\documentclass{article}

\usepackage{epubtk}
\usepackage{graphicx}
\usepackage{amssymb}
\usepackage{amsmath}
\usepackage{multirow}
\usepackage[normalem]{ulem}

\begin{document}

\title{Binary Neutron Star Mergers}

\author{\epubtkAuthorData{Joshua A.\ Faber}{%
 Center for Computational Relativity and Gravitation and School of Mathematical Sciences\\
Rochester Institute of Technology, 85 Lomb Memorial Drive, Rochester NY 14623, USA}{%
jafsma@rit.edu}{%
http://ccrg.rit.edu/~jfaber}%
\and
\epubtkAuthorData{Frederic A.\ Rasio}{%
Center for Interdisciplinary Exploration and Research in Astrophysics,\\
 and Department of Physics \& Astronomy,\\
Northwestern University, 2145 Sheridan Road, Evanston IL 60208, USA}{%
rasio@northwestern.edu}{%
http://ciera.northwestern.edu/rasio/}
}

\date{}
\maketitle

\begin{abstract}
We review the current status of studies of the coalescence of binary
neutron star systems.  We begin with a discussion of the formation channels
of merging binaries and we discuss the most recent theoretical predictions for 
merger rates.  Next, we
turn to the quasi-equilibrium formalisms that are used to study
binaries prior to the merger phase and to generate initial data for fully
dynamical simulations. The quasi-equilibrium approximation has played a key 
role in developing our understanding of the physics of binary coalescence 
and, in particular, of the orbital instability processes
that can drive binaries to merger at the end of their lifetimes.
We then turn to the numerical techniques used in dynamical simulations,
including relativistic formalisms, (magneto-)hydrodynamics,
gravitational-wave extraction techniques, and nuclear
microphysics treatments.  This is followed by a summary of the
simulations performed across the field to date, including the most
recent results from both fully relativistic and
microphysically detailed simulations.  Finally, we discuss the likely
directions for the field as we transition from the first to the second generation of
gravitational-wave interferometers and while supercomputers reach the petascale 
frontier.
\end{abstract}

\epubtkKeywords{general relativity}

\newpage 

\section{Introduction}

Binaries composed of neutron stars (NSs) and black holes (BHs) have long been of interest to astrophysicists.  They provide many important constraints for models of massive star evolution and compact object formation, 
and are among the leading potential sources for detection by gravitational-wave (GW) observatories.  While it remains uncertain whether mergers of compact binaries are an important contributor to the production of r-process elements, they are now thought to be the leading candidate to explain short-duration, hard-spectrum gamma-ray bursts (often abbreviated to ``short-hard'' GRBs, or merely SGRBs).  

The first neutron star--neutron star (NS--NS) binary to be observed was PSR~B1913+16, in which a radio pulsar was found to be in close orbit around another NS~\cite{Hulse:1974eb}.   In the decades since its discovery, the decay of the orbit of PSR~B1913+16 at exactly the rate predicted by Einstein's general theory of relativity (see, e.g.,~\cite{Taylor:1989sw,Weisberg:2010zz}) 
has provided strong indirect evidence that gravitational radiation exists and is indeed correctly described by general relativity (GR).  
This measurement led to the 1993 Nobel Prize in physics for Hulse and Taylor.

According to the lowest-order dissipative contribution from GR, which arises at the 2.5PN level (post-Newtonian; where the digit indicates the expansion order in $[v/c]^2$ in the Taylor expansion term), and assuming that both NSs may be approximated as point masses, a circular binary orbit decays at a rate $da/dt=- a/\tau_{\mathrm{GW}}$ where $a$ is the binary separation and the gravitational radiation merger timescale $\tau_{\mathrm{GW}}$ is given by
\begin{eqnarray}
\tau_{\mathrm{GW}}&=&\frac{5}{64}\frac{a^4}{\mu M^2} = \frac{5}{64}\frac{a^4}{q(1+q)M_1^3}\nonumber\\
&=&2.2\times 10^8 q^{-1}(1+q)^{-1} \left(\frac{a}{R_\odot}\right)^4\left(\frac{M_1}{1.4\,M_{\odot}}\right)^{-3}\mathrm{\ yr},\label{eq:taugw}
\end{eqnarray}
where $M_1$, $M_2$, and $M\equiv M_1+M_2$ are the individual NS masses and the total mass of the binary, $\mu=M_1M_2/M$ is the reduced mass, $q=M_2/M_1$ is the binary mass ratio, and we assume geometrized units where $G=c=1$ (as we do throughout this paper, unless otherwise noted).  The timescale for an elliptical orbit is shorter, and it can be shown that eccentricity is reduced over time by GW emission, leading to a circularization of orbits as they decay.  A quick integration shows that the time until merger is given by $\tau_{\mathrm{merge}}=\tau_{\mathrm{GW}}/4$.

The luminosity of such systems in gravitational radiation is
\begin{eqnarray}
L_{\mathrm{GW}}=-\frac{dE_{\mathrm{GW}}}{dt} &=& \frac{32}{5}\frac{\mu^2M^3}{a^5}=\frac{32}{5}\frac{m_1^2m_2^2(m_1+m_2)}{a^5}\nonumber\\
&=&5.34\times 10^{32}q^2(1+q)\left(\frac{M_1}{1.4\,M_{\odot}}\right)^5\left(\frac{a}{R_{\odot}}\right)^{-5}\mathrm{\ erg/s}\nonumber\\
&=&8.73\times 10^{51}q^2(1+q)\left(\frac{M_1}{1.4\,M_{\odot}}\right)^5\left(\frac{a}{100\mathrm{\ km}}\right)^{-5}\mathrm{\ erg/s},
\end{eqnarray}
which, at the end of a binary's lifetime, when the components have approached to within a few NS radii of each other, is comparable to the luminosity of all the visible matter in the universe ($\sim 10^{53}$~erg/s).  The resulting strain amplitude observed at a distance $D$ from the source (assumed to be oriented face-on) is given approximately by 
\begin{eqnarray}
h=\frac{4M_1M_2}{aD}=5.53\times 10^{-23}q\left(\frac{M_1}{1.4\,M_{\odot}}\right)^2\left(\frac{a}{100\mathrm{\ km}}\right)^{-1}\left(\frac{D}{100\mathrm{\ Mpc}}\right)^{-1},
\end{eqnarray}
at a characteristic frequency
\begin{eqnarray}
f_{GW}=2f_{\mathrm{orb}} = \frac{1}{\pi}\sqrt{\frac{M}{a^3}} = 194\left(\frac{M}{2.8\,M_{\odot}}\right)^{1/2}\left(\frac{a}{100\mathrm{\ km}}\right)^{-3/2}\mathrm{\ Hz}.\label{eq:fgw}
\end{eqnarray}

The first measurement that will likely be made with direct GW observations is the orbital decay rate, with the period evolving (for the circular case) according to the relation
\begin{equation}
\frac{dT}{dt}=-\frac{192\pi}{5}\left(\mathcal{M}_c \omega\right)^{5/3},
\end{equation}
where $T$ is the orbital period and $\omega$ the angular frequency, and thus the ``chirp mass,'' 
\begin{equation}\mathcal{M}_c\equiv \mu^{3/5}M^{2/5} = M_1^{3/5}M_2^{3/5}(M_1+M_2)^{-1/5},\label{eq:chirp}
\end{equation}
is likely to be the easiest parameter to determine from GW observations.

Several NS--NS systems are now known, including PSR J0737--3039~\cite{Burgay:2003jj}, a binary consisting of two observed pulsars, which allows for the prospect of even more stringent tests of GR~\cite{Kramer:2008ARA&A..46..541K}.  
Even with the handful of observed sources to date, one may use this sample to place empirical limits on the expected rate of NS--NS mergers~\cite{Kim:2002uw} and to constrain the many parameters that enter into population synthesis calculations~\cite{O'Shaughnessy:2007fb}.  
With regard to the former, the very short merger timescale for J0737, $\tau_{\mathrm{merge}}$~=85~Myr, makes it especially important for estimating the overall rate of NS--NS mergers since it is a priori very unlikely to detect a system with such a short lifetime.

Although black hole-neutron star (BH--NS) binaries are expected to form through the same processes as NS--NS binaries, none has been detected to date. This is generally thought to reflect their lower probability of detection in current surveys, in addition to intrinsically smaller numbers compared to NS--NS systems~\cite{Belczynski:2009nx}. BH--NS systems are an expected byproduct of binary stellar evolution, and properties of the population may be inferred from population synthesis studies calibrated to the observed NS--NS sample (see, e.g.,~\cite{Belczynski:2005mr}).

In this review, we will summarize the current state of research on relativistic mergers, beginning in Section~\ref{sec:population} with a description of the astrophysical processes that produce merging binaries and the expected parameters of these systems.  The phases of the merger are briefly described in Section~\ref{sec:phases}.  In Section~\ref{sec:initialdata}, we discuss the numerical techniques used to generate quasi-equilibrium (QE) sequences of NS--NS configurations, and we summarize the QE calculations that have been performed.  These sequences yield a lot of information about NS physics, particularly with regard to the nuclear matter equation of state (EOS).  They also serve as initial data for dynamical merger calculations, which we discuss next, focusing in turn on the numerical hydrodynamics techniques used to compute mergers and the large body of results that has been generated, in Sections~\ref{sec:dyntech} and \ref{sec:dynsim}, respectively. We pay particular attention to how numerical studies have taken steps toward answering a number of questions about the expected GW and electromagnetic (EM) emission from merging binaries, and we discuss briefly the possibility that they may be the progenitors of SGRBs and a source of r-process elements.  We close with conclusions and a look to the future in Section~\ref{sec:conclusions}.  

While most of this review focuses on NS--NS mergers, many of the methods used to study NS--NS binaries are also used to evolve BH--NS binaries, and it has become clear that both merger types may produce similar observational signatures as well.  For a review focusing on BH--NS merger calculations, we encourage the reader to consult the recent work by Shibata and Taniguchi~\cite{ST_LRR}.

\newpage
\section{Evolutionary Channels and Population Estimates}
\label{sec:population}

Merging NS--NS and BH--NS binaries, i.e., those for which the merger timescale is smaller than the Hubble time, are typically formed through similar evolutionary channels in stellar field populations of galaxies~\cite{Belczynski:2005mr} (both may also be formed through dynamical processes in the high-density cores of some star clusters, but the overall populations are smaller and more poorly constrained; see~\cite{Sadowski:2007dz} for a review).  It is difficult to describe the evolutionary pathways that form NS--NS binaries without discussing BH--NS binaries as well, and it is important to note that the joint distribution of parameters such as merger rates and component masses that we could derive from simultaneous GW and EM observations will constrain the underlying physics of binary stellar evolution much more tightly than observing either source alone.

Population synthesis calculations for both merging NS--NS and BH--NS binaries typically favor the standard channel in which the first-born compact object goes through a common-envelope (CE) phase,
although other models have been proposed, including recent ones where the progenitor binary is assumed to have very nearly-equal mass components that leave the main sequence and enter a CE phase prior to either undergoing a supernova~\cite{Bethe:2005ju,Brown:2008cn}.  
Simulations of this latter process have shown that close NS--NS systems could indeed be produced by twin giant stars with core masses $\gtrsim 0.15\,M_{\odot}$, though twin main sequence stars typically merge during the contact phase~\cite{Lombardi:2010xh}.

In the standard channel (see, e.g.,~\cite{Bhattacharya91,Lorimer:2008se}, and Figure~\ref{fig:evol} for an illustration of the process), the progenitor system is a high-mass binary (with both stars of mass $M\gtrsim 8-10\,M_{\odot}$ to ensure a pair of supernovae). The more massive primary evolves over just a few million years before it leaves the main sequence, passes through its giant phase, and undergoes a Type Ib, Ic, or II supernova, leaving behind what will become the heavier compact object (CO): the BH in a BH--NS binary or the more massive NS in an NS--NS binary.  The secondary then evolves off the main sequence in turn,  triggering a CE phase when it reaches the giant phase and overflows its Roche lobe. Dynamical friction shrinks the binary separation dramatically, until sufficient energy is released to expel the envelope.  Without this step, binaries would remain too wide to merge through the emission of GWs within a Hubble time.  Eventually, the exposed, Helium-rich core of the secondary  undergoes a supernova, either unbinding the system or leaving behind a tight binary, depending on the magnitude and orientation of the supernova kick.  

\epubtkImage{}{%
\begin{figure}[!ht]
\centerline{\includegraphics[width=\textwidth]{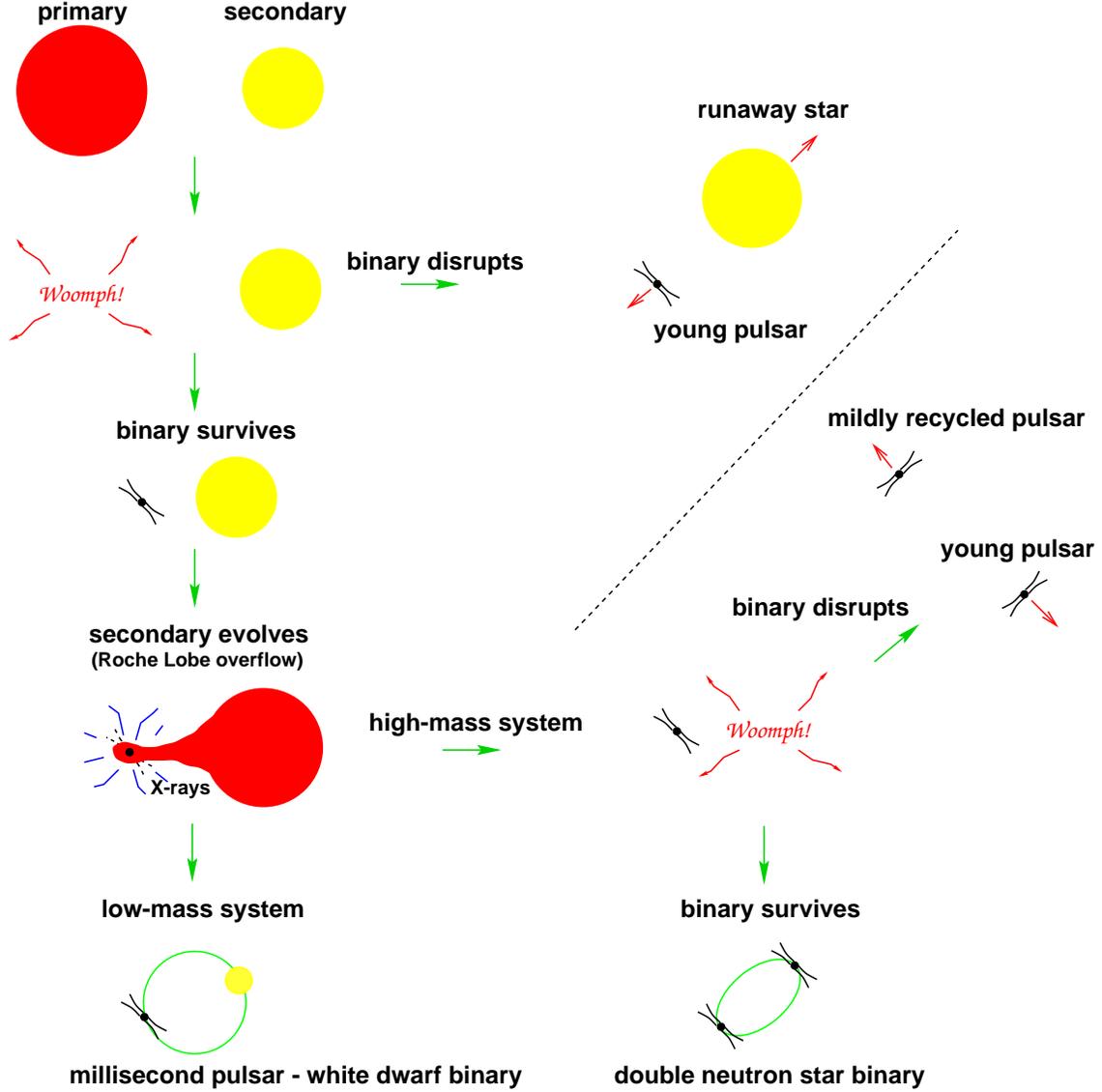}}
\caption{Cartoon showing standard formation channels for close NS--NS binaries through binary stellar evolution, taken from~\cite{Lorimer:2008se}.}
\label{fig:evol}
\end{figure}}

This evolutionary pathway has important effects on the physical parameters of NS--NS and BH--NS binaries, leading to preferred regions in phase space.  The primary, which can accrete some matter during the CE phase, or during an episode of stable mass transfer from  the companion Helium star, should be spun up to rapid rotation  (see~\cite{Lorimer:2008se} for a review).  In NS--NS binaries, we expect that this process will also reduce the magnetic field of the primary down to levels seen in ``recycled'' pulsars, typically up to four orders of magnitude lower than for young pulsars~\cite{Lovelace:2005gr,Cumming:2001kr}.  The secondary NS, which never undergoes accretion, is likely to spin down rapidly from its nascent value, but is likelier to maintain a stronger magnetic field.

While this evolutionary scenario has been well studied for several decades, many aspects remain highly uncertain.  In particular: 
\begin{itemize}
\item The CE efficiency, 
which helps to determine the expected range of binary separations and the mass of the primary compact object after its accretion phase, remains very poorly constrained~\cite{Postnov:2007jv,Belczynski:2005mr,Davis:2009rz,Zorotovic:2010da}.  If too much matter is accreted by the NS, it may undergo accretion-induced collapse to a BH~\cite{Bethe:2005ju}, though the growing body of observed NS--NS systems does help place constraints on the allowed range of accretion-related parameters.
\item The exact relation between a star's initial mass and the eventual compact object mass is better understood, but significant theoretical uncertainties remain, and the relation is sensitive to the metallicity (often unknown), mainly through the effects of mass loss in stellar winds, and to the details of the explosion itself~\cite{Zhang:2007nw,O'Connor:2010tk}.
\item The maximum allowed NS mass will affect whether the primary remains a NS or undergoes accretion-induced collapse to a BH; its value is dependent upon  the as-yet undetermined nuclear matter EOS.  At present, the strongest limit is set by the binary millisecond pulsar PSR~J1614--2230, for which a mass of $M_{\mathrm{NS}}=1.97\pm 0.04\,M_{\odot}$ was determined by Shapiro time delay measurements~\cite{Demorest:2010bx}.  Determining the NS EOS from GW observations may eventually provide stronger constraints~\cite{Read:2008iy,Ozel:2009da,Markakis:2011vd}, including a determination of whether supernova remnants are indeed classical hadronic NS or instead have cores consisting of some form of strange quark matter or other elementary particle condensate~\cite{Pandharipande:1975NuPhA.237..507P,Glendenning:1984jr,Prakash:1995uw,Glendenning:1997ak,Balberg:1997yw,Alford:2004pf}. 
\item The supernova kick velocity distribution is only partially understood, leading to uncertainties as to which systems will become unbound after the second explosion.~\cite{Hobbs:2005yx,Wang:2005jg,O'Shaughnessy:2006wh,Kuranov:2009pp}
\end{itemize}

Given all these uncertainties, it is reassuring that most estimates of the NS--NS and BH--NS merger rate, expressed either as a rate of mergers per Myr per ``Milky Way equivalent galaxy'' or as a predicted detection rate for LIGO (the Laser Interferometer Gravitational-Wave Observatory) and Virgo (see Section~\ref{sec:gw} below), agree to within 1\,--\,2 orders of magnitude, which is comparable to the typical uncertainties that remain once all possible sources of error are folded into a population synthesis model.  In Table~\ref{table:popsynth}, we show the predicted detection rates of NS--NS and BH--NS mergers for both the first generation LIGO detectors (``LIGO''), which ran at essentially their design specifications~\cite{Abbott:2009tt}, and the Advanced LIGO (``AdLIGO'') configuration due to go online in 2015~\cite{Smith:2009bx}.  We note that the methods used to generate these results varied widely.  In~\cite{Kim:2002uw}, the authors used the observed parameters of close binary pulsar systems to estimate the Galactic NS--NS merger rate empirically (such results do not constrain the BH--NS merger rate).  In~\cite{Nakar:2005bs,Guetta:2008qw}, the two groups independently estimated the binary merger rate from the observed statistics of SGRBs. 
 In these cases, one does not get an independent prediction for the NS--NS and BH--NS merger rate, but rather some linear combination of the two. In both cases, the authors estimated that, if NS--NS and BH--NS mergers are roughly equal contributors to the observed SGRB sample, LIGO will detect about an order of magnitude more BH--NS mergers since their higher mass allows them to be seen over a much larger volume of the Universe.  As they both noted, should either type of system dominate the SGRB sample, we would expect a doubling of LIGO detections for that class, and lose our ability to constrain the rate of the other using this method.  Many population synthesis models have attempted to understand binary evolution within our Galaxy by starting from a basic parameter survey of the various assumptions made about CE evolution, supernova kick distributions, and other free parameters.  In~\cite{Voss:2003ep,deFreitasPacheco:2005ub}, population synthesis models are normalized by estimates of the star formation history of the Milky Way.  In~\cite{Kalogera:2006uj,O'Shaughnessy:2009ft}, parameter choices are judged based on their ability to reproduce the observed Galactic binary pulsar sample, which allows posterior probabilities to be applied to each model in a Bayesian framework.  A review by the LIGO collaboration of this issue may be found in~\cite{Abadie:2010cf}.

\begin{table}
\caption[Estimated initial and advanced LIGO rates for BH--NS and NS--NS
  mergers from population synthesis calculations and other
  methods.]{Estimated initial and advanced LIGO rates for BH--NS and
  NS--NS mergers from population synthesis calculations and other
  methods.  The methods used are, in order, empirical constraints from
  the observed sample of binary pulsars (`Empirical'), constraints
  on the combined NS--NS/BH--NS merger rate assuming that they are the
  progenitors of short-hard gamma-ray bursts (`SGRBs'), population
  synthesis models calibrated to the star formation rate in the Milky
  Way (`Pop.\ Synth.\ -- SFR'), and population synthesis calibrated
  against the observed Galactic binary pulsar sample (`Pop.\ Synth.\ --
  NS--NS').  We note that observations of binary pulsars  do not yield
  constraints for BH--NS binaries.  SGRB observations may produce
  constraints on NS--NS merger rates, BH--NS, merger rates, or both,
  depending on which sources are the true progenitors, but this remains
  unclear. Therefore, the table
  quotes results assuming a roughly equal split between the two.  The
  official review of these results and their implications by the
  LIGO/Virgo Scientific Collaborations may be found
  in~\cite{Abadie:2010cf}.}
\label{table:popsynth}
\centering
\begin{tabular}{l|cc|cc|c}
Author & \multicolumn{2}{c}{NS--NS} & \multicolumn{2}{c}{BH--NS} & Method\\
& LIGO  & AdLIGO & LIGO & AdLIGO &\\\hline
Kim et al. \cite{Kim:2002uw} & 5e-3 & 27 & & & Empirical\\\hline
Nakar et al. \cite{Nakar:2005bs} & & $\sim$2 & & $\sim$20.0 & SGRBs \\
Guetta \& Stella \cite{Guetta:2008qw} & 7.0e-3 & 22 & 7.0e-2 & 220 & SGRBs\\\hline
Voss \& Tauris \cite{Voss:2003ep} & 6.0e-4 & 2.0 & 1.2e-3 & 4.0 & Pop. Synth. - SFR\\
de Freitas Pacheco et al. \cite{deFreitasPacheco:2005ub} & 8.0e-4 & 6.0 & & &  Pop. Synth. - SFR\\\hline
Kalogera et al. \cite{Kalogera:2006uj} & 1.0e-2 & 35 & 4.0e-3 & 20 & Pop. Synth. - NS--NS\\
O'Shaughnessy et al. \cite{O'Shaughnessy:2009ft} & 1.0e-2  & 10 & 1.0e-2 & 10 & Pop. Synth. - NS--NS\\\hline
\end{tabular}
\end{table}

Should the next generation of GW interferometers begin to detect a statistically significant number of merger events including NSs, it should be possible to constrain several astrophysical parameters describing binary evolution much more accurately.  These include
\begin{itemize}
\item \emph{The relative numbers of BH--NS and NS--NS mergers:} Interferometric detections are sensitive to a binary's ``chirp mass'' (see Eq.~\ref{eq:chirp}), and to the binary mass ratio as well~\cite{Buonanno:2005pt,Babak:2006ty,vanderSluys:2008qx,VanDenBroeck:2009gd} if the signal-to-noise ratio is sufficiently high.  Even if the merger signal takes place at frequencies too high to fall within the LIGO band, it should still be possible in most cases to determine whether the primary's mass exceeds the maximum mass of a NS.

\item \emph{The mass ratio probability distribution for merging binary systems:}  If both binary components' masses are determined, we will be able to constrain both the BH mass distribution in merging binaries and the NS mass ratio distribution.  Knowledge of the former would determine, e.g., whether the current low estimates for the mass accreted onto the primary CO core during the CE phase are correct~\cite{Belczynski:2007xg}, as this model predicts that BH masses in close BH--NS binaries should cluster narrowly around $M_{\mathrm{BH}}=10\,M_{\odot}$.  Previous calculations assuming larger accreted masses typically favored mass ratios closer to unity, since the primaries often began as NSs and underwent ``accretion-induced collapse'' to a BH during the CE phase.  The NS--NS mass distribution will allow for tests of whether ``twins'', i.e., systems whose zero-age main sequence (or ``ZAMS'') masses are so close that they both leave the main sequence before either undergoes a supernova, play a significant role in the merging NS--NS population~\cite{Bethe:2005ju}.
\end{itemize}

Interestingly enough, while it has generally been assumed that Advanced LIGO or another interferometer will make the first direct observations of NS--NS mergers and their immediate aftermath, such systems may have already been observed and simply not interpreted correctly. Indeed, while virtually all NS--NS merger calculations performed to date that consider EM emission have looked at the high-energy prompt emission, Nakar and Piran suggest that the mass ejection from mergers should yield an observable radio afterglow~\cite{Nakar:2011cw}. Moreover, they argue that such post-merger afterglows may have already been observed as a $\sim$month-long radio burst occurring in the outer regions of a galaxy.  While such an outburst could also result from a supernova, the luminosity required would be an order of magnitude larger than those previously observed.  Given the length and timescales characterizing radio bursts, no NS--NS simulation has been able to address the model directly, but it certainly seems plausible that the time-variable magnetic fields within a stable hypermassive remnant could generate enough EM energy to power the resulting radio burst~\cite{Shibata:2011fj}.  These potential EM observations of mergers are likely to spur further research into the amount and velocity of merger ejecta, which could then be coupled to a larger-scale astrophysical simulation of the potential radio afterglow.

\newpage
\section{Stages of a Binary Merger}
\label{sec:phases}

The qualitative evolution of NS--NS mergers, or indeed any compact binary merger, has long been understood, and may be divided roughly into inspiral, merger, and ringdown phases, each of which presents a distinct set of challenges for numerical modeling and detection.  
As a visual aid, we include a cartoon summary in Figure~\ref{fig:thorne}, originally intended to describe black hole -- black hole (BH--BH) mergers, and attributed to Kip Thorne.  Drawn before the advent of the supercomputer simulations it envisions, we note that merger waveforms for all compact binary mergers have proven to be much smoother and simpler than shown here. To adapt it to NS--NS mergers, we note that NSs are generally assumed to be essentially nonspinning, and that the ``ringdown'' phase may describe either a newly formed BH or a NS that survives against gravitational collapse.  

\epubtkImage{}{%
\begin{figure}[!ht]
\centerline{\includegraphics[width=0.8\textwidth]{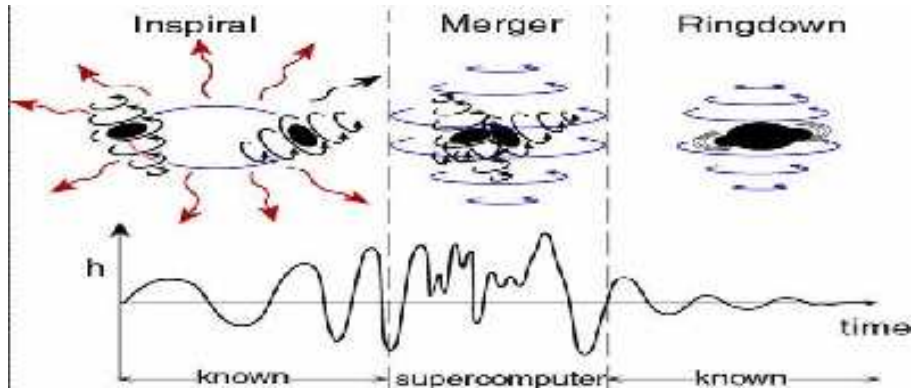}}
\caption{Cartoon picture of a compact binary coalescence, drawn for a BH--BH merger but applicable to NS--NS mergers as well (although NSs are generally assumed to be nonspinning). }
\label{fig:thorne}
\end{figure}}

Summarizing the evolution of the system through the three phases:
\begin{itemize}
\item After a pair of supernovae yields a relatively tight NS--NS binary, the orbital separation decays over long timescales through GW emission, a phase that takes up virtually all of the lifetime of the binary except the last few milliseconds.  During the \emph{inspiral} phase, binary systems may be accurately described by QE formalisms, up until the point where the gravitational radiation timescale becomes comparable to the dynamical timescale.  The evolution in time is well-described by PN expansions, currently including all terms to 3PN \cite{Blanchet:2006zz}, though small deviations can arise because of finite-size effects, especially at smaller separations (see Eqs.~\ref{eq:taugw} -- \ref{eq:chirp} for the lowest-order 2.5PN expressions for circular inspirals).
\item Once the binary separation becomes no more than a few times the radii of the two NSs, binaries rapidly become unstable.  The stars plunge together, following the onset of dynamical instability, and enter the \emph{merger} phase, requiring full GR simulations to understand the complicated hydrodynamics that ensues.  According to all simulations to date, if the NS masses are nearly equal, the merger resembles a slow collision, while if the primary is substantially more massive than the secondary the latter will be tidally disrupted during the plunge and will essentially accrete onto the primary. Since the NSs are most likely irrotational just prior to merging, there is a substantial velocity discontinuity at the surface of contact, leading to rapid production of vortices.  Meanwhile, some fraction of the mass may be lost through the outer Lagrange points of the system to form a disk around the central remnant. 
This phase yields the maximum GW amplitude predicted by numerical simulations, but with a signal much simpler and more quasi-periodic than in the original cartoon version.  GW emission during the merger encodes important information about the NS EOS, particularly the GW frequency $f_{\rm cut}$ at which the binary orbit becomes unstable (see Eq.~\ref{eq:fgw}) resulting in a characteristic cutoff in GW emission at those frequencies.  Meanwhile, the merger itself may generate the thermal energy that eventually powers a SGRB, which occurs when the neutrinos and anti-neutrinos produced by shock-heated material annihilate around the remnant to produce high-energy photons.
\item Finally, the system will eventually settle into a new, dynamically stable configuration through a phase of \emph{ringdown}, with a particular form for the GW signal that depends on the remnant's mass and rotational profile.
 If the remnant is massive enough, it will be gravitationally unstable and collapse promptly to form a spinning BH. Otherwise it must fall into one of three classes depending on its total mass.  Should the remnant mass be less than the maximum mass $M_{\mathrm{iso}}$ supported by the nuclear matter EOS for an isolated, non-rotating configuration, it will clearly remain stable forever. Instead a remnant that is \emph{``supramassive,''} i.e., with a mass above the isolated stationary mass limit but below that allowed for a \emph{uniformly rotating} NS (typically $\lesssim 1.2\,M_{\mathrm{iso}}$, with very weak dependence on the EOS; see, e.g.,~\cite{Komatsu:1989zz,Cook92,Cook:1993qr}, and references therein) may become unstable.  Supramassive remnants are stable against gravitational collapse unless angular momentum losses, either via pulsar-type emission or magnetic coupling to the outer disk, can drive the angular velocity below the critical value for stability.
  If the remnant has a mass above the supramassive limit, it may fall into the \emph{hypermassive} 
  regime, where it is supported against gravitational collapse by rapid differential rotation.  Hypermassive NS (HMNS) remnants can have significantly larger masses, depending on the EOS (see, e.g.,~\cite{Baumgarte:1999cq,Shibata:2005mz,Duez:2005cj,Stephens:2006cn,Galeazzi:2011nn}),
 and will survive for timescales much longer than the dynamical time, undergoing a wide variety of oscillation modes.  They can emit GWs should a triaxial configuration yield a significant quadrupole moment, and potentially eject mass into a disk around the remnant.  Eventually, some combination of radiation reaction and magnetic and viscous dissipation will dampen the differential rotation and lead the HMNS to collapse to a spinning BH, again with the possibility that it may be surrounded by a massive disk that could eventually accrete.
The energy release during HMNS collapse provides the possibility for a ``delayed'' SGRB, in which the peak GW emission occurs during the initial merger event, but the gamma-ray emission, powered by the collapse of the HMNS to a BH, occurs significantly later.

Most calculations indicate that a geometrically thick, lower-density, gravitationally bound disk of material will surround whatever remnant is formed.  
Such disks, which are geometrically thick, are widely referred to as ``tori'' throughout the literature, though there is no clear distinction between the two terms, and we will use ``disk'' throughout this paper to describe generically the bound material outside a central merger remnant.
Such disks are expected to heat up significantly, and much of the material will eventually accrete onto the central remnant, possibly yielding observable EM emission.  Given the low densities and relatively axisymmetric configuration expected, disks are not significant GW emitters.
There may be gravitationally unbound outflow from mergers as well, though dynamical simulations neither confirm nor deny this possibility yet.  Such outflows, which can be the sites of exotic nuclear reactions, are frequently discussed in the context of r-process element formation, but their inherently low densities make them difficult phenomena to model numerically with high accuracy.
\end{itemize}

\subsection{Comparison to BH--NS mergers}

Since this three-phase picture is applicable to BH--NS mergers as well, it is worthwhile to compare the two merger processes at a qualitative level to understand the key similarities and differences.  Inspiral for BH--NS mergers is also well-described by PN expansions up until shortly before the merger, but the parameter space is fundamentally different.  First, since BHs are heavier than NSs, the dynamics can be quite different.  Also, since BHs may be rapidly-spinning (i.e., have dimensionless spin angular momenta as large as $J/M^2\sim 1$), spin-orbit couplings can play a very important role in the orbital dynamics of the binary, imprinting a large number of oscillation modes into the GW signal (see, e.g.,~\cite{Grandclement:2003ck,Buonanno:2005pt}).  From a practical standpoint, the onset of instability in BH--NS mergers should be easier to detect for LIGO, Virgo, and other interferometers, since the larger mass of BH--NS binaries implies that instability occurs at lower GW frequencies (see Eq.~\ref{eq:fgw}, noting that the separation $a$ at which mass-transfer begins scales roughly proportionally to the BH mass).

The onset of instability of a BH--NS binary is determined by the interplay of the binary mass-ratio, NS compactness, and BH spin, with the first of these playing the largest role (see Figures~13\,--\,15 of~\cite{Taniguchi:2007aq} and the summary in~\cite{ST_LRR}).  In general, systems with high BH masses and/or more compact NSs tend to reach a minimum in the binding energy as the radius increases, leading to a dynamical orbital instability that occurs near the classical innermost stable circular orbit (ISCO).  In these cases, the NS plunges toward the BH before the onset of tidal disruption, and is typically ``swallowed whole''. leaving behind almost no material to form a disk.  The GW emission from such systems
is sharply curtailed after the merger event, yielding only a low-amplitude, high-frequency, rapidly-decaying ``ringdown'' signal (see, e.g.,~\cite{Kyutoku:2010zd}).  Numerical calculations have shown that even in borderline cases between dynamical instability and mass-shedding the NS is essentially swallowed whole, especially in cases where the BH in either non-spinning or spinning in the retrograde direction, which pushes the ISCO out to larger radii (see, e.g.,~\cite{Shibata:2006bs,Shibata:2006ks,Shibata:2007zm,Etienne:2007jg,Etienne:2008re}).

A richer set of phenomena occurs when the BH--NS mass ratio is closer to unity, the NS is less compact, the BH has a prograde spin direction, or more generally, some combination of those factors.  In such cases, the NS will reach the mass-shedding limit prior to dynamical instability, and be tidally disrupted.  Unlike what was described in semi-analytic Newtonian models (see, e.g.,~\cite{Clark77,PortegiesZwart:1998xm,Davies:2004pu}) and seen in some early Newtonian and quasi-relativistic simulations (see, e.g.,~\cite{Lee:1998qk,Lee:1999kcb,Janka:1999qu}, \emph{stable} mass transfer, in which angular momentum transfer via mass-shedding halts the inspiral, has never been seen in full GR calculations, nor even in approximate GR models (see the discussion in~\cite{Faber:2005yg}).
Even so, unstable mass transfer can produce a substantial disk around the BH, though in every GR simulation to data the substantial majority of the NS matter is accreted promptly by the BH (see~\cite{ST_LRR} for a detailed summary of all current results).  The exact structure of the disk and its projected lifetime depend on the binary system parameters, with the binary mass ratio and spin both important in determining the disk mass and the BH spin orientation critical for determining both the disk's vertical structure and its thermodynamic state given the shock heating that occurs during the NS disruption. In general, the mass and temperature of the post-merger disks are comparable to those seen in some NS--NS mergers, and inasmuch as either is a plausible SGRB progenitor candidate then both need to be viewed as such.  
To date, no calculation performed in full GR has found any bound and self-gravitating NS remnant left over after the merger, including both NS cores that survive the initial onset of mass transfer by recoiling outward (predicted for cases in which stable mass transfer was thought possible, as noted above) or those in which bound objects form through fragmentation of the circum-BH disk.  Motivated by observations of wide-ranging timescales for X-ray flares in both long and short GRBs \cite{Perna:2005tv}, the latter channel  has been suggested to occur in collapsars \cite{Piro:2006ja} and mentioned in the context of mergers \cite{Rosswog:2006rh}, possibly on longer timescales than current simulations permit.
Even so, there is no analogue to the HMNS state that may result from NS--NS mergers, nor any mechanism for a delayed SGRB as provided by HMNS collapse.

\subsection{Qualitative numerical results}
Constructing QE sequences for a given set of NS parameters requires sophisticated numerical schemes, but not supercomputer-scale resources, as we discuss in Section~\ref{sec:initialdata} below, focusing first on the numerical techniques used to construct QE binary data in GR, and the astrophysical information contained in the GW emission during the inspiral phase.  Merger and ringdown, on the other hand, typically require large-scale numerical simulations, including some of the largest calculations performed at major supercomputer centers, as we discuss in detail Section~\ref{sec:dyntech} and \ref{sec:dynsim} below. 

To illustrate the various physical processes that occur during NS--NS mergers, we show
the evolution of three different NS--NS merger simulations in Figures~\ref{fig:sim_1}, \ref{fig:sim_2}, and \ref{fig:sim_3}, taken from Figures~4\,--\,6 of~\cite{Kiuchi:2009jt}.  In Figure~\ref{fig:sim_1}, we see the merger of two equal-mass NSs, each of mass $M_{\mathrm{NS}}=1.4\,M_{\odot}$, described by the APR (Akmal--Pandharipande--Ravenhall) EOS~\cite{Akmal:1998cf}.  In the second panel, clear evidence of ``tidal lags'' is visible shortly after first contact, leading to a slightly off-center collision pattern.  By the third panel, an ellipsoidal HMNS has been formed, surrounded by a disk of material of lower density, which gradually relaxes to form a more equilibrated HMNS by the final panel.  In Figure~\ref{fig:sim_2}, we see a merger of two slightly heavier equal-mass NSs with $M_{\mathrm{NS}}=1.5\,M_{\odot}$.  In this case, the deeper gravitational potential limits the amount of mass that goes into the disk, and once a BH is formed (with a horizon indicated by the dashed blue circle in the final panel) it accretes virtually all of the rest mass initially present in the two NSs, with only $0.004\%$ of the total remaining outside the horizon.

\epubtkImage{}{%
\begin{figure}[!ht]
\centerline{
\includegraphics[width=0.5\textwidth]{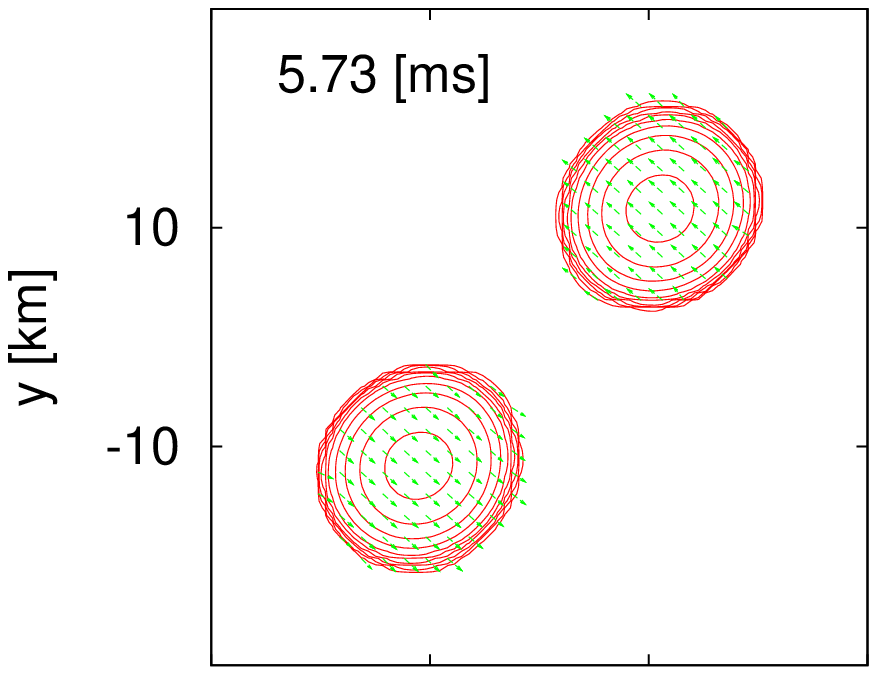}
\includegraphics[width=0.5\textwidth]{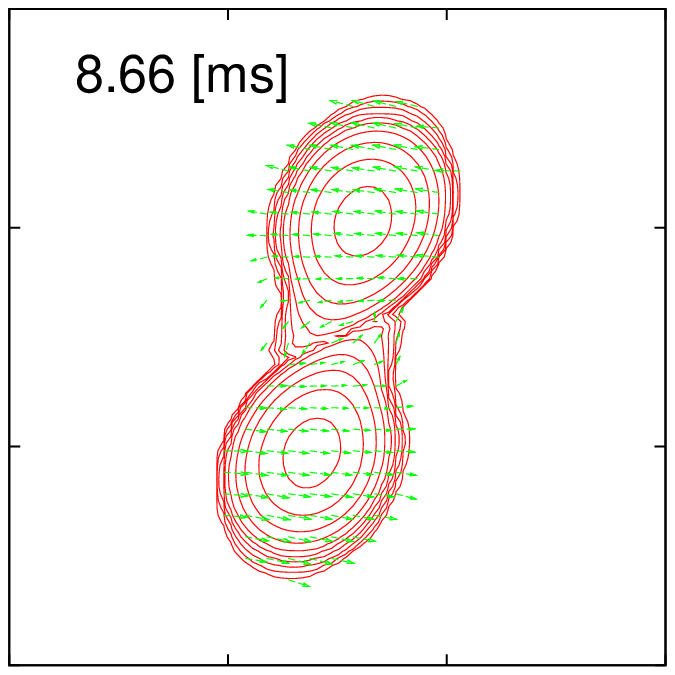}\\
}
\centerline{
\includegraphics[width=0.5\textwidth]{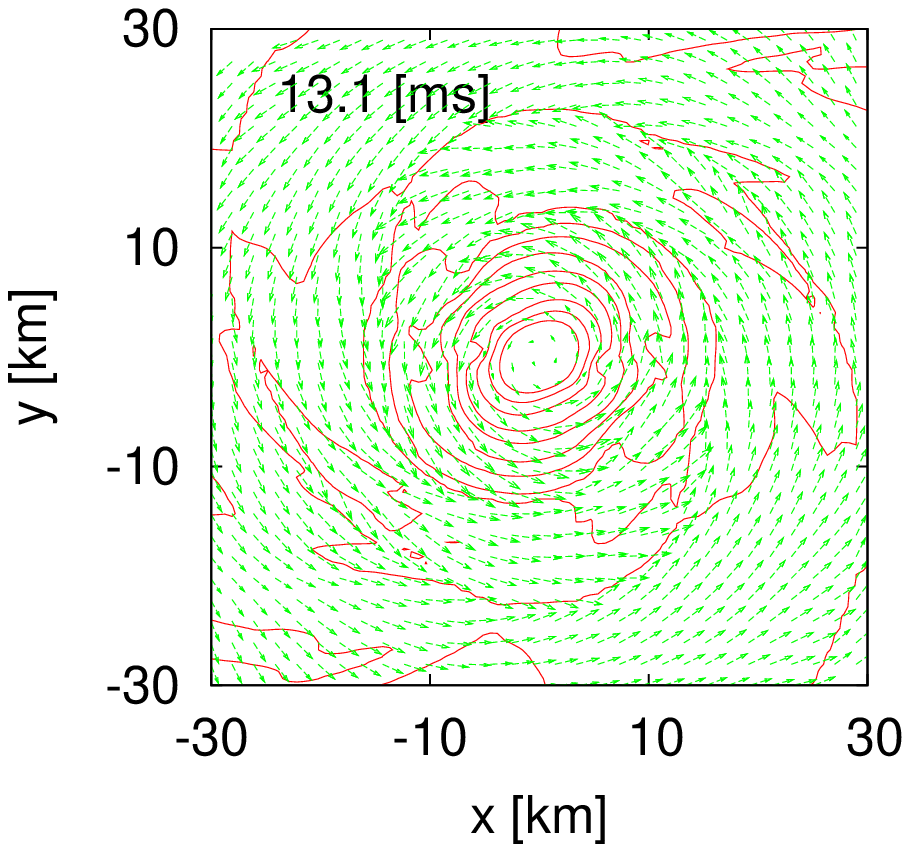}
\includegraphics[width=0.5\textwidth]{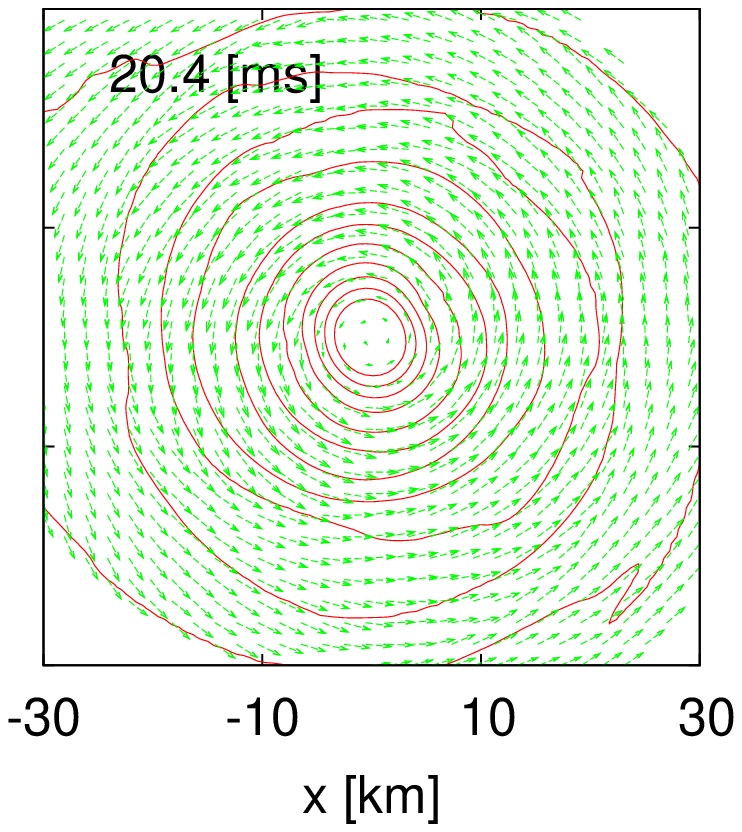}
}
\caption{Isodensity contours and velocity profile in the equatorial plane for a merger of two equal-mass NSs with $M_{\mathrm{NS}}=1.4\,M_{\odot}$ assumed to follow the APR model~\cite{Akmal:1998cf} for the NS EOS, taken from Figure~4 of~\cite{Kiuchi:2009jt}.  The hypermassive  merger remnant survives until the end of the numerical simulation.}
\label{fig:sim_1}
\end{figure}}

\epubtkImage{}{%
\begin{figure}[!ht]
\centerline{
\includegraphics[width=0.5\textwidth]{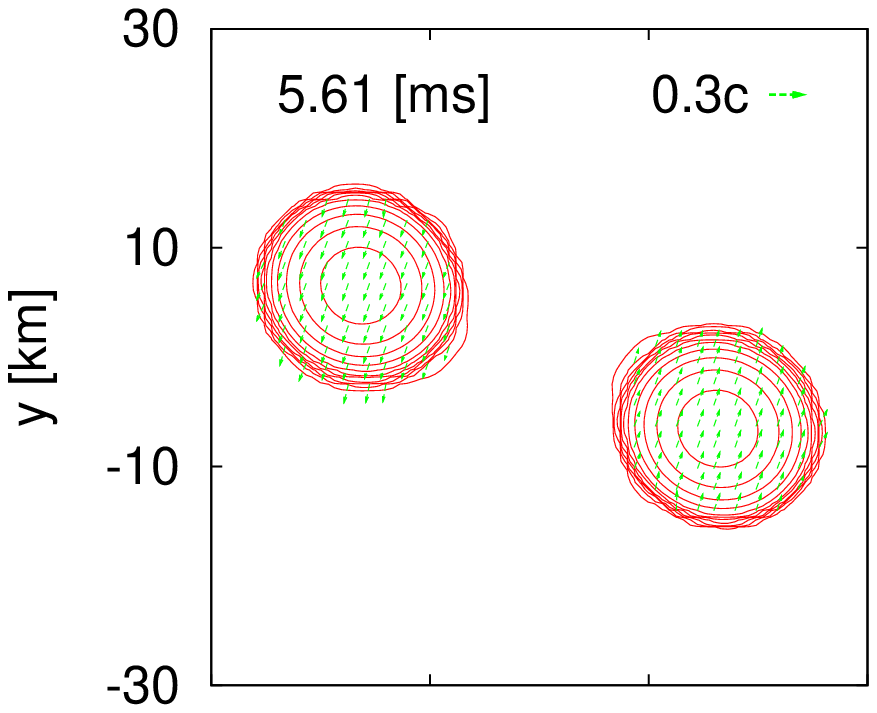}
\includegraphics[width=0.5\textwidth]{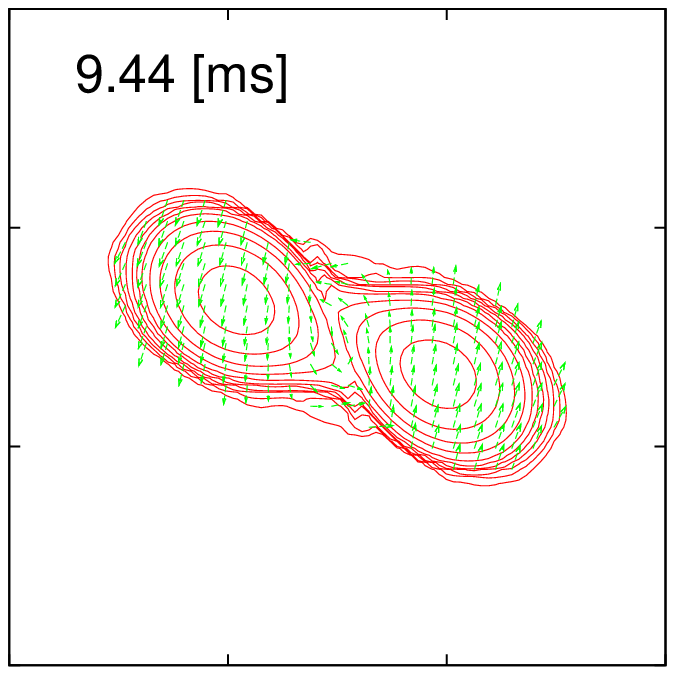}\\
}
\centerline{
\includegraphics[width=0.5\textwidth]{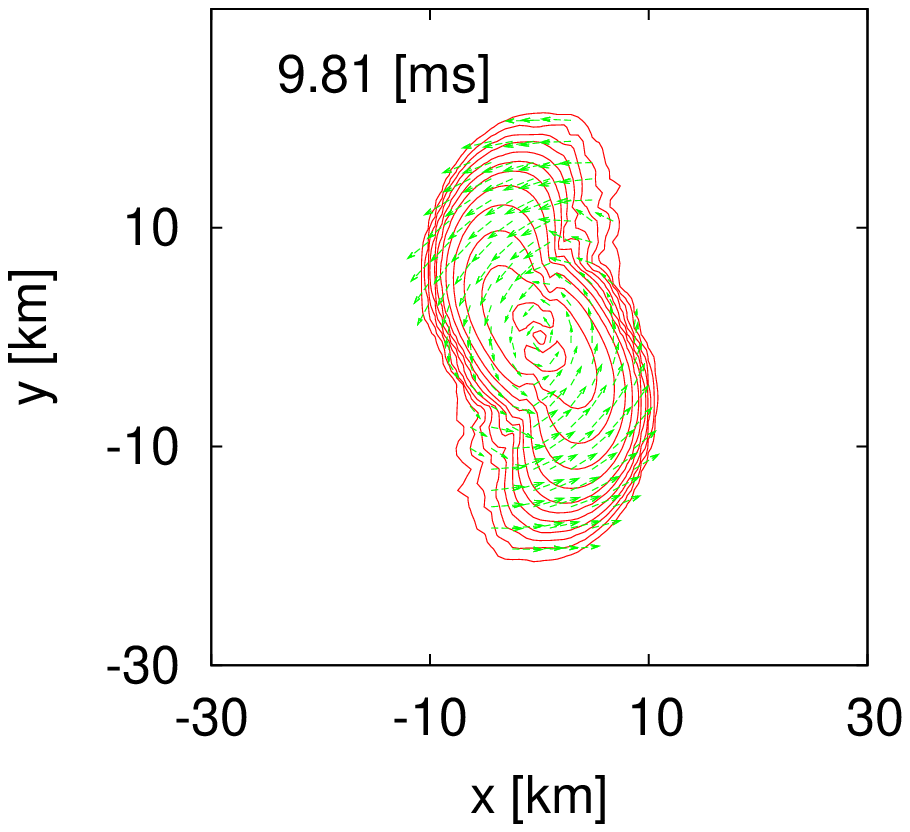}
\includegraphics[width=0.5\textwidth]{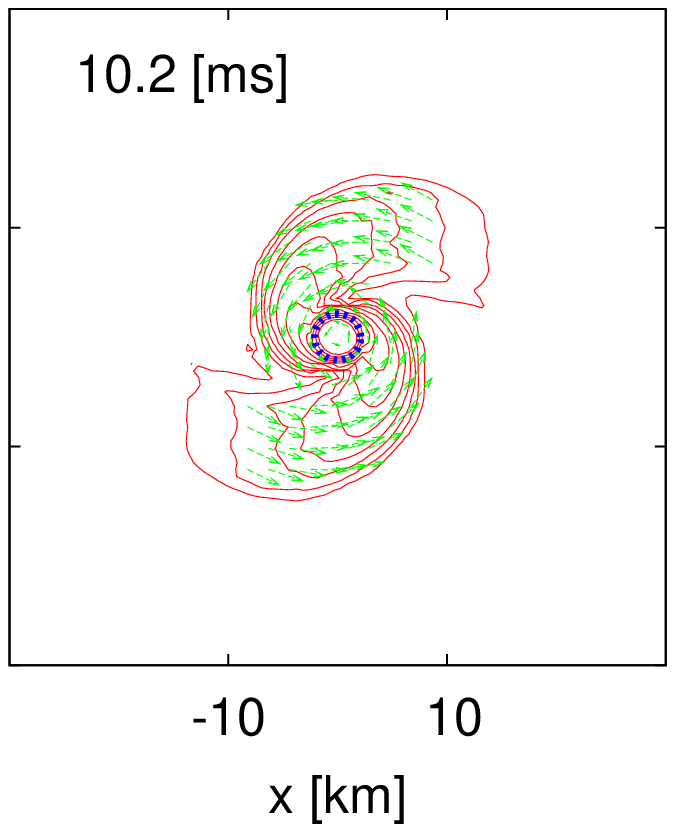}
}
\caption{Isodensity contours and velocity profile in the equatorial plane for a merger of two equal-mass NSs with $M_{\mathrm{NS}}=1.5\,M_{\odot}$ assumed to follow the APR model~\cite{Akmal:1998cf} for the NS EOS, taken from Figure~5 of~\cite{Kiuchi:2009jt}.  With a higher mass than the remnant shown in Figure~\ref{fig:sim_1}, the remnant depicted here collapses promptly to form a BH, its horizon shown by the dashed blue circle, absorbing all but 0.004\% of the total rest mass from the original system.}
\label{fig:sim_2}
\end{figure}}

In Figure~\ref{fig:sim_3}, we see the merger of an unequal-mass binary, with masses $M_1=1.3\,M_{\odot}$ and $M_2=1.6\,M_{\odot}$.  In this case, the heavier NS partially disrupts the lighter NS prior to merger, leading to the secondary NS being accreted onto the primary.  In this case, a much more massive disk is formed and, even after a BH forms in the center of the remnant, a substantial amount of matter, representing 0.85\% of the total mass, remains outside the horizon. Later accretion of this material could potentially release the energy required to power a SGRB.

\epubtkImage{}{%
\begin{figure}[!ht]
\centerline{
\includegraphics[width=0.39\textwidth]{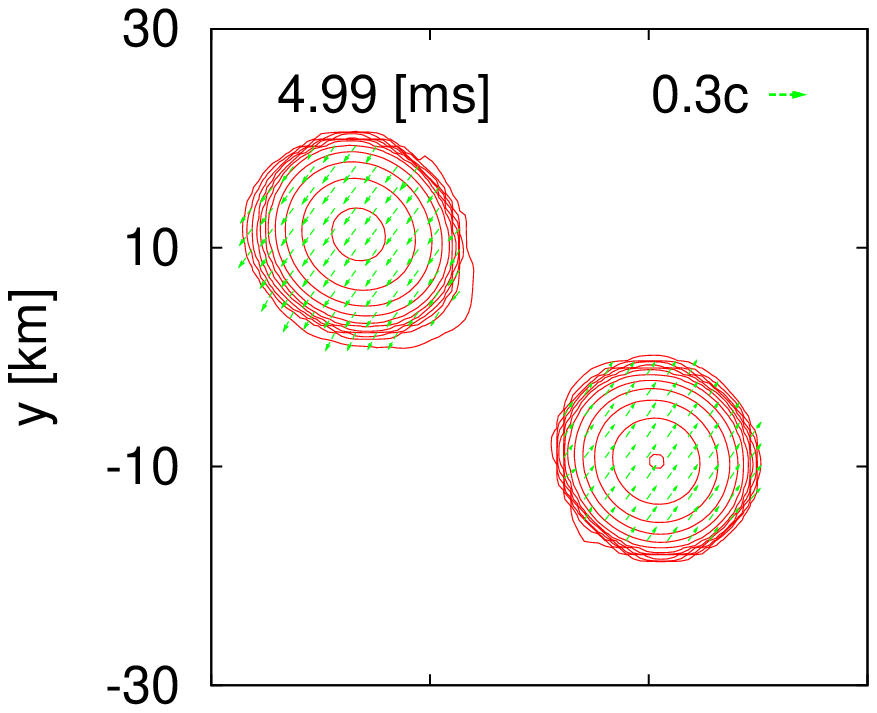}
\includegraphics[width=0.2925\textwidth]{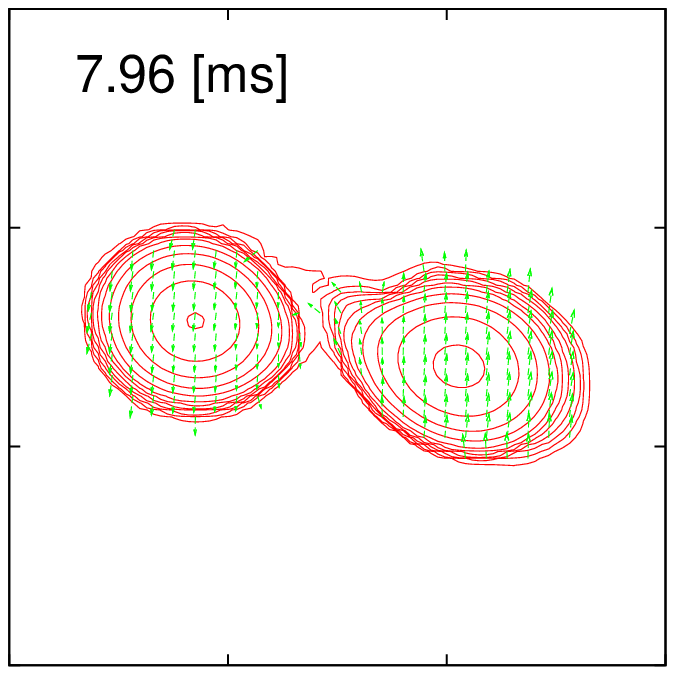}
\includegraphics[width=0.2925\textwidth]{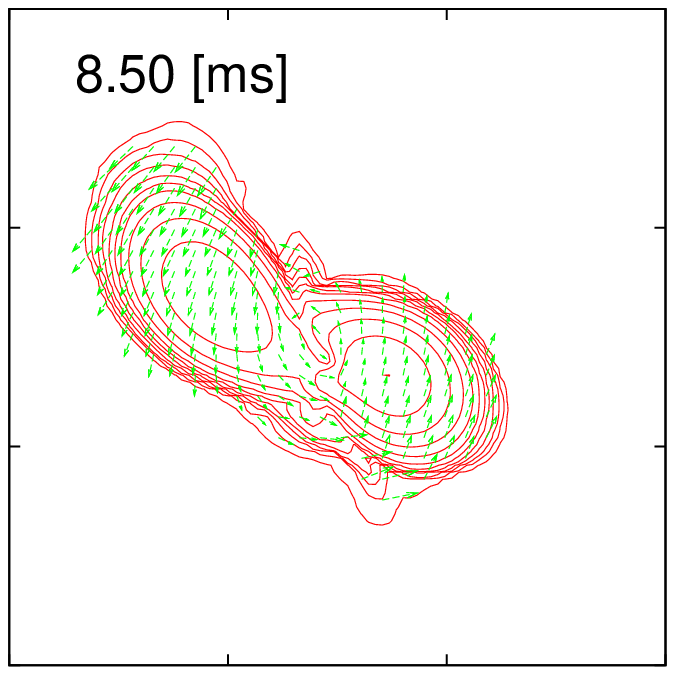}\\
}
\centerline{
\includegraphics[width=0.39\textwidth]{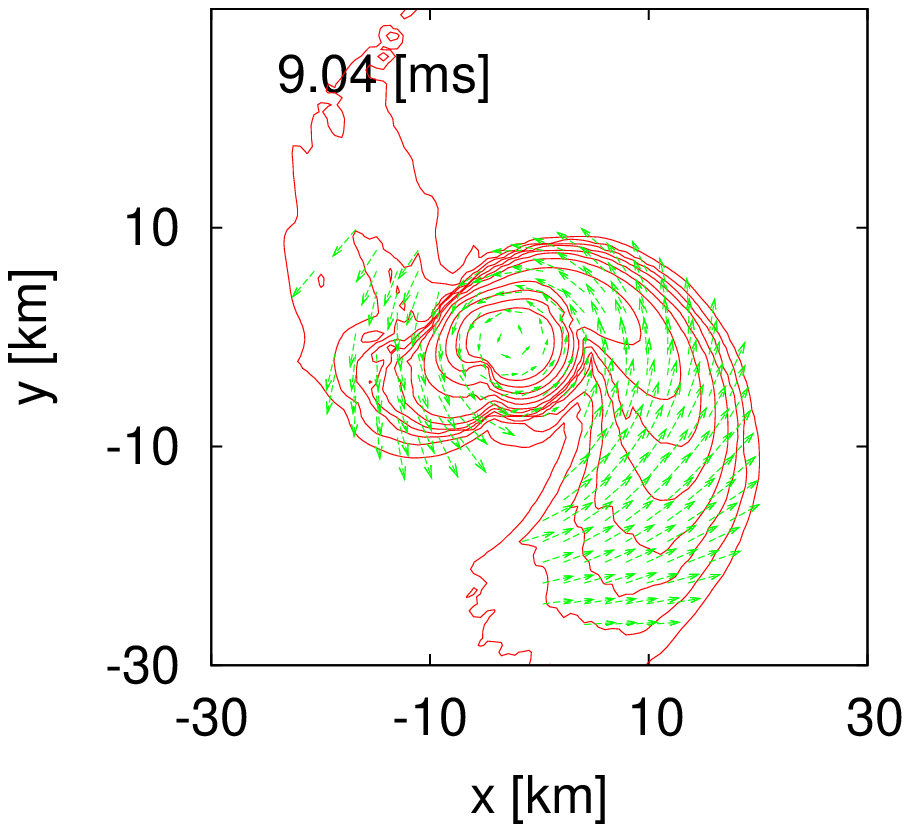}
\includegraphics[width=0.2925\textwidth]{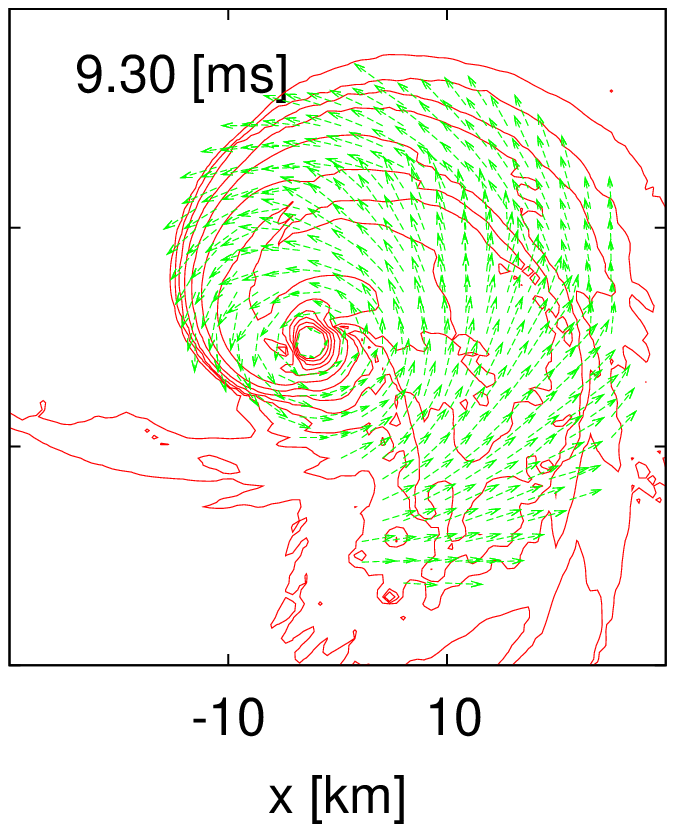}
\includegraphics[width=0.2925\textwidth]{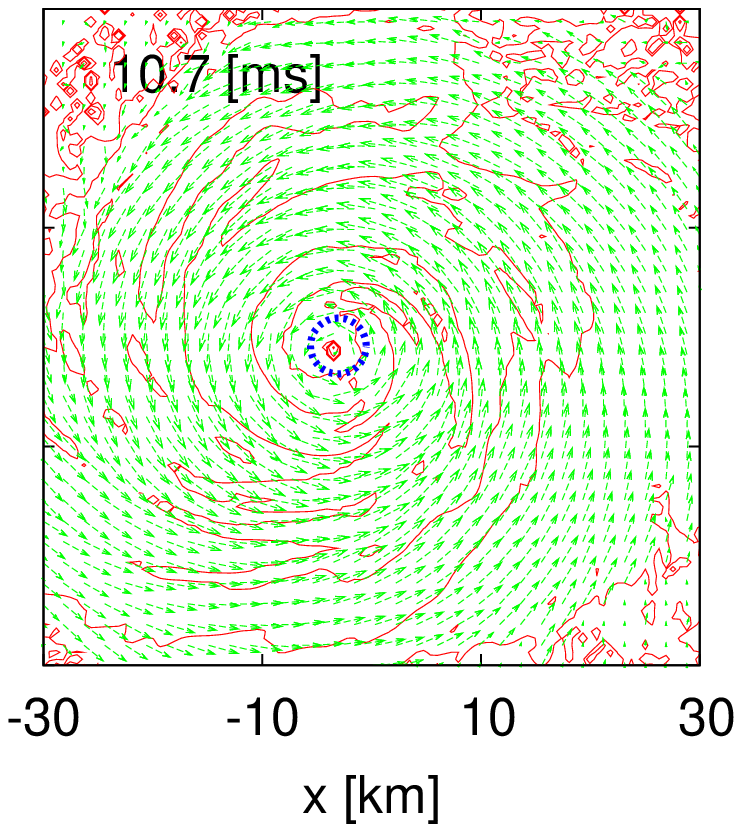}
}
\caption{Isodensity contours and velocity profile in the equatorial plane for a merger of two unequal-mass NSs with $M_1=1.3\,M_{\odot}$ and $M_2=1.6\,M_{\odot}$, with both assumed to follow the APR model~\cite{Akmal:1998cf} for the NS EOS, taken from Figure~6 of~\cite{Kiuchi:2009jt}. In unequal-mass mergers, the lower mass NS is tidally disrupted during the merger, forming a disk-like structure around the heavier NS.  In this case, the total mass of the remnant is sufficiently high for prompt collapse to a BH, but 0.85\% of the total mass remains outside the BH horizon at the end of the simulation, which is substantially larger than  for equal-mass mergers with prompt collapse (see Figure~\ref{fig:sim_2}).} 
\label{fig:sim_3}
\end{figure}}

\newpage
\section{Initial Data and Quasi-Equilibrium Results}
\label{sec:initialdata}

\subsection{Overview}

While dynamical calculations are required to understand the GW and EM emission from BH--NS and NS--NS mergers, some of the main qualitative features of the signals may be derived directly from QE sequences.
From the variation of total system energy with binary angular velocity along a given sequence, it is possible to construct an approximate GW energy spectrum $dE_{\mathrm{GW}}/{df}$ immediately from QE results, essentially by performing a numerical derivative (see Figure~\ref{fig:QE_binding}).  Doing so for a number of different sequences makes it possible to identify key frequencies where tidal effects may become measurable and to identify these with binary parameters such as the system mass ratio and NS radius.  Similarly, since QE sequences should indicate whether a binary begins to shed mass prior to passage through the ISCO (see Figure~\ref{fig:QE_massshed}), one may be able to classify observed signals into mass-shedding and non-mass-shedding events, and to use the critical point dividing those cases to help constrain the NS EOS.  Single-parameter estimates have been derived for NS--NS binaries using QE sequences~\cite{Faber:2002zn} (and for BH--NS binaries using QE~\cite{Taniguchi:2007xm} and dynamical calculations~\cite{Shibata:2007zm}).
NS--NS binaries typically approach instability at frequencies $f_{\mathrm{GW}}\gtrsim 1$ kHz, where laser shot noise is severely degrading the sensitivity of an interferometer detector.  To observe ISCO-related effects with higher signal-to-noise, it may be necessary to operate GW observatories using narrow-band signal recycling modes, in which the sensitivity in a narrow range of frequencies is enhanced at the cost of lower sensitivity to broadband signals~\cite{Buonanno:2003fw}.

\epubtkImage{}{%
\begin{figure}[!ht]
\centerline{\includegraphics[width=0.8\textwidth]{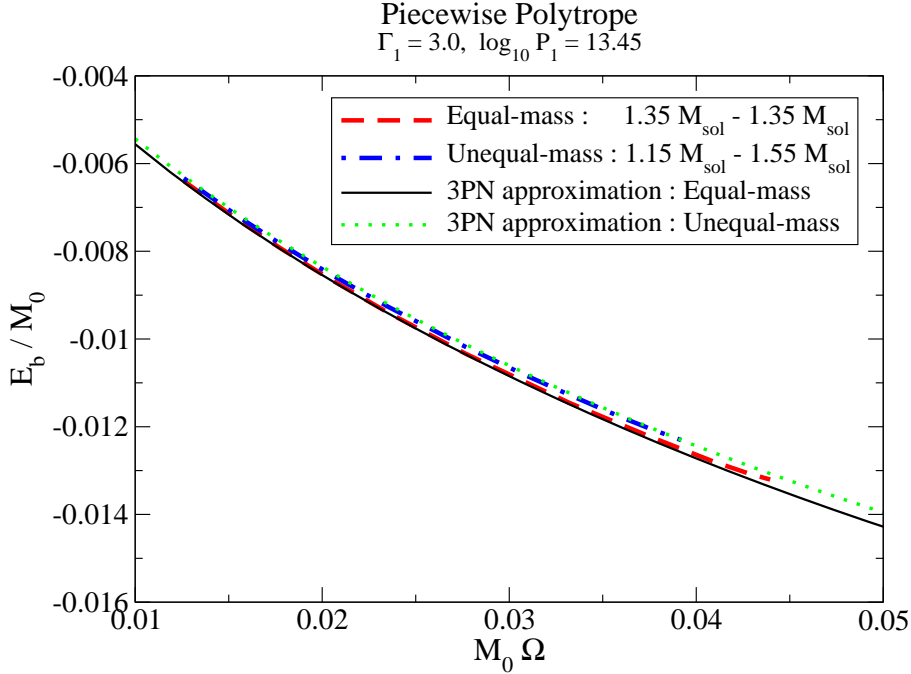}}
\caption{Dimensionless binding energy $E_b/M_0$ vs.\ dimensionless orbital frequency $M_0\Omega$, where $M_0$ is the total ADM (Arnowitt--Deser--Misner) mass of the two components at infinite separation, for two QE NS--NS sequences that assume a piecewise polytropic NS EOS, taken from Figure~16 of~\cite{Taniguchi:2010kj}.  The equal-mass case assumes $M_{NS}=1.35\,M_{\odot}$ for both NSs, while the unequal-mass case assumes $M_1=1.15\,M_{\odot}$ and $M_2=1.55\,M_{\odot}$.  The thick curves are the numerical results, while the thin curves show the results from the 3PN approximation. The lack of any minimum suggests that instability for these configurations occurs at the onset of mass shedding, and not through a secular orbital instability.}
\label{fig:QE_binding}
\end{figure}}

\epubtkImage{}{%
\begin{figure}[!ht]
\centerline{\includegraphics[width=0.8\textwidth]{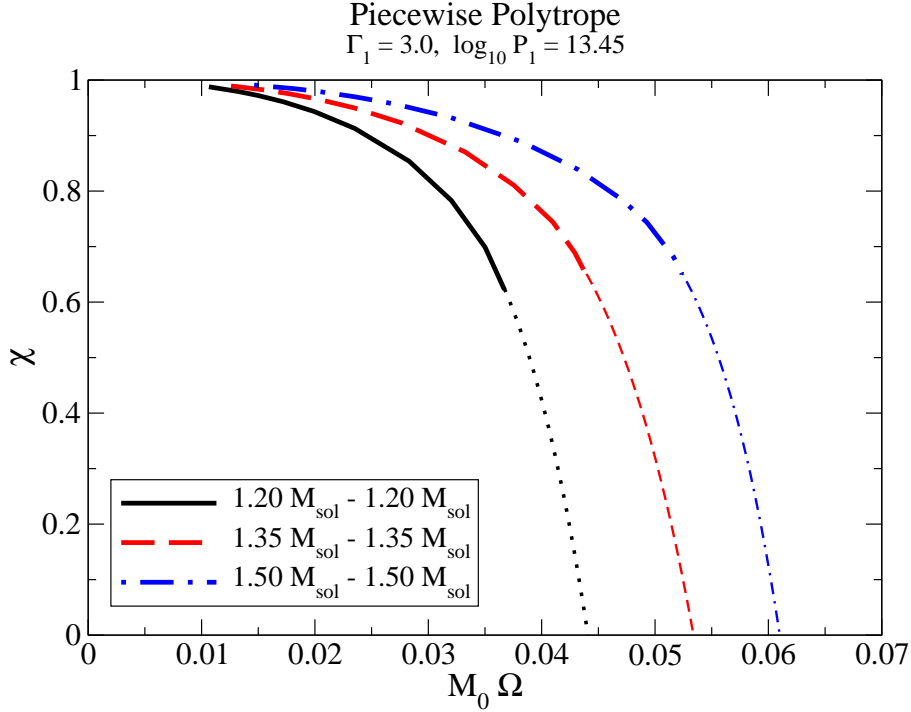}}
\caption{Mass-shedding indicator $\chi\equiv \left(\frac{\partial (\ln h)}{\partial r}\right)_{\mathrm{eq}}/\left(\frac{\partial (\ln h)}{\partial r}\right)_{\mathrm{pole}}$ vs.\ orbital frequency $M_0\Omega$, where $h$ is the fluid enthalpy and the derivative is measured at the NS surface in the equatorial plane toward the companion and toward the pole in the direction of the angular momentum vector, for a series of QE NS--NS sequences assuming equal-mass components, taken from Figure~19 of~\cite{Taniguchi:2010kj}. Here 
$\chi=1$ corresponds to a spherical NS, while $\chi=0$ indicates the onset of mass shedding.  More massive NSs are more compact, and thus able to reach smaller separations and higher angular frequencies before mass shedding gets underway.}
\label{fig:QE_massshed}
\end{figure}}

It is important to note that, while the potential parameter space for NS EOS models is still very large, a much smaller set may serve to classify models for comparison with the first generation of GW detections.  Indeed, by breaking up the EOS into piecewise polytropic segments, one may use as few as four parameters to roughly approximate all known EOS models, including standard nuclear models as well as models with kaon or other condensates~\cite{Read:2008iy}. 
 To illustrate this, we show in Figure~\ref{fig:QE}  four different QE models for NS--NS configurations with different EOS, taken from~\cite{Taniguchi:2010kj}; all have $M_1=1.15\,M_{\odot}$ and $M_2=1.55\,M_{\odot}$, and they correspond to the closest separation for which the QE code still finds a convergent result.

\epubtkImage{}{%
\begin{figure}[!ht]
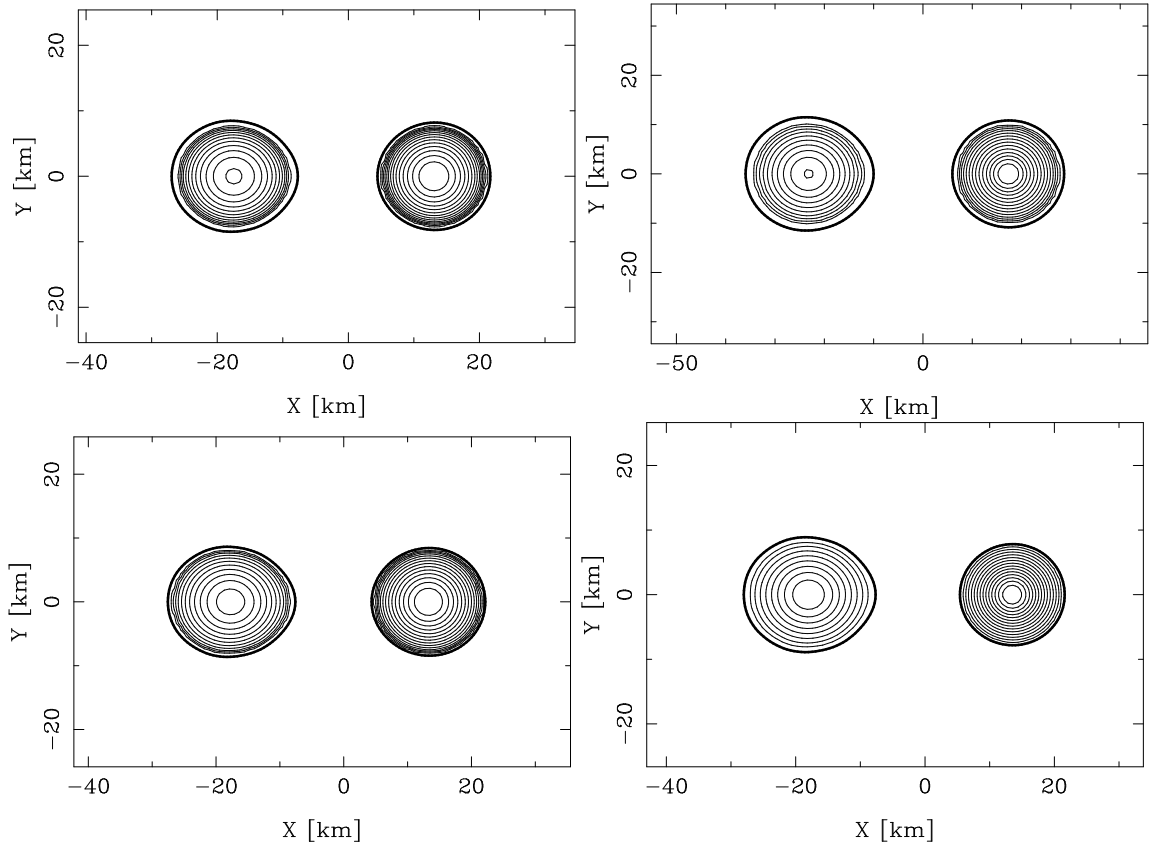

\centerline{
\includegraphics[width=0.5\textwidth]{NSNS_QE1}
\includegraphics[width=0.5\textwidth]{NSNS_QE2}\\
}
\centerline{
\includegraphics[width=0.5\textwidth]{NSNS_QE3}
\includegraphics[width=0.5\textwidth]{NSNS_QE4}
}
\caption{Isodensity contours for QE models of NS--NS binaries, taken from Figures~9\,--\,12 of~\cite{Taniguchi:2010kj}.  In each case, the two NSs have masses $M_1=1.15\,M_{\odot}$ (left) and $M_2=1.55\,M_{\odot}$ (right), and the center-of-mass separation is as small as the QE numerical methods allow while able to find a convergent result.  The models assume different EOS, resulting in different central concentrations and tidal deformations.}
\label{fig:QE}
\end{figure}}

The inspiral of NS--NS binaries may yield complementary information about the NS structure beyond what can be gleaned from QE studies of tidal disruption.  NSs have a wide variety of oscillation modes, including f-modes, g-modes, and r-modes, any of which may be excited by resonances with the orbital frequency as the latter sweeps upward.  Should a particular oscillation mode be excited resonantly, it can then serve briefly as an energy sink for the system, potentially changing the phase evolution of the binary.  For example, in a rapidly spinning NS, excitation of the $m=1$ r-mode can be significant, yielding a change of over 100 radians for the pre-merger GW signal phase in the case of a millisecond spin period~\cite{Lai:2006pr}.  For NS--NS mergers in the field, this would require one of the NSs to be a young pulsar that has not yet spun down significantly, which is unlikely because of the difficulty in obtaining such an extremely small binary separation after the second supernova.  
Other modes, such as the $l=2$ f-mode, may be excited in less extreme circumstances, also yielding information about NS structure parameters~\cite{Flanagan:2007ix}. 

\subsection{Quasi-equilibrium formalisms}

It has long been known that the GW emission from eccentric binaries is very efficient at radiating away angular momentum relative to the radiated energy~\cite{Peters:1963ux}; as a result, the orbital eccentricity decreases as a binary inspirals, so that orbits should be very nearly circular long before they enter the detection range of ground-based interferometers. The only exception could be from a dynamical capture process that would create a binary with a significant eccentricity and very small orbital separation.  Such eccentric binaries have been predicted to form in the nuclear cluster of our Galaxy (see, e.g., \cite{O'Leary:2008xt}) and in core-collapsed globular clusters \cite{Grindlay:2005ym,Lee:2009ca}. However, at present, no formalism exists to construct initial data for such systems, besides superposing the individual components with sufficiently large initial separations to minimize constraint violations.  

Using this circularity of primordial binaries as a starting point, one may use the constraint equations of GR, along with an assumption of quasi-circularity, to derive sets of elliptic equations describing compact binary configurations.
For both QE and dynamical calculations, most groups typically make use of the Arnowitt-Deser-Misner (ADM) 3+1 splitting of the metric~\cite{Arnowitt:1962hi}, which foliates the metric into a set of three-dimensional hypersurfaces by introducing a time coordinate.  The resulting form of the metric, which is completely general, is written
\begin{equation}
g_{\mu\nu}\equiv (-\alpha^2+\beta_i\beta^i)dt^2+2\beta_i dt~dx^i+\gamma_{ij} dx^idx^j,\label{eq:adm}
\end{equation}
where $\alpha$ is known as the lapse function, $\beta_i$ the shift vector, and $\gamma_{ij}$ the spatial three-metric intrinsic to the hypersurface.  We are following the standard relativistic notation here where Greek indices correspond to four-dimensional quantities and Latin indices to spatial three-dimensional quantities.  Thus, the shift vector is a 3-vector, raised and lowered with the spatial 3-metric $\gamma_{ij}$ rather than the spacetime 4-metric $g_{\mu\nu}$.  To simplify matters, one typically defines a conformal factor $\psi$ that factors out the determinant of the 3-metric, such that
\begin{equation}
\psi\equiv \left[\det(\gamma_{ij})\right]^{1/12}.\label{eq:confpsi}
\end{equation}
introducing the conformal 3-metric $\tilde{\gamma}_{ij}\equiv \psi^{-4}\gamma_{ij}$ with unit determinant.
While the 3-metric is a fundamental component of the geometric structure of the spacetime, the lapse function and shift vector are \emph{gauge quantities} that simply reflect our choice of coordinates.  Thus, while one often determines the lapse and shift in order to construct a appropriately ``stationary'' solution in the relevant coordinates between neighboring time slices, their values are often replaced to initialize dynamical runs with more convenient choices and thus different assumptions about coordinate evolution in time.

The field equations of general relativity take the deceptively simple form
\begin{equation}
G_{\mu\nu}\equiv R_{\mu\nu}-\frac{1}{2}g_{\mu\nu} R  = 8\pi T_{\mu\nu}\label{eq:einstein}
\end{equation}
where $G_{\mu\nu}$ is the Einstein tensor, $R_{\mu\nu}$ and $R$ the Ricci curvature tensor and the curvature scalar, and $T_{\mu\nu}$ the stress-energy tensor that accounts for the presence of matter, electromagnetic fields, and other physical effects that contribute to the mass-energy of the spacetime.
Since GR is a second-order formulation, valid initial data must include not only the metric but also its first time derivative.  It generally proves most convenient to introduce the time derivative of the metric after subtracting away the Lie derivative with respect to the shift, yielding a quantity known as the extrinsic curvature, $K_{ij}$:
\begin{equation}
(\partial_t-\mathcal{L}_\beta)\gamma_{ij}\equiv -2\alpha K_{ij}.
\end{equation}
Both the 3-metric and extrinsic curvature are symmetric tensors with six free parameters.

For systems containing NSs, one must consider the effects of nuclear matter through its presence in the stress-energy tensor $T^{\mu\nu}$.  It is common to assume that the matter has the EOS describing a perfect, isotropic fluid, for which the stress energy tensor is given by
\begin{equation}T^{\mu\nu}\equiv (\rho_0 + \rho_0\varepsilon +
P)u^{\mu}u^{\nu} + P g^{\mu\nu},\end{equation}
where $\rho_0$, $\varepsilon$, $P$ and $u^\mu$ are the fluid's rest-mass
density, specific internal energy, pressure, and 4-velocity, respectively.  Many calculations further assume that the NS EOS is described by an adiabatic polytrope, for which
\begin{equation}
P=(\Gamma-1)\rho_0\varepsilon = k\rho_0^\Gamma, \label{eq:polytrope}
\end{equation}
where $\Gamma$ is the adiabatic index of the gas and $k$ a constant, though a number of models designed to incorporate nuclear physics and/or strange matter condensates have also been constructed and studied (see Sections~\ref{sec:QEsim} and \ref{sec:dynsim} below).

The problem in constructing initial data is not so much producing solutions that are self-consistent within GR, but rather to specify a sufficient number of assumptions
to fully constrain a solution.  Indeed, there are only four constraints imposed by the equations of GR, known as the Hamiltonian and momentum constraints.  The Hamiltonian constraint is found by projecting Einstein's equations twice along the direction defined by a normal observer, and describes the way stress-energy leads to curvature in the metric (see, e.g.,~\cite{Baumgarte:2002jm} for a thorough review):
\begin{equation}
R+K^2-K_{ij}K^{ij}=16\pi\rho,
\end{equation}
where $R$ is the scalar curvature of the 4-metric, $K=K^i_i$ is the trace of the extrinsic curvature,  and
\begin{equation}
\rho\equiv {\mathbf n}\cdot{\mathbf T}\cdot{\mathbf n}=\alpha^2 T^{00}=\rho h (\alpha u^0)^2- P
\end{equation} 
is the total energy density seen by a normal observer. 
The third term indicates that the total energy density is found by projecting the stress-energy tensor in the direction of the unit-length timelike normal vector $n$, whose components are given by 
\begin{equation}n_\mu=(-\alpha,0,0,0).\label{eq:unitnormal}\end{equation} 
In the final expression $h\equiv 1+\varepsilon+P/\rho_0$ is the specific enthalpy of the fluid, and the combination $\Gamma_n\equiv \alpha u^0$ represents the Lorentz factor of the matter seen by an inertial observer.  The notation here makes use of the standard summation convention, in which repeated indices are summed over.  

Projecting Einstein's equations in the space and time directions leads to the vectorial momentum constraint
\begin{equation}
D_iK^i_j-D_i K=8\pi j_i,
\end{equation}
where $D_i$ represents a three-dimensional covariant derivative and $j_i\equiv \rho_0 h\Gamma_n u_i$ is the total momentum seen by a normal observer.

\subsubsection{The Conformal Thin Sandwich formalism}\label{sec:CTS}

In order to specify all the free variables that remain once the
Hamiltonian and momentum constraints are satisfied, two different techniques have been widely employed throughout the numerical relativity community.  One, known as the Conformal Transverse-Traceless (CTT) decomposition, underlies the Bowen-York \cite{Bowen:1980yu} solution for black holes with known spin and/or linear momentum that is widely used in the ``moving puncture'' approach.  To date, however, the CTT formalism has not been used to generate NS--NS initial data, and we refer readers to \cite{ST_LRR,Centrella:2010mx} for descriptions of the CTT formalism applied to BH--NS and BH--BH initial data, respectively.

To date, most groups have
used the Conformal Thin Sandwich (CTS) formalism to generate QE NS--NS data
(see~\cite{Baumgarte:2002jm} for a
review,~\cite{Baierlein:1962zz,Isenberg:2007zg} for the initial steps
in the formulation,
and~\cite{Wilson:1989fnr..book..306W,Wilson:1995uh,York:1998hy,Cook:2000vr}
for derivations of the form in which it is typically used today) .
One first specifies that the conformal 3-metric is spatially flat,
i.e., $\tilde{\gamma}_{ij}=\delta_{ij}$, where $\delta_{ij}$ is the
Kronecker delta function.  Under this assumption, the only remaining
parameter defining the spatial metric is the conformal factor $\psi$,
which serves the role of a gravitational potential. Indeed, in the
limit of weak sources, it is linearly related to the standard
Newtonian potential.  Next, one specifies that there exists a helical
Killing vector, so that, as the configuration advances forward in
time, all quantities remain unchanged when properly rotated at
constant angular velocity in the azimuthal direction.  This is
sufficient to fix all but the trace of the extrinsic curvature, with
the other components forced to satisfy the relation
\begin{equation}
K^{ij}=-\frac{1}{2\alpha\psi^4}\left[\nabla^i\beta^j+\nabla^j\beta^i-\frac{2}{3}\gamma^{ij}\nabla_k\beta^k\right].
\end{equation}

The trace of the extrinsic curvature $K$ remains a free parameter in this approach.  While one may choose arbitrary  prescriptions to fix it, most implementations choose a maximal slicing of the spatial hypersurfaces by setting $K=\partial_t K=0$.  Under these assumptions the Hamiltonian and momentum constraints, along with the trace of the Einstein equations, yield five elliptic equations for the lapse, shift vector, and conformal factor:
\begin{eqnarray}
\nabla^2\psi&=&-\psi^5\left(\frac{1}{8}K^{ij}K_{ij}-2\pi \rho\right),\label{eq:ellippsi}\\
\nabla^2(\alpha\psi)&=&\alpha\psi^5(\frac{7}{8}\psi^4K_{ij}K^{ij}+2\pi\psi^4(\rho+2S),\label{eq:ellipapsi}\\
\nabla^2\beta^i+\frac{1}{3}\nabla^j\nabla_j\beta^i&=&2\alpha\psi^4K^{ij}\nabla_j(\alpha\psi^{-6})+16\alpha\psi^4j^i\label{eq:ellipshift},
\end{eqnarray}
where 
\begin{equation}
S=(g_{\mu\nu}+n_\mu n_\nu)T^{\mu\nu}=S^j_j=3P+(\rho+P)\left[1-\Gamma_n^{-2}\right]
\end{equation} 
is the trace of the stress-energy tensor projected twice in the spatial direction.  While these five equations are linked and the right-hand sides are non-linear, they are amenable to solution using iterative methods.
Boundary conditions are set by assuming asymptotic flatness: at large radii, the metric takes on the Minkowski form so $\alpha\rightarrow 1$, $\psi\rightarrow 1$, and $\beta^i_{rot}\rightarrow \Omega\times \vec{r}$.  We note that a purely corotating shift term yields zero when we apply the left-hand side of Eq.~\ref{eq:ellipshift}, so we may subtract it away and solve the equation with a boundary condition of zero instead.

The breakdown in Eqs.~\ref{eq:ellippsi}, \ref{eq:ellipapsi}, and \ref{eq:ellipshift} is not unique.  The Meudon group~\cite{Gourgoulhon:2000nn,lorene_web}, to pick one example, has often chosen to define $\nu\equiv\ln\alpha$ and $\beta\equiv\ln(\alpha\psi^2)$, and replace Eqs.~\ref{eq:ellippsi} and \ref{eq:ellipapsi} with the equivalent pair
\begin{eqnarray*}
\nabla^2\nu&=&\psi^4K^{ij}K_{ij}-\nabla_i\nu\nabla^i\beta+4\pi\psi^4(\rho+S),\\
\nabla^2\beta&=&\frac{3}{4}\psi^4K^{ij}K_{ij}-\frac{1}{2}(\nabla_i\nu\nabla^i\nu+\nabla_i\beta\nabla^i\beta)+4\pi\psi^4S.
\end{eqnarray*}

This approach is sufficient to define the field component of the configuration, but one still needs to solve for the matter quantities as well.  One starts by assuming that there is a known prescription for reconstructing the density, internal energy, and pressure from the enthalpy $h$.  Next, one has to assume some model for the spin of the NS.  While corotation is often a simpler choice, since the velocity field of the matter is zero in the corotating frame, the more physically reasonable condition is {\em irrotational} flow. 
Indeed a realistic NS viscosity is never sufficiently large to tidally lock the NS to its companion during inspiral~\cite{Bildsten:1992my,Kochanek:1992wk}. 
 If we define the co-momentum vector $w_i=hu_i$, irrotational flow implies the vanishing of its curl:
\begin{equation}
\nabla\times w = \nabla_\mu w_\nu-\nabla_\nu w_\mu = \partial_{\mu}w_\nu-\partial_\nu w_\mu=0,
\end{equation}
which allows us to define a velocity potential $\Psi$ such that $w\equiv \nabla\Psi$.
Using these quantities, one may write down the integrated Euler equation
\begin{equation}
h\alpha \Gamma_n/(1-\gamma_{ij}U^iU_0^j)= \mathrm{\ const.},
\end{equation}
where the 3-velocity $U^i$ of the fluid with respect to an Eulerian observer is given by
\begin{equation}U^i\equiv \frac{u^i+\beta^iu^0}{\Gamma_n}, \end{equation} 
and the orbital 3-velocity $U^i_0$ with respect to the same observer by $U_0^i = \beta^i/\alpha$.
For details on the ways in which one may construct an elliptic equation for the velocity potential, we refer to the derivation in~\cite{Gourgoulhon:2000nn}.

To date, all QE sequences and dynamical runs in the literature have assumed that NSs are either irrotational or synchronized, but it is possible to construct the equations for arbitrary NS spins so long as they are aligned~\cite{Baumgarte:2009fw,Tichy:2011gw}.  While suggestions are also given there on how to construct QE sequences with intermediate spins using the new formalism, none have yet appeared in the literature.  Similarly, a formalism to add magnetic fields self-consistently to QE sequences has been constructed~\cite{Uryu:2010su}, as current dynamical simulations typically begin from data assuming either zero magnetic fields or those that only contribute via magnetic pressure.

\subsubsection{Other formalisms}

The primary drawback of the CTS system is the lack of generality in assuming the spatial metric to be conformally flat, which introduces several problems.  The Kerr metric, for example, is known not to be conformally flat, and conformally flat attempts to model Kerr BHs inevitably include spurious GW content.  The same problem affects binary initial data: in order to achieve a configuration that is instantaneously time-symmetric, one actually introduces spurious gravitational radiation into the system, which can affect both the measured parameters of the initial system as well as any resulting evolution.

Other numerical formalisms to specify initial data configurations in GR have been derived using different assumptions.  Usui and collaborators derived an elliptic set of equations by allowing the azimuthal component of the 3-metric to independently vary from the radial and longitudinal components~\cite{Usui:1999eu,Usui:2002fx}, finding good agreement with the other methods discussed above.
A number of techniques have been developed to construct helically symmetric spacetimes in which one actually solves Einstein's equations to evaluate the non-conformally-flat component of the metric, which are typically referred to as ``waveless'' or ``WL'' formalisms~\cite{Schaefer:2003ra,Bonazzola:2003dm,Shibata:2004qz}.  In terms of the fundamental variables, rather than specifying the components of the conformal spatial metric by ansatz, one specifies instead the time derivative of the extrinsic curvature using a physically motivated prescription.
These methods are designed to match the proper asymptotic behavior of the metric at large distances, and may be combined with techniques designed to enforce  helical symmetry of the metric and gauge in the near zone (the near zone helical symmetry, or ``NHS'' formalism) to produce a global solution~\cite{Uryu:2005vv,Yoshida:2006cn,Uryu:2009ye}.   QE sequences generated using this formalism \cite{Uryu:2005vv} have shown that the resulting conformal metric is indeed non-flat, with deviations of approximately $1\%$ for the metric components, and similar differences in the system's binding energy when compared to equivalent CTS results.  They suggest \cite{Uryu:2009ye} that underestimates in the quadrupole deformations of NS prior to merger may result in total phase accumulation errors of a full cycle, especially for more compact NS models.

QE formalisms reflect the assumption that binaries will be very nearly circular, since GW emission acting over very long timescales damps orbital eccentricity to negligible values for primordial NS--NS binaries between their formation and final merger.  Binaries formed by tidal capture and other dynamical processes, which may be created with much smaller initial separations, are more likely to maintain significant eccentricities all the way to merger (see, e.g., \cite{O'Leary:2008xt} for a discussion of such processes for BH--BH binaries)  and it has been suggested based on simple analytical models that such mergers, likely occurring in or near dense star clusters, may account for a significant fraction of the observed SGRB sample \cite{Lee:2009ca}. However, more detailed modeling is required to work out accurate estimates of merger rates given the complex interplay between dynamics and binary star evolution that determines the  evolution of dense star clusters, and given the large uncertainties in the distributions of star cluster properties in galaxies throughout the universe.  No initial data have ever been constructed in full GR  for merging NS--NS binaries with eccentric orbits since the systems are then highly time-dependent, while the calculations performed to evolve them generally use a superposition of two stationary NS configurations~\cite{Gold:2011df} .

\subsection{Numerical implementations}

There are a number of numerical techniques that have been used to solve these elliptic systems.  The first calculations of NS--NS QE sequences, in both cases for synchronized binaries, were performed by Wilson, Mathews, and Marronetti~\cite{Wilson:1995uh,Wilson:1996ty,Mathews:1999km} and Baumgarte et al.~\cite{Baumgarte:1997xi,Baumgarte:1997eg}. The former used a finite differencing scheme, and centered different quantities at cells, vertices, and faces in order to construct a system of equations that was solved using fast matrix inversion techniques, while
the latter used a Cartesian multigrid scheme, restricted to an octant to increase computational efficiency.  

After a  formalism for evaluating QE irrotational NS--NS sequences was developed~\cite{Bonazzola:1998yq,Uryu:1999uu}, some of the first results were obtained by Ury\=u and Eriguchi, who developed a finite-differencing code in spherical coordinates allowing for the solution of relativistic NS--NS binaries using Green's functions\cite{Uryu:1999uu,Uryu:2000dw}.  Their method extended the self-consistent field (SCF) work of~\cite{Komatsu:1989zz}, which had previously been applied to axisymmetric configurations.
Irrotational configurations were also generated by Marronetti et al.~\cite{Marronetti:1999ya}, using the same finite difference scheme as found in the work on synchronized binaries.

The most widely used direct grid-based solver in numerical relativity is the {\tt Bam\_Elliptic} solver~\cite{Bruegmann:1999we},  which solves elliptic equations on single rectangular grids or multigrid configurations.  It is included within the {\tt Cactus} code, which is widely used in 3-D numerical relativity~\cite{cactus_web}. In particular it has been used to initiate a number of single and binary BH simulations, including one of the original breakthrough binary puncture works~\cite{Campanelli:2005dd}.

Lagrangian methods, typically based on smoothed
particle hydrodynamics (SPH)
\cite{Lucy:1977zz,Gingold:1977sh,Monaghan:1992rr}) have been used to
generate both synchronized and irrotational configurations for
PN~\cite{Ayal:1999wn,Faber:1999gj,Faber:2000uf,Faber:2002cg} and
conformally flat (CF)
\cite{Oechslin:2001km,Oechslin:2002vy,Faber:2003sb,Oechslin:2004yj,Oechslin:2005mw,Oechslin:2006uk,Oechslin:2007gn,Bauswein:2008gx,Bauswein:2009im,Bauswein:2010dn}
calculations of NS--NS mergers, but they have not yet been extended to
fully GR calculations, in part because of the difficulties
in evolving the global spacetime metric.

The most widely used data for numerical calculations are those generated by the Meudon group (see Section~\ref{sec:QEsim} below for details on their calculations and~\cite{Gourgoulhon:2000nn} for a detailed description of their methods).  The code they developed, {\tt Lorene}~\cite{lorene_web}, uses multidomain spectral methods to solve elliptic equations (while the code has been used primarily for relativistic stellar and binary configurations, it can be used as a more general solver).  Around each star, one creates a set of nested, contiguous grids, with points arrayed in the radial and angular directions.  The innermost grid has spheroidal geometry, and the surrounding grids are annular.  The outermost grid may be allowed to extend to spatial infinity through a compactification transformation of the radial coordinate.  To solve elliptic equations for various field quantities, one breaks each into a sum of two components, each of whose source terms are concentrated in one NS or the other.  Similarly, the source terms themselves are split into two pieces, ideally so each component is well-described by spheroidal spectral coefficients centered around each star.  Using the spectral expansion, one may pass values from one star to the other and then recalculate spectral coefficients for the other grid configuration.
This scheme has several efficiency advantages over direct grid-based methods, which helps to explain its popularity.  First, the domain geometry may be chosen to fit to a NS surface, which eliminates Gibbs phenomenon-related errors and allows for exponential convergence with respect to the number of grid points, rather than the geometric convergence that characterizes finite difference-based grid codes.  Second, the use of spectral methods requires much less computer memory than grid-based codes, and, as a result, {\tt Lorene} is a serial code that can run easily on any off-the-shelf PC, rather than requiring a supercomputer platform.

\subsection{Quasi-equilibrium and pre-merger simulations}
\label{sec:QEsim}
NS--NS binaries may be well approximated by QE configurations up until they reach separations comparable to the sizes of the binary components themselves, that latter phase lasting a fraction of a second after an inspiral of millions of years or more.  The eventual merger will occur after the binary undergoes one of two possible orbital instabilities.  If the total binary energy and angular momenta reach a minimum at some separation, which defines the ISCO, the binary becomes dynamically unstable and plunges toward merger.  Alternately, if the NS fills its Roche Lobe (typically the lower density NS) mass will transfer onto the primary and the secondary will be tidally disrupted.  The parameters of some NS--NS systems could technically allow for stable mass transfer, in which mass loss from a lighter object to a heavier one leads to a widening of the binary separation.  This does occur for some binaries containing white dwarfs, but every dynamical calculation to date using full GR or even approximate GR has found that the rapid inspiral rate leads to inevitably unstable mass transfer and the prompt merger of a binary.

Many of the results later confirmed using relativistic QE sequences were originally derived in Newtonian and PN calculations, particularly as explicit extensions of Chandrasekhar's body of work (see~\cite{Chandrasekhar:1987QB410.C47}).  Chandrasekhar's studies of incompressible fluids were first extended to compressible binaries by Lai, Rasio, and Shapiro~\cite{Lai:1993ApJ...406L..63L,Lai:1993ve,Lai:1993pa,Lai:1993rs,Lai:1994hf}, who used an energy variational method with an ellipsoidal treatment for polytropic NSs.  They established,  among other results, the magnitude of the rapid inspiral velocity near the dynamical stability limit~\cite{Lai:1993ApJ...406L..63L},  the existence of a critical polytropic index ($n\approx 2$) separating binary sequences undergoing the two different terminal instabilities~\cite{Lai:1993ve}, the role played by the NS spin and viscosity and magnitude of finite-size effects in relation to 1PN terms~\cite{Lai:1993pa,Lai:1993rs}, and the development of tidal lag angles as the binary approaches merger~\cite{Lai:1994hf}.  They also determined that for most reasonable EOS models and non-extreme mass ratios, as would pertain to NS--NS mergers, an energy minimum is inevitably reached before the onset of mass transfer through Roche lobe overflow.  The general results found in those works were later confirmed by~\cite{New:1997xi}, who used a SCF technique~\cite{Hachisu:1986ApJS...61..479H,Hachisu:1986ApJS...62..461H}, finding similar locations for instability points as a function of the adiabatic index of polytropes, but a small positive offset in the radius at which instability occurred.  Similar results were also found by~\cite{Uryu:1997zc,Uryu:1998kq}, but with a slight modification in the total system energy and decrease in the orbital frequency at the onset of instability.

The first PN ellipsoidal treatments were developed by Shibata and collaborators using self-consistent fields~\cite{Shibata:1995sb,Shibata:1996ci,Shibata:1997dr,Shibata:1997xn,Taniguchi:1998wi} and by Lombardi, Rasio, and Shapiro~\cite{Lombardi:1997aw}.  Both groups found that the nonlinear gravitational effects imply a decrease in the orbital separation (increase in the orbital frequency) at the instability point for more compact NS. 
 This result reflects a fairly universal principle in relativistic binary simulations: as gravitational formalisms incorporate more relativistic effects, moving from Newtonian gravity to 1PN and on to CF approximations and finally full GR, the strength of the gravitational interaction inevitably becomes stronger. 
  The effects seen in fully dynamical calculations will be discussed in Section~\ref{sec:dynsim}, below.

The first fully relativistic CTS QE data for synchronized NS--NS binaries were constructed by Baumgarte et al.~\cite{Baumgarte:1997eg,Baumgarte:1997xi}, using a grid-based elliptic solver.  Their results demonstrated that the maximum allowed mass of NSs in close binaries was \emph{larger} than that of isolated NSs with the same (polytropic) EOS, clearly disfavoring the ``star-crushing'' scenario that had been suggested by~\cite{Wilson:1995uh,Mathews:1997pf} using a similar CTS formalism (but see also the error in these latter works addressed in~\cite{Flanagan:1998zt}, discussed in Section~\ref{sec:approxrel} below).  Baumgarte et al.\  also identified how varying the NS radius affects the ISCO frequency, and thus might be constrained by GW observations.  Using a multigrid method, Miller et al. \cite{Miller:2003vc} showed that while conformal flatness remained valid until relatively near the ISCO, the assumption of sycnronized rotation broke down much earlier.  Usui et al.~\cite{Usui:1999eu} used the Green's function approach with a slightly different formalism to compute relativistic sequences and determined that the CTS conditions were valid up until extremely relativistic binaries were considered.

The first relativistic models of physically realistic irrotational NS--NS binaries were constructed by the Meudon group~\cite{Bonazzola:1998yq} using the {\tt Lorene}  multi-domain pseudo-spectral method code.  Since then, the Meudon group and collaborators have constructed a wide array of NS--NS initial data, including polytropic NS models~\cite{Gourgoulhon:2000nn,Taniguchi:2002ns,Taniguchi:2003hx}, as well as physically motivated NS EOS models~\cite{Bejger:2004zx} or quark matter condensates~\cite{Limousin:2004vc}.  Irrotational models have also been constructed by Ury\=u and collaborators~\cite{Uryu:1999uu,Uryu:2000dw} for use in dynamical calculations, and nuclear/quark matter configurations have been generated by Oechslin and collaborators~\cite{Oechslin:2004yj,Oechslin:2006uk}.  A large compilation of QE CTS sequences constructed using physically motivated EOS models including FPS (Friedman--Pandharipande)~\cite{Pandharipande:1989nmhi.conf..103P}, SLy (Skyrme Lyon)~\cite{Douchin:2001sv}, and APR~\cite{Akmal:1998cf}  models, along with piecewise polytropes designed to model more general potential cases (see~\cite{Read:2008iy}), was published in~\cite{Taniguchi:2010kj}.   

The most extensive set of results calculated using the waveless/near-zone helical symmetry condition appear in~\cite{Uryu:2009ye}, with equal-mass NS--NS binary models constructed for the FPS, Sly, and APR EOS in addition to $\Gamma=3$ polytropes.  Results spanning all of these QE techniques are summarized in Table~\ref{table:qe}.

\begin{table}
\caption[A summary of various studies focusing on QE sequences of NS--NS
  binaries.]{A summary of various studies focusing on QE sequences of
  NS--NS binaries.  Please refer to Section~\ref{sec:dynsim} for a
  discussion of papers that focus on dynamical simulations instead.
  Gravitational schemes include Newtonian gravity (`Newt.'), lowest-order post-Newtonian theory (`PN'), conformal thin sandwich (`CTS') including modified forms of the spatial metric (`Mod.\ CTS'), and waveless/near-zone helical symmetry techniques.  Numerical methods include ellipsoidal formalisms (`Ellips.'), self-consistent fields (`SCF'), numerical grids (`Grid'), multigrids, and multipatch, Green's function techniques (`Green's'), spectral methods (`Spectral'), or SPH relaxation (`SPH').   With regard to EOS models, `WD' refers to the exact white dwarf
  EOS assuming a cold degenerate electron gas
  \cite{Chandrasekhar:1967aits.book}.  The `Physical' EOS models
  include the FPS~\cite{Pandharipande:1989nmhi.conf..103P},
  SLy~\cite{Douchin:2001sv}, and APR~\cite{Akmal:1998cf} nuclear EOS
  models, along with their parameterized approximations and other
  physically motivated models.  The compactness ${\mathcal C}=M/R$
  refers to the value for a NS in isolation before it is placed in a
  binary, and plays no role in Newtonian physics.  The mass ratio
  $q=M_2/M_1$ is defined to be less than unity, and `spin' refers to
  either synchronized or irrotational configurations.}\label{table:qe}
\centering
\begin{tabular}{ll|ll|llll}
\hline
Author & Ref. & Grav. & Method & EOS & Compactness & Mass ratio & Spin\\\hline
Lai & \cite{Lai:1993ve} & Newt. & Ellips. & $ \Gamma=\frac{7}{5},\frac{5}{3},2,\infty$ & N/A & 1.0 & Syn.\\
Lai & \cite{Lai:1993pa} & Newt. & Ellips. & $ \Gamma=\frac{5}{3},2,3$ & N/A & 1.0 & Syn./Irr.\\
Lai & \cite{Lai:1993rs} & Newt. & Ellips. & $ \Gamma=\frac{7}{5},\frac{5}{3},3,\infty$ & N/A & 0.2\,--\,1.0 & Syn./Irr.\\
New & \cite{New:1997xi} & Newt. & SCF & $\Gamma=\frac{5}{3},2,3$,WD & N/A & 1.0 & Syn.\\
Ury\=u & \cite{Uryu:1998kq} & Newt. & SCF & $\Gamma=\frac{5}{3},2,\frac{17}{7},2,3,\infty$ & N/A & 1.0 & Irr.\\
\hline
Shibata & \cite{Shibata:1995sb} & PN & Grid & $\Gamma=2$ & ${\mathcal C}=0.08-0.12$ & 1.0 & Syn.\\
Shibata & \cite{Shibata:1997dr} & PN & Grid & $\Gamma=3$ &  ${\mathcal C}=0.03$ & 1.0 & Syn.\\
Shibata & \cite{Shibata:1997xn} & PN & Ellips. & $\Gamma=\frac{5}{3},2,\frac{7}{3},3,5,\infty$ & ${\mathcal C}=0-0.03$ & 1.0 & Syn.\\
Lombardi & \cite{Lombardi:1997aw} & PN & Ellips. & $\Gamma=2,3$ & ${\mathcal C}=0.12-0.25$ & 1.0 & Syn./Irr.\\ 
\hline
Baumgarte & \cite{Baumgarte:1997xi} & CTS & Multigrid & $\Gamma=2$ & ${\mathcal C} = 0.05-0.2$ & 1.0 & Syn.\\
Usui & \cite{Usui:1999eu} & Mod. CTS & Green's & $\Gamma=2,3,\infty$ & ${\mathcal C}=0.05-0.25$ & 1.0 & Syn.\\
Ury\=u & \cite{Uryu:1999uu} & CTS & Green's & $\Gamma=2$ & ${\mathcal C} = 0.1-0.19$ & 1.0 & Syn./Irr.\\
Ury\=u & \cite{Uryu:2000dw} & CTS & Green's & $\Gamma=\frac{9}{5},2,2.25,2.5,3$ & ${\mathcal C} = 0.1-0.19$ & 1.0 & Irr.\\
Bonazzola & \cite{Bonazzola:1998yq} & CTS & Spectral & $\Gamma=2$ & ${\mathcal C}=0.14$ & 1.0 & Syn/Irr.\\
Taniguchi & \cite{Taniguchi:2002ns} & CTS & Spectral & $\Gamma=2$ & ${\mathcal C}=0.12-0.18$ & 0.9\,--\,1.0 & Syn./Irr.\\
Taniguchi & \cite{Taniguchi:2003hx} & CTS & Spectral & $\Gamma=1.8,2.25,2.5$ & ${\mathcal C}=0.08-0.18$ & 0.83\,--\,1.0 & Syn./Irr.\\
Miller & \cite{Miller:2003vc} & CTS & Multigrid & $\Gamma=2$ & ${\mathcal C} = 0.15$ & 1.0 & Syn.\\
Bejger & \cite{Bejger:2004zx} & CTS & Spectral & Phys.  & ${\mathcal C}=0.14-0.19$ & 1.0 & Irr. \\
Limousin & \cite{Limousin:2004vc} & CTS & Spectral & Quark & ${\mathcal C}=0.19$ & 1.0 & Syn./Irr.\\
Oechslin & \cite{Oechslin:2004yj} & CTS & SPH & Quark & ${\mathcal C} = 0.12-0.20$ & 1.0 & Irr.\\
Taniguchi & \cite{Taniguchi:2010kj} & CTS & Spectral & Physical & ${\mathcal C}=0.1-0.3$ & 0.7\,--\,1.0 & Irr.\\
Ury\=u & \cite{Uryu:2009ye} & WL/NHS & Multipatch & $\Gamma=3$, Physical  & ${\mathcal C}=0.13-0.22$ & 1.0 & Irr.\\\hline
\end{tabular}
\end{table}

\newpage
\section{Dynamical Calculations: Numerical Techniques}
\label{sec:dyntech}

\subsection{Overview: General relativistic (magneto-)hydrodynamics and microphysical treatments}

NS--NS binaries are highly relativistic systems, numerous groups now run codes that evolve both  GR metric fields and fluids self-consistently, with some groups also incorporating an ideal magnetohydrodynamic evolution scheme that assumes infinite conductivity.  The codes that evolve the GR hydrodynamics or magnetohydrodynamics (GRHD and GRMHD, respectively) equations are many and varied, incorporating different spatial meshes, relativistic formalisms, and numerical techniques, and we will summarize the leading variants here.  All full GR codes now make use of the significant insight gained from  BH--BH merger calculations, but much work on these systems predates the three 2005 ``breakthrough''  papers by Pretorius~\cite{Pretorius:2005gq}, Goddard~\cite{Baker:2005vv}, and the group then at UT Brownsville (now at RIT)~\cite{Campanelli:2005dd}, with the first successful NS--NS merger calculations announced already in 1999~\cite{Shibata:1999wm}.  A list of the groups that have performed NS--NS merger calculations using full GR is presented below; note that many of these groups have also performed BH--NS simulations, as discussed in the review by Shibata and Taniguchi \cite{ST_LRR}.

Of the full GR codes used to evolve NS--NS binaries, almost all are grid-based  and make use of some form of adaptive mesh refinement. The one exception is the {\tt SpEC} code developed by the SXS collaboration,  formed originally by Caltech and Cornell, which has used a hybrid spectral-method field solver with grid-based hydrodynamics.   Most make use of the BSSN formalism for evolving Einstein's equations (see Sec.~\ref{sec:formalism} below), while the {\tt HAD} code uses the alternate Generalized Harmonic  Gauge (GHG) approach.  This technique is also used by the SXS collaboration and the Princeton group, who have both performed simulations of merging BH--NS binaries (see Sec.~\ref{sec:bhns}) but have yet to report any results on NS--NS mergers.  Three groups have reported results for NS--NS mergers including MHD (HAD, Whisky, and UIUC), while the KT (Kyoto/Tokyo) group has reported magnetized evolutions of HMNS remnants (see \cite{Shibata:2011fj} and references therein for a discussion of their work and that of other numerical relativity groups), but have yet to use that code for a NS--NS merger calculation.

While full GR codes were being developed to study NS--NS binaries, a parallel and rather independent track  developed to study detailed microphysical effects in binary mergers using approximate relativistic schemes.  This includes codes like that developed by the MPG  group that accurately track the production of neutrinos and antineutrinos and their annihilation during a merger, as well as post-processing routines  that use extensive nuclear chains to track the production of rare high-atomic number r-process elements in merger ejecta \cite{Goriely:2011vg}.  Meanwhile, the Bremen group's SPH code includes variable-temperature physically motivated equations of state \cite{Rosswog:2007ue} and magnetohydrodynamics \cite{Price:2006fi}, and has been used with a multi-group flux-limited diffusion neutrino code to generate expected neutrino signatures from merger calculations \cite{Dessart:2008zd}.  A summary of groups performing NS--NS merger calculations is presented in Table~\ref{table:collab}.

\begin{table}
\caption[A summary of groups  reporting  calculations of NS--NS mergers]{A summary of groups  reporting NS--NS merger calculation results.  The asterisk for the KT collaboration's MHD column indicates that they have used an MHD-based code for other projects, but not yet for NS--NS merger simulations.  Gravitational formalisms include full GR, assumed to be implemented using the BSSN decomposition except for the HAD collaborations's GHG approach, the CF approximation, or Newtonian gravity.  Microphysical treatments include physically motivated EOS models or quark-matter EOS and neutrino leakage schemes.}\label{table:collab}
\centering
\begin{tabular}{ll|ll|llll}
Abbrev. & Refs. & Grav. & MHD & Microphysics \\\hline
KT & \protect\cite{Hotokezaka:2011dh,Kiuchi:2009jt,Kiuchi:2010ze,Sekiguchi:2011zd,Sekiguchi:2011mc,Shibata:1999wm,Shibata:2002jb} & GR & $*$ & Phys. EOS, $\nu$-leakage\\
& \protect\cite{Shibata:2003ga,Shibata:2005ss,Shibata:2006nm,Yamamoto:2008js}  & & & \\
HAD & \protect\cite{Anderson:2007kz,Anderson:2008zp} & GR (GHG)  & Y & N\\
Whisky & \protect\cite{Baiotti:2008ra,Baiotti:2009gk,Baiotti:2010xh,Baiotti:2011am,Giacomazzo:2009mp,Giacomazzo:2010bx,Rezzolla:2010fd,Rezzolla:2011da} & GR & Y & N\\
UIUC & \protect\cite{Liu:2008xy} & GR & Y & N\\
Jena & \protect\cite{Thierfelder:2011yi} & GR & N & N\\ \hline
MPG & \protect\cite{Bauswein:2008gx,Bauswein:2009im,Bauswein:2010dn,Bauswein:2011tp,Goriely:2011vg,Oechslin:2006uk,Oechslin:2007gn,Stergioulas:2011gd} & CF & N & Quark, Phys. EOS, $\nu$-leakage\\
 Bremen & \protect\cite{Dessart:2008zd,Price:2006fi,Rosswog:2007ue} & Newt & Y & Phys. EOS , $\nu$-leakage\\
 \end{tabular}
\end{table}

\subsection{GR Numerical techniques}

\subsubsection{GR formalisms and gauge choice}\label{sec:formalism}

There are two distinct schemes used in all binary merger calculations performed to date, the BSSN (Baumgarte--Shapiro--Shibata--Nakamura) 
\cite{Shibata:1995we,Baumgarte:1998te} and Generalized Harmonic formalisms.   For general reviews of these formalisms, as well as other developments in numerical relativity, we refer readers to two recent books on numerical relativity\cite{Alcubierre:book,BaumgarteShapiro:book}. Here we only briefly summarize these two schemes.

The BSSN formalism was adapted from the 3+1 ADM  approach, with quantities defined as in Eqs.~\ref{eq:adm} and \ref{eq:confpsi}.
While the original ADM scheme proved to be numerically unstable,  defining the auxiliary quantities $\tilde\Gamma^i =-{\tilde{\gamma}^{ij}}_{,j}$ and treating these expressions as independent variables stabilized the system and allowed for long-term evolutions.  
While slight variants exist, the 19 evolved variables are typically either the conformal factor $\psi$ or its logarithm $\phi$, the conformal
3-metric $\tilde{\gamma}_{ij}$, the trace $K$ of the extrinsic curvature,
the trace free extrinsic curvature $A_{ij}$ and the conformal connection
functions $\tilde{\Gamma}^i$.   The evolution equations themselves are given in Appendix~\ref{app:field}.

The BSSN scheme was used in the binary merger calculations of the KT collaboration~\cite{Shibata:1999wm,Shibata:2002jb,Shibata:2003ga,Shibata:2005ss}, the first completely successful NS--NS calculations ever performed in full GR.  Ever since the UTB/RIT~\cite{Campanelli:2005dd} and Goddard~\cite{Baker:2005vv} groups showed simultaneously that a careful choice of gauge allows BHs to be evolved in BSSN schemes without the need to excise the singularity, these ``puncture gauges''  have gained widespread hold, and have been used to evolve  NS--NS binaries (and in some cases, BH--NS binaries) by the KT collaboration~\cite{Yamamoto:2008js}, UIUC~\cite{Liu:2008xy}, and Whisky~\cite{Baiotti:2008ra}. 

The  generalized harmonic formalism, developed over about two decades from initial theoretical suggestions up to its current numerical implementation \cite{Friedrich:1985aa,Garfinkle:2001ni,Pretorius:2004jg,Gundlach:2005eh,Pretorius:2005gq} was used to perform the first calculations of merging BH--BH binaries by Pretorius~\cite{Pretorius:2005gq}, and has since been applied to NS--NS binaries by the HAD collaboration~\cite{Anderson:2008zp} and to BH--NS mergers by HAD \cite{Chawla:2010sw}, SXS~\cite{Duez:2008rb,Duez:2009yy,Foucart:2010eq}, and the Princeton group \cite{Stephens:2011as,East:2011xa}.
It involves introducing a set of auxiliary quantities denoted $h^\mu$ representing the action of the wave operator on the spacetime coordinates themselves
\begin{equation}
H^{\mu}\equiv \Gamma^\mu\equiv  g^{\alpha\beta}\Gamma^{\mu}_{\alpha\beta} = -\frac{1}{\sqrt{-g}}\partial_\nu (\sqrt{-g}g^{\mu\nu}) = g^{\alpha\beta}\nabla_\alpha\nabla_\beta x^\mu = \Box~x^\mu\label{eq:GH_H}
\end{equation}
which are treated as independent gauge variables whose evolution equation must be specified.  Current first-order formulations \cite{Lindblom:2005qh,Anderson:2008zp} evolve equations for the spacetime metric $g_{\mu\nu}$ along with its spatial derivative, $\Phi_{i\mu\nu} = \partial_i g_{\mu\nu}$ and projected time derivative $\Pi_{\mu\nu} = -n^\alpha\partial_\alpha g_{\mu\nu}$, subject to consistency constraints on the spatial derivatives.
 The first BH--NS merger calculations in the GH formalism used a first-order reduction \cite{Lindblom:2005qh} of the Einstein equations and specified the source functions to damp to zero exponentially in time, while the first binary NS merger work \cite{Anderson:2008zp} used a similar first-order reduction and chose harmonic coordinates with $H^\mu=0$.

In both formalisms, most groups employ grid-based finite differencing to evaluate spatial derivatives. 
While finite differencing operators may be easily written down to arbitrary orders of accuracy, there is a trade-off between the computational efficiency achievable by using smaller, second-order stencils and the higher accuracy that can be attained using larger, higher-order ones.  For the moment, many groups are now moving to at least fourth-order accurate differencing techniques, with a high likelihood that at least the field sector of NS--NS merger calculations will soon be performed at comparable order to BH--BH calculations, at either sixth~\cite{Husa:2007hp} or eighth-order~\cite{Lousto:2007rj} accuracy, if not higher.
The main limitation to date involves the complexity of shock capturing using higher-order schemes, as we discuss in Section~\ref{sec:shock} below.

Imposing physically realistic and accurate boundary conditions remains a difficult task for numerical codes. Ideally, one wishes to impose boundary conditions at large distances that preserve the GR constraints and yield a well-posed initial-boundary value problem.  On physical grounds, the boundary should only permit outgoing waves, preventing the reflection of spurious waves back into the numerical grid.  Otherwise, reflections may be avoided by placing the outer boundaries so far away from the matter that they remain causally disconnected from the merging objects for the full duration of the simulation.  Building upon previous work (see, e.g.,~\cite{Friedrich:1998xt,Kreiss:2006mi,Babiuc:2006ik,Kreiss:2007zz,Ruiz:2007hg,Rinne:2008vn}, Winicour~\cite{Winicour:2009dr} derived boundary conditions that satisfy all of the desired conditions for the generalized harmonic formalism.
No such conditions have been derived for the full BSSN formalism, though progress has been made (see, e.g., \cite{Beyer:2004sv,Gundlach:2006tw}) so that we may now construct well-posed BCs in the weak-gravity linearized limit of BSSN~\cite{Nunez:2009wn} and for related first-order gravitational formalisms~\cite{Bona:2010wn}.

\subsubsection{Grid decompositions}

While unigrid schemes are extremely convenient, they tend to be extremely inefficient, since one must resolve small-scale features in the central regions of a merger but also extend grids out to the point where the GW signal has taken on its roughly asymptotic form  Thus, nearly every code incorporates some means of focusing resolution on the high-density regions via some form of mesh refinement.  A simple approach, for instance,  is to use ``fisheye'' coordinates that represent  a continuous radial deformation of a grid~\cite{Liu:2008xy}, a technique that had previously been used successfully, e.g., for BH--BH mergers~\cite{Baker:2001sf,Campanelli:2006gf}.

While fixed mesh refinement offers the chance for greater computational efficiency and accuracy, much current work focuses on \emph{adaptive} mesh refinement, in which the level of refinement of the grid is allowed to evolve dynamically to react to the evolving binary configuration.  Several codes, both public and private, now implement this functionality.  The publicly available {\tt Carpet} computational toolkit~\cite{Schnetter:2003rb,carpet_web}, which is distributed to the community as part of the open source {\tt Einstein~Toolkit} \cite{einsteintoolkit_web,Loffler:2011ay} uses a ``moving boxes'' approach, and has been designed to be compatible with the widely used and publicly available {\tt Cactus} framework.  It has been successfully implemented by the Whisky group~\cite{Baiotti:2008ra} to perform NS--NS mergers, by UIUC for their BH--NS mergers \cite{Etienne:2008re}, and a host of other groups for BH--BH mergers and additional problems.  The KT code {\tt SACRA} also implements an adaptive mesh refinement (AMR) scheme for NS--NS and BH--NS mergers~\cite{Yamamoto:2008js}, as does the most recent version of the HAD collaboration's code~\cite{Anderson:2008zp}, which is based on the publicly available infrastructure of the same name~\cite{had_web,Anderson:2006ay}, and the {\tt BAM} code employed by the Jena group \cite{Thierfelder:2011yi,Bernuzzi:2011aq,Gold:2011df}.  The Princeton group also has an AMR code, which has been used to perform BH--NS mergers to date \cite{Stephens:2011as,East:2011xa,East:2011aa}

One drawback of employing rectangular grids is that memory costs scale like $N^3$, where $N$ is the number of grid cells across a side, and total computational effort like $N^4$ once Courant conditions are figured in.  Since one does not necessarily need high angular resolution at large radii, there is great current interest in developing schemes that use some form of spheroidal grid, for which the memory scaling is merely $\propto N$.  A group at LSU has implemented a multi-patch method~\cite{Zink:2007xn}, in which space is broken up into a number of non-overlapping domains in such a way that the six outermost regions (projections of the faces of a cube onto spheres of constant radius), maintain constant angular resolution and thus produce linear dependence of the total number of grid points on the number of radial points.  To date, it has been used primarily for vacuum spacetimes~\cite{Pazos:2009vb} and hydrodynamics on a fixed background.  The SXS collaboration, begun at Caltech and Cornell and now including members at CITA and Washington State, has used a spectral evolution code with multiple domains to evolve BH--NS binaries, which achieves the same scaling by expanding the metric fields in modes rather than in position space~\cite{Duez:2008rb}.  Their first published results on NS--NS binaries are currently in preparation (see~\cite{Kaplan10} for a brief summary of work to date).

While all of these grid techniques produce tremendous advantages in computational efficiency, each required careful study since deformations of a grid or the introduction of multiple domains can introduce inaccuracies and non-conservative effects.  As an example, in AMR schemes, one must deal with the same reflection issues at refinement boundaries that are present at the physical boundaries of the grid, as discussed above, though the interior nature of the boundaries allows for additional techniques, such as tapered grid boundaries \cite{Lehner:2005vc}, to be used to minimize reflections there.
The study of how to minimize spurious effects in these schemes continues, and will represent an important topic for years to come, especially as evolution schemes become more complicated by including more physical effects.  

\subsubsection{Hydrodynamics, MHD, and high-resolution shock capturing}
\label{sec:shock}

Fluids cannot be treated in the same way as the spacetime metric because finite differencing operators do not return meaningful results when placed across  discontinuities induced by shocks.  Instead, GR(M)HD calculations must include some form of shock-capturing that accounts for these jumps.  These are typically implemented in conservative schemes, in which the fluid variables are transformed from the standard ``primitive'' set $\vec{P}$, which includes the fluid density, pressure, and velocity (and magnetic field in MHD evolutions), into a new set $\vec{U}$ so that the evolution equations may be written  in the form
\begin{equation}
\partial_t(\vec{U})=\nabla\cdot\vec{F}+\vec{S},
\end{equation}
where the \emph{flux functions} $\vec{F}(\vec{P})$ and \emph{source terms} $\vec{S}(\vec{P})$ can be expressed in terms of the primitive variables but not their derivatives.  These schemes allow one to evolve the resulting MHD set of equations so that numerical fluxes are conserved to numerical precision across cell walls as the fluid evolves in time.   One widely used scheme, often referred to as the Valencia formulation \cite{Marti:1991wi}, is described in Appendix~\ref{app:hydro}..

There are important mathematical reasons for casting the GRHD/GRMHD system in conservative form, primarily since the mathematical techniques describing Godunov methods may be called into play \cite{Godunov:1959}.  In such methods, we assume that the evolution of the quantities $\vec{U}$ may be expressed in the form
\begin{equation}
\vec{U}(x=x_i,t=t_{n+1}) = \vec{U}(x=x_i,t=t_n) + \frac{\Delta t}{\Delta x}\left(\vec{F}(x=x_{i-1/2})- \vec{F}(x=x_{i+1/2})\right)
\end{equation}
where the points have spatial coordinates $x_i\equiv x_0+i\Delta x$ and the timesteps satisfy $t_n = t_0+n\Delta t$.  The fluxes must be determined by solving the Riemann problem at each cell face (thus the half-integer indices), either exactly or approximately.   It can be shown that solutions constructed this way, when convergent, must converge to a solution of the original problem, even when shocks are present \cite{LaxWendroff:1960}.

First one \emph{reconstructs} the primitives from the conserved variables on both sides of an interface, using interpolation schemes designed to respect the presence of shocks.  Simple schemes involving limiters yield second-order accuracy in general but first-order accuracy at shocks, while higher-order methods such as PPM (piecewise parabolic method) and essentially non-oscillatory (ENO) schemes such as CENO (central ENO) and WENO (weighted ENO) yield higher accuracy but at much higher computational cost.  Once primitives are interpolated to the cell interfaces, flux terms are evaluated there and one solves the Riemann problem describing the evolution of two distinct fluid configurations placed on either side of a membrane (see~\cite{Font:2007zz} for a description).  While complete solutions of the Riemann problem are painstaking to evolve, a number of approximation techniques exist and do not reduce the order of accuracy of the code.  Finally, one must take the conservative solution, now advanced forward in time, and recover the underlying primitive variables, a process that requires solving as many as eight simultaneous equations in the case of GRMHD or five for GRHD systems.  A number of methods to do this have been widely studied~\cite{Noble:2005gf}, and simplifying techniques are known for specific cases (for the case of polytropic EOSs in GRHD evolutions, one need only invert a single non-analytic expression and the remaining variables can then be derived analytically).

The inclusion of magnetic fields in hydrodynamic calculations adds another layer of complexity beyond shock capturing.  Magnetic fields must be evolved in such a way that they remain divergence-free, much in the same way that relativistic evolutions must satisfy the Hamiltonian and momentum constraints.  Brute force attempts to subtract away any spurious divergence often lead to instabilities, so more intricate schemes have been developed.  ``Divergence cleaning'' schemes typically introduce a new field representing the magnetic field divergence and use parabolic/hyperbolic equations to damp the divergence away while moving it off the computational domain; the approach is relatively simple to implement but prone to small-scale numerical errors~\cite{Balsara:2003ui}.
``Constrained transport'' schemes stagger the grids on which different physical terms are calculated to enforce the constraints (see, e.g.,~\cite{Toth:2000JCoPh.161..605T} for a particular implementation), and have been applied widely to many different physical configurations.  Recently, a new scheme in which the vector potential is used rather than the magnetic field was introduced by Etienne, Liu, and Shapiro~\cite{Etienne:2010ui,Etienne:2011re}, and found to yield successful results for a variety of physical configurations including NSs and BHs.

\subsection{Microphysical numerical techniques}

\subsubsection{Neutron star physics and equations of state}

One of the largest uncertainties in the input physics of NS--NS merger simulations is the true behavior of the nuclear matter EOS.  To date, EM observations have yielded relatively weak constraints on the NS mass-radius relation, with the most precise simultaneous measurement of both as of now resulting from observations of Type 1 X-ray bursts from accreting NSs in three different sources~\cite{Ozel:2010fw}.  In each case, the NS mass was found to lie in the  range $1.3\,M_{\odot}\lesssim M_{\mathrm{NS}}\lesssim 2\,M_{\odot}$ and the radius $8\mathrm{\ km}\lesssim R_{\mathrm{NS}}\lesssim 12\mathrm{\ km}$, implying a NS compactness 
\begin{equation}
\mathcal{C}\equiv\frac{GM_{\mathrm{NS}}}{R_{\mathrm{NS}}c^2} = 0.1476\left(\frac{M_{\mathrm{NS}}}{\,M_{\odot}}\right)\left(\frac{R_{\mathrm{NS}}}{10\mathrm{\ km}}\right)^{-1}\approx 0.16-0.37.
\end{equation}
A more stringent constraint on the NS EOS is provided by observations of the Shapiro time delay in the binary millisecond pulsar PSR~J1614--2230, which was found to have a mass $M_{\mathrm{NS}}=1.97\pm 0.04\,M_{\odot}$~\cite{Demorest:2010bx}, which would rule out extremely soft EOS models incapable of supporting such a massive NS against collapse.  As we discuss in more detail below, GW observations are likely to eventually yield tighter constraints than our current EM-based ones, though BH--NS mergers, which can undergo stronger tidal disruptions than NS--NS mergers at frequencies closer to LIGO and other GW observatories' maximum frequency sensitivity band, may prove to be more useful for the task than NS--NS mergers.

Given the large theoretical uncertainties in describing the proper
physical NS EOS, many groups have chosen the simplest possible
parameterization: a polytrope (see Eq.~\ref{eq:polytrope}).  Under this choice, the enthalpy $h$ takes the particularly simple form
\begin{equation}
h\equiv 1+\varepsilon+P/\rho = 1+\Gamma\varepsilon.\label{eq:idealfluid}
\end{equation}
Initial data are generally assumed to follow the relation 
\begin{equation}P=K\rho^\Gamma,\label{eq:polyfluid}
\end{equation}
where $K$ is constant across the fluid.  In the presence of shocks, the value of $K$ for a particular fluid element will increase with time.  We note that the Whisky group~\cite{Baiotti:2008ra,Baiotti:2009gk} uses the term ``polytropic'' to refer to simulations in which Eq.~\ref{eq:polyfluid} is enforced throughout, which implies adiabatic evolution without shock heating, and use the term ``ideal fluid'' to describe an EOS that includes the effects of shock heating and enforces Eq.~\ref{eq:idealfluid}.  

Since the temperatures of NSs typically yield thermal energies per baryon substantially below the Fermi energy, one may treat nearly all NSs as effectively cold, except for the most recently born ones.  During the merger process for NS--NS binaries, the matter will remain cold until the two NSs are tidally disrupted and a disk forms, at which point the thermal energy input and substantially reduced fluid densities require a temperature evolution model to properly model the underlying physics.  In light of these results, some groups adopt a two-phase model for the NS EOS (see, e.g.,~\cite{Shibata:2005ss}), where a cold, zero-temperature EOS, evaluated as a function of the density only, encodes as much information about as we possess about the NS EOS, and the hot phase depends on both the density and internal energy, typically in a polytropic way
\begin{equation}
P(\rho,\varepsilon) = P_{\mathrm{cold}}(\rho)+P_{\mathrm{hot}}(\rho,\varepsilon)~~~~[P_{\rm hot}(\rho,\varepsilon) = (\Gamma-1)\rho(\varepsilon-\varepsilon_{\mathrm{cold}})].
\label{eq:2phaseEOS}
\end{equation}

There are a number of physically motivated EOS models that have been implemented for merger simulations, whose exact properties vary depending on the assumptions of the underlying model. These include models for which the pressure is tabulated as a function of the density only: FPS~\cite{Pandharipande:1989nmhi.conf..103P}, SLy~\cite{Douchin:2001sv}, and APR~\cite{Akmal:1998cf}; as well as models including a temperature dependence: Shen~\cite{Shen:1998by,Shen:1998gq} and Lattimer--Swesty~\cite{Lattimer:1991nc}.  A variety of models have been used  to study the effects of quarks, kaons, and other condensates, which typically serve to soften the EOS, leading to reduced maximum masses and more compact NSs~\cite{Pandharipande:1975NuPhA.237..507P,Glendenning:1984jr,Prakash:1995uw,Glendenning:1997ak,Balberg:1997yw,Alford:2004pf}.  

Given the variance among even the physically motivated EOS models, it has proven useful to parameterize known EOS models  with a much more restricted set of parameters.  In a series of works, a Milwaukee/Tokyo collaboration determined that essentially all current EOS models could be fit using four parameters, so that their imprint on GW signal properties could be easily analyzed ~\cite{Read:2008iy,Read:2009yp,Markakis:2011vd}.  Their method assumes that the SLy EOS describes NS matter at low densities, and that the EOS at higher densities can be described by a piecewise polytropic fit with breaks at $\rho=10^{14.7}$ and $10^{15}\mathrm{\ g/cm^3}$.  The four resulting parameters are $P_1=P\left(\rho=10^{14.7}\right)$, the pressure at the first breakpoint density, which normalizes the overall density scale, as well as $\Gamma_1, \Gamma_2, \Gamma_3$, the adiabatic exponents in the three regions.  Their results indicate that advanced LIGO should be able to determine the NS radius to approximately 1~km at an effective distance of 100~Mpc, which would place tight constraints on the value of $P_1$ in particular.

\subsection{Electromagnetic and neutrino signature modeling}

Motivated by the evidence that SGRBs frequently appear in galaxies with very low star formation rates~\cite{Berger:2005rv,Fox:2005kv}, astronomers have suggested that their progenitors are likely to be mergers of either NS--NS and/or BH--NS binaries.  While soft-gamma repeaters (SGRs) have been confirmed as an SGRB source from observations of the system SGR~1806--20, they make up no more than approximately 15\% of the total observed SGRB fraction according to the leading population estimates~\cite{Lazzati:2005vn,Nakar:2005hs}. There has been much interest in predicting the EM signatures of NS--NS and BH--NS mergers, along with the associated neutrino emission.  The simplest models estimate a local radiation cooling rate for the matter but do not attempt to follow the paths of the photons and/or neutrinos after they are emitted, instead calculating the time-dependent luminosity assuming free streaming.  Such models have been used in non-GR simulations of binary mergers going back more than a decade~\cite{Ruffert:1995fs,Rosswog:2003rv}, and recently such schemes have been used to perform full GR NS--NS mergers \cite{Sekiguchi:2011zd}, including a self-consistent evolution of the electron fraction of the material $Y_e$, rather than a passive advection approach.

More complicated flux-limited diffusion schemes, in which the neutrino fluxes for given species and energies are given by explicit formulae that limit to the correct values for zero optical depth (free-streaming) and very large optical depth (diffusion), have been used as a post-processing tool to investigate the merger remnants in Newtonian NS--NS mergers~\cite{Dessart:2008zd}, but have yet to be applied to full GR simulations.  Finally,
radiation transport schemes to evolve EM and neutrino fluxes passing through fluid configurations have been implemented in numerical GR codes~\cite{DeVilliers:2008ut,Farris:2008fe}, but have yet to be used in binary merger simulations.

\subsection{GW signal modeling}
\label{sec:gw}

Measuring the GW signal from a dynamical merger calculation is a rather difficult task.  One must determine, using a method unaffected by gauge effects, the perturbations at asymptotically large distances from a source by extrapolating various quantities measured at large but finite distances from the merger itself.  

In the early days of numerical merger simulations, most groups typically assumed Newtonian and/or quasi-Newtonian gravitation, for which there is no well-defined dynamical spacetime metric.  GW signals were typically calculated using the quadrupole formalism, which technically only applies for slow-moving, non-relativistic sources (see~\cite{Misner:1973grav.book} for a thorough review of the theory).  Temporarily reintroducing physical constants, the strains of the two polarizations for signals emitted in the $z$-direction are
\begin{eqnarray*}
h_{+}&=&\frac{G}{rc^4}(\ddot{\mbox{\sout{I}}}_{xx}-\ddot{\mbox{\sout{I}}}_{yy}),\\
h_\times&=&\frac{2G}{rc^4}\ddot{\mbox{\sout{I}}}_{xy},
\end{eqnarray*}
where $r$ is the distance from the source to the observer and $\ddot{\mbox{\sout{I}}}_{ij}$ it the traceless quadrupole moment of the system.  
The GW luminosity and angular momentum loss rate of the system are given, respectively, by
\begin{eqnarray*}
\left(\frac{dE}{dt}\right)_{\mathrm{GW}} &=& \frac{G}{5c^5}\dddot{\mbox{\sout{I}}}_{ij}\dddot{\mbox{\sout{I}}}_{ij},\\
\left(\frac{dL_k}{dt}\right)_{\mathrm{GW}} &=& \frac{2G}{5c^5}\epsilon_{ijk}\ddot{\mbox{\sout{I}}}_{il}\dddot{\mbox{\sout{I}}}_{lj}.
\end{eqnarray*}
While only approximate, the quadrupole formulae do yield equations that are extremely straightforward to implement in both grid and particle-based codes using standard integration techniques.

Quadrupole methods were adopted for later PN and CF simulations, again because the metric was assumed either to be static or  artificially constrained in such a way that made self-consistent determination of the GW signal impossible.  
One important development from this period was the introduction of a simple method to calculate the GW energy spectrum $dE/df$ from the GW time-series through Fourier transforming into the frequency domain~\cite{Xing:1994ak}.  GW signals analyzed in the frequency domain allowed for direct comparison with the LIGO noise curve, making it much easier to determine approximate distances at which various GW sources would be detectable and the potential signal-to-noise ratio that would result from a template search.  To constrain the nuclear matter EOS, one can examine where a GW merger spectrum deviates in a measurable way from the quadrupole point-mass form,
\begin{equation}
\left(\frac{dE}{df}\right)_{\mathrm{GW}} = \frac{\pi G m_1m_2}{3}\left(\pi G(m_1+m_2)f_{\mathrm{GW}}\right)^{-1/3};~~~f_{\mathrm{GW}}\equiv 2f_{\mathrm{orb}}=\frac{\omega_{\mathrm{orb}}}{\pi},
\end{equation}
because of finite-size effects, and then link the deviation to the properties of the NS~\cite{Faber:2002zn}, as we show in Figure~\ref{fig:GW_dedf}.

\epubtkImage{}{%
\begin{figure}[!ht]
\centerline{\includegraphics[width=\textwidth]{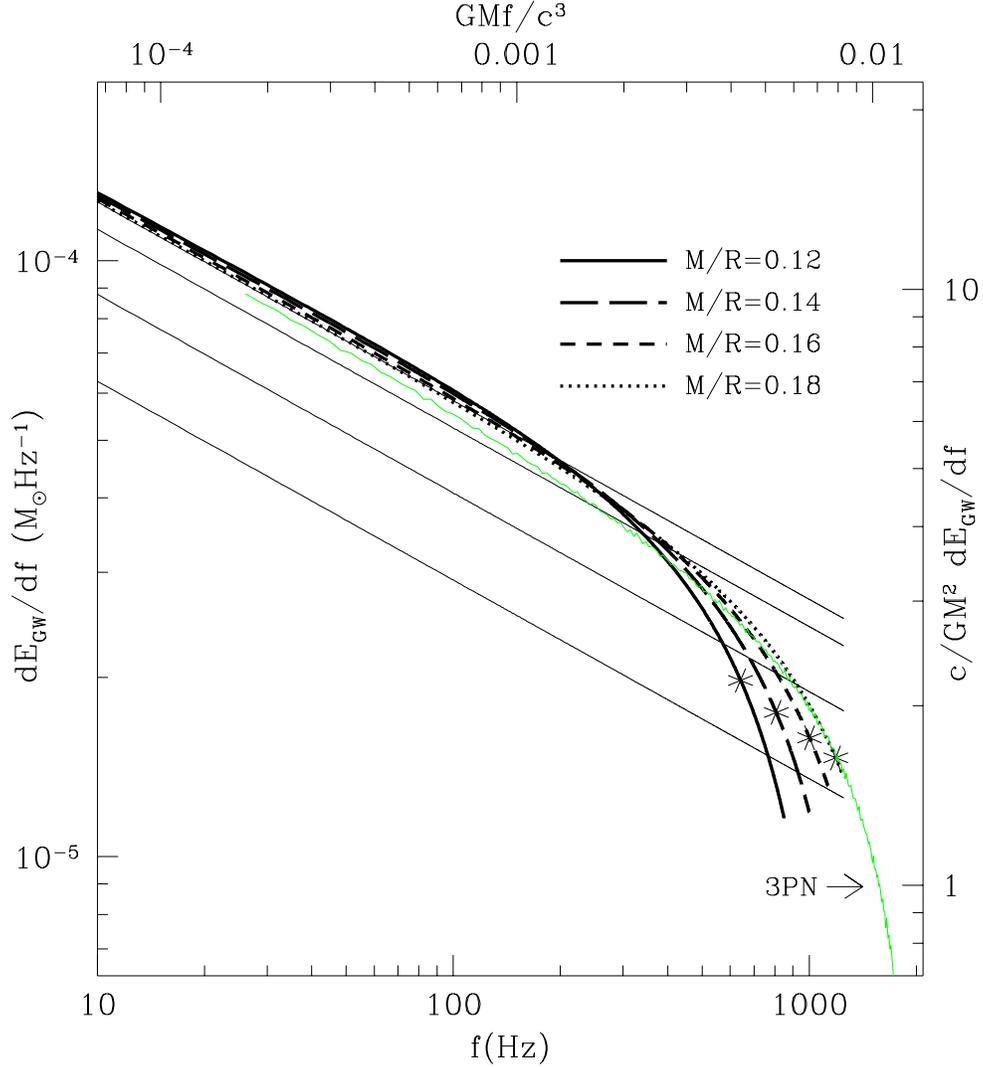}}
\caption{Approximate energy spectrum $dE_{\mathrm{GW}}/df$ derived from QE sequences of equal-mass NS--NS binaries with isolated ADM masses $M_{\mathrm{NS}}=1.35\,M_{\odot}$ and a $\Gamma=2$ EOS, but varying compactnesses (denoted $M/R$ here), originally described in~\cite{Taniguchi:2002ns}.  The diagonal lines show the energy spectrum corresponding to a point-mass binary, as well as values with 90\%, 75\%, and 50\% of the power at a given frequency (see Figure~2 of~\cite{Faber:2002zn}).  Asterisks indicate the onset of mass-shedding, beyond which QE results are no longer valid.}
\label{fig:GW_dedf}
\end{figure}}

Full GR dynamical calculations, in which the metric is evolved according to the Einstein equations, generally use one  of two approaches to calculate the GW signal from the merger, if not both.  The first method, developed first by by Regge and Wheeler~\cite{Regge:1957td} and  Zerilli~\cite{Zerilli:1971wd} and written down in a gauge-invariant way by Moncrief~\cite{Moncrief:1974am} involves analyzing perturbations of the metric away from a Schwarzschild background.  The second uses the Newman--Penrose formalism~\cite{Newman:1961qr} to calculate the Weyl scalar $\psi_4$, a contraction of the Weyl curvature tensor, to represent the outgoing wave content on a specially constructed null tetrad that may be calculated approximately~\cite{Campanelli:1998jv}.  The two methods are complementary since they incorporate different metric information and require different numerical integrations to produce a GW time series.  Regardless of the method used to calculate the GW signal, results are often presented by calculating the dominant $s=-2$ spin-weighted spherical harmonic mode.  For circular binaries, the $l=2$, $m=2$ mode generally carries the most energy, followed by other harmonics; in cases where the components of the binary have nearly equal masses and the orbit is circular, the falloff is typically quite rapid, while extreme mass ratios can pump a significant amount of the total energy into other harmonics.  For elliptical orbits, other modes can dominate the signal, e.g., a $3:1$ ratio in power for the $l=2,~m=0$ mode to the $l=2$, $m=2$ mode observed for high-ellipticity close orbits in~\cite{Gold:2011df}.   A thorough summary of both methods and their implementation may be found in~\cite{Ruiz:2007yx}.  

\newpage
\section{Dynamical Calculations}
\label{sec:dynsim}

NS--NS merger simulations address a broad set of questions, which can be roughly summarized as follows (note the same questions apply to BH--NS mergers as well):
\begin{enumerate}
\item What is the final fate of the system, assuming a given set of initial parameters?  Do we get a prompt collapse to a BH or the formation of a HMNS supported against collapse by differential rotation? Other outcomes are disfavored, at least for pre-merger NSs with masses $M_{\mathrm{NS}}\gtrsim 1.4\,M_{\odot}$ since the supramassive limit is at most 20\% larger than that of a non-rotating NS~\cite{Cook92,Cook:1993qr}, and even for the stiffest EOS  these values are typically less than $2.8\,M_{\odot}$.
\item  What is the GW signal from the merger, and how does it inform us about the initial pre-merger parameters of the system?  
\item What fraction of the system mass is left in a disk around the central BH or HMNS?  While deriving exact EM emission profiles from a hydrodynamical configuration remains a challenge for the future, minimum conditions that would allow for the energy release observed in SGRBs have been established based on scaling arguments.
\item What is the neutrino and EM emission from the system, in both the time and energy domains?  Obviously, the answer to this question and those that follow depend critically on the answers above.
\item What role do B-fields play in the GW, EM, and neutrino emission, and how does that tie in with other models suspected of having the same disk/jet geometry and gamma-ray emission like active galactic nuclei, pre-main sequence stars, etc.?
\item Do mergers produce a cosmologically significant quantity of r-process elements, or do those likely get produced by other astrophysical events instead?
\end{enumerate}

The influence of the gravitational formalism used in a numerical simulation on the answer one finds for the questions above differs item by item.  Determining the final fate of a merging system is highly dependent on the gravitational formalism; NS--NS merger remnants only undergo collapse in quasi-relativistic and fully GR schemes.  Moreover, orbital dynamics at separations comparable to the ISCO and even somewhat larger depend strongly on the gravitational scheme.  In particular, mass loss rates  into a disk are often suppressed by orders of magnitude in GR calculations when compared to CF simulations, and even more so in comparison to PN and Newtonian calculations.  
EM emission profiles from a disk 
are difficult to calculate accurately without the use of full GR for this reason.  
On the other hand, while GR is required to calculate the exact GW signal from a merger, even early Newtonian simulations predicted many of the qualitative GW emission features correctly, and PN and CF schemes yielded results with some degree of quantitative accuracy about the full wavetrain.  B-fields have only begun to be explored, but it already seems clear that they will affect the hydrodynamical evolution primarily after the merger in cases where differential rotation in a HMNS or disk winds up magnetic field strengths up to energy equipartition levels, vastly stronger than those found in pre-merger NSs.  For such configurations, non-relativistic calculations can often reproduce the basic physical scenario but full GR is required to properly understand the underlying dynamics.  Finally, the production of r-process elements, which depends sensitively on the thermodynamic evolution of the merger, seems to generally disfavor binary mergers as a significant source of the observed stellar abundances since the temperature and thus the electron fraction of the fluid remains too small~\cite{Ruffert:2001gf}, regardless of the nature of the gravitational treatment used in the calculations.  This picture may need to be revised if significant mass loss occurs from the hot accretion disk that forms around the central post-merger object, but numerical calculations do not currently predict sufficient mass loss to match observations~\cite{Surman:2008qf}.  We will address each of these topics in greater detail in the sections below.

Since the first NS--NS merger calculations, there have been two main directions for improvements: more accurate relativistic gravitation, resulting in the current codes that operate using a self-consistent fully GR approach, and the addition of microphysical effects, which now include treatments of magnetic fields and neutrino/EM radiation.  Noting that several of the following developments overlapped in time, e.g., the first full GR simulations by Shibata and Ury\=u~\cite{Shibata:1999wm} are coincident with the first PN SPH calculations, and predate the first CF SPH calculations, we consider in turn the original Newtonian calculations, those performed using approximate relativistic schemes, the calculations performed using full GR, and finally those that have included more advanced microphysical treatments.  

\subsection{Quasi-equilibrium and semi-analytic methods vs fully dynamical results}

Before reviewing fully dynamical calculations of NS--NS mergers, it is worthwhile to ask how much information can already be deduced from QE calculations, which may be performed at much smaller computational cost, as well as from semi-analytic PN treatments and related approximate techniques.  Clearly, the details of the merger and ringdown phases fall outside the QE regime, so only dynamical calculations can yield reliable information about the stability of remnants, properties of ejecta, or other processes that arise during the merger itself or in its aftermath.  Thus, the primary point of comparison is the  GW signal just prior to merger, which is also easier to detect (for first and second generation interferometers).

The strength of QE calculations lies in their ability to model self-consistently finite-size effects not captured in PN treatments (which always assume two orbiting point masses).  The increased tidal interaction between the objects typically results in a more rapid phase advance of the binary orbit, which is important for constructing template waveforms that cover the entire NS--NS inspiral, merger, and ringdown.  While QE sequences potentially offer a wealth of information about well-separated binaries and can help fix the phase evolution of the inspiraling binary, they do have two weaknesses arising as the binary approaches the stability limit.  First, most QE methods, including the CTS formalism described in Sec.~\ref{sec:CTS}, are time-symmetric, and assume that the NS possess a symmetry plane perpendicular to the direction of motion (i.e., a front-back symmetry whose axis is perpendicular to the orbital angular momentum and the binary separation vector).  In reality, \emph{tidal lags} develop prior to final plunge, with the innermost edge of each NS rotating forward and the outer edge backwards.  This effect has been captured in analytic and semi-analytic approaches (see, e.g., \cite{Lai:1994hf} for an early example), and is clearly seen in dynamical calculations (see Fig.~\ref{fig:sim_1}), but is not captured in CTS-based schemes (tidal lags also develop in BH--NS merger calculations when the BH has a non-zero spin, since this breaks the front-back symmetry; see \cite{Taniguchi:2005fr} for an example).  

A second weakness of QE methods is the treatment of the ISCO, particularly its importance as a characteristic point along an evolutionary sequence that, in theory, could encode information about the NS EOS.  Simple estimates of the infall trajectory derived solely from QE sequences predict a very sudden and rapid infall near the ISCO, i.e., the point where the binding energy reaches a minimum along the sequence (see, e.g., the argument in \cite{Faber:2002zn}). However, this is clearly an oversimplification.  In reality, binaries transition more gradually to the merger phase, and the inward plunge may occur significantly before reaching the formal ISCO; this in turns leads to more rapidly growing deviations from the QE approximation.  Looking at the GW energy spectrum, one typically sees minor deviations from the point-mass predictions at frequencies below those characterizing the ISCO, but substantially more power at frequencies above it.  Equivalently, the cutoff frequency for GW emission $f_{\rm cut}$, where the spectrum starts deviating strongly from the point-mass prediction, is usually higher than the QE frequency near the ISCO, $f_{\rm ISCO}$, while simple QE estimates assume these two frequencies to coincide.

To date, most attempts to generate waveforms extended back to arbitrarily early starting points involve numerically matching PN signals, typically generated using the Taylor T4  approach \cite{Boyle:2007ft}, onto the early stages of numerically generated waveforms, with some form of maximum overlap method used to provide the most physical transition from one to the other.  These approaches may be improved by adding tidal effects to the evolution, typically parameterized by the tidal Love numbers that describe how tidal gravity fields induce quadrupole deformations \cite{Flanagan:2007ix}.  Tidal effects can be placed into a relativistic framework \cite{Binnington:2009bb,Damour:2009vw}, which may be included within the effective one-body (EOB) formalism to produce high-accuracy waveforms \cite{Damour:2009wj}.  In the EOB approach \cite{Buonanno:1998gg}, resummation methods are used to include higher-order PN effects, though some otherwise unfixed parameters need to be set by comparing to numerical simulations.

Work is in its early stages to compare {\em directly\/} the GW spectra inferred from QE sequences of NS--NS binaries  with those generated in numerical relativity simulations, but this comparison has been discussed at some length with regard to BH--NS mergers.  Noting that NS--NS mergers generally correspond more closely to the BH--NS cases in which an ISCO is reached prior to the onset of tidal disruption,  the KT collaboration~ \cite{Shibata:2007zm,Shibata:2009cn} concluded that the cutoff frequency marking significant deviations from PN point-mass behavior is roughly $30\%$ higher than that marking emission near the classical ISCO for BH--NS systems ($f_{\rm cut}\simeq 1.3\, f_{\rm ISCO}$).  

A more detailed study has now been performed comparing EOB methods to numerical evolutions.  By comparing to long-term simulations of NS--NS mergers, Baiotti et al. \cite{Baiotti:2011am} find that EOB models may be tuned, via careful choices of their unfixed parameters, to reproduce the GW phases and amplitudes seen in NR evolutions up until the onset of merger.  They further suggest that the EOB approach seems to cover a wider range of phase space than the Taylor T4 approach, presumable because of a more consistent representation of tidal effects, and offers the best route forward for construction of more accurate NS--NS inspiral templates.

\subsection{Early dynamical calculations}

The earliest NS--NS merger calculations were performed in Newtonian gravity, sometimes with the addition of lowest-order 2.5PN radiation reaction forces, and typically assumed that the NS EOS was polytropic.  Both Eulerian grid codes~\cite{Oohara:1989cb,Nakamura:1989jk,Oohara:1990dp,Nakamura:1991rt,Shibata:1992py,Ruffert:1996qu,New:1997xi,Swesty:1999ke} and Lagrangian SPH codes~\cite{Rasio:1992ApJ...401..226R,Rasio:1994bd,Rasio:1994uw,Davies:1993zn,Xing:1994ak,Xing:1996sr}   were employed,  and GW signals were derived under the assumptions of the quadrupole formalism.
Configurations in Newtonian gravity cannot collapse, so a stable (possibly hypermassive) remnant was always formed.  For polytropic EOS models with adiabatic indices larger than the classical minimum for production of a Jacobi ellipsoid, $\Gamma\gtrsim 2.6$, remnants were typically triaxial and maintained a significant-amplitude GW signal until the end of the simulation.  For simulations using smaller values of $\Gamma$, remnants rapidly relaxed to spheroidal configurations, quickly damping away the resulting GW signal.  Mass loss from the central remnant was often quite significant, with thick accretion disks or completely unbound material comprising up to 10\,--\,20\% of the total system mass

Mass loss was suppressed in numerical simulations by constructing irrotational, rather than synchronized, initial data. Irrotational flow is widely thought to be the more physically realistic case, since viscous forces are much too weak to synchronize a NS prior to merger~\cite{Bildsten:1992my,Kochanek:1992wk}.  When irrotational NSs (which are  counter-rotating in the corotating frame of the binary) first make contact, a vortex sheet forms.  Since the low-density fluid layers at the contact surface are surrounded at first contact by the denser fluid layers located originally within each NS, the configuration is well understood to be Kelvin--Helmholtz unstable, resulting in rapid mixing through vortex production.  Meanwhile, mass loss through the outer Lagrange points is hampered by the reduced rotational velocity along the outer halves of each NS.  

The GW emission from these mergers is composed of a ``chirp,'' increasing in frequency and amplitude as the NSs spiral inward, followed by a ringdown signal once the stars collide and merge.  In~\cite{Xing:1994ak,Xing:1996sr}, a procedure to calculate the energy spectrum in the frequency band was laid out, with the resulting signal following the quadrupole, point-mass power-law form
 up to GW frequencies characterizing the beginning of the plunge.  Above the plunge frequency,  a sharp drop in the GW energy was seen, followed in some cases by spikes at kHz frequencies representing coherent emission during the ringdown phase.  

\subsection{Approximate relativistic schemes}
\label{sec:approxrel}

The first steps toward approximating the effects of GR included the use of 1PN dynamics or the CF approximation.  Using a formalism derived by Blanchet, Damour, and Schaefer~\cite{Blanchet:1989fg}, the 1PN equations of motion require the solution of eight Poisson-like equations in the form
\begin{equation}
\nabla^2\psi = f(\vec{x}),
\end{equation}
where the source terms $f(\vec{x})$ are compactly supported, and thus
the fields $\psi$ may be determined using the same techniques already
in place to find the  Newtonian potential.  Adding in
the lowest-order dissipative radiation reaction effects requires
solution of a ninth Poisson equation for a reaction potential. The 1PN
formalism was implemented in both grid-based~\cite{Oohara:1992dm} and SPH
codes~\cite{Ayal:1999wn,Faber:1999gj,Faber:2000uf,Faber:2002cg}.
Unfortunately, physically realistic NSs are difficult to model using a
PN expansion, since the characteristic NS compactness ${\mathcal
  C}\gtrsim 0.15$, leads to first order ``corrections'' that often
rival Newtonian terms in magnitude.  To deal with this problem Ayal et
al.~\cite{Ayal:1999wn} considered large ($R\approx 30\mathrm{\ km}$),
low-mass ($<1\,M_{\odot}$) NSs, allowing them to study relativistic
effects but  making results more difficult to interpret for physically
realistic mergers.  In~\cite{Faber:1999gj,Faber:2000uf,Faber:2002cg},
a dual speed of light approach was used, in which all 1PN effects were
scaled down by a constant factor to yield smaller quantities while
Newtonian and radiation reaction terms were included at full-strength.
Both SPH groups found that the GW signal in PN mergers is strongly
modulated, whereas Newtonian merger calculations typically yielded smooth, 
either monotonically decreasing or nearly constant-amplitude ringdown signals. 
Even reduced 1PN effects were shown to suppress
mass loss by a factor of 2\,--\,5 for initially synchronized cases,
and disk formation was seen to be virtually non-existent for initially
irrotational, equal-mass NSs with a stiff ($\Gamma=3$ polytropic)
EOS~\cite{Faber:2000uf}. 

Moving beyond the linearized regime, several groups explored the CF approximation, which incorporates many of the nonlinear effects of GR into an \emph{elliptic}, rather than hyperbolic, evolution scheme.  While non-linear elliptic solvers are expensive computationally, they typically yield stable evolution schemes since field solutions are always calculated instantaneously from the given matter configuration.  Summarized quickly, the CF approach involves solving the CTS field equations, Eqs.~\ref{eq:ellippsi}, \ref{eq:ellipapsi}, and \ref{eq:ellipshift}, at every timestep, and evolving the matter configuration forward in time.  The metric fields act like potentials, with various gradients appearing in the Euler and energy equations.  While the CTS formalism remains the most widely used method to construct NS--NS (and BH--NS) initial data, it does not provide a completely consistent dynamical solution to the GR field equations.  In particular, while it reproduces spherically symmetric configurations like the Schwarzschild solution exactly, it cannot describe more complicated configurations, including Kerr BHs. Moreover, because the CF approximation is time-symmetric, it also does not allow one to consistently predict the GW signal from a merging configuration.  As a result, most dynamical calculations are performed by adding the lowest-order dissipative radiation reaction terms, either in the quadrupole limit or via the radiation reaction potential introduced in~\cite{Blanchet:1989fg}.

The CTS equations themselves were originally written down in essentially complete form by Isenberg in the 1970s, but his paper was rejected and only published after a delay of nearly 30 years~\cite{Isenberg:2007zg}.  In the intervening years, Wilson, Mathews, and Marronetti~\cite{Wilson:1995uh,Wilson:1996ty,Mathews:1997vw,Mathews:1997pf} independently re-derived the entire formalism and used it to perform the first non-linear calculations of NS--NS mergers (as a result, the formalism is often referred to as the ``Wilson--Mathews'' or ``Isenberg--Wilson--Mathews'' formalism).  The key result in~\cite{Wilson:1995uh,Wilson:1996ty,Mathews:1997vw,Mathews:1997pf} was the existence of a ``collapse instability,'' in which the deeper gravitational wells experienced by the NSs as they approached each other prior to merger could force one or both to collapse to BHs prior to the orbit itself becoming unstable. Unfortunately, their results were affected by an error, pointed out in~\cite{Flanagan:1998zt}, which meant that much of the observed compression was spurious.  While their later calculations still found some increase in the central density as the NSs approached each other~\cite{Mathews:1999km}, these results have been contradicted by other QE sequence calculations (see, e.g.,~\cite{Uryu:1999uu}).  Furthermore, using a ``CF-like'' formalism in which the non-linear source terms for the field equations are ignored, dynamical calculations demonstrated the maximum allowed mass for a NS actually increases in response to the growing tidal stress~\cite{Shibata:1998sg}. 

The CF approach was adapted into a Lagrangian scheme for SPH calculations by the same groups that had investigated PN NS--NS mergers,  with Oechslin, Rosswog, and Thielemann~\cite{Oechslin:2001km} using a multigrid scheme  and Faber, Grandcl\'ement, and Rasio~\cite{Faber:2003sb} a spectral solver based on the {\tt Lorene} libraries~\cite{lorene_web}.  The effects of non-linear gravity were immediately evident in both sets of calculations.  In~\cite{Oechslin:2001km}, NS--NS binaries consisting of initially synchronized NSs merged without appreciable mass loss, with no more that $\sim 10^{-4}$ of the total system mass ejected, strikingly different from previous Newtonian and PN simulations.  When evolving initially irrotational systems,~\cite{Faber:2003sb} found no appreciable developments of ``spiral arms'' whatsoever, indicating a complete lack of mass loss through the outer Lagrange points.  Both groups also found strong emission from remnants for a stiff EOS, as the triaxial merger remnant produced an extended period of strong ringdown emission.  Neither set of calculations indicated that the remnant should collapse promptly to form a BH, but given the high spin of the remnant it was noted that conformal flatness would have already broken down for those systems.

\subsection{Full GR calculations}

\begin{table}\label{table:fullgr}
\caption[A summary of Full GR NS--NS merger calculations]{A summary of Full GR NS--NS merger calculations.  EOS models include polytropes, piecewise polytropes (PP), as well as physically motivated models including cold SLy~\cite{Douchin:2001sv}, FPS~\cite{Pandharipande:1989nmhi.conf..103P}, and APR~\cite{Akmal:1998cf} models to which one adds an ideal-gas hot component to reflect shock heating, as well as the Shen~\cite{Shen:1998by,Shen:1998gq} finite temperature model and EOS that include Hyperonic contributions~\cite{Sekiguchi:2011mc}.  ``Co/Ir'' indicates that both corotating and irrotational models were considered; ``BHB'' indicates that BH binary mergers were also presented, including both BH--BH and BH--NS types, ``$\nu$-leak'' indicates a neutrino leakage scheme was included in the calculation, ``GH'' indicates calculations were performed using the GHG formalism rather than BSSN, ``non-QE'' indicates superposition initial data were used, including cases where eccentric configurations were studied (``Eccen.'');  ``MHD'' indicates MHD was used to evolve the system.}
\centering
\begin{tabular}{ll|lll|l}
\hline
Group & Ref. & NS EOS & Mass ratio & $\mathcal{C}$ & notes\\\hline \hline
KT & \cite{Shibata:1999wm} & $\Gamma=2$ &  1 & 0.09--0.15 & Co/Ir\\ 
-- & \cite{Shibata:2002jb} & $\Gamma=2,2.25$ & 0.89--1 & 0.1--0.17 &\\
-- & \cite{Shibata:2003ga} & $\Gamma=2$ & 0.85--1 & 0.1--0.12 &\\
-- & \cite{Shibata:2005ss} & SLy,FPS+Hot & 0.92--1 & 0.1--0.13 &\\
-- & \cite{Shibata:2006nm} & SLy,APR+Hot & 0.64--1 & 0.11--0.13 &\\ 
-- & \cite{Yamamoto:2008js} & $\Gamma=2$ & 0.85--1 & 0.14--0.16 &BHB \\ 
-- & \cite{Kiuchi:2009jt} & APR+Hot & 0.8--1 & 0.14--0.18 & \\
-- & \cite{Kiuchi:2010ze} & APR,SLy,FPS+Hot & 0.8--1.0 & 0.16-0.2 & \\
-- & \cite{Sekiguchi:2011zd} & Shen & 1 & 0.14--0.16 & $\nu$-leak \\
-- & \cite{Hotokezaka:2011dh} & PP+hot & 1 & 0.12--0.17 & \\
-- & \cite{Sekiguchi:2011mc} & Shen, Hyp & 1.0 & 0.14--0.16 & $\nu$-leak \\
\hline
HAD & \cite{Anderson:2007kz} & $\Gamma=2$ & 1.0 & 0.08 & GH, non-QE\\
-- & \cite{Anderson:2008zp} & $\Gamma=2$ & 1.0 & 0.08 & GH, non-QE, MHD\\ \hline
Whisky & \cite{Baiotti:2008ra} & $\Gamma=2$ & 1.0 & 0.14--0.18 &\\
--& \cite{Baiotti:2009gk} & $\Gamma=2$ & 1.0 & 0.20 &\\
--& \cite{Giacomazzo:2009mp} & $\Gamma=2$ & 1.0 & 0.14--0.18 & MHD\\
--& \cite{Giacomazzo:2010bx} & $\Gamma=2$ & 1.0 & 0.14--0.18 & MHD\\
--& \cite{Rezzolla:2010fd} & $\Gamma=2$ & 0.70--1.0 & 0.09--0.17 &\\
--& \cite{Baiotti:2010xh,Baiotti:2011am} & $\Gamma=2$ & 1.0 & 0.12--0.14 &\\
--& \cite{Rezzolla:2011da} & $\Gamma=2$ &1.0 &0.18 & MHD\\ \hline
UIUC & \cite{Liu:2008xy} & $\Gamma=2$ & 0.85--1 & 0.14--0.18 &MHD\\ \hline
Jena & \cite{Thierfelder:2011yi,Bernuzzi:2011aq} & $\Gamma=2$ & 1.0 & 0.14 & \\
-- & \cite{Gold:2011df} & $\Gamma=2$ & 1.0 & 1.4 & Eccen.\\  \hline
\end{tabular}
\end{table}

A summary of full GR calculations of NS--NS mergers is presented in Table~\ref{table:fullgr}.
The KT collaboration was responsible for the only full GR calculations of NS--NS mergers that predate the breakthrough calculations of numerically stable binary BH evolutions \cite{Pretorius:2005gq,Baker:2005vv,Campanelli:2005dd}, which have since transformed the field of GR hydrodynamics and MHD in addition to vacuum relativistic evolutions (Miller et al. \cite{Miller:2003vc} performed NS--NS inspiral calculations in full GR, but were not able to follow binaries through to merger).  
The first calculations of NS--NS mergers using a completely self-consistent treatment of GR were performed by Shibata and collaborators in the KT collaboration 
using a grid based code and the BSSN formalism~\cite{Shibata:1999wm}.  CTS initial data consisting of equal-mass NSs described by a $\Gamma=2$ polytropic EOS were constructed  via SCF techniques~\cite{Uryu:1999uu}, for both synchronized and irrotational configurations.  The hyperbolic system was evolved on a grid, with an approximate maximal slicing condition that results in a parabolic equation for the lapse~\cite{Shibata:1999va} and an approximate minimal distortion condition for the shift vector requiring the solution of an elliptic equation at every time step~\cite{Shibata:1999wi}.  The shift vector gauge condition was found to fail when BHs were produced in the merger remnant, a well-known problem that had long bedeviled simulations involving binary and even single BH evolutions, so modifications were introduced to extend the stability of the algorithm as far as possible.  Among the key results from this early work was a clear differentiation between mergers of moderately low-compactness NSs ($\mathcal{C}\gtrsim 0.11$), 
where the remnant collapsed promptly to a BH, and very low-compactness models, which yielded hypermassive 
remnants stabilized against gravitational collapse by differential rotation.  Virtually all the NS matter was contained within the remnant for initially irrotational models, which served as evidence against equal-mass NSs mergers being a leading source of r-process elements in the universe through ejection.  The lack of significant mass loss in equal-mass mergers, together with insignificant shock-heating of the material, also argued against the likelihood of such mergers as progenitors for SGRBs if the gamma-ray emission was assumed to be coincident with the GW burst; instead a delayed burst following the collapse of a HMNS to a BH appeared more likely.

Later works, in particular  a paper by Shibata, Taniguchi, and Ury\=u~\cite{Shibata:2003ga}, introduced several new techniques to perform dynamical calculations that most codes at present still include in nearly the same or lightly modified form.  These included the use of a high-resolution shock-capturing scheme for the hydrodynamics, as well as a Gamma-driver shift condition closely resembling the moving puncture gauge conditions that later proved instrumental in allowing for long-term BH evolution calculations.  
In the series of papers that followed their original calculations, the KT group established a number of results about NS--NS mergers that form the basis for much of our thinking about their hydrodynamic evolution:
\begin{itemize}
\item By varying the EOS model for the NS as well as the mass ratio, it was possible to constrain the binary parameters separating cases that form a HMNS rather than producing prompt collapse to a BH, and  it was quickly determined that
the total system mass as a proportion of the maximum allowed mass for an isolated NS is the key parameter, with only weak dependence on the binary mass ratio.

  For polytropic EOS models, the critical compactness values leading to prompt collapse for equal-mass binary mergers were found to be ${\mathcal C}=0.14$ for $\Gamma=2$ and ${\mathcal C}=0.16$ for $\Gamma=2.25$ \cite{Shibata:2002jb}.  As a rough rule, collapse occurred for polytropic EOS when the total system rest-mass was at least $1.7\,M_{\max}$, where $M_{\max}$ is the maximum mass of an isolated non-rotating NS for the given EOS.  
For physically motivated EOS models~\cite{Shibata:2005ss}, the critical mass  was significantly smaller; indeed, the critical NS mass was found to be $\sim 1.35\,M_{\max}$ for the SLy EOS~\cite{Douchin:2001sv}  (i.e., collapse for $M_{\mathrm{tot}} \ge 2.7 \,M_{\odot}$ with $M_{\max}=2.04 \,M_{\odot}$) and $\sim 1.39\,M_{\max}$ for the FPS EOS~\cite{Pandharipande:1989nmhi.conf..103P} (collapse for $M_{\mathrm{tot}}\ge 2.5 \,M_{\odot}$ with $M_{\max}=1.8 \,M_{\odot}$). 
This was not a complete surprise, since for the physically motivated EOS the NS radius is nearly independent of the mass across much of the parameter space, limiting the ability of the HMNS to expand in response to the extra mass absorbed during the merger.

\item The mass ratio was found to  play a critical role in the evolution of the remnant/disk configuration, since unequal-mass cases are better characterized as disruptions of the smaller secondary followed by its accretion onto the primary, rather than a true merger between the two NSs.  Disk masses from full GR calculations~\cite{Shibata:2003ga} are generically smaller  than those predicted from non-GR calculations.  For polytropic EOS, disks contain approximately 4\% of the total system mass for mass ratios $q\simeq 0.8$, varying roughly $\propto (1-q)$ for a fixed total mass, with the disk mass decreasing for heavier binaries (and thus larger compactnesses) given the stronger gravity of the central remnant.  Using the stiffer APR EOS~\cite{Akmal:1998cf}, the dependence on the mass ratio was seen to be much steeper for a physical EOS than for polytropes, scaling like $(1-q)^p$, where $p\simeq 3-4$~\cite{Shibata:2006nm}.

\item With respect to GW emission, it was determined in \cite{Shibata:2002jb} that in low-mass cases in which a HMNS was formed, stiffer polytropic EOS models were able to support the development of a bar-mode instability, leading to transient spiral arm formation from the remnant and an extended period of strong GW emission, in the characteristic modulated form that results from differentially rotating ellipsoids (see, e.g.,~\cite{Lai:1994ke,Shibata:2000jt,Saijo:2000qt,Baiotti:2006wn}).  Unequal-mass cases typically yielded one high frequency peak at roughly $f_{\mathrm{GW}}\approx 2$kHz corresponding to non-axisymmetric oscillations, and equal-mass cases yielded multiple peaks including those associated with quasi-radial oscillations as well \cite{Shibata:2003ga}.   For physical EOS models \cite{Shibata:2005ss}, mass loss into a disk is reduced relative to the polytropic case given the higher compactness of the central region, and GW oscillation peaks, while very strong, occur at correspondingly higher frequencies.
  By contrast, prompt formation of a BH led to a  ringdown signal with rapidly decreasing amplitude becoming negligible within a few dynamical times. 
    
  The GW signals were evaluated under the gauge-dependent assumption of transverse tracelessness, and energy and angular momentum loss rates into each spherical harmonic mode were computed using the gauge-invariant Zerilli-Moncrief formalism~\cite{Regge:1957td,Zerilli:1971wd,Moncrief:1974am} in much the same way that is used by some groups in numerical relativity today (many BH--BH and hydrodynamics simulations report GW signals derived from the alternate $\psi_4$ Weyl scalar formulation~\cite{Newman:1961qr,Campanelli:1998jv}, or use both methods).

\item   In \cite{Shibata:2005ss}, it was concluded that mergers of NSs with comparable masses made poor SGRB progenitor candidates, assuming prompt emission (because of  the lack of energy available for neutrino annihilation), but that the energy budget in the HMNS case is orders of magnitude larger.  Remarkably, this discussion from 2005  predates the first identifications of SGRBs with older populations, which greatly improved our theoretical understanding of compact object mergers as their likely progenitors. In the first work that followed the initial localizations of SGRBs, mergers of binaries with relatively small mass ratios,  $q\approx 0.7$,  were seen to form sufficiently hot and massive disk to power a SGRB, albeit a relatively brief, low-luminosity one.  It was  suggested that the more likely SGRB progenitor is indeed a HMNS, since dissipative effects within the remnant can boost temperatures up to $\sim 10^{11}$~K. 

Further approximate relativistic investigations of NS--NS mergers, along with BH--NS mergers, as potential SGRB sources quickly swept through the community after the initial localizations of SGRBs, with several groups using a wide variety of methods all concluding that mergers were plausible progenitors, but finding it extremely difficult to constrain the scenario in quantitative ways given the extremely complicated microphysics ultimately responsible for powering the burst (see, e.g.,~\cite{Oechslin:2005mw,Oechslin:2006vt} who investigated potential disk energies;~\cite{Rosswog:2006rh}, who modeled the fall-back accretion phase onto a BH; and~\cite{Setiawan:2005ah}, who considered the thermodynamic and nuclear evolution of disks around newly-formed BHs produced by mergers).  We will return to this topic below in light of recent GRMHD Simulations.
\end{itemize}

In the past few years, five groups have reported results from NS--NS mergers in full GR; KT, HAD, Whisky, UIUC, and Jena.  Much of the work of the HAD and Whisky  groups, developers respectively of the code of those names, began at Louisiana State University (HAD) and the Albert-Einstein Institut in Potsdam (Whisky), though both efforts now include several other collaborating institutions.  Two other groups, the SXS collaboration that originated at Caltech and Cornell, and the Princeton group, have reported BH--NS merger results and are actively studying NS--NS mergers as well, but have yet to publish their initial papers about the latter.  All of the current groups use AMR-based Eulerian grid codes, with four evolving Einstein's equations using the BSSN formalism and the HAD collaboration making use of the GHG method instead.  HAD, Whisky, and UIUC have all reported results about magnetized NS--NS mergers (the KT collaboration has used a GRMHD code to study the evolution of magnetized HMNS, but not complete NS--NS mergers).  The KT collaboration has considered a wide range of EOS models, including finite-temperature physical models such as the Shen EOS, and have also implemented a neutrino leakage scheme, while all other results reported to date have assumed a $\Gamma=2$ polytropic EOS model.  

Given the similarities of the various codes used to study NS--NS mergers, it is worthwhile to ask whether they do produce consistent results.  A comparison paper between the {\tt Whisky} code and the KT collaboration's {\tt SACRA} codes~\cite{Baiotti:2010ka} found that both codes performed well for conservative global quantities, with global extrema such as the maximum rest-mass density in agreement to within 1\% and waveform amplitudes and frequencies differing by no more than 10\% throughout a full simulation, and typically much less.  

Several of the the groups listed above have also been leaders in the field of BH--NS simulations: the  KT, HAD, and UIUC groups have all presented BH--NS merger results, as have the SXS collaboration \cite{Duez:2008rb,Duez:2009yy,Foucart:2010eq},  and Princeton group \cite{Stephens:2011as,East:2011xa} (see \cite{ST_LRR} for a thorough review).

We discuss the current understanding of NS--NS mergers in light of all these calculations below.

\subsubsection{HMNS and BH remnant properties}

Using their newly developed {\tt SACRA} code~\cite{Yamamoto:2008js}, the KT group~\cite{Kiuchi:2009jt}, found that when a hybrid EOS is used to model the NS, in which the cold part is described by the APR EOS and the thermal component as a $\Gamma=2$ ideal gas, the critical total binary mass for prompt collapse to a BH is $M_{\mathrm{tot}}=2.8-2.9\,M_{\odot}$, independent of the initial binary mass ratio, a result consistent with previous explorations of other polytropic and physically motivated NS EOS models (see above).  In all cases, the BH was formed with a spin parameter $a\approx 0.78$ depending very weakly on the total system mass and mass ratio.

They further classified the critical masses for a number of other physical EOS in~\cite{Hotokezaka:2011dh}, finding that binaries with total masses $M_{\mathrm{tot}}\lesssim 2.7\,M_{\odot}$ should yield long-lived HMNSs ($>$~10~ms) and substantial disk masses with $M_{\mathrm{disk}}>0.04\,M_{\odot}$ assuming that the current limit on the heaviest observed NS, $M=1.97\,M_{\odot}$~\cite{Demorest:2010bx} is correct.  In Figure~\ref{fig:finalfate}, we show the final fate of the merger remnant as a function of the total pre-merger mass of the binary.  ``Type~I'' indicates a prompt collapse of the merger remnant to a BH, ``Type~II'' a short-lived HMNS, which lasts for less than 5~ms after the merger until its collapse, and ``Type~III'' a long-lived HMNS which survives for at least 5~ms.  See~\cite{Hotokezaka:2011dh} for an explanation of the EOS used in each simulation.

\epubtkImage{}{%
\begin{figure}[!ht]
\centerline{\includegraphics[width=0.6\textwidth]{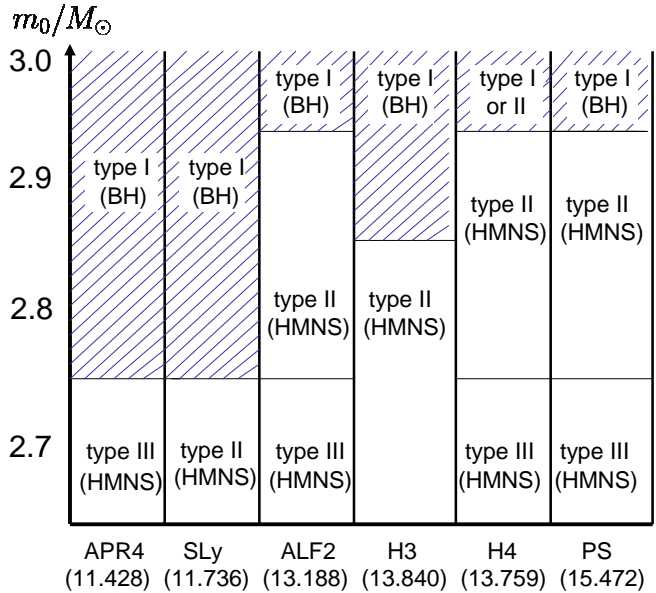}}
\caption{Type of final remnant corresponding to different EOS models (see Figure~3 of~\cite{Hotokezaka:2011dh}).  The vertical axis shows the total mass of two NSs. The horizontal axis shows the EOSs together with the corresponding NS radii for $M_{\mathrm{NS}}=1.4\,M_{\odot}$.}
\label{fig:finalfate}
\end{figure}}

While all of the above results incorporated shock heating, the addition  of both finite-temperature effects in the EOS and neutrino emission modifies the numerically determined critical masses separating HMNS formation from prompt collapse.
Adding in a neutrino leakage scheme for a NS--NS merger performed using the relatively stiff finite-temperature Shen EOS, the KT collaboration reports in~\cite{Sekiguchi:2011zd} that HMNSs will form generically for binary masses $\lesssim 3.2\,M_{\odot}$, not because they are centrifugally supported but rather because they are pressure-supported, with a remnant temperature in the range 30\,--\,70~MeV.  Since they are not supported by differential rotation, these HMNSs were predicted to be stable until neutrino cooling, with luminosities of $\sim 3-10\times 10^{53}$~erg/s, can remove the pressure support.   Even for cases where the physical effects of  hyperons were included, which effectively soften the EOS and reduce the maximum allowed mass for an isolated NS to $1.8M\odot$, 
the KT collaboration~\cite{Sekiguchi:2011mc} still finds that thermal support can stabilize HMNS with masses up to $2.7M_\odot$.

Using a {\tt Carpet/Cactus}-based hydrodynamics code called  {\tt Whisky}~\cite{Baiotti:2010zf} that works within the BSSN formalism (a version of which has been publicly released as {\tt GRHydro} within the {\tt Einstein Toolkit}~\cite{einsteintoolkit_web}),  the Whisky collaboration has analyzed the dependence of disk masses on binary parameters in some detail.  For mass ratios $q=0.7-1.0$~\cite{Rezzolla:2010fd},  they found that bound disks with masses of up to $0.2\,M_{\odot}$ can be formed, with the disk mass following the approximate form
\begin{equation}
M_{\mathrm{disk}} = 0.039 (M_{\max}-M_{\mathrm{tot}}) + 1.115(1-q)(M_{\max}-M_{\mathrm{tot}});~~~~~M_{\max} = 1.139(1+q)M_*~,\label{eq:mdisk}
\end{equation}
where $M_{\max}$ the maximum mass of a binary system for a given EOS ($\Gamma=2$ ideal gas for these calculations), 
$M_*$ is the maximum mass of an isolated non-rotating NS for the EOS, and $M_{\mathrm{tot}}$ the mass of the binary, with all masses here defined as baryonic. 
The evolution of the total rest mass present in the computational domain for a number of simulations is shown in Figure~\ref{fig:diskmass_q}.

\epubtkImage{}{%
\begin{figure}[!ht]
\centerline{\includegraphics[width=0.5\textwidth]{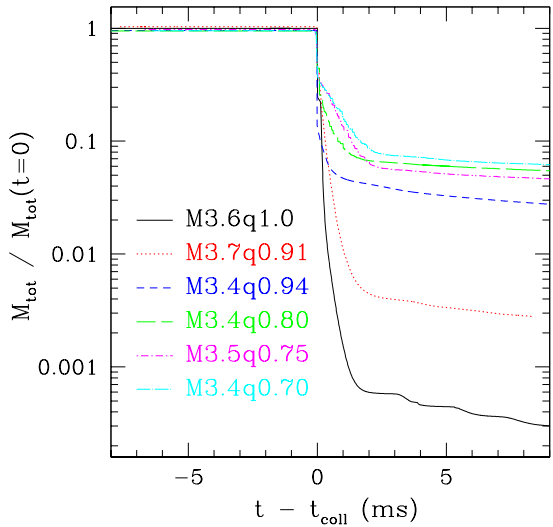}}
\caption{Evolution of the total rest mass $M_{\mathrm{tot}}$ of the remnant disk (outside the BH horizon) normalized to the initial value for NS--NS mergers using a $\Gamma=2$ polytropic EOS with differing mass ratios and total masses (see Figure~5 of~\cite{Rezzolla:2010fd}). The order of magnitude of the mass fraction in the disk can be read off the logarithmic mass scale on the vertical axis. The curves referring to different models have been shifted in time to coincide at $t_{\mathrm{coll}}$.}
\label{fig:diskmass_q}
\end{figure}}

\subsubsection{Magnetized NS--NS mergers}

Using the {\tt HAD} code described in~\cite{Anderson:2006ay} that evolves the GHG system on an AMR-based grid with CENO 
reconstruction techniques, Anderson et al.\ ~\cite{Anderson:2007kz}  performed the first study of magnetic effects in full GR NS--NS mergers~\cite{Anderson:2008zp}.  Beginning from spherical NSs with extremely strong poloidal magnetic fields ($9.6\times 10^{15}$G, as is found in magnetars), their merger simulations showed that magnetic repulsion can delay merger by 1--2 orbits and lead to the formation of magnetically buoyant cavities at the trailing end of each NS as contact is made (see Figure~\ref{fig:buoyancy}), although the latter may be affected by the non-equilibrium initial data.  Both effects would have been greatly reduced if more realistic magnetic fields strengths had been considered.  Magnetic fields in the HMNS remnant, which can be amplified through dynamo effects regardless of their initial strengths, helped to distribute angular momentum outward via the magneto-rotational instability (MRI), leading to a less differentially rotating velocity profile and a more axisymmetric remnant.  The GW emission in the magnetized case was seen to occur at lower characteristic frequencies and amplitudes as a result.

The UIUC group was among the first to produce fully self-consistent GRMHD results~\cite{Duez:2005sf}. Using a newly developed {\tt Cactus}-based code, they  performed the first studies of unequal-mass magnetized NS--NS mergers~\cite{Liu:2008xy}.  Using poloidal, magnetar-level initial magnetic fields, Liu et al.\  found that magnetic effects are essentially negligible prior to merger, but can  increase the mass in a disk around a newly formed BH moderately, from $1.3\%$ to $1.8\%$ of the total system mass for mass ratios of $q=0.85$ and $\Gamma=2$.  They point out that MHD effects can efficiently channel outflows away from the system's center after collapse \cite{Stephens:2008hu}, and may be important for the late-stage evolution of the system.

 In~\cite{Giacomazzo:2009mp}, the Whisky group performed simulations of magnetized mergers with field strengths ranging from $10^{12}$ to $10^{17}$G.  Agreeing with the UIUC work that magnetic field strengths would have essentially no effect on the GW emission during inspiral, they note that magnetic  effects  become significant for the HMNS, since differential rotation can amplify B-fields, with marked deviations in the GW spectrum appearing at frequencies of $f_{\mathrm{GW}}\gtrsim 2$kHz.  They also point out that high-order MHD reconstruction schemes, such as third-order PPM, can produce significantly more accurate results that second-order limiter-based schemes.  A follow-up paper~\cite{Giacomazzo:2010bx} showed that a plausible way to detect the effect of physically realistic magnetic fields on the GW signal from a merger was through a significant shortening of the timescale for a HMNS to collapse, though a third-generation GW detector could perhaps observe differences in the kHz emission of the HMNS as well.
 
 More recently, they have used very long-term simulations to focus attention on the magnetic field strength and geometry found after the remnant collapses to a BH~\cite{Rezzolla:2011da}  They find that the large, turbulent magnetic fields ($B\sim 10^{12}$~G) present in the initial binary configuration are boosted exponentially in time up to a poloidal field of strength $10^{15}$~G in the remnant disk, with the field lines maintaining a half-opening angle of $30^\circ$ along the BH spin axis, a configuration thought to be extremely promising for producing a SGRB.  The resulting evolution, shown in Figure~\ref{fig:bfield}, is perhaps the most definitive result indicating that NS--NS mergers should produce SGRBs for some plausible range of initial parameters.

\epubtkImage{}{%
\begin{figure}[!ht]
\centerline{
\includegraphics[width=0.45\textwidth]{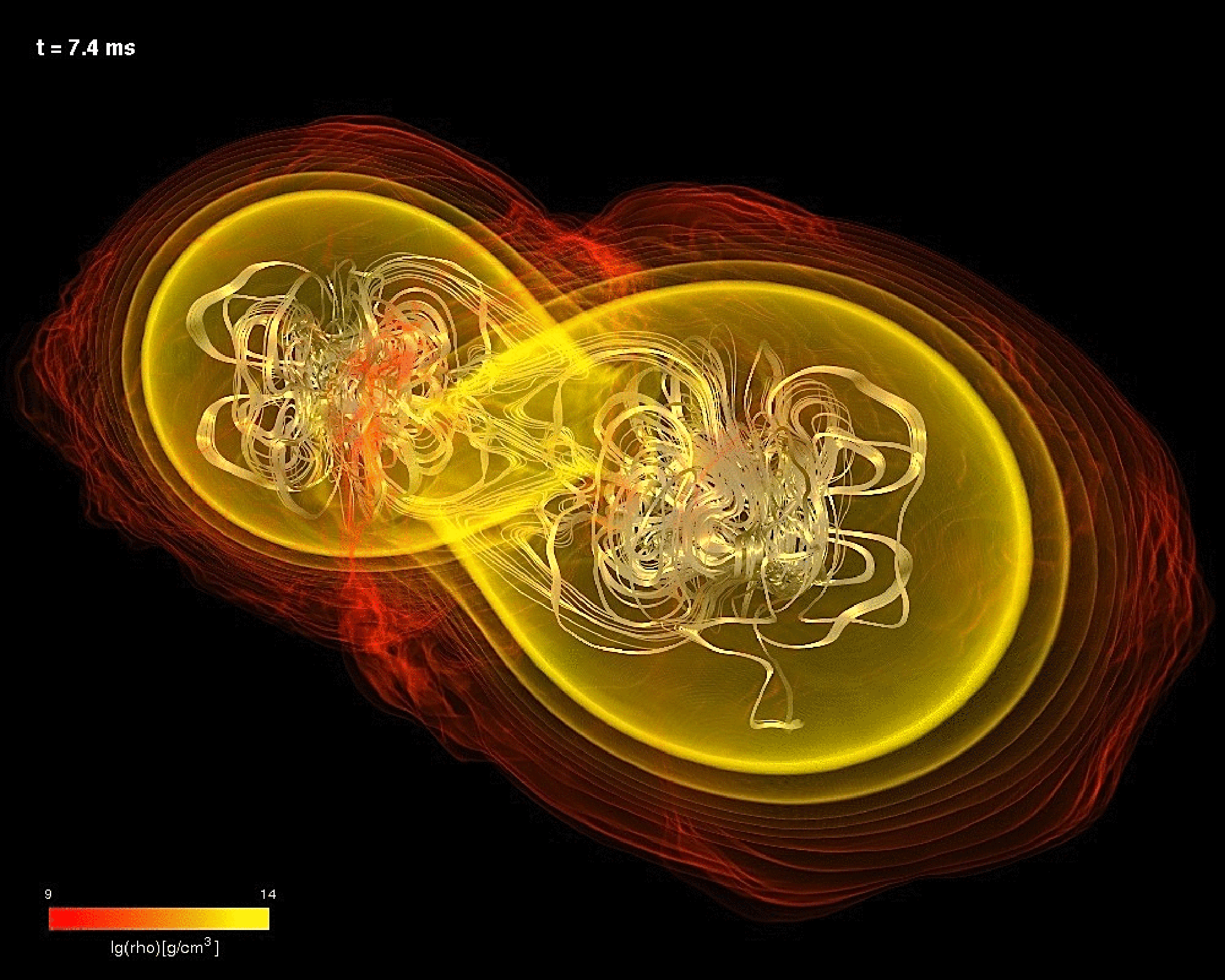}
\includegraphics[width=0.45\textwidth]{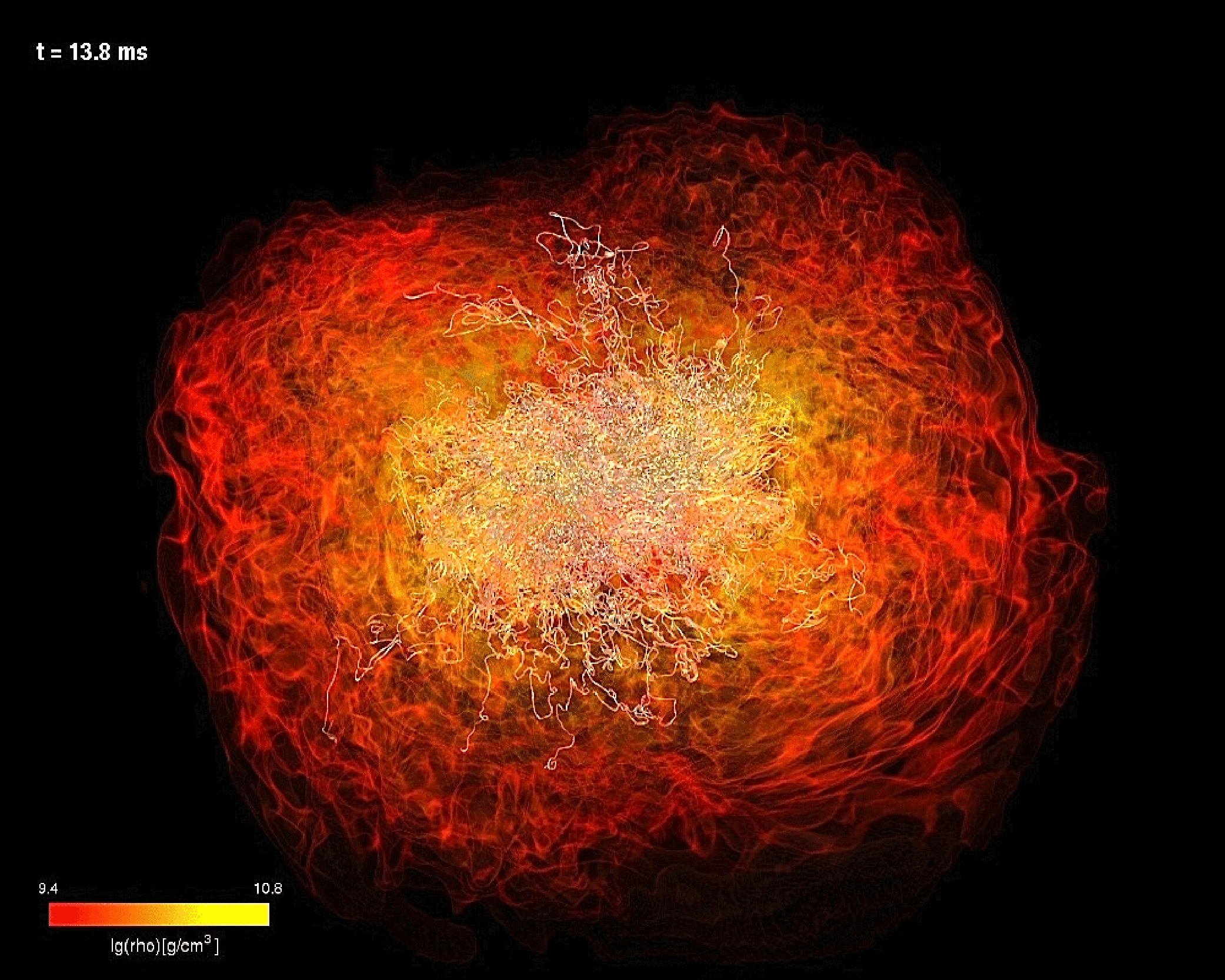}\\
}
\centerline{
\includegraphics[width=0.45\textwidth]{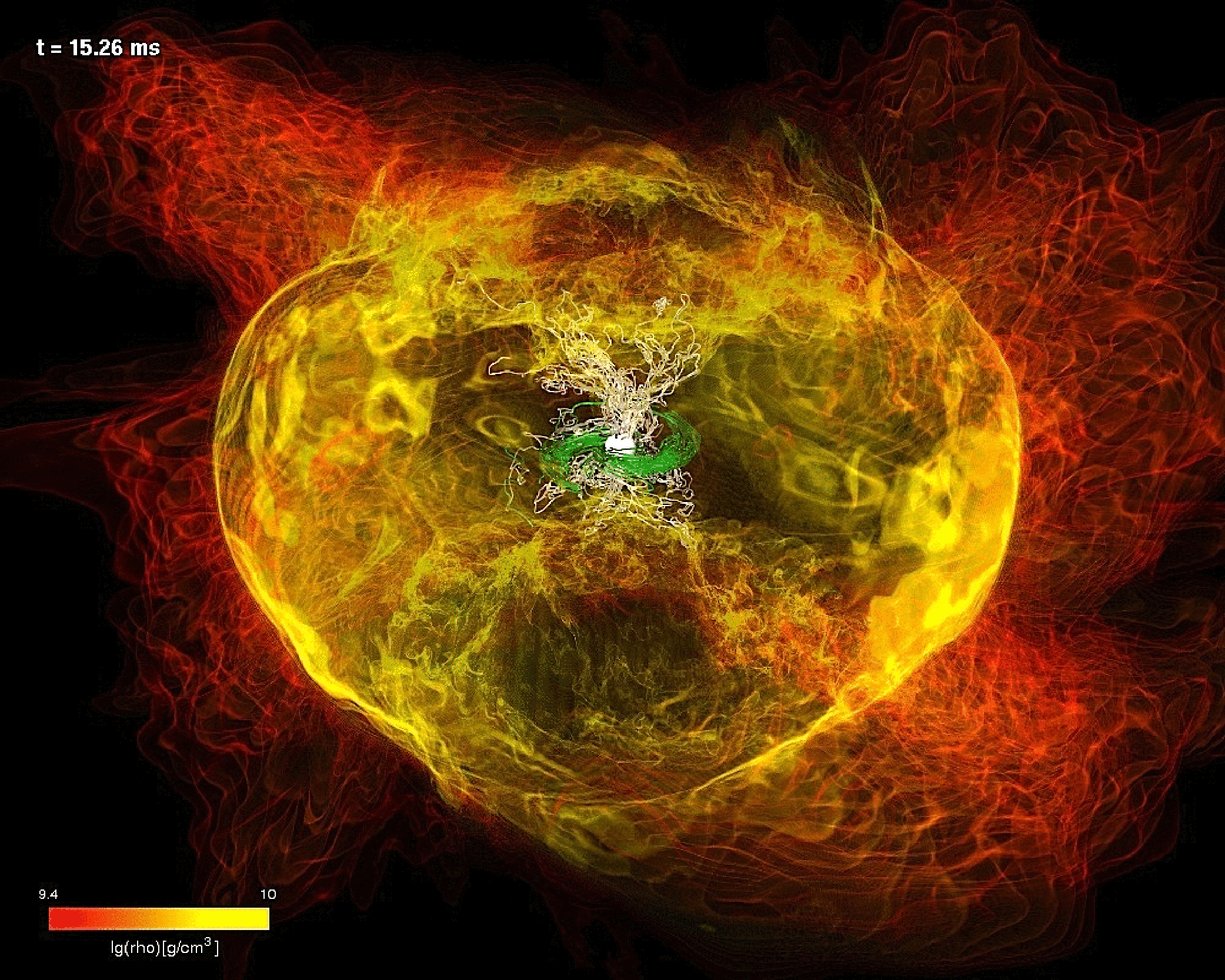}
\includegraphics[width=0.45\textwidth]{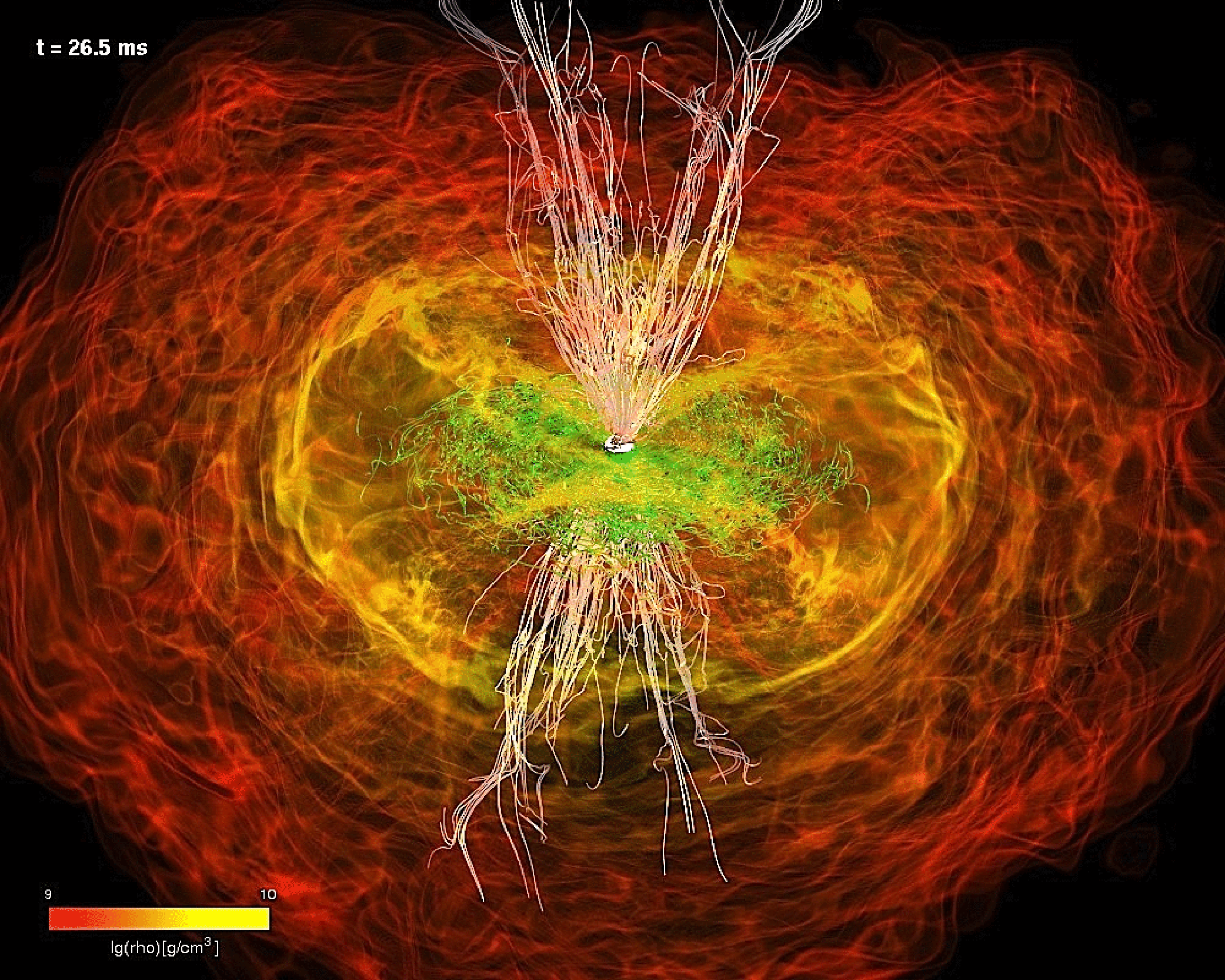}
}
\caption{Evolution of the density in a NS--NS merger, with magnetic field lines superposed, taken from Figure~1 of~\cite{Rezzolla:2011da}.  The first panel shows the binary shortly after contact, while the second shows the short-lived HMNS remnant shortly before it collapses.  In the latter two panels, a BH has already formed, and the disk around it winds up the magnetic field to a poloidal geometry of extremely large strength, $\sim 10^{15}$G, with an half-opening angle of $30^\circ$, consistent with theoretical SGRB models.}
\label{fig:bfield}
\end{figure}}

\epubtkImage{}{%
\begin{figure}[!ht]
\includegraphics[width=\textwidth]{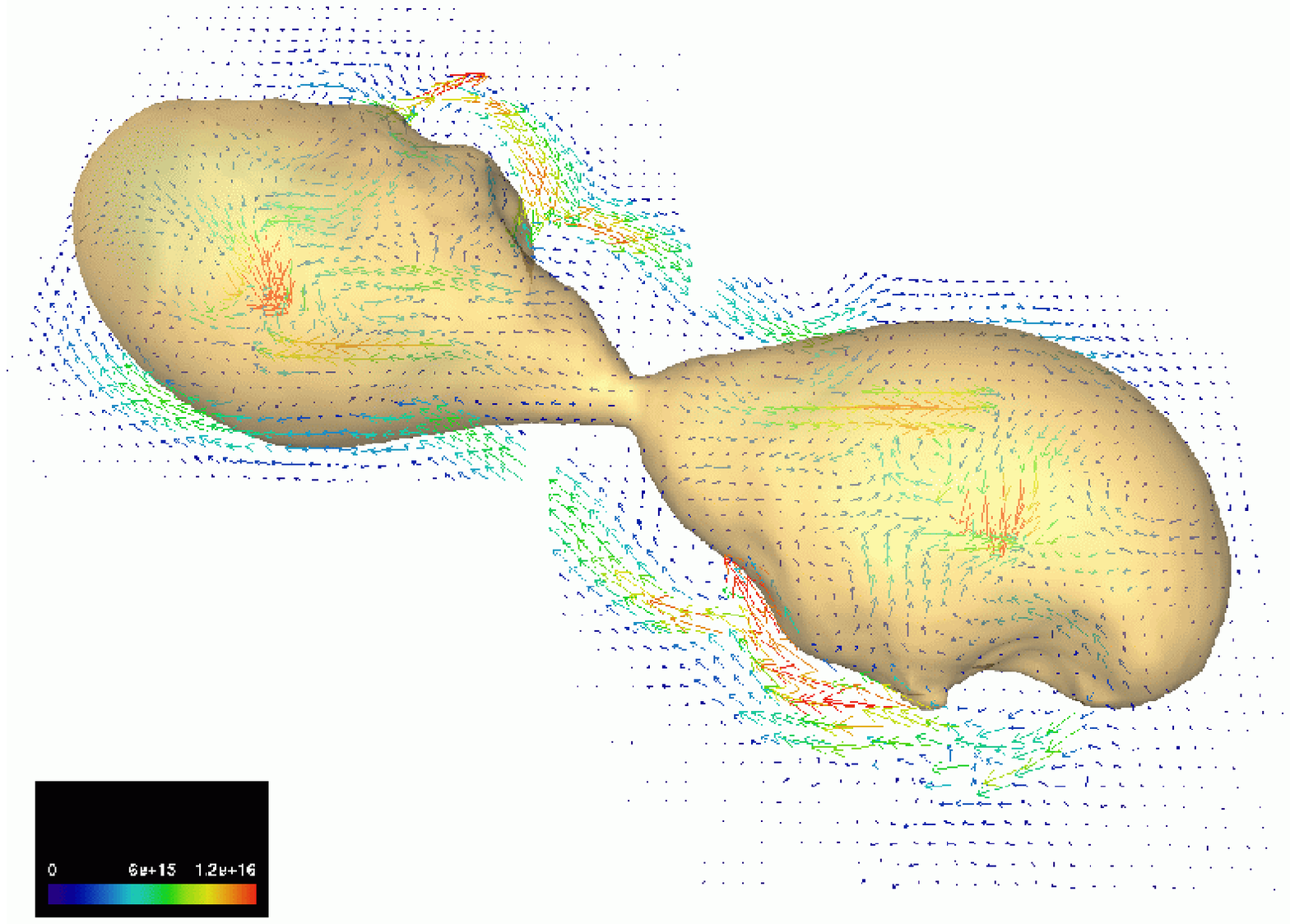}
\caption{Fluid density iscontours and magnetic field distribution (in a plane slightly above the equator) immediately after first contact for a magnetized merger simulation (see Figure~1 of~\cite{Anderson:2008zp}).  The cavities at both trailing edges are attributed to magnetic pressure inducing buoyancy.}
\label{fig:buoyancy}
\end{figure}}

It is worth noting that all magnetized NS--NS merger calculations that have been attempted to date have made use of unphysically large magnetic fields.  This is not merely a convenience designed to enhance the role of magnetic effects during the merger, though it does have that effect.  Rather, magnetic fields are boosted in HMNS remnants by the MRI, whose fastest growing unstable mode depends roughly linearly on the Alfven speed, and thus the magnetic field strength.  In order to move to physically reasonable magnetic field values, one would have to resolve the HMNS at least a factor of 100 times better in each of three dimensions, which is beyond the capability of even the largest supercomputers at present, and likely will be for some time to come.  

\subsubsection{GW emission}

In~\cite{Kiuchi:2010ze}, the KT collaboration found a nearly linear relationship between the GW spectrum cutoff frequency $f_{\rm cut}$  and the NS compactness, independent of the EOS,  as well as  a relationship between the disk mass and the width of the kHz hump seen in the GW energy spectrum.   While $f_{\rm cut}$ is a somewhat crude measure of the NS compactness, it occurs at substantially lower frequencies than any emission process associated with merger remnants, and thus is the parameter most likely to be accessible to GW observations with a second generation detector.

The qualitative form of the high-frequency components of the GW spectrum is primarily determined by the type of remnant formed.  In Figures~\ref{fig:GWsignal} and \ref{fig:GWSpectra}, we show $h(t)$ and $\tilde{h}(f)$, respectively, for four of the runs calculated by the KT collaboration and described in~\cite{Hotokezaka:2011dh}.  Type~I collapses are characterized by a rapid decrease in the GW amplitude immediately after the merger, yielding relatively low power at frequencies above the cutoff frequency.  Type~II and III mergers yield longer periods of GW emission after the merger, especially the latter, with the remnant oscillation modes leading to clear peaks at GW frequencies $f_{\mathrm{GW}}$~=~2\,--\,4~khz that should someday be detectable by third generation detectors like the Einstein Telescope, or possibly even by advanced LIGO should the source be sufficiently close ($D\lesssim 20$Mpc) and the high-frequency peak of sufficiently high quality~\cite{Sekiguchi:2011zd}.

\epubtkImage{}{%
\begin{figure}[!ht]
\centerline{
\includegraphics[width=0.5\textwidth]{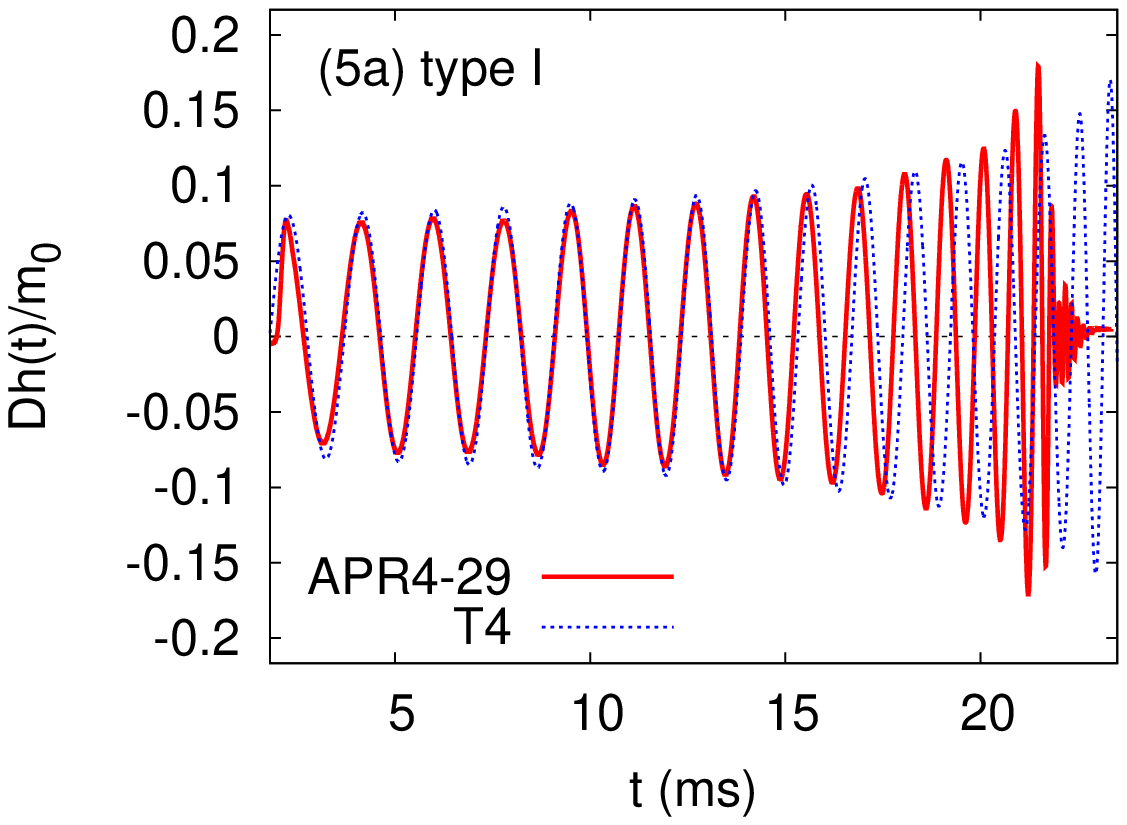}
\includegraphics[width=0.5\textwidth]{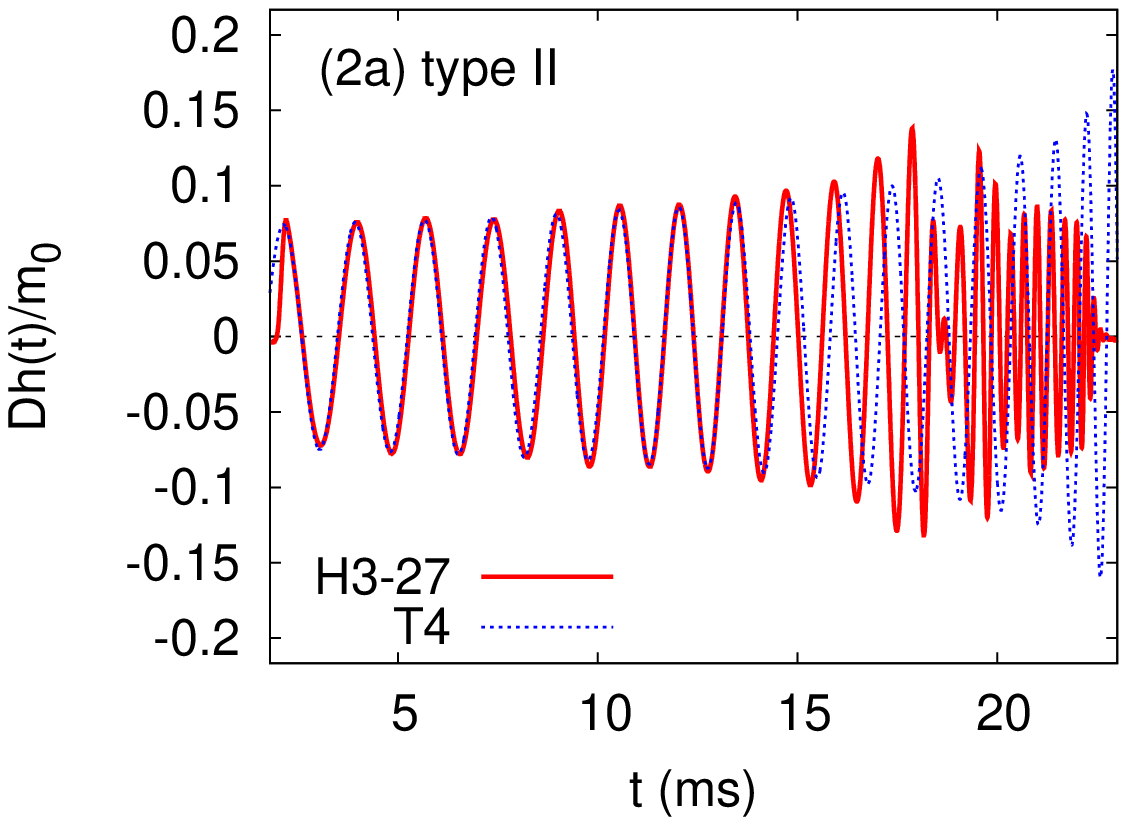}\\
}
\centerline{
\includegraphics[width=0.5\textwidth]{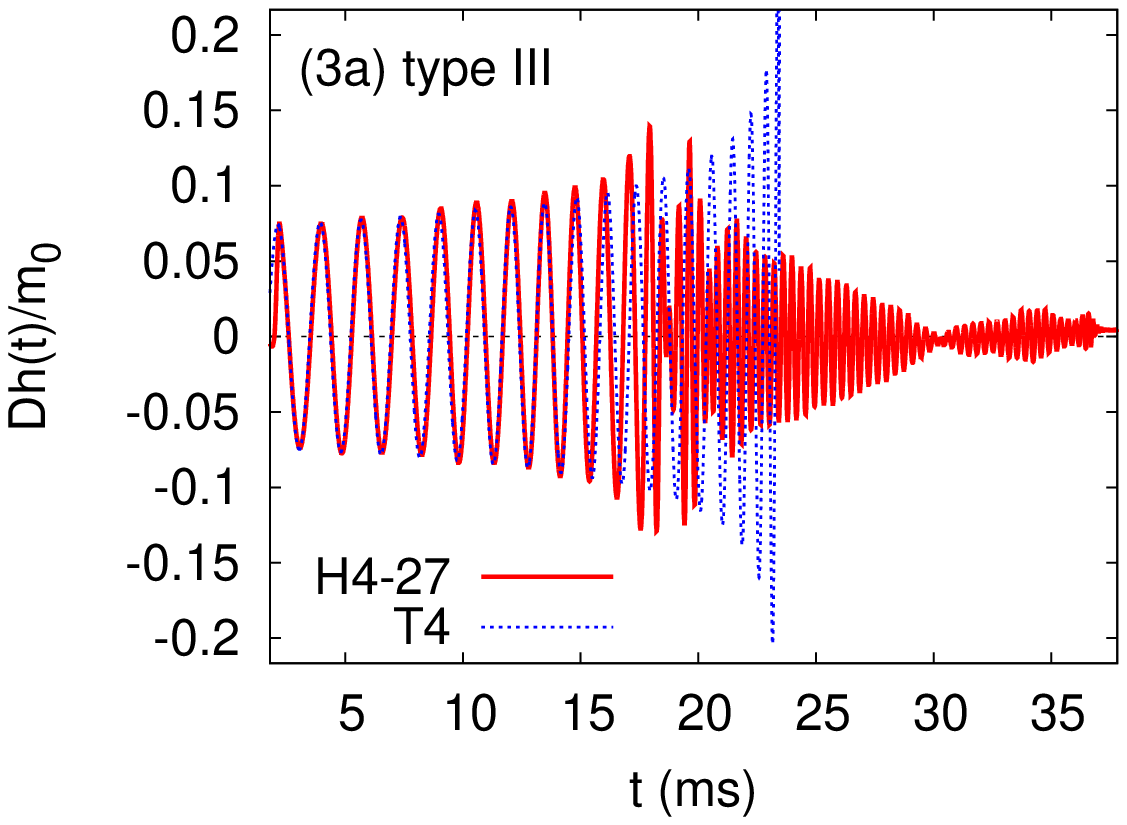}
\includegraphics[width=0.5\textwidth]{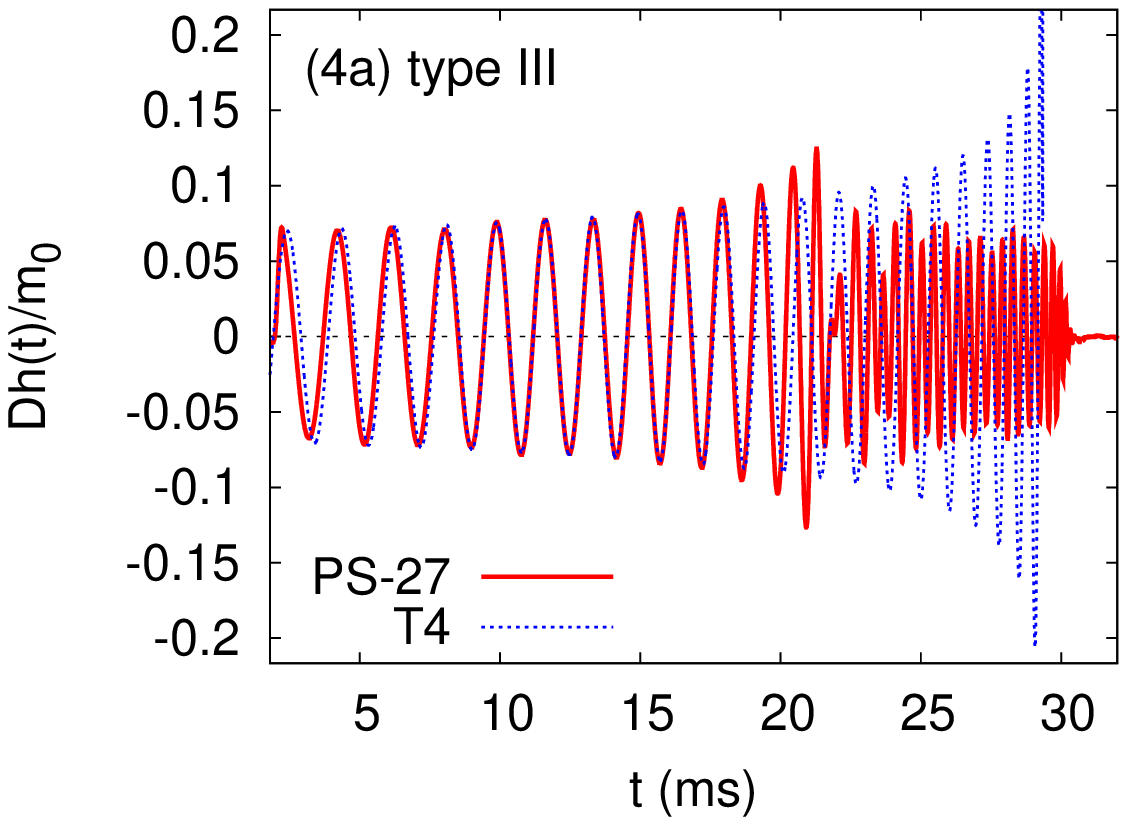}
}
\caption{Dimensionless GW strain $Dh/m_0$, where $D$ is the distance to the source and $m_0$ the total mass of the binary, versus time for four different NS--NS merger calculations, taken from Figures~5 and~6 of~\cite{Hotokezaka:2011dh}.  The different merger types become apparent in the post-merger GW signal, clearly indicating how BH formation rapidly drives the GW signal down to negligible amplitudes.}  \label{fig:GWsignal}
\end{figure}}

\epubtkImage{}{%
\begin{figure}[!ht]
\centerline{
\includegraphics[width=0.5\textwidth]{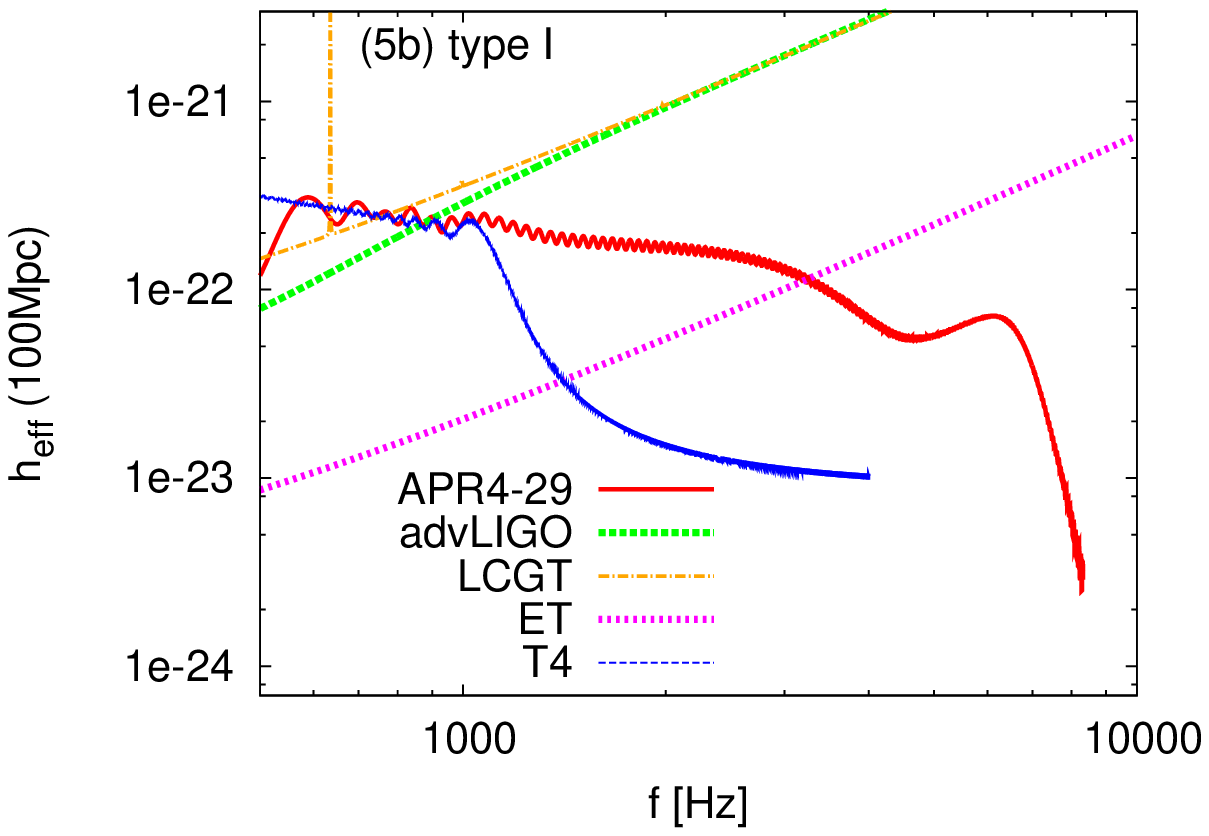}
\includegraphics[width=0.5\textwidth]{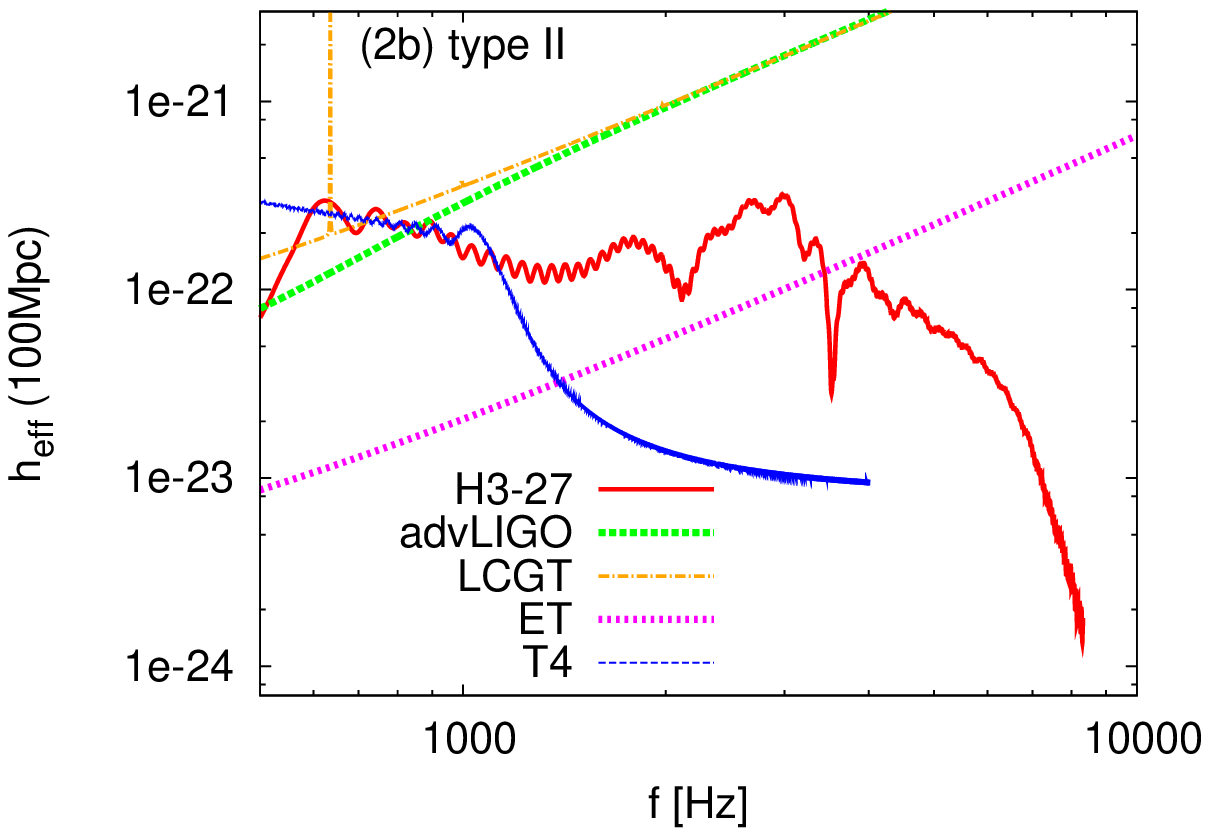}\\
}
\centerline{
\includegraphics[width=0.5\textwidth]{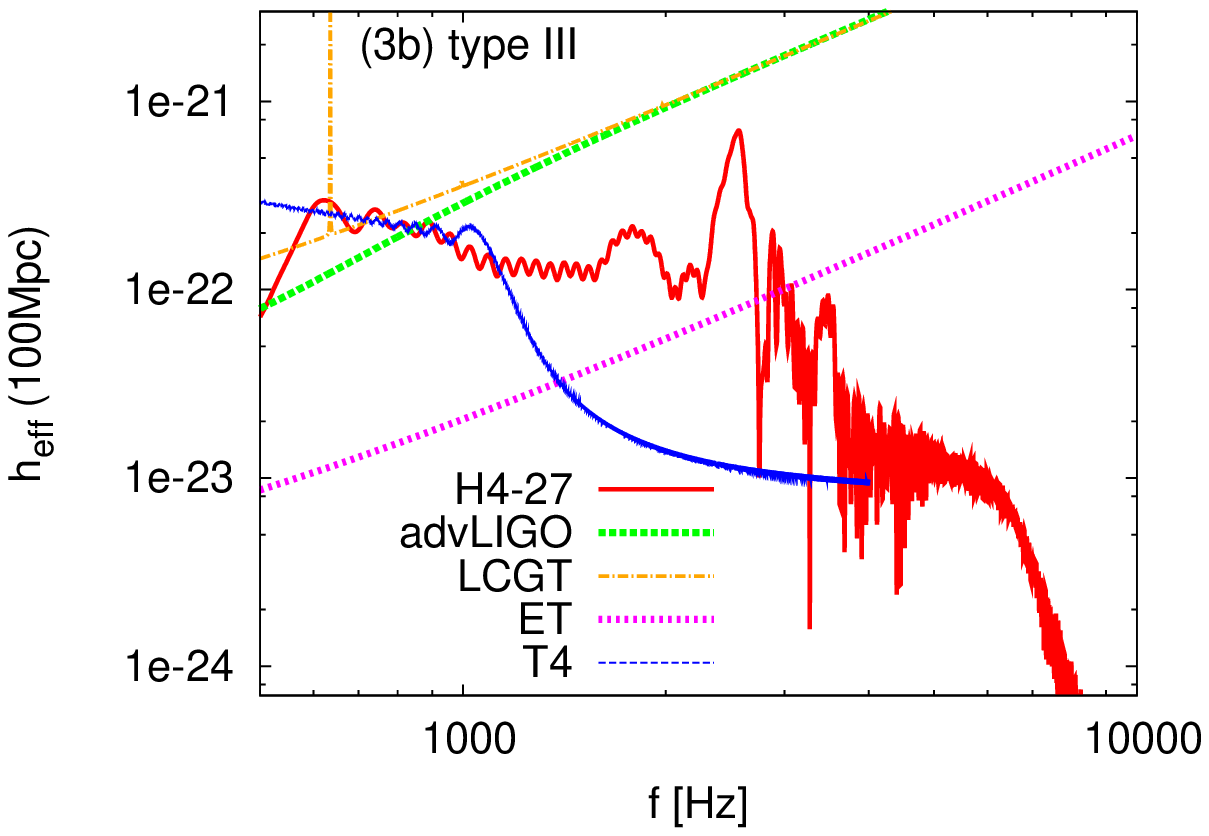}
\includegraphics[width=0.5\textwidth]{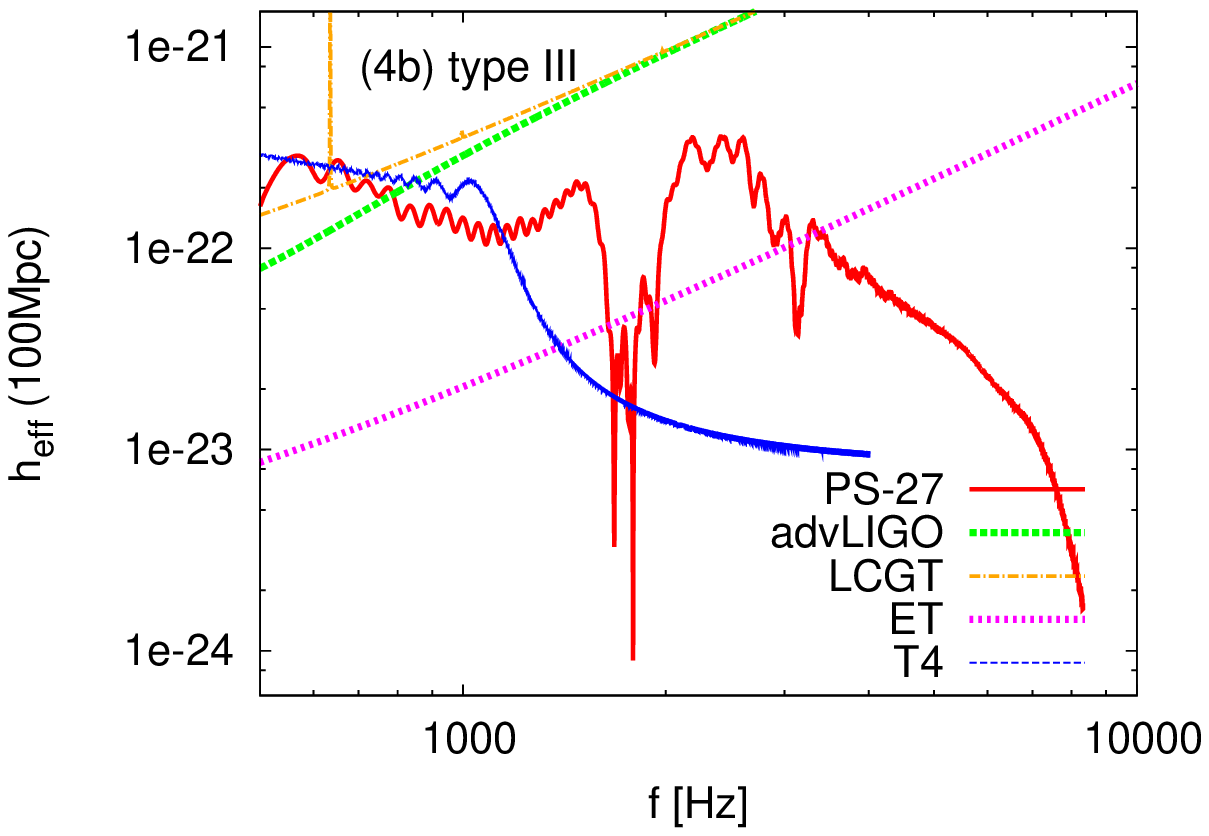}
}
\caption{Effective strain at a distance of 100~Mpc shown as a function of the GW frequency (solid red curve) for the same four merger calculations depicted in~\ref{fig:GWsignal}, taken from Figures~5 and 6 of~\cite{Hotokezaka:2011dh}.   Post-merger quasi-periodic oscillations are seen as broad peaks in the GW spectrum at frequencies $f_{\mathrm{GW}}$~=~2\,--\,4~kHz. The blue curve shows the Taylor T4 result, which represents a particular method of deducing the signal from a 3PN evolution. The thick green dashed curve and orange dot-dashed curves depict the sensitivities of the second-generation Advanced LIGO and LCGT (Large Scale Cryogenic Gravitational Wave Telescope) detectors, 
respectively, while the maroon dashed curve shows the sensitivity of a hypothetical third-generation Einstein Telescope.}
\label{fig:GWSpectra}
\end{figure}}

Using new multi-orbit simulations of NS--NS mergers, Baiotti et al.~\cite{Baiotti:2010xh,Baiotti:2011am} showed that the semi-analytic effective one-body (EOB) formalism severely underestimates high-order relativistic corrections even when lowest-order finite-size tidal effects were included.  As a result, phase errors of almost a quarter of a radian can develop, although these may be virtually eliminated by introducing a second-order ``next-to-next-to-leading order'' (NNLO) correction term and fixing the coefficient to match numerical results.  The excellent agreement between pre-merger numerical waveforms and the revised semi-analytic EOB approximant is shown in Figure~\ref{fig:GWvsEOB}.

\epubtkImage{}{%
\begin{figure}[!ht]
\centerline{\includegraphics[width=0.9\textwidth]{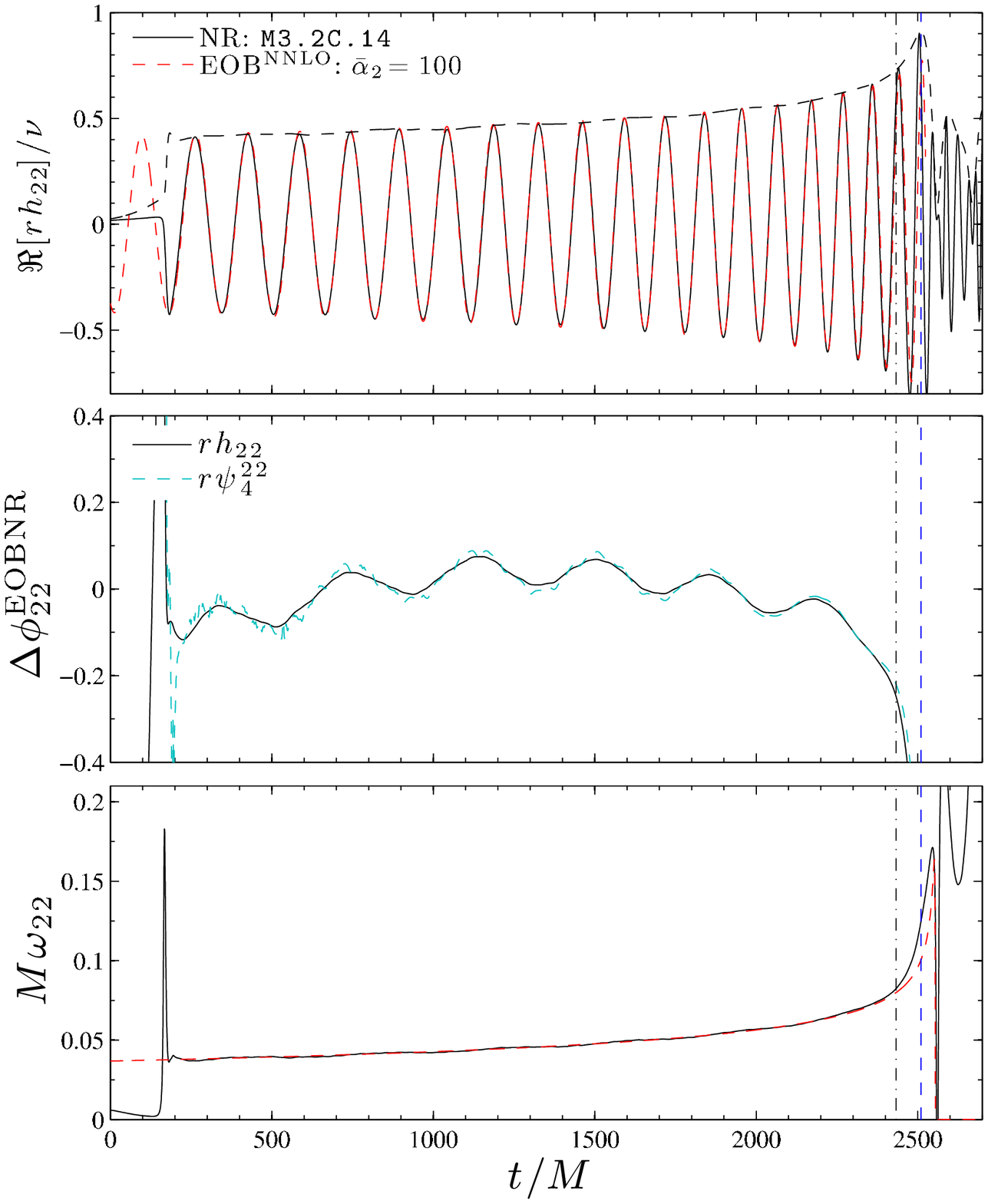}}
\caption{Comparison between numerical waveforms, shown as a solid black line, and semi-analytic NNLO EOB waveforms, shown as a red dashed line (top panel), taken from Figure~14 of~\cite{Baiotti:2011am}.  The top panels show the real parts of the EOB and numerical relativity waveforms, and the middle panels display the corresponding phase differences between waveforms generated with the two methods. There is excellent agreement between with the numerical waveform almost up to the time of the merger as shown by the match of the orbital frequencies (bottom panel).}
\label{fig:GWvsEOB}
\end{figure}}

\subsubsection{Binary eccentricity}

The effects of binary eccentricity on NS--NS mergers was recently studied by the Jena group~\cite{Gold:2011df}.  Such systems, which would indicate dynamical formation processes rather than the long-term evolution of primordial binaries, evolve differently  in several fundamental ways from binaries that merger from circular orbits.  For nearly head-on collisions, they found prompt BH formation and negligible disk mass production, with only a single GW burst at frequencies comparable to the quasi-normal mode of the newly formed BH.  For a collision in which mass transfer occurred at the first passage but two orbits were required to complete the merger and form a BH, a massive disk was formed, containing $8\%$ of the total system mass, more than  half of which was accreted by the BH within 100M of its formation.  Between the first close passage and the second, during which the two NS merged, the GW signal was seen to be quasi-periodic, and a a frequency comparable to the fundamental oscillation mode of the two NS, a result that was duplicated in a calculation for which the periastron fell outside the Roche limit and the eccentric binary survived for the full duration of the run, comprising several orbits.

\subsection{Simulations including microphysics}

In parallel to efforts in full GR, there has also been great progress in numerical simulations that include approximate relativistic treatments but a more detailed approach to microphysical issues.
The first simulations to use a realistic EOS for NS--NS mergers were performed by Ruffert, Janka, and collaborators~\cite{Ruffert:1995fs,Janka:1995cq,Ruffert:1996by}, who assumed the Lattimer-Swesty EOS for their Newtonian PPM-based Eulerian calculations.  They were able to determine a physically meaningful temperature for NS--NS merger remnants of 30\,--\,50~MeV, an overall neutrino luminosity of roughly $10^{53}$~erg/s for tens of milliseconds, and a corresponding annihilation rate of $2-5\times 10^{50}$~erg/sec given the computed annihilation efficiencies of a few parts in a thousand  This resulted in an  energy loss of $2-4\times 10^{49}$~erg over the lifetime of the remnant~\cite{Janka:1995cq}, a value later confirmed in multigrid simulations that replaced the newly formed HMNS by a Newtonian or quasi-relativistic BH surrounded by the bound material making up a disk~\cite{Ruffert:1998qg}.  The temperatures in the resulting neutron-rich ($Y_e\approx 0.05-0.2$) remnant were thought to be encouraging for the production of r-process elements~\cite{Ruffert:1996by}, although numerical resolution  of the low-density ejecta limited the ability to make quantitatively accurate estimates of its exact chemical distribution.  Further calculations, some of which involved unequal-mass binaries, indicated that the temperatures and electron fractions in the ejecta were likely not sufficient to produce solar abundances of r-process elements~\cite{Ruffert:2001gf}, with electron fractions in particular smaller than those set by hand in the r-process production model that appears in~\cite{Freiburghaus:1999ApJ...525L.121F,Rosswog:2000qm}. More recently, it was suggested~\cite{Goriely:2011vg} that the decompression of matter 
originally located in the inner crust of a NS and ejected during a merger has a nearly solar elemental distribution for heavy r-process elements ($A>140$). This indicates that NS--NS mergers may be the source of the observed cosmic r-process elements should there be sufficient mass loss per merger event , $M_{ej}\sim 3-5\times 10^{-5}\,M_{\odot}$, although these amounts have yet to be observed in full GR simulations which have often admittedly been performed using cruder microphysical treatments.

 In~\cite{Rosswog:2001fh}, Rosswog and Davies included a detailed neutrino leakage scheme in their calculations and also adopted the Shen EOS for several calculations, finding in a later paper~\cite{Rosswog:2002rt} that the gamma-ray energy release is roughly $10^{48}$erg, in line with previous results from other groups, but noting that the values would be significantly higher if temperatures in the remnant were higher, since the neutrino luminosity scales like a very high power of the temperature.  These calculations also identified NS--NS mergers as likely SGRB candidates given the favorable geometry~\cite{Rosswog:2003ts}, and the possibility that the MRI in a HMNS remnant could dramatically boost magnetic fields on the sub-second timescales characterizing SGRBs~\cite{Rosswog:2003tn}.  Rosswog and Liebend\"orfer~\cite{Rosswog:2003rv} found that electron antineutrinos $\bar{\nu}_e$ dominate the emission, as had Ruffert and Janka~\cite{Ruffert:2001gf}, though the exact thermodynamic and nuclear profiles were found to be somewhat sensitive to the properties of the EOS model.  More recently, using the {\tt VULCAN} 
2-dimensional multi-group flux-limited-diffusion radiation hydrodynamics code~\cite{Livne:2003ai} to evaluate slices taken from SPH calculations, Dessart et al.~\cite{Dessart:2008zd} found that neutrino heating of the remnant material will eject roughly $10^{-4}\,M_{\odot}$ from the system.

Price and Rosswog~\cite{Price:2006fi,Rosswog:2007ue} performed the first MHD simulation of merging NS--NS binaries using an SPH code that included magnetic field effects, finding that the Kelvin-Helmholtz unstable vortices formed at the contact surface between the two NSs could boost magnetic fields rapidly up to $\sim 10^{17}G$.  These results were not seen in GRMHD simulations, where  gains in the magnetic field strength generated by dynamos were limited by the swamping of the  vortex sheet at the surface of contact by rapidly infalling NS material that went on to form the eventual HMNS or BH~\cite{Giacomazzo:2010bx}.  Longer-term simulations did note that shearing instabilities were able to support power-law, or perhaps even exponential, growth of the magnetic fields on longer timescales ($\sim$~10~s of ms), which augurs well for NS--NS mergers as the central engines of SGRBs~\cite{Rezzolla:2011da}.

An effort to identify potential observational differences between NSs and COs with quark-matter interiors has been led by Oechslin and collaborators.  Using an SPH code with CF gravity, Oechslin et al.~\cite{Oechslin:2002vy,Oechslin:2004yj} considered mergers of NSs with quark cores described by the MIT bag model~\cite{Chodos:1974je,Farhi:1984qu,Kettner:1994zs}, which have significantly smaller maximum masses than traditional NSs.  They found the hybrid nuclear-quark EOS yielded higher ISCO frequencies for NSs with masses $\gtrsim 1.5\,M_{\odot}$ and slightly larger GW oscillation frequencies for any resulting merger remnant compared to purely hadronic EOS.  Bauswein et al.~\cite{Bauswein:2008gx} followed up this work by investigating whether ``strangelets'', or small lumps of strange quark matter, would be ejected in sufficient amounts throughout the interstellar medium to begin the phase transition that would convert traditional hadronic NSs into strange stars.  They determined that the total rate of strange matter ejection in NS--NS mergers could be as much as $10^{-8}\,M_{\odot}$ per year per galaxy or essentially zero 
depending on the parameters input into the MIT bag model, with the upper values clearly detectable by orbiting magnetic spectrometers such as the AMS-02 detector that was recently installed on the International Space Station~\cite{Madsen:2004vw,Kounine:2010js}.  Further calculations concluded that the mergers of strange stars produce a much more tenuous halo than traditional NS mergers, more rapid formation of a BH, and higher frequency ringdown emission~\cite{Bauswein:2009im}, as we show in Figure~\ref{fig:SS}.

\epubtkImage{}{%
\begin{figure}[!ht]
\centerline{
\includegraphics[width=0.5\textwidth]{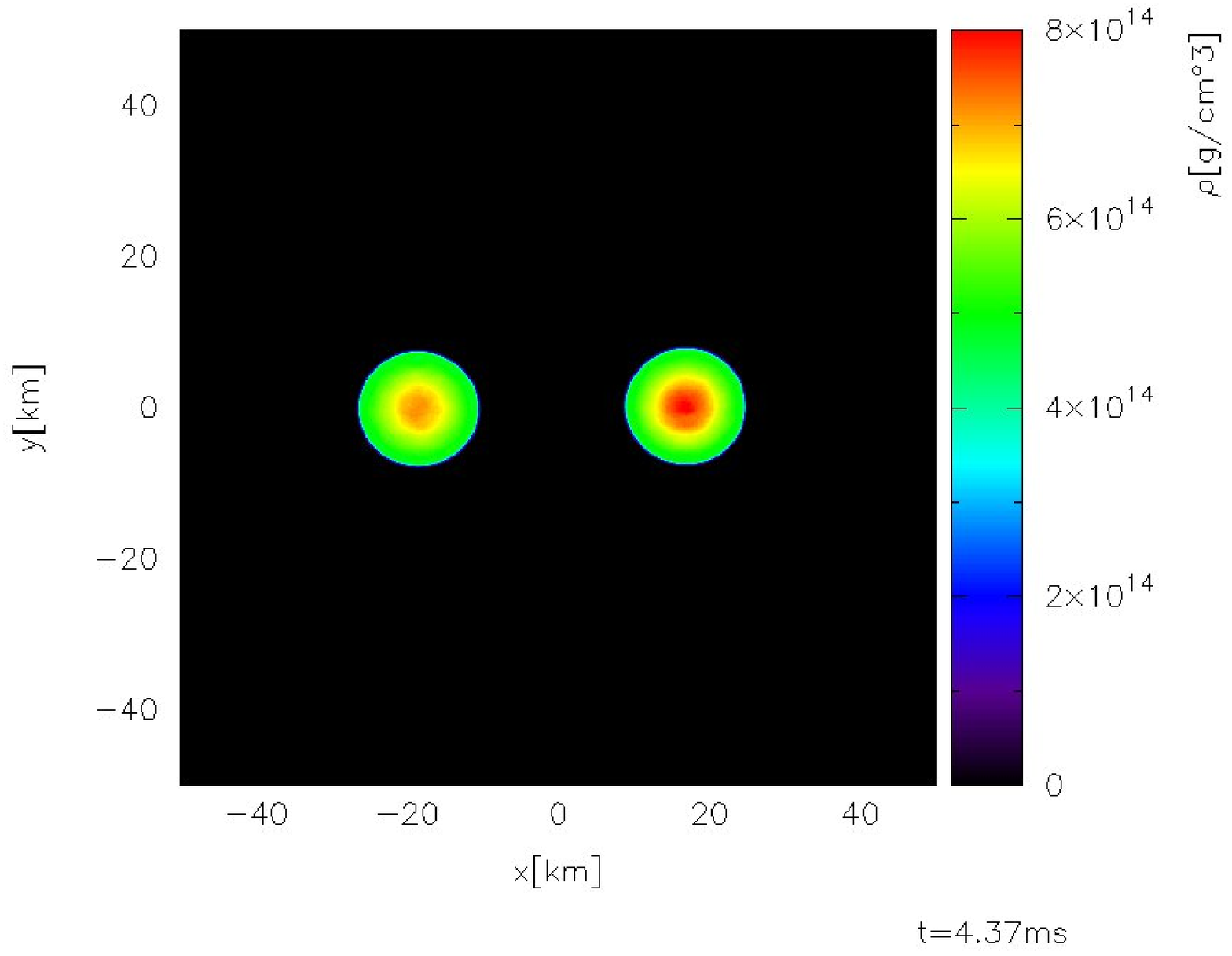}
\includegraphics[width=0.5\textwidth]{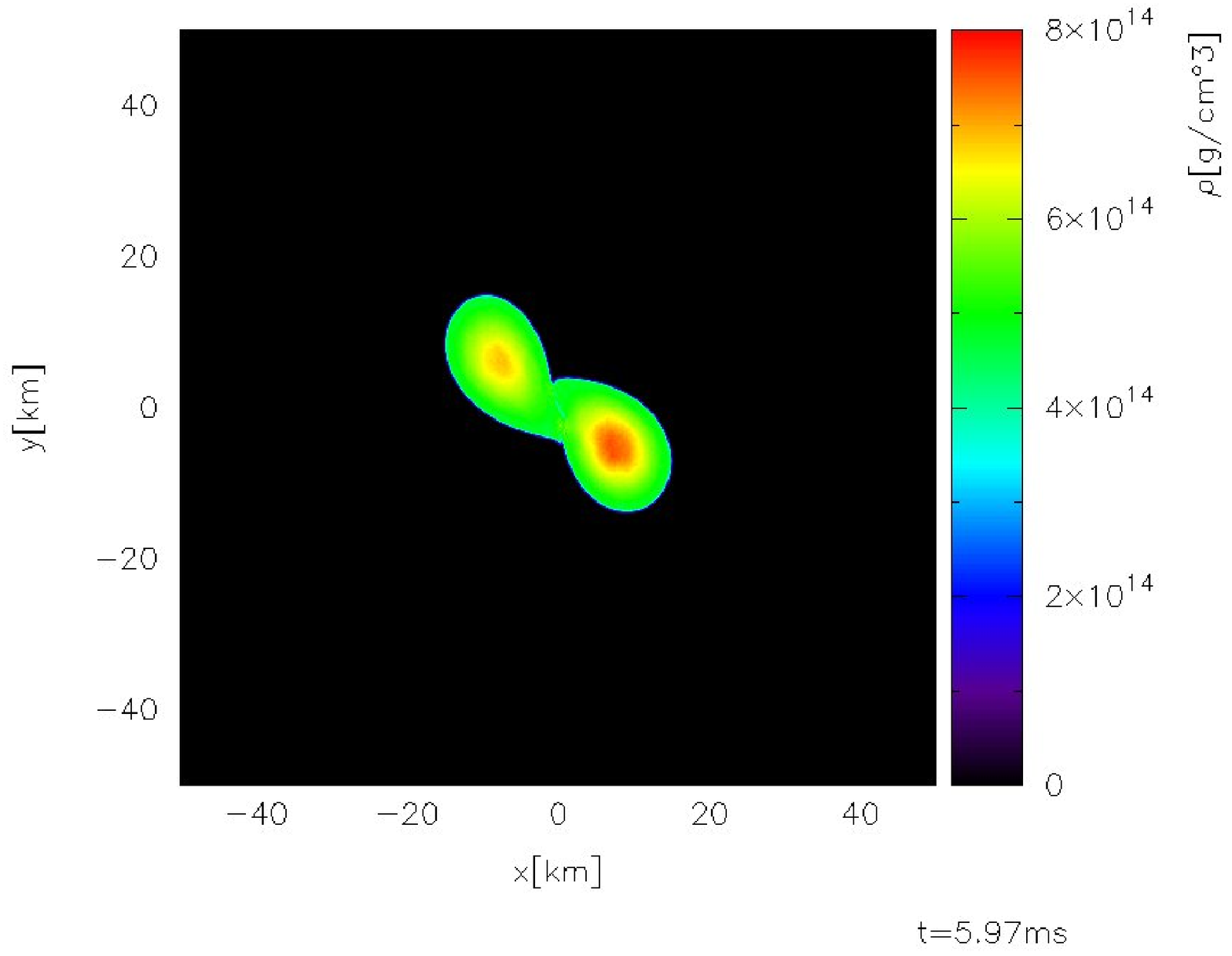}\\
}
\centerline{
\includegraphics[width=0.5\textwidth]{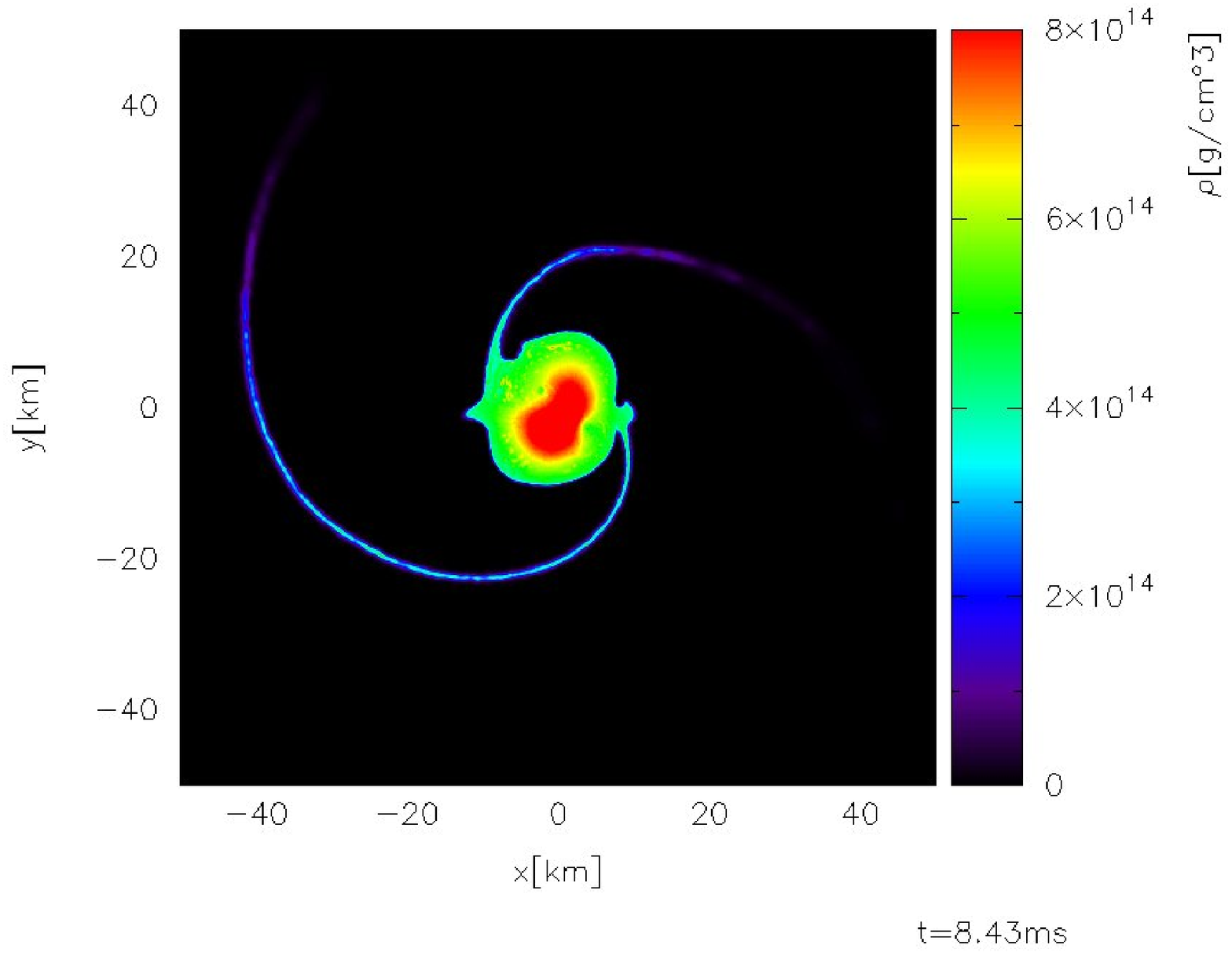}
\includegraphics[width=0.5\textwidth]{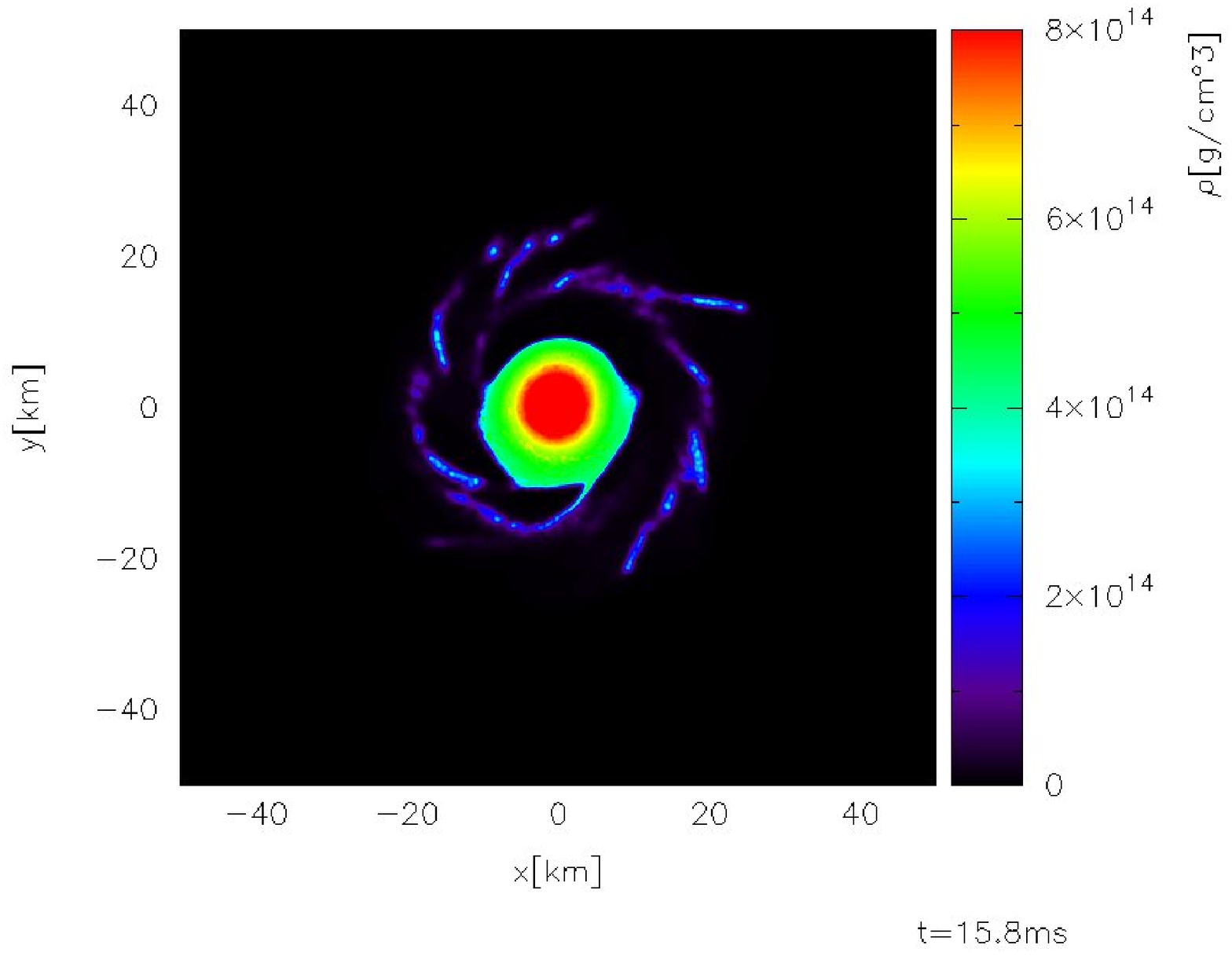}
}
\caption{Evolution of a binary strange star merger performed using a CF SPH evolution, taken from Figure~4 of~\cite{Bauswein:2009im}.  The ``spiral arms'' representing mass loss through the outer Lagrange points of the system are substantially narrower than those typically seen in CF calculations of NS--NS mergers with typical nuclear EOS models.}
\label{fig:SS}
\end{figure}}

Oechslin, Janka, and Marek also analyzed a wide range of EOS models using their CF SPH code, finding that matter in spiral arms was typically cold and that the dynamics of the disk formed around a post-merger BH depends on the initial temperature assumed for the pre-merger NS~\cite{Oechslin:2006uk}. 
They also determined that the kHz GW emission peaks produced by HMNSs may help to constrain various parameters of the original NS EOS, especially it's high-density behavior~\cite{Oechslin:2007gn}, with further updates to the prediction provided by Bauswein and Janka~\cite{Bauswein:2011tp}.  Most recently, 
Stergioulas et al.~\cite{Stergioulas:2011gd}
studied the effect of non-linear mode couplings in HMNS oscillations, leading to the prediction of a triplet peak of frequencies being present or low mass ($M_{\mathrm{NS}}=1.2-1.35\,M_{\odot}$) systems in the kHz range.

\subsection{Comparison to BH--NS merger results}\label{sec:bhns}

While there is a history of Newtonian, quasi-relativistic, post-Newtonian, and CF gravitational formalisms being used to perform BH--NS merger simulations, their results are nearly always quantitatively, if not qualitatively, different than full GR simulations, and we focus here on the latter (see~\cite{ST_LRR} for a more thorough historical review).
Most of the groups that have performed full GR NS--NS merger calculations have also published results on BH--NS mergers, including Whisky (for head-on collisions)~\cite{Loffler:2006nu},
KT~\cite{Shibata:2006bs,Shibata:2006ks,Shibata:2007zm,Yamamoto:2008js,Shibata:2009cn,Kyutoku:2010zd,Kyutoku:2011vz}, UIUC~\cite{Etienne:2007jg,Etienne:2008re,Etienne:2011ea}, HAD~\cite{Chawla:2010sw}, as well as the SXS collaboration~\cite{Duez:2008rb,Duez:2009yy,Foucart:2010eq,Foucart:2011mz} and Princeton (for elliptical mergers)~\cite{Stephens:2011as,East:2011xa}.  Summarizing the results of these works, we get a rather coherent picture, which we describe below.

The GW signal from BH--NS mergers is somewhat ``cleaner'' than that from
NS--NS mergers, since the disruption of the NS and its accretion by the
BH rapidly terminate the GW emission.  In general, 3PN estimates model
the signal well until tidal effects become important.  The more
compact the NS, the higher the dimensionless ``cutoff frequency''
$M_{\mathrm{tot}}f_{\rm cut}$ at which the GW energy spectrum
plummets, with direct plunges in which the NS is swallowed whole
typically yielding excess power near the frequency maximum from the
final pre-merger burst.  For increasingly prograde BH spins, there is
more excess power over the 3PN prediction at lower frequencies, but
also a lower cutoff frequency marking the plunge (see the discussion
in~\cite{ST_LRR}).  From an observational standpoint, the deviations
from point-mass form become more visible for a higher mass BH--NS
system, because frequencies scale characteristically like the inverse
of the total mass.  The distinction is particularly important for
Advanced LIGO, as systems with $M_{\mathrm{BH}}\gtrsim 3\,M_{\odot}$
typically yield cutoff frequencies within the advanced LIGO band at
source distances of $D\sim 100\mathrm{\ Mpc}$, while for lower-mass
systems the cutoff occurs at or just above the upper end of the
frequency band.  This is significantly different than the situation
for NS--NS mergers, in which the characteristic frequencies
corresponding to the merger itself typically fall at frequencies above
the advanced LIGO high-frequency sensitivity limit, and those
corresponding to remnant oscillations in the range 2-4kHz, which will
prove a challenge even for third-generation GW detectors.

Disk masses for BH--NS mergers were found to be extremely small in the first calculations, all performed using non-spinning BHs~\cite{Shibata:2006bs,Shibata:2006ks,Etienne:2007jg}, but have since been corrected to larger values once more sophisticated grid-based schemes and atmosphere treatments were added to those codes.  More recent results indicate disk masses for reasonable physical parameters can be as large as $0.4 \,M_{\odot}$, for  highly-spinning ($a_{\mathrm{BH}}/M=0.9$) mergers~\cite{Foucart:2010eq}, with values of $0.035-0.05\,M_{\odot}$ characterizing non-spinning models with mass ratios $q\approx 1/5$ \cite{Kyutoku:2011vz}.  Mass loss into a disk is suppressed by misaligned spins, especially for highly-inclined BHs, so the aligned cases should currently be interpreted as upper limits for the disk mass when alignment is varied~\cite{Foucart:2010eq}.
 Overall, disk masses for BH--NS merger remnants are comparable to those from NS--NS merger remnants, and may not be clearly distinguishable from them based solely on the emission properties of the disk.  For BH--NS mergers with mass ratios $q=1/3$ and prograde spins of dimensionless magnitude $a_{\mathrm{BH}}/M=0.5$, the disk parameters found after a run performed with the inclusion of a finite-temperature NS EOS~\cite{Duez:2009yy} indicated that the neutrino luminosity from the disk might be as high as $10^{53}$ erg/s.  While NS--NS merger simulations have led to predictions of neutrino luminosities a few times larger than this, the result does indicate that BH--NS mergers are also plausible SGRB progenitor candidates, possibly with lower characteristic luminosities than bursts resulting from NS--NS mergers.

The role of magnetic fields in BH--NS mergers has only been investigated recently~\cite{Chawla:2010sw,Etienne:2011ea}, in simulations that apply an initially poloidal magnetic field to the NSs in the binary.  Magnetic fields were found to have very little effect on the resulting GW signal and the mass accretion rate for the BH for physically reasonable magnetic field strengths, with visible divergences appearing only for $B\sim 10^{17}$G \cite{Etienne:2011ea}, which is not particularly surprising.  Just as in NS--NS mergers, magnetic fields play very little role during inspiral, and unlike the case of NS--NS mergers there is no opportunity to boost fields at a vortex sheet that forms when the binary makes contact, nor in a HMNS via differential rotation.  While the MRI may be important in determining the thermal evolution and mass accretion rate in a post-merger disk, such effects will likely be observable primarily on longer timescales.

Just as full-GR NS--NS simulations do not  indicate that such mergers are likely sources of the r-process elements we observe in the universe, BH--NS simulations in full GR make the same prediction:  no detectable mass loss from the system whatsoever, at least in the calculations performed to date.  The picture may change when even larger prograde spins are modeled, since this should lead to maximal disk production, or if more detailed microphysical treatments indicate that a significant wind can be generated from either a HMNS or BH disk and unbind astrophysically interesting amounts of material, but neither has been seen in the numerical results to date.

As is seen in NS--NS mergers, the pericenter distance plays a critical role in the evolution of eccentric BH--NS mergers as well.  Large disk masses containing up to $0.3M_\odot$, with an unbound fraction of roughly $0.15M_\odot$, can occur when the periastron separation is located just outside the classical ISCO, with GW signals taking on the characteristic zoom-whirl form predicted for elliptical orbits~\cite{Stephens:2011as}.  In between pericenter passages, radial oscillations of the neutron star produce GW emission at frequencies corresponding to the f-mode for the NS as well\cite{East:2011xa}.

\newpage
\section{Summary and Likely Future Directions}
\label{sec:conclusions}

Returning to the questions posed in the previous section, we can now provide the current state of the field's best answers, though this remains a very active area of research and new results will certainly continue to modify this picture. 

\begin{enumerate}
\item With regard to the final fate of the merger remnant, calculations using full GR are required, but the details of the microphysics do not seem to play a very strong role.  It is now possible to determine whether or not a pair of NSs with given parameters and specified EOS will form a BH or HMNS promptly after merger, and to estimate whether a HMNS will collapse on a dynamical timescale or one of the longer dissipative timescales  (see, e.g.,~\cite{Hotokezaka:2011dh}).  
For NS--NS binaries with sufficiently small masses, it is also possible to  determine quickly whether the remnant mass is below the supramassive limit for which a NS is stabilized against collapse by uniform rotation alone, and thus would be unlikely to collapse, barring a significant amount of fallback accretion, unless pulsar emission or magnetic field coupling to the outer disk reduced the rotation rate below the critical value.  This scenario likely applies only for mergers where the total system mass is relatively small $M_{\mathrm{tot}}\lesssim 2.5-2.6\,M_{\odot}$~\cite{Cook92}, even taking into account the current maximum observed NS mass of $M_{\mathrm{NS}}=1.97\,M_{\odot}$~\cite{Demorest:2010bx}.
Based on the wide arrays  of EOS models already considered, it is entirely possible to infer the likely fate for sets of parameters and/or EOS models that have not yet been simulated, although no one has yet published a ``master equation'' that summarizes all of the current work into a single global form.   While  magnetic fields with realistic magnitudes are unlikely to affect the BH versus HMNS question~\cite{Liu:2008xy,Giacomazzo:2010bx}, finite-temperature effects might play a nontrivial role should NSs be sufficiently hot prior to merger~\cite{Sekiguchi:2011zd} (and see also~\cite{Oechslin:2006uk}).  In the end, by the time the second generation of GW detectors make the first observations of mergers, the high-frequency shot-noise cutoff will prove to be a bigger obstacle to determining the fate of the remnant than any numerical uncertainty.  A schematic diagram showing the possible final fates for 
a NS--NS merger along with the potential EM emission (see Figure~21 of~\cite{Shibata:2006nm}) is shown in Figure~\ref{fig:remnant}. 

\epubtkImage{}{%
\begin{figure}[!ht]
\centerline{\includegraphics[width=0.8\textwidth]{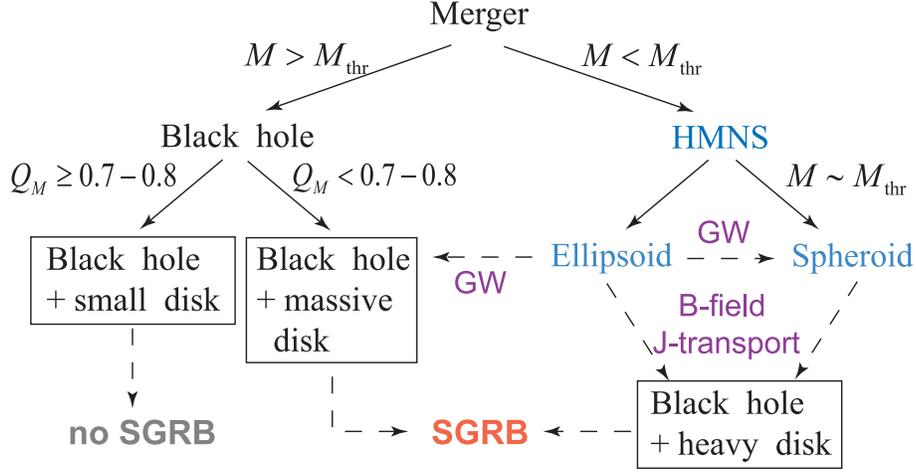}}
\caption{Summary of potential outcomes from NS--NS mergers, taken from~\cite{Shibata:2006nm}.  Here $M_{\mathrm{thr}}$ is the threshold mass (given the EOS) for collapse of a HMNS to a BH, and $Q_M$ is the binary mass ratio.  `Small', `massive', and `heavy' disks imply total disk masses $M_{\mathrm{disk}} \ll 0.01\,M_{\odot}$, $0.01\,M_{\odot}\lesssim M_{\mathrm{disk}}\lesssim 0.03\,M_{\odot}$, and $M_{\mathrm{disk}}\gtrsim0.05\,M_{\odot}$, respectively.  `B-field' and `J-transport' indicate potential mechanisms for the HMNS to eventually lose its differential rotation support and collapse: magnetic damping and angular momentum transport outward into the disk.  Spheroids are likely formed only for the APR and other stiff EOS models that can support remnants with relatively low rotational kinetic energies against collapse.}
\label{fig:remnant}
\end{figure}}

\item GW emission during merger is also well-understood, though there are a few gaps that need to be filled, with full GR again a vital requirement.  While the PN inspiral signal prior to merger is very well understood, finite-size tidal effects introduce complications beyond those seen in BH--BH mergers, yet the longest calculations performed to date~\cite{Baiotti:2011am} encompass fewer orbits prior to merger than the longest BH--BH runs~\cite{Scheel:2008rj}.  As noted in~\cite{Baiotti:2011am} and elsewhere, longer calculations will likely appear over time, helping to refine the prediction for the NS--NS merger GW signal as the binary transitions from a PN phase into one that can only be simulated using full GR, and teasing out the NS physics encoded in the GW signal.
It seems clear from the published work that the emission during the onset of the merger is well-understood, as is the very rapid decay that occurs once the remnant collapses to a BH, either promptly or following some delay.  GW emission from HMNSs has been investigated widely, and there have been correlations established  between properties of the initial binary and the late-stage high-frequency emission (see, e.g.,~\cite{Kiuchi:2010ze,Giacomazzo:2010bx}), but given that magnetic fields, neutrino cooling, and other microphysical effects seem to be important, a great deal of work remains to be done.  Perhaps more importantly, since HMNSs emit radiation at frequencies well beyond the shot-noise limit of even second-generation GW detectors, while the final inspiral occurs near peak sensitivity, it is likely that the first observations of NSs will constrain the nuclear EOS (or perhaps the quark matter EOS~\cite{Bauswein:2009im}) primarily via the detection of small finite-size effects during inspiral.
Since QE calculations are computationally inexpensive compared to numerical merger simulations, there should be much more numerical data available about the inspiral stage than other phases of NS--NS mergers, which should help optimize the inferences to be drawn from future observations.

\item Determining the mass of the thick disk that forms around a NS--NS merger remnant remains a very difficult challenge, since its density is much lower and harder to resolve using either grid-based or particle-based simulations.  The parameterization given by Eq.~\ref{eq:mdisk}~\cite{Rezzolla:2010fd} is generally consistent with the GR calculations of other groups (see, e.g.,~\cite{Hotokezaka:2011dh}), and seems to reflect a current consensus.  It is also clear that disk masses around HMNSs (up to $0.2\,M_{\odot}$) are significantly larger than those forming around prompt collapses, which are limited to about $0.05\,M_{\odot}$. 

\item It is likely that several orders of magnitude more mass-energy are present in the remnant disk than is observed in EM radiation from a SGRB.  Modeling the emission from the disk (and possibly a HMNS) remains extremely challenging.  
Neutrino leakage schemes have been applied in both approximate relativistic calculations ~\cite{Ruffert:2001gf,Rosswog:2003rv} and full GR~\cite{Sekiguchi:2011zd}, and a more complex flux-limited diffusion scheme has been applied to the former as a post-processing step~\cite{Dessart:2008zd}, but there are no calculations that follow in detail the neutrinos as they flow outward, annihilate, and produce observable EM emission.  At present, nuclear reactions are typically not followed in detail; rather, the electron fraction of the nuclear material, $Y_e$, is evolved, and used to calculate neutrino emission and absorption rates.

\item Magnetic fields, on the other hand, are starting to be much better understood.  B-fields do seem to grow quite large through winding effects, even during the limited amount of physical time that can currently be modeled numerically~\cite{Anderson:2008zp,Liu:2008xy,Yamamoto:2008js,Rezzolla:2011da}, with some calculations indicating exponential growth rates.  
The resulting  geometries seem likely to produce the disk/jet structure observed throughout astrophysics when magnetized objects accrete material, which span scales from stellar BHs or pre-main sequence stars all the way up to active galactic nuclei~\cite{Rezzolla:2011da}.

\item While recent numerical simulations have strengthened the case for NS--NS mergers as SGRB progenitors, full GR calculations have not generated much support for the same events yielding significant amounts of r-process elements.  Noting the standard caveat that low-density ejecta are difficult to model, there is still tension between CF calculations producing ejecta with the proper temperatures and masses to reproduce the observed cosmic r-process abundances (see, e.g.,~\cite{Goriely:2011vg}), and full GR calculations that produce almost no measurable unbound material whatsoever.
\end{enumerate}

While NS--NS merger calculations have seen tremendous progress in the past decade, the future remains extremely exciting.  Between the addition of more accurate and realistic physical treatments, the exploration of the full phase space of models, and the linking of numerical relativity to astrophysical observations and GW detection, there remain many unsolved problems that will be attacked over the course of the next decade and beyond.

\section{Acknowledgements}

J.A.F.\ acknowledges support from NASA under  award 08-ATFP-0093 and from NSF Grant PHY-0903782.
F.A.R.\ acknowledges support from NSF Grant PHY-0855592 and thanks the Aspen Center for Physics, supported
by NSF Grant PHY-1066293, for hospitality while this work was being completed.  We thank the referees for their careful reading and helpful suggestions.

\newpage
\appendix
\section{Field evolution equations}\label{app:field}

In the BSSN evolution system, we define the following variables in terms of the
standard ADM 4-metric $g_{ij}$, 3-metric $\gamma_{ij}$, and extrinsic
curvature $K_{ij}$:
\begin{eqnarray}
  \phi & \equiv &\frac{1}{12} \log \left[  \det \gamma_{ij} \right] = \log\psi\,,
  \\
  \tilde\gamma_{ij} & \equiv & e^{-4\phi}\; \gamma_{ij}\,,
  \\
  K & \equiv & g^{ij} K_{ij}\,,
  \\
  \tilde A_{ij} & \equiv & e^{-4\phi} \left[ K_{ij} - \frac{1}{3} g_{ij} K\,,
    \right]
  \\
  \tilde\Gamma^i & \equiv & \tilde\gamma^{jk} \tilde\Gamma^i_{jk} .
\end{eqnarray}
The evolution system consists of 15 equations for the various field terms,
\begin{eqnarray}
  \partial_0 K & = & -\gamma^{ij} \tilde{D}_i \tilde{D}_j \alpha
 + \alpha\left( \tilde{A}^{ij} \tilde{A}_{ij} + \frac{1}{3} K^2 \right) +4\pi \left(E+\gamma^{ij}S_{ij}\right)
  \\
  \partial_0 \phi & = & -\frac{1}{6}\, \left(\alpha K -\partial_k\beta^k\right)
  \\
  \partial_0 \tilde{\gamma}_{ij} & = & -2\alpha\tilde{A}_{ij} 
  + 2\tilde{\gamma}_{k(i}\partial_{j)}\beta^k 
  - \frac{2}{3}\tilde{\gamma}_{ij}\partial_k\beta^k
  \\
  \partial_0 \tilde{A}_{ij} & = & e^{-4\phi}\left[ 
    \alpha\tilde{R}_{ij} + \alpha R^\phi_{ij} - \tilde{D}_i\tilde{D}_j\alpha 
    \right]^{TF}
  \nonumber\\
  & & {} + \alpha K\tilde{A}_{ij} - 2\alpha\tilde{A}_{ik}\tilde{A}^k_{\; j}
  + 2\tilde{A}_{k(i}\partial_{j)}\beta^k 
  - \frac{2}{3}\tilde{A}_{ij}\partial_k\beta^k
  - 8\pi \alpha e^{-4\phi} S^{TF}_{ij}
  \\
  \partial_0\tilde{\Gamma}^i & = & 
   - 2\tilde{A}^{ij}\partial_j\alpha 
  + 2\alpha\left[ \tilde{\Gamma}^i_{\; kl}\tilde{A}^{kl} +
    6\tilde{A}^{ij}\partial_j\phi - \frac{2}{3}\tilde{\gamma}^{ij} K_{,j}
  \right]\nonumber\\
 &&-\tilde{\Gamma}^j\partial_j\beta^i+\frac{2}{3}\tilde{\Gamma}^i\partial_j\beta^j+\frac{1}{3}\tilde{\gamma}^{ik}\beta^j_{,jk}+\tilde{\gamma}^{jk}\beta^i_{,jk}-16\pi\alpha \tilde{\gamma}^{ik}S_k
\end{eqnarray}
where the matter source terms contain various projections of the stress-energy tensor, defined through the relations 
\begin{eqnarray}
  E & \equiv &n^\mu n^\nu T_{\mu\nu} = \frac{1}{\alpha^2} \left( T_{00} - 2 \beta^i T_{0i} +
  \beta^i \beta^j T_{ij} \right),
  \\
   S_i & \equiv & -n^\mu \gamma^\nu_i T_{\mu\nu} = - \frac{1}{\alpha} \left( T_{0i} - \beta^j T_{ij} \right) ,
\\ 
S_{ij} & \equiv & \gamma_i^\mu \gamma_j^\nu T_{\mu\nu}.
\end{eqnarray}
We have introduced the notation $\partial_0 = \partial_t -
\beta^j\partial_j$. All quantities with a tilde involve
the conformal 3-metric $\tilde{\gamma}_{ij}$, which is used to
raise and lower indices. In particular, $\tilde{D}_i$ and
$\tilde{\Gamma}^k_{ij}$ refer to the covariant derivative and the
Christoffel symbols with respect to $\tilde{\gamma}_{ij}$.  Parentheses indicate symmetrization of indices, and the
expression $[ \cdots ]^{TF}$ denotes the trace-free part of the
expression inside the brackets.  In the BSSN approach, the Ricci tensor is typically split into two pieces, whose respective contributions are given by
\begin{eqnarray}
\tilde{R}_{ij} 
 &=& -\frac{1}{2} \tilde{\gamma}^{kl}\partial_k\partial_l\tilde{\gamma}_{ij} 
  + \tilde{\gamma}_{k(i}\partial_{j)}\tilde{\Gamma}^k
  - \partial_k\tilde{\gamma}_{l(i}\partial_{j)}\tilde{\gamma}^{kl}+\frac{1}{2}\tilde{\Gamma}^l\tilde{\gamma}_{ij,l}-\tilde{\Gamma}^l_{ik}\tilde{\Gamma}^k_{jl}
\\
R^\phi_{ij} &=& -2\tilde{D}_i\tilde{D}_j\phi 
  - 2\tilde{\gamma}_{ij} \tilde{D}^k\tilde{D}_k\phi
  + 4\tilde{D}_i\phi\, \tilde{D}_j\phi 
  - 4\tilde{\gamma}_{ij}\tilde{D}^k\phi\, \tilde{D}_k\phi .
\end{eqnarray}

These equations must be supplemented with gauge conditions that determine the evolution of the lapse function $\alpha$ and shift vector $\beta^i$.  Noting that some groups introduce slight variants of these, the moving puncture gauge conditions that have become popular for all GR merger calculations involving BHs and NSs typically take the form
\begin{eqnarray} 
\partial_0\alpha&=&-2\alpha K,\label{eq:lapseevol}\\
\partial_t\beta^i &=& \frac{3}{4} B^i,\label{eq:shiftevol1}\\
\partial_t B^i&=& \partial_t {\tilde{\Gamma}}^i -\eta B^i.\label{eq:shiftevol2}
\end{eqnarray}
where $B^i$ is an intermediate quantity used to convert the second-order ``Gamma-driver'' shift condition into a pair of first-order equations, and $\eta$ is a user-prescribed term used to control dissipation in the simulation.  Note that it is possible to replace  the three instances of $\partial_t$ in the shift evolution equations Eqs.~\ref{eq:shiftevol1} and~\ref{eq:shiftevol2} by the shift-advected time derivative $(\partial_t-\beta^j\partial_j)$ without changing the stability or hyperbolicity properties of the evolution scheme; in both cases moving punctures translate smoothly across a grid over long time periods \cite{vanMeter:2006vi} and both systems are strongly hyperbolic so long as the shift vector does not grow too large within the simulation domain \cite{Gundlach:2006tw}.

The generalized harmonic formulation involves recasting the Einstein field equations, Eq.~\ref{eq:einstein} in the form
\begin{equation}
R_{\mu\nu} = 8\pi\left(T_{\mu\nu}-\frac{1}{2}g_{\mu\nu}T\right)
\end{equation} and, after some tensor algebra, rewriting the Ricci tensor in the form
\begin{equation}
-2R_{\mu\nu} = g^{\gamma\delta}g_{\alpha\beta,\gamma\delta}+{g^{\gamma\delta}}_{,\beta}g_{\alpha\delta,\gamma}+g^{\gamma\delta}_{,\alpha}g_{\beta\delta,\gamma} +2H_{(\alpha,\beta)}-2H_\delta \Gamma^{\delta}_{\alpha\beta}+2\Gamma^{\gamma}_{\delta\beta}\Gamma^{\delta}_{\gamma\alpha}\label{eq:GHRicci}
\end{equation}
The Christoffel coefficients are calculated from the full 4-metric,
\begin{equation}
\Gamma^{\alpha}_{\beta\gamma}\equiv \frac{1}{2}g^{\alpha\delta}\left[g_{\beta\delta,\gamma}+g_{\gamma\delta,\beta} -g_{\beta\gamma,\delta}\right]
\end{equation} 
and the gauge source terms $H^\mu$ are defined in Eq.~\ref{eq:GH_H}.

Given well-posed initial data for the metric and its first time derivative (since the system is second-order in time according to Eq.~\ref{eq:GHRicci}), the evolution of the system may be treated by a first-order reduction that specifies the evolution of
the four functions $H^{\mu}$ along with the spacetime metric $g_{\mu\nu}$, its projected time derivatives $\Pi_{\mu\nu} = -n^\alpha\partial_\alpha g_{\mu\nu}$, and its spatial derivatives
\begin{equation}
\Phi_{i\mu\nu} =    \partial_i g_{\mu\nu}
\end{equation}
subject to a constraint specifying that the derivative terms $\Phi_{i\mu\nu}$ remain consistent with the metric $g_{\mu\nu}$ in time.
In practice, one typically introduces a constraint for the source functions, defining
\begin{equation}
C^\mu = H^\mu - \Box x^\mu
\end{equation}
and then modifies the evolution equation by appending a constraint damping term to the RHS of the stress energy-term (following \cite{Gundlach:2005eh,Pretorius:2005gq}
\begin{equation}
-8\pi(2T_{\mu\nu}-g_{\mu\nu}T)~~\Rightarrow~~ -8\pi(2T_{\mu\nu}-g_{\mu\nu}T) - \kappa(n_{\mu}C_\nu + n_\nu C_\mu - g_{\mu\nu} n^{\delta}C_{\delta})
\end{equation}
where $n_\mu$ is the unit normal vector to the hypersurface (see Eq.~\ref{eq:unitnormal}).
The gauge conditions used in the first successful simulations of merging BH binaries \cite{Pretorius:2005gq} consisted of the set
\begin{equation}
\Box H_t = -\xi_1\frac{\alpha-1}{\alpha^\eta}+\xi_2 H_{t,\nu} n^\nu;~~~H_i=0
\end{equation}
with $\xi_1, \xi_2,\eta$ constant parameters used to tune the evolution.  The first one drives the coordinates toward the ADM form and the latter provides dissipation.  The binary NS--NS merger work of \cite{Anderson:2008zp}  chose harmonic coordinates with $H^\mu=0$.

\section{GR Hydrodynamical and MHD equations}\label{app:hydro}

In what follows, we will adopt the stress energy tensor of an ideal relativistic fluid,
\begin{equation}
\label{eq:Tmunu}
T^{\mu\nu} = \rho h u^\mu u^\nu + g^{\mu\nu} P\,\,,
\end{equation}
where $\rho$, $P$, and $u^\mu$ are the rest mass density, pressure, and fluid 4-velocity, respectively, and 
\begin{equation}
h = 1 + \epsilon + P/\rho\label{eq:enthalpy}
\end{equation}
 is the relativistic specific enthalpy, with $\epsilon$ the specific internal energy of the fluid.

The equations of ideal GR hydrodynamics~\cite{Marti:1991wi} may be 
derived from the local GR conservation laws of mass and
energy-momentum:
\begin{equation}
  \nabla_{\!\mu} J^\mu = 0, \qquad \nabla_{\!\mu} T^{\mu \nu} = 0\,\,,
  \label{eq:equations_of_motion_gr}
\end{equation}
where $ \nabla_{\!\mu} $ denotes the covariant derivative with respect
to the 4-metric, and $ J^{\,\mu} = \rho u^{\,\mu} $ is the mass current.

The 3-velocity $v^i$ can be calculated in the form
\begin{equation}
v^i = \frac{u^i}{W} + \frac{\beta^i}{\alpha}\,\,,
\label{eq:vel}
\end{equation}
where 
\begin{equation}
W \equiv \alpha u^0 = (1-v^i v_i)^{-1/2}\label{eq:Lorentz}
\end{equation}
 is the Lorentz factor.  The contravariant 4-velocity is then given by:
\begin{equation}
u^0  = \frac{W}{\alpha}\,,\qquad
u^i = W \left( v^i - \frac{\beta^i}{\alpha}\right)\,\,,
\end{equation}
and the covariant 4-velocity is:
\begin{equation}
u_0  = W(v^i \beta_i - \alpha)\,,\qquad
u_i = W v_i\,\,.
\end{equation}

To cast the equations of GR hydrodynamics as a first-order hyperbolic
flux-conservative system for the conserved variables
$D$, $S^i$, and $\tau$,  defined in terms of the primitive
variables $\rho, \epsilon, v^i$, we define
\begin{eqnarray}
  D &=& \sqrt{\gamma} \rho W,\label{eq:p2c1}\\
  S^i &=& \sqrt{\gamma} \rho h W^{\,2} v^i,\label{eq:p2c2}\\
  \tau &=& \sqrt{\gamma} \left(\rho h W^{\,2} - P\right) - D\label{eq:p2c3}\,,
\end{eqnarray}
where $ \gamma $ is the determinant of $\gamma_{ij} $.
The evolution system then becomes
\begin{equation}
  \frac{\partial \mathbf{U}}{\partial t} +
  \frac{\partial \mathbf{F}^{\,i}}{\partial x^{\,i}} =
  \mathbf{S}\,\,,
  \label{eq:conservation_equations_gr}
\end{equation}
with
\begin{eqnarray}
  \mathbf{U} & = & [D, S_j, \tau], \nonumber\\
  \mathbf{F}^{\,i} & = & \alpha
  \left[ D \tilde{v}^{\,i}, S_j \tilde{v}^{\,i} + \delta^{\,i}_j P,
  \tau \tilde{v}^{\,i} + P v^{\,i} \right]\!, \nonumber \\
  \mathbf{S} & = & \alpha
  \bigg[ 0, T^{\mu \nu} \left( \frac{\partial g_{\nu j}}{\partial x^{\,\mu}} - 
  \Gamma^{\,\lambda}_{\mu \nu} g_{\lambda j} \right), \nonumber\\
  & &\qquad\alpha \left( T^{\mu 0}
  \frac{\partial \ln \alpha}{\partial x^{\,\mu}} -
  T^{\mu \nu} \Gamma^{\,0}_{\mu \nu} \right) \bigg]\,.
\end{eqnarray}%
Here, $ \tilde{v}^{\,i} = v^{\,i} - \beta^i / \alpha $ and $
\Gamma^{\,\lambda}_{\mu \nu} $ are the 4-Christoffel symbols.  

Magnetic fields may be included in the formalism, in the ideal MHD limit under which we assume infinite conductivity, by adding three new evolution equations and modifying those above to include magnetic stress-energy contributions of the form
\begin{equation}
T_{\mu\nu}^{EM} = b^2\left(u^\mu u^\nu+\frac{1}{2}g^\mu\nu\right)-b^\mu b^\nu
\end{equation}
where the magnetic field seen by a comoving observer, $b^\mu$ is defined in terms of the dual faraday tensor $^*\!F^{\nu\mu}$ by the condition
\begin{equation}
b^\mu = ^*\!F^{\nu\mu} u_\nu
\end{equation}
where $b^2 = b^\mu b_\mu$ represents twice the magnetic pressure.   With magnetic terms included, we may rewrite the stress-energy tensor in a familiar form by introducing magnetically modified pressure and enthalpy contributions:
\begin{equation}
T^{\mu\nu} = \rho h^* u^\mu u^\nu + P^* g^{\mu\nu};~~h^*\equiv 1+\epsilon+\frac{P+b^2}{\rho},~~P^* = P+\frac{b^2}{2}
\end{equation}
and redefine the conserved momentum and energy variables $S^i$ and $\tau$ accordingly:
\begin{eqnarray}
 S^i &=& \sqrt{\gamma} \rho h^* W^{\,2} v^i -\alpha b^0 b^i\\
  \tau &=& \sqrt{\gamma} \left(\rho h^* W^{\,2} - P^*-(\alpha b^0)^2\right) - D
\end{eqnarray}
Defining the (primitive) magnetic field 3-vector as 
\begin{equation}
B^i = - ^*\!F^{\mu i} n_\mu = \alpha~^*\!F^{0i}
\end{equation}
and the conserved variable $\mathcal{B}^i=\sqrt{\gamma} B^i$, which are related to the comoving magnetic field 4-vector $b^\mu$
through the relations 
\begin{eqnarray}
b^0&=&\frac{WB^kv_k}{\alpha}\\
b^i&=&\frac{B^i}{W}+W(B^kv_k)\left(v^i-\frac{\beta^i}{\alpha}\right)\\
b_i&=&\frac{B_i}{W}+W(B^kv_k) v_i\\
b^2&=&\frac{B^iB_i}{W^2}+(B^iv_i)^2,\\
\end{eqnarray}
we may rewrite the conservative evolution scheme in the form
\begin{eqnarray}
  \mathbf{U}  ~=~ & &[D, S_j, \tau,\mathcal{B}^k], \nonumber\\
  \mathbf{F}^{\,i} ~ = ~&& \alpha\times
  \left[\begin{array}{c} D \tilde{v}^{\,i},\\ S_j \tilde{v}^{\,i} + \delta^{\,i}_j P^* - b_j\mathcal{B}^i/W,\\
  \tau \tilde{v}^{\,i} + P^* v^{i}-\alpha b^0 B^i/W,\\ \mathcal{B}^k\tilde{v}^i-\mathcal{B}^i\tilde{v}^k\end{array} \right]\!,  \\
  \mathbf{S} ~ =~ && \alpha\times\left[\begin{array}{c}
  0, \\ T^{\mu \nu} \left( \frac{\partial g_{\nu j}}{\partial x^{\,\mu}} - 
  \Gamma^{\,\lambda}_{\mu \nu} g_{\lambda j} \right), \\
  \qquad\alpha \left( T^{\mu 0}
  \frac{\partial \ln \alpha}{\partial x^{\,\mu}} -
  T^{\mu \nu} \Gamma^{\,0}_{\mu \nu} \right),\\ \vec{0} \end{array}\right]\, ,
\end{eqnarray}
where the magnetic field evolution equation is just the relativistic version of the induction equation.  An external mechanism to enforce the divergence-free nature of the magnetic field, $\partial_i \mathcal{B}^i$ must also be applied.

\bibliography{refs}

\begin{thebibliography}{100}

\bibitem{Abadie:2010cf}
Abadie, J. et~al. (LIGO Scientific Collaboration, Virgo Collaboration),
  ``Predictions for the Rates of Compact Binary Coalescences Observable by
  Ground-based Gravitational-wave Detectors'', {\em Class. Quantum Grav.}, {\bf
  27}, 173001, (2010).
  {\small[\href{http://dx.doi.org/10.1088/0264-9381/27/17/173001}{DOI}]},
  {\small[\href{http://adsabs.harvard.edu/abs/2010CQGra..27q3001A}{ADS}]},
  {\small[\href{http://arxiv.org/abs/1003.2480}{{arXiv:1003.2480
  {\small[astro-ph.HE]}}}]}.

\bibitem{Abbott:2009tt}
Abbott, B.P. et~al. (LIGO Scientific Collaboration), ``Search for Gravitational
  Waves from Low Mass Binary Coalescences in the First Year of LIGO's S5
  Data'', {\em Phys. Rev. D}, {\bf 79}, 122001, (2009).
  {\small[\href{http://dx.doi.org/10.1103/PhysRevD.79.122001}{DOI}]},
  {\small[\href{http://adsabs.harvard.edu/abs/2009PhRvD..79l2001A}{ADS}]},
  {\small[\href{http://arxiv.org/abs/0901.0302}{{arXiv:0901.0302
  {\small[gr-qc]}}}]}.

\bibitem{Akmal:1998cf}
Akmal, A., Pandharipande, V.R.  and Ravenhall, D.G., ``The Equation of state of
  nucleon matter and neutron star structure'', {\em Phys. Rev. C}, {\bf 58},
  1804--1828, (1998).
  {\small[\href{http://dx.doi.org/10.1103/PhysRevC.58.1804}{DOI}]},
  {\small[\href{http://adsabs.harvard.edu/abs/1998PhRvC..58.1804A}{ADS}]},
  {\small[\href{http://arxiv.org/abs/nucl-th/9804027}{{arXiv:nucl-th/9804027
  {\small[nucl-th]}}}]}.

\bibitem{Alcubierre:book}
{Alcubierre}, M., {\em {Introduction to 3+1 Numerical Relativity}}, (Oxford
  University Press, Oxford, UK, 2008).
  {\small[\href{http://adsabs.harvard.edu/abs/2008itnr.book.....A}{ADS}]}.

\bibitem{Alford:2004pf}
Alford, M., Braby, M., Paris, M.W.  and Reddy, S., ``Hybrid stars that
  masquerade as neutron stars'', {\em Astrophys. J.}, {\bf 629}, 969--978,
  (2005). {\small[\href{http://dx.doi.org/10.1086/430902}{DOI}]},
  {\small[\href{http://adsabs.harvard.edu/abs/2005ApJ...629..969A}{ADS}]},
  {\small[\href{http://arxiv.org/abs/nucl-th/0411016}{{arXiv:nucl-th/0411016
  {\small[nucl-th]}}}]}.

\bibitem{Anderson:2006ay}
Anderson, M., Hirschmann, E., Liebling, S.L.  and Neilsen, D., ``Relativistic
  MHD with Adaptive Mesh Refinement'', {\em Class. Quantum Grav.}, {\bf 23},
  6503--6524, (2006).
  {\small[\href{http://dx.doi.org/10.1088/0264-9381/23/22/025}{DOI}]},
  {\small[\href{http://adsabs.harvard.edu/abs/2006CQGra..23.6503A}{ADS}]},
  {\small[\href{http://arxiv.org/abs/gr-qc/0605102}{{arXiv:gr-qc/0605102
  {\small[gr-qc]}}}]}.

\bibitem{Anderson:2008zp}
{Anderson}, M., {Hirschmann}, E.~W., {Lehner}, L., {Liebling}, S.~L., {Motl},
  P.~M., {Neilsen}, D., {Palenzuela}, C.  and {Tohline}, J.~E., ``Magnetized
  Neutron Star Mergers and Gravitational Wave Signals'', {\em Phys. Rev.
  Lett.}, {\bf 100}, 191101, (2008).
  {\small[\href{http://dx.doi.org/10.1103/PhysRevLett.100.191101}{DOI}]},
  {\small[\href{http://adsabs.harvard.edu/abs/2008PhRvL.100s1101A}{ADS}]},
  {\small[\href{http://arxiv.org/abs/0801.4387}{{arXiv:0801.4387
  {\small[gr-qc]}}}]}.

\bibitem{Anderson:2007kz}
{Anderson}, M., {Hirschmann}, E.~W., {Lehner}, L., {Liebling}, S.~L., {Motl},
  P.~M., {Neilsen}, D., {Palenzuela}, C.  and {Tohline}, J.~E., ``Simulating
  binary neutron stars: Dynamics and gravitational waves'', {\em Phys. Rev. D},
  {\bf 77}, 024006, (2008).
  {\small[\href{http://dx.doi.org/10.1103/PhysRevD.77.024006}{DOI}]},
  {\small[\href{http://adsabs.harvard.edu/abs/2008PhRvD..77b4006A}{ADS}]},
  {\small[\href{http://arxiv.org/abs/0708.2720}{{arXiv:0708.2720
  {\small[gr-qc]}}}]}.

\bibitem{Arnowitt:1962hi}
Arnowitt, R., Deser, S.  and Misner, C.W., ``The dynamics of general
  relativity'', in Witten, L., ed., {\em Gravitation: An Introduction to
  Current Research}, pp. 227--265, (Wiley, New York; London, 1962).
  {\small[\href{http://dx.doi.org/10.1007/s10714-008-0661-1}{DOI}]},
  {\small[\href{http://adsabs.harvard.edu/abs/2008GReGr..40.1997A}{ADS}]},
  {\small[\href{http://arxiv.org/abs/gr-qc/0405109}{{arXiv:gr-qc/0405109
  {\small[gr-qc]}}}]}.

\bibitem{Ayal:1999wn}
Ayal, S., Piran, T., Oechslin, R., Davies, M.B.  and Rosswog, S.,
  ``PostNewtonian SPH'', {\em Astrophys. J.}, {\bf 550}, 846--859, (2001).
  {\small[\href{http://dx.doi.org/10.1086/319769}{DOI}]},
  {\small[\href{http://adsabs.harvard.edu/abs/2001ApJ...550..846A}{ADS}]},
  {\small[\href{http://arxiv.org/abs/astro-ph/9910154}{{arXiv:astro-ph/9910154
  {\small[astro-ph]}}}]}.

\bibitem{Babak:2006ty}
Babak, S., Balasubramanian, R., Churches, D., Cokelaer, T.  and Sathyaprakash,
  B.S., ``A Template bank to search for gravitational waves from inspiralling
  compact binaries. I. Physical models'', {\em Class. Quantum Grav.}, {\bf 23},
  5477--5504, (2006).
  {\small[\href{http://dx.doi.org/10.1088/0264-9381/23/18/002}{DOI}]},
  {\small[\href{http://adsabs.harvard.edu/abs/2006CQGra..23.5477B}{ADS}]},
  {\small[\href{http://arxiv.org/abs/gr-qc/0604037}{{arXiv:gr-qc/0604037
  {\small[gr-qc]}}}]}.

\bibitem{Babiuc:2006ik}
Babiuc, M.C., Kreiss, H.-O.  and Winicour, J., ``Constraint-preserving
  Sommerfeld conditions for the harmonic Einstein equations'', {\em Phys. Rev.
  D}, {\bf 75}, 044002, (2007).
  {\small[\href{http://dx.doi.org/10.1103/PhysRevD.75.044002}{DOI}]},
  {\small[\href{http://adsabs.harvard.edu/abs/2007PhRvD..75d4002B}{ADS}]},
  {\small[\href{http://arxiv.org/abs/gr-qc/0612051}{{arXiv:gr-qc/0612051
  {\small[gr-qc]}}}]}.

\bibitem{Baierlein:1962zz}
Baierlein, R.F., Sharp, D.H.  and Wheeler, J.A., ``Three-Dimensional Geometry
  as Carrier of Information about Time'', {\em Phys. Rev.}, {\bf 126},
  1864--1865, (1962).
  {\small[\href{http://dx.doi.org/10.1103/PhysRev.126.1864}{DOI}]},
  {\small[\href{http://adsabs.harvard.edu/abs/1962PhRv..126.1864B}{ADS}]}.

\bibitem{Baiotti:2010xh}
Baiotti, L., Damour, T., Giacomazzo, B., Nagar, A.  and Rezzolla, L.,
  ``Analytic modelling of tidal effects in the relativistic inspiral of binary
  neutron stars'', {\em Phys. Rev. Lett.}, {\bf 105}, 261101, (2010).
  {\small[\href{http://dx.doi.org/10.1103/PhysRevLett.105.261101}{DOI}]},
  {\small[\href{http://adsabs.harvard.edu/abs/2010PhRvL.105z1101B}{ADS}]},
  {\small[\href{http://arxiv.org/abs/1009.0521}{{arXiv:1009.0521
  {\small[gr-qc]}}}]}.

\bibitem{Baiotti:2011am}
Baiotti, L., Damour, T., Giacomazzo, B., Nagar, A.  and Rezzolla, L.,
  ``Accurate numerical simulations of inspiralling binary neutron stars and
  their comparison with effective-one-body analytical models'', {\em Phys. Rev.
  D}, {\bf 84}, 024017, (2011).
  {\small[\href{http://dx.doi.org/10.1103/PhysRevD.84.024017}{DOI}]},
  {\small[\href{http://adsabs.harvard.edu/abs/2011PhRvD..84b4017B}{ADS}]},
  {\small[\href{http://arxiv.org/abs/1103.3874}{{arXiv:1103.3874
  {\small[gr-qc]}}}]}.

\bibitem{Baiotti:2006wn}
Baiotti, L., De~Pietri, R., Manca, G.M.  and Rezzolla, L., ``Accurate
  simulations of the dynamical barmode instability in full General
  Relativity'', {\em Phys. Rev. D}, {\bf 75}, 044023, (2007).
  {\small[\href{http://dx.doi.org/10.1103/PhysRevD.75.044023}{DOI}]},
  {\small[\href{http://adsabs.harvard.edu/abs/2007PhRvD..75d4023B}{ADS}]},
  {\small[\href{http://arxiv.org/abs/astro-ph/0609473}{{arXiv:astro-ph/0609473
  {\small[astro-ph]}}}]}.

\bibitem{Baiotti:2008ra}
Baiotti, L., Giacomazzo, B.  and Rezzolla, L., ``Accurate evolutions of
  inspiralling neutron-star binaries: prompt and delayed collapse to black
  hole'', {\em Phys. Rev. D}, {\bf 78}, 084033, (2008).
  {\small[\href{http://dx.doi.org/10.1103/PhysRevD.78.084033}{DOI}]},
  {\small[\href{http://adsabs.harvard.edu/abs/2008PhRvD..78h4033B}{ADS}]},
  {\small[\href{http://arxiv.org/abs/0804.0594}{{arXiv:0804.0594
  {\small[gr-qc]}}}]}.

\bibitem{Baiotti:2009gk}
Baiotti, L., Giacomazzo, B.  and Rezzolla, L., ``Accurate evolutions of
  inspiralling neutron-star binaries: assessment of the truncation error'',
  {\em Class. Quantum Grav.}, {\bf 26}, 114005, (2009).
  {\small[\href{http://dx.doi.org/10.1088/0264-9381/26/11/114005}{DOI}]},
  {\small[\href{http://adsabs.harvard.edu/abs/2009CQGra..26k4005B}{ADS}]},
  {\small[\href{http://arxiv.org/abs/0901.4955}{{arXiv:0901.4955
  {\small[gr-qc]}}}]}.

\bibitem{Baiotti:2010zf}
Baiotti, L., Hawke, I., Montero, P.J.  and Rezzolla, L., ``A new
  three-dimensional general-relativistic hydrodynamics code'', {\em Mem. Soc.
  Astron. Ital. Supplement}, {\bf 1}, S210--219, (2003).
  {\small[\href{http://adsabs.harvard.edu/abs/2003MSAIS...1..210B}{ADS}]},
  {\small[\href{http://arxiv.org/abs/1004.3849}{{arXiv:1004.3849
  {\small[gr-qc]}}}]}.

\bibitem{Baiotti:2010ka}
Baiotti, L., Shibata, M.  and Yamamoto, T., ``Binary neutron-star mergers with
  Whisky and SACRA: First quantitative comparison of results from independent
  general-relativistic hydrodynamics codes'', {\em Phys. Rev. D}, {\bf 82},
  064015, (2010).
  {\small[\href{http://dx.doi.org/10.1103/PhysRevD.82.064015}{DOI}]},
  {\small[\href{http://adsabs.harvard.edu/abs/2010PhRvD..82f4015B}{ADS}]},
  {\small[\href{http://arxiv.org/abs/1007.1754}{{arXiv:1007.1754
  {\small[gr-qc]}}}]}.

\bibitem{Baker:2001sf}
{Baker}, J., {Campanelli}, M.  and {Lousto}, C.~O., ``{The Lazarus project: A
  Pragmatic approach to binary black hole evolutions}'', {\em Phys. Rev. D},
  {\bf 65}, 044001, (2002).
  {\small[\href{http://dx.doi.org/10.1103/PhysRevD.65.044001}{DOI}]},
  {\small[\href{http://adsabs.harvard.edu/abs/2002PhRvD..65d4001B}{ADS}]},
  {\small[\href{http://arxiv.org/abs/gr-qc/0104063}{{arXiv:gr-qc/0104063
  {\small[gr-qc]}}}]}.

\bibitem{Baker:2005vv}
Baker, J.G., Centrella, J., Choi, D.-I., Koppitz, M.  and van Meter, J.,
  ``Gravitational wave extraction from an inspiraling configuration of merging
  black holes'', {\em Phys. Rev. Lett.}, {\bf 96}, 111102, (2006).
  {\small[\href{http://dx.doi.org/10.1103/PhysRevLett.96.111102}{DOI}]},
  {\small[\href{http://adsabs.harvard.edu/abs/2006PhRvL..96k1102B}{ADS}]},
  {\small[\href{http://arxiv.org/abs/gr-qc/0511103}{{arXiv:gr-qc/0511103
  {\small[gr-qc]}}}]}.

\bibitem{Balberg:1997yw}
Balberg, S.  and Gal, A., ``An Effective equation of state for dense matter
  with strangeness'', {\em Nucl. Phys. A}, {\bf 625}, 435--472, (1997).
  {\small[\href{http://dx.doi.org/10.1016/S0375-9474(97)81465-0}{DOI}]},
  {\small[\href{http://adsabs.harvard.edu/abs/1997NuPhA.625..435B}{ADS}]},
  {\small[\href{http://arxiv.org/abs/nucl-th/9704013}{{arXiv:nucl-th/9704013
  {\small[nucl-th]}}}]}.

\bibitem{Balsara:2003ui}
Balsara, D.S.  and Kim, J., ``A Comparison between Divergence-Cleaning and
  Staggered-Mesh Formulations for Numerical Magnetohydrodynamics'', {\em
  Astrophys. J.}, {\bf 602}, 1079--1090, (2004).
  {\small[\href{http://dx.doi.org/10.1086/381051}{DOI}]},
  {\small[\href{http://adsabs.harvard.edu/abs/2004ApJ...602.1079B}{ADS}]},
  {\small[\href{http://arxiv.org/abs/astro-ph/0310728}{{astro-ph/0310728}}]}.

\bibitem{Baumgarte:1997xi}
Baumgarte, T.W., Cook, G.B., Scheel, M.A., Shapiro, S.L.  and Teukolsky, S.A.,
  ``Binary neutron stars in general relativity: Quasiequilibrium models'', {\em
  Phys. Rev. Lett.}, {\bf 79}, 1182--1185, (1997).
  {\small[\href{http://dx.doi.org/10.1103/PhysRevLett.79.1182}{DOI}]},
  {\small[\href{http://adsabs.harvard.edu/abs/1997PhRvL..79.1182B}{ADS}]},
  {\small[\href{http://arxiv.org/abs/gr-qc/9704024}{{arXiv:gr-qc/9704024
  {\small[gr-qc]}}}]}.

\bibitem{Baumgarte:1997eg}
Baumgarte, T.W., Cook, G.B., Scheel, M.A., Shapiro, S.L.  and Teukolsky, S.A.,
  ``General relativistic models of binary neutron stars in quasiequilibrium'',
  {\em Phys. Rev. D}, {\bf 57}, 7299--7311, (1998).
  {\small[\href{http://dx.doi.org/10.1103/PhysRevD.57.7299}{DOI}]},
  {\small[\href{http://adsabs.harvard.edu/abs/1998PhRvD..57.7299B}{ADS}]},
  {\small[\href{http://arxiv.org/abs/gr-qc/9709026}{{arXiv:gr-qc/9709026
  {\small[gr-qc]}}}]}.

\bibitem{Baumgarte:1998te}
Baumgarte, T.W.  and Shapiro, S.L., ``On the numerical integration of
  Einstein's field equations'', {\em Phys. Rev. D}, {\bf 59}, 024007, (1999).
  {\small[\href{http://dx.doi.org/10.1103/PhysRevD.59.024007}{DOI}]},
  {\small[\href{http://adsabs.harvard.edu/abs/1999PhRvD..59b4007B}{ADS}]},
  {\small[\href{http://arxiv.org/abs/gr-qc/9810065}{{arXiv:gr-qc/9810065
  {\small[gr-qc]}}}]}.

\bibitem{Baumgarte:2002jm}
Baumgarte, T.W.  and Shapiro, S.L., ``Numerical relativity and compact
  binaries'', {\em Phys. Rept.}, {\bf 376}, 41--131, (2003).
  {\small[\href{http://dx.doi.org/10.1016/S0370-1573(02)00537-9}{DOI}]},
  {\small[\href{http://adsabs.harvard.edu/abs/2003PhR...376...41B}{ADS}]},
  {\small[\href{http://arxiv.org/abs/gr-qc/0211028}{{arXiv:gr-qc/0211028
  {\small[gr-qc]}}}]}.

\bibitem{Baumgarte:2009fw}
Baumgarte, T.W.  and Shapiro, S.L., ``A Formalism for the construction of
  binary neutron stars with arbitrary circulation'', {\em Phys. Rev. D}, {\bf
  80}, 064009, (2009).
  {\small[\href{http://dx.doi.org/10.1103/PhysRevD.80.064009}{DOI}]},
  {\small[\href{http://adsabs.harvard.edu/abs/2009PhRvD..80f4009B}{ADS}]},
  {\small[\href{http://arxiv.org/abs/0909.0952}{{arXiv:0909.0952
  {\small[gr-qc]}}}]}.

\bibitem{Baumgarte:1999cq}
Baumgarte, T.W., Shapiro, S.L.  and Shibata, M., ``On the maximum mass of
  differentially rotating neutron stars'', {\em Astrophys. J. Lett.}, {\bf
  528}, L29--L32, (2000).
  {\small[\href{http://dx.doi.org/10.1086/312425}{DOI}]},
  {\small[\href{http://adsabs.harvard.edu/abs/2000ApJ...528L..29B}{ADS}]},
  {\small[\href{http://arxiv.org/abs/astro-ph/9910565}{{arXiv:astro-ph/9910565
  {\small[astro-ph]}}}]}.

\bibitem{BaumgarteShapiro:book}
{Baumgarte}, T.~W.  and {Shapiro}, S.~L., {\em {Numerical Relativity: Solving
  Einstein's Equations on the Computer}}, (Cambridge University Press, New
  York, 2010).
  {\small[\href{http://adsabs.harvard.edu/abs/2010nure.book.....B}{ADS}]}.

\bibitem{Bauswein:2011tp}
Bauswein, A.  and Janka, H.-T., ``Measuring neutron-star properties via
  gravitational waves from binary mergers'', {\em Phys. Rev. Lett.}, {\bf
  108}(1), 011101, 011101, (January 2012).
  {\small[\href{http://dx.doi.org/10.1103/PhysRevLett.108.011101}{DOI}]},
  {\small[\href{http://adsabs.harvard.edu/abs/2012PhRvL.108a1101B}{ADS}]},
  {\small[\href{http://arxiv.org/abs/1106.1616}{{arXiv:1106.1616
  {\small[astro-ph.SR]}}}]}.

\bibitem{Bauswein:2010dn}
Bauswein, A., Janka, H.-T.  and Oechslin, R., ``Testing Approximations of
  Thermal Effects in Neutron Star Merger Simulations'', {\em Phys. Rev. D},
  {\bf 82}, 084043, (2010).
  {\small[\href{http://dx.doi.org/10.1103/PhysRevD.82.084043}{DOI}]},
  {\small[\href{http://adsabs.harvard.edu/abs/2010PhRvD..82h4043B}{ADS}]},
  {\small[\href{http://arxiv.org/abs/1006.3315}{{arXiv:1006.3315
  {\small[astro-ph.SR]}}}]}.

\bibitem{Bauswein:2008gx}
{Bauswein}, A., {Janka}, H.-T., {Oechslin}, R., {Pagliara}, G., {Sagert}, I.,
  {Schaffner-Bielich}, J., {Hohle}, M.~M.  and {Neuh{\"a}user}, R., ``Mass
  Ejection by Strange Star Mergers and Observational Implications'', {\em Phys.
  Rev. Lett.}, {\bf 103}, 011101, (2009).
  {\small[\href{http://dx.doi.org/10.1103/PhysRevLett.103.011101}{DOI}]},
  {\small[\href{http://adsabs.harvard.edu/abs/2009PhRvL.103a1101B}{ADS}]},
  {\small[\href{http://arxiv.org/abs/0812.4248}{{arXiv:0812.4248
  {\small[astro-ph]}}}]}.

\bibitem{Bauswein:2009im}
Bauswein, A., Oechslin, R.  and Janka, H.-T., ``Discriminating Strange Star
  Mergers from Neutron Star Mergers by Gravitational-Wave Measurements'', {\em
  Phys. Rev. D}, {\bf 81}, 024012, (2010).
  {\small[\href{http://dx.doi.org/10.1103/PhysRevD.81.024012}{DOI}]},
  {\small[\href{http://adsabs.harvard.edu/abs/2010PhRvD..81b4012B}{ADS}]},
  {\small[\href{http://arxiv.org/abs/0910.5169}{{arXiv:0910.5169
  {\small[astro-ph.SR]}}}]}.

\bibitem{Bejger:2004zx}
Bejger, M., Gondek-Rosinska, D., Gourgoulhon, E., Haensel, P., Taniguchi, K.
  and Zdunik, J.L., ``Impact of the nuclear equation of state on the last
  orbits of binary neutron stars'', {\em Astron. Astrophys.}, {\bf 431},
  297--306, (2005).
  {\small[\href{http://dx.doi.org/10.1051/0004-6361:20041441}{DOI}]},
  {\small[\href{http://adsabs.harvard.edu/abs/2005A%26A...431..297B}{ADS}]},
  {\small[\href{http://arxiv.org/abs/astro-ph/0406234}{{arXiv:astro-ph/0406234
  {\small[astro-ph]}}}]}.

\bibitem{Belczynski:2005mr}
Belczynski, K., Kalogera, V., Rasio, F.A., Taam, R.E., Zezas, A., Bulik, T.,
  Maccarone, T.J.  and Ivanova, N., ``Compact Object Modeling with the
  StarTrack Population Synthesis Code'', {\em Astrophys. J. Suppl. Ser.}, {\bf
  174}, 223--260, (2008).
  {\small[\href{http://dx.doi.org/10.1086/521026}{DOI}]},
  {\small[\href{http://adsabs.harvard.edu/abs/2008ApJS..174..223B}{ADS}]},
  {\small[\href{http://arxiv.org/abs/astro-ph/0511811}{{arXiv:astro-ph/0511811%
}}]}.

\bibitem{Belczynski:2007xg}
Belczynski, K., Taam, R.E., Rantsiou, E.  and van~der Sluys, M., ``Black Hole
  Spin Evolution: Implications for Short-hard Gamma Ray Bursts and
  Gravitational Wave Detection'', {\em Astrophys. J.}, {\bf 682}, 474--486,
  (2008).
  {\small[\href{http://adsabs.harvard.edu/abs/2008ApJ...682..474B}{ADS}]},
  {\small[\href{http://arxiv.org/abs/astro-ph/0703131}{{arXiv:astro-ph/0703131
  {\small[astro-ph]}}}]}.

\bibitem{Belczynski:2009nx}
Belczynski, K.  and Ziolkowski, J., ``On the Apparent Lack of Be X-ray Binaries
  with Black Holes'', {\em Astrophys. J.}, {\bf 707}, 870--877, (2009).
  {\small[\href{http://dx.doi.org/10.1088/0004-637X/707/2/870}{DOI}]},
  {\small[\href{http://adsabs.harvard.edu/abs/2009ApJ...707..870B}{ADS}]},
  {\small[\href{http://arxiv.org/abs/0907.4990}{{arXiv:0907.4990
  {\small[astro-ph.GA]}}}]}.

\bibitem{Berger:2005rv}
Berger, E. {et~al.}, ``The afterglow and elliptical host galaxy of the short
  gamma-ray burst GRB 050724'', {\em Nature}, {\bf 438}, 988--990, (2005).
  {\small[\href{http://dx.doi.org/10.1038/nature04238}{DOI}]},
  {\small[\href{http://adsabs.harvard.edu/abs/2005Natur.438..988B}{ADS}]},
  {\small[\href{http://arxiv.org/abs/astro-ph/0508115}{{arXiv:astro-ph/0508115
  {\small[astro-ph]}}}]}.

\bibitem{Bernuzzi:2011aq}
{Bernuzzi}, S., {Thierfelder}, M.  and {Bruegmann}, B., ``{Accuracy of
  numerical relativity waveforms from binary neutron star mergers and their
  comparison with post-Newtonian waveforms}'', (2011).
  {\small[\href{http://adsabs.harvard.edu/abs/2011arXiv1109.3611B}{ADS}]},
  {\small[\href{http://arxiv.org/abs/1109.3611}{{arXiv:1109.3611
  {\small[gr-qc]}}}]}.

\bibitem{Bethe:2005ju}
Bethe, H.A., Brown, G.E.  and Lee, C.-H., ``Evolution and merging of binaries
  with compact objects'', {\em Phys. Rept.}, {\bf 442}, 5--22, (2007).
  {\small[\href{http://dx.doi.org/10.1016/j.physrep.2007.02.004}{DOI}]},
  {\small[\href{http://adsabs.harvard.edu/abs/2007PhR...442....5B}{ADS}]},
  {\small[\href{http://arxiv.org/abs/astro-ph/0510379}{{arXiv:astro-ph/0510379
  {\small[astro-ph]}}}]}.

\bibitem{Beyer:2004sv}
Beyer, H.R.  and Sarbach, O., ``{On the well posedness of the
  Baumgarte-Shapiro-Shibata-Nakamura formulation of Einstein's field
  equations}'', {\em Phys. Rev. D}, {\bf 70}, 104004, (2004).
  {\small[\href{http://dx.doi.org/10.1103/PhysRevD.70.104004}{DOI}]},
  {\small[\href{http://adsabs.harvard.edu/abs/2004PhRvD..70j4004B}{ADS}]},
  {\small[\href{http://arxiv.org/abs/gr-qc/0406003}{{arXiv:gr-qc/0406003
  {\small[gr-qc]}}}]}.

\bibitem{Bhattacharya91}
Bhattacharya, D.  and van~den Heuvel, E.P.J., ``Formation and evolution of
  binary and millisecond radio pulsars'', {\em Phys. Rept.}, {\bf 203}, 1--124,
  (1991). {\small[\href{http://dx.doi.org/10.1016/0370-1573(91)90064-S}{DOI}]},
  {\small[\href{http://adsabs.harvard.edu/abs/1991PhR...203....1B}{ADS}]}.

\bibitem{Bildsten:1992my}
Bildsten, L.  and Cutler, C., ``Tidal interactions of inspiraling compact
  binaries'', {\em Astrophys. J.}, {\bf 400}, 175--180, (1992).
  {\small[\href{http://dx.doi.org/10.1086/171983}{DOI}]},
  {\small[\href{http://adsabs.harvard.edu/abs/1992ApJ...400..175B}{ADS}]}.

\bibitem{Binnington:2009bb}
Binnington, T.  and Poisson, E., ``{Relativistic theory of tidal Love
  numbers}'', {\em Phys. Rev. D}, {\bf 80}, 084018, (2009).
  {\small[\href{http://dx.doi.org/10.1103/PhysRevD.80.084018}{DOI}]},
  {\small[\href{http://adsabs.harvard.edu/abs/2009PhRvD..80h4018B}{ADS}]},
  {\small[\href{http://arxiv.org/abs/0906.1366}{{arXiv:0906.1366
  {\small[gr-qc]}}}]}.

\bibitem{Blanchet:2006zz}
Blanchet, L., ``Gravitational radiation from post-Newtonian sources and
  inspiralling compact binaries'', {\em Living Rev. Relativity}, {\bf 9},
  lrr-2006-4, (2006).
  {\small[\href{http://adsabs.harvard.edu/abs/2006LRR.....9....4B}{ADS}]}. URL
  (accessed 30 March 2012):
  \newline\url{http://www.livingreviews.org/lrr-2006-4}.

\bibitem{Blanchet:1989fg}
Blanchet, L., Damour, T.  and Sch\"afer, G., ``Post-Newtonian hydrodynamics and
  post-Newtonian gravitational wave generation for numerical relativity'', {\em
  Mon. Not. R. Astron. Soc.}, {\bf 242}, 289--305, (1990).
  {\small[\href{http://adsabs.harvard.edu/abs/1990MNRAS.242..289B}{ADS}]}.

\bibitem{Bona:2010wn}
Bona, C.  and Bona-Casas, C., ``Constraint-preserving boundary conditions in
  the 3+1 first-order approach'', {\em Phys. Rev. D}, {\bf 82}, 064008, (2010).
  {\small[\href{http://dx.doi.org/10.1103/PhysRevD.82.064008}{DOI}]},
  {\small[\href{http://adsabs.harvard.edu/abs/2010PhRvD..82f4008B}{ADS}]},
  {\small[\href{http://arxiv.org/abs/1003.3328}{{arXiv:1003.3328
  {\small[gr-qc]}}}]}.

\bibitem{Bonazzola:2003dm}
Bonazzola, S., Gourgoulhon, E., Grandcl{\'{e}}ment, P.  and Novak, J., ``A
  Constrained scheme for Einstein equations based on Dirac gauge and spherical
  coordinates'', {\em Phys. Rev. D}, {\bf 70}, 104007, (2004).
  {\small[\href{http://dx.doi.org/10.1103/PhysRevD.70.104007}{DOI}]},
  {\small[\href{http://adsabs.harvard.edu/abs/2004PhRvD..70j4007B}{ADS}]},
  {\small[\href{http://arxiv.org/abs/gr-qc/0307082}{{arXiv:gr-qc/0307082
  {\small[gr-qc]}}}]}.

\bibitem{Bonazzola:1998yq}
Bonazzola, S., Gourgoulhon, E.  and Marck, J.-A., ``Numerical models of
  irrotational binary neutron stars in general relativity'', {\em Phys. Rev.
  Lett.}, {\bf 82}, 892--895, (1999).
  {\small[\href{http://dx.doi.org/10.1103/PhysRevLett.82.892}{DOI}]},
  {\small[\href{http://adsabs.harvard.edu/abs/1999PhRvL..82..892B}{ADS}]},
  {\small[\href{http://arxiv.org/abs/gr-qc/9810072}{{arXiv:gr-qc/9810072
  {\small[gr-qc]}}}]}.

\bibitem{Bowen:1980yu}
{Bowen}, J.~M.  and {York}, Jr., J.~W., ``{Time asymmetric initial data for
  black holes and black hole collisions}'', {\em Phys. Rev. D}, {\bf 21},
  2047--2056, (1980).
  {\small[\href{http://dx.doi.org/10.1103/PhysRevD.21.2047}{DOI}]},
  {\small[\href{http://adsabs.harvard.edu/abs/1980PhRvD..21.2047B}{ADS}]}.

\bibitem{Boyle:2007ft}
{Boyle}, M., {Brown}, D.~A., {Kidder}, L.~E., {Mrou{\'e}}, A.~H., {Pfeiffer},
  H.~P., {Scheel}, M.~A., {Cook}, G.~B.  and {Teukolsky}, S.~A.,
  ``{High-accuracy comparison of numerical relativity simulations with
  post-Newtonian expansions}'', {\em Phys. Rev. D}, {\bf 76}, 124038, (2007).
  {\small[\href{http://dx.doi.org/10.1103/PhysRevD.76.124038}{DOI}]},
  {\small[\href{http://adsabs.harvard.edu/abs/2007PhRvD..76l4038B}{ADS}]},
  {\small[\href{http://arxiv.org/abs/0710.0158}{{arXiv:0710.0158
  {\small[gr-qc]}}}]}.

\bibitem{Brown:2008cn}
Brown, G.E., Lee, C.-H.  and Rho, M., ``Kaon Condensation, Black Holes and
  Cosmological Natural Selection'', {\em Phys. Rev. Lett.}, {\bf 101}, 091101,
  (2008).
  {\small[\href{http://dx.doi.org/10.1103/PhysRevLett.101.091101}{DOI}]},
  {\small[\href{http://adsabs.harvard.edu/abs/2008PhRvL.101i1101B}{ADS}]},
  {\small[\href{http://arxiv.org/abs/0802.2997}{{arXiv:0802.2997
  {\small[hep-ph]}}}]}.

\bibitem{Bruegmann:1999we}
Bruegmann, B., ``Numerical relativity in (3+1)-dimensions'', {\em Ann. Phys.},
  {\bf 9}, 227--246, (2000).
  {\small[\href{http://adsabs.harvard.edu/abs/2000AnP...512..227B}{ADS}]},
  {\small[\href{http://arxiv.org/abs/gr-qc/9912009}{{arXiv:gr-qc/9912009
  {\small[gr-qc]}}}]}.

\bibitem{Buonanno:2005pt}
Buonanno, A., Chen, Y., Pan, Y., Tagoshi, H.  and Vallisneri, M., ``Detecting
  gravitational waves from precessing binaries of spinning compact objects. II.
  Search implementation for low-mass binaries'', {\em Phys. Rev. D}, {\bf 72},
  084027, (2005).
  {\small[\href{http://dx.doi.org/10.1103/PhysRevD.72.084027}{DOI}]},
  {\small[\href{http://adsabs.harvard.edu/abs/2005PhRvD..72h4027B}{ADS}]},
  {\small[\href{http://arxiv.org/abs/gr-qc/0508064}{{arXiv:gr-qc/0508064
  {\small[gr-qc]}}}]}.

\bibitem{Buonanno:2003fw}
Buonanno, A.  and Chen, Y.-B., ``Improving the sensitivity to gravitational
  wave sources by modifying the input output optics of advanced
  interferometers'', {\em Phys. Rev. D}, {\bf 69}, 102004, (2004).
  {\small[\href{http://dx.doi.org/10.1103/PhysRevD.69.102004}{DOI}]},
  {\small[\href{http://adsabs.harvard.edu/abs/2004PhRvD..69j2004B}{ADS}]},
  {\small[\href{http://arxiv.org/abs/gr-qc/0310026}{{arXiv:gr-qc/0310026
  {\small[gr-qc]}}}]}.

\bibitem{Buonanno:1998gg}
Buonanno, A.  and Damour, T., ``{Effective one-body approach to general
  relativistic two-body dynamics}'', {\em Phys. Rev. D}, {\bf 59}, 084006,
  (1999). {\small[\href{http://dx.doi.org/10.1103/PhysRevD.59.084006}{DOI}]},
  {\small[\href{http://adsabs.harvard.edu/abs/1999PhRvD..59h4006B}{ADS}]},
  {\small[\href{http://arxiv.org/abs/gr-qc/9811091}{{arXiv:gr-qc/9811091
  {\small[gr-qc]}}}]}.

\bibitem{Burgay:2003jj}
Burgay, M. {et~al.}, ``An Increased estimate of the merger rate of double
  neutron stars from observations of a highly relativistic system'', {\em
  Nature}, {\bf 426}, 531--533, (2003).
  {\small[\href{http://dx.doi.org/10.1038/nature02124}{DOI}]},
  {\small[\href{http://adsabs.harvard.edu/abs/2003Natur.426..531B}{ADS}]},
  {\small[\href{http://arxiv.org/abs/astro-ph/0312071}{{arXiv:astro-ph/0312071
  {\small[astro-ph]}}}]}.

\bibitem{Campanelli:1998jv}
Campanelli, M.  and Lousto, C.O., ``Second order gauge invariant gravitational
  perturbations of a Kerr black hole'', {\em Phys. Rev. D}, {\bf 59}, 124022,
  (1999). {\small[\href{http://dx.doi.org/10.1103/PhysRevD.59.124022}{DOI}]},
  {\small[\href{http://adsabs.harvard.edu/abs/1999PhRvD..59l4022C}{ADS}]},
  {\small[\href{http://arxiv.org/abs/gr-qc/9811019}{{arXiv:gr-qc/9811019
  {\small[gr-qc]}}}]}.

\bibitem{Campanelli:2005dd}
Campanelli, M., Lousto, C.O., Marronetti, P.  and Zlochower, Y., ``Accurate
  evolutions of orbiting black-hole binaries without excision'', {\em Phys.
  Rev. Lett.}, {\bf 96}, 111101, (2006).
  {\small[\href{http://dx.doi.org/10.1103/PhysRevLett.96.111101}{DOI}]},
  {\small[\href{http://adsabs.harvard.edu/abs/2006PhRvL..96k1101C}{ADS}]},
  {\small[\href{http://arxiv.org/abs/gr-qc/0511048}{{arXiv:gr-qc/0511048
  {\small[gr-qc]}}}]}.

\bibitem{Campanelli:2006gf}
Campanelli, M., Lousto, C.O.  and Zlochower, Y., ``The Last orbit of binary
  black holes'', {\em Phys. Rev. D}, {\bf 73}, 061501, (2006).
  {\small[\href{http://dx.doi.org/10.1103/PhysRevD.73.061501}{DOI}]},
  {\small[\href{http://adsabs.harvard.edu/abs/2006PhRvD..73f1501C}{ADS}]},
  {\small[\href{http://arxiv.org/abs/gr-qc/0601091}{{arXiv:gr-qc/0601091
  {\small[gr-qc]}}}]}.

\bibitem{carpet_web}
``CarpetCode: A mesh refinement driver for Cactus''. URL (accessed 30 March
  2012): \newline\url{http://www.carpetcode.org}.

\bibitem{Centrella:2010mx}
{Centrella}, J., {Baker}, J.~G., {Kelly}, B.~J.  and {van Meter}, J.~R.,
  ``{Black-hole binaries, gravitational waves, and numerical relativity}'',
  {\em Rev. Mod. Phys.}, {\bf 82}, 3069, (2010).
  {\small[\href{http://dx.doi.org/10.1103/RevModPhys.82.3069}{DOI}]},
  {\small[\href{http://adsabs.harvard.edu/abs/2010RvMP...82.3069C}{ADS}]},
  {\small[\href{http://arxiv.org/abs/1010.5260}{{arXiv:1010.5260
  {\small[gr-qc]}}}]}.

\bibitem{Chandrasekhar:1967aits.book}
Chandrasekhar, S., {\em An introduction to the study of stellar structure},
  (Dover, New York, 1967).
  {\small[\href{http://adsabs.harvard.edu/abs/1967aits.book.....C}{ADS}]}.

\bibitem{Chandrasekhar:1987QB410.C47}
Chandrasekhar, S., {\em Ellipsoidal figures of equilibrium}, (Dover, New York,
  1987).
  {\small[\href{http://adsabs.harvard.edu/abs/1987QB410.C47......}{ADS}]}.

\bibitem{Chawla:2010sw}
{Chawla}, S., {Anderson}, M., {Besselman}, M., {Lehner}, L., {Liebling}, S.~L.,
  {Motl}, P.~M.  and {Neilsen}, D., ``Mergers of Magnetized Neutron Stars with
  Spinning Black Holes: Disruption, Accretion and Fallback'', {\em Phys. Rev.
  Lett.}, {\bf 105}, 111101, (2010).
  {\small[\href{http://dx.doi.org/10.1103/PhysRevLett.105.111101}{DOI}]},
  {\small[\href{http://adsabs.harvard.edu/abs/2010PhRvL.105k1101C}{ADS}]},
  {\small[\href{http://arxiv.org/abs/1006.2839}{{arXiv:1006.2839
  {\small[gr-qc]}}}]}.

\bibitem{Chodos:1974je}
Chodos, A., Jaffe, R.L., Johnson, K., Thorn, C.B.  and Weisskopf, V.F., ``A New
  Extended Model of Hadrons'', {\em Phys. Rev. D}, {\bf 9}, 3471--3495, (1974).
  {\small[\href{http://dx.doi.org/10.1103/PhysRevD.9.3471}{DOI}]},
  {\small[\href{http://adsabs.harvard.edu/abs/1974PhRvD...9.3471C}{ADS}]}.

\bibitem{Clark77}
Clark, J.P.A.  and Eardley, D.M., ``Evolution of close neutron star binaries'',
  {\em Astrophys. J.}, {\bf 215}, 311--322, (1977).
  {\small[\href{http://dx.doi.org/10.1086/155360}{DOI}]},
  {\small[\href{http://adsabs.harvard.edu/abs/1977ApJ...215..311C}{ADS}]}.

\bibitem{Cook:2000vr}
Cook, G.B., ``Initial data for numerical relativity'', {\em Living Rev.
  Relativity}, {\bf 3}, lrr-2000-5, (2000).
  {\small[\href{http://adsabs.harvard.edu/abs/2000LRR.....3....5C}{ADS}]},
  {\small[\href{http://arxiv.org/abs/gr-qc/0007085}{{arXiv:gr-qc/0007085
  {\small[gr-qc]}}}]}. URL (accessed 30 March 2012):
  \newline\url{http://www.livingreviews.org/lrr-2000-5}.

\bibitem{Cook92}
Cook, G.B., Shapiro, S.L.  and Teukolsky, S.A., ``Spin-up of a rapidly rotating
  star by angular momentum loss - Effects of general relativity'', {\em
  Astrophys. J.j}, {\bf 398}, 203--223, (1992).
  {\small[\href{http://dx.doi.org/10.1086/171849}{DOI}]},
  {\small[\href{http://adsabs.harvard.edu/abs/1992ApJ...398..203C}{ADS}]}.

\bibitem{Cook:1993qr}
Cook, G.B., Shapiro, S.L.  and Teukolsky, S.A., ``Rapidly rotating neutron
  stars in general relativity: Realistic equations of state'', {\em Astrophys.
  J.}, {\bf 424}, 823--845, (1994).
  {\small[\href{http://dx.doi.org/10.1086/173934}{DOI}]},
  {\small[\href{http://adsabs.harvard.edu/abs/1994ApJ...424..823C}{ADS}]}.

\bibitem{Cumming:2001kr}
Cumming, A., Zweibel, E.G.  and Bildsten, L., ``Magnetic screening in accreting
  neutron stars'', {\em Astrophys. J.}, {\bf 557}, 958--966, (2001).
  {\small[\href{http://dx.doi.org/10.1086/321658}{DOI}]},
  {\small[\href{http://adsabs.harvard.edu/abs/2001ApJ...557..958C}{ADS}]},
  {\small[\href{http://arxiv.org/abs/astro-ph/0102178}{{arXiv:astro-ph/0102178
  {\small[astro-ph]}}}]}.

\bibitem{Damour:2009vw}
Damour, T.  and Nagar, A., ``{Relativistic tidal properties of neutron
  stars}'', {\em Phys. Rev. D}, {\bf 80}, 084035, (2009).
  {\small[\href{http://dx.doi.org/10.1103/PhysRevD.80.084035}{DOI}]},
  {\small[\href{http://adsabs.harvard.edu/abs/2009PhRvD..80h4035D}{ADS}]},
  {\small[\href{http://arxiv.org/abs/0906.0096}{{arXiv:0906.0096
  {\small[gr-qc]}}}]}.

\bibitem{Damour:2009wj}
{Damour}, T.  and {Nagar}, A., ``{Effective One Body description of tidal
  effects in inspiralling compact binaries}'', {\em Phys. Rev. D}, {\bf 81},
  084016, (2010).
  {\small[\href{http://dx.doi.org/10.1103/PhysRevD.81.084016}{DOI}]},
  {\small[\href{http://adsabs.harvard.edu/abs/2010PhRvD..81h4016D}{ADS}]},
  {\small[\href{http://arxiv.org/abs/0911.5041}{{arXiv:0911.5041
  {\small[gr-qc]}}}]}.

\bibitem{Davies:1993zn}
Davies, M.B., Benz, W., Piran, T.  and Thielemann, F.K., ``Merging neutron
  stars. I. Initial results for coalescence of noncorotating systems'', {\em
  Astrophys. J.}, {\bf 431}, 742--753, (1994).
  {\small[\href{http://dx.doi.org/10.1086/174525}{DOI}]},
  {\small[\href{http://adsabs.harvard.edu/abs/1994ApJ...431..742D}{ADS}]},
  {\small[\href{http://arxiv.org/abs/astro-ph/9401032}{{arXiv:astro-ph/9401032
  {\small[astro-ph]}}}]}.

\bibitem{Davies:2004pu}
Davies, M.B., Levan, A.J.  and King, A.R., ``The Ultimate outcome of black hole
  - neutron star mergers'', {\em Mon. Not. R. Astron. Soc.}, {\bf 356}, 54--58,
  (2005).
  {\small[\href{http://dx.doi.org/10.1111/j.1365-2966.2004.08423.x/abs/}{DOI}]%
}, {\small[\href{http://adsabs.harvard.edu/abs/2005MNRAS.356...54D}{ADS}]},
  {\small[\href{http://arxiv.org/abs/astro-ph/0409681}{{arXiv:astro-ph/0409681
  {\small[astro-ph]}}}]}.

\bibitem{Davis:2009rz}
Davis, P.J., Kolb, U.  and Willems, B., ``A comprehensive population synthesis
  study of post-common envelope binaries'', {\em Mon. Not. R. Astron. Soc.},
  {\bf 403}, 179--195, (2010).
  {\small[\href{http://dx.doi.org/10.1111/j.1365-2966.2009.16138.x}{DOI}]},
  {\small[\href{http://adsabs.harvard.edu/abs/2010MNRAS.403..179D}{ADS}]},
  {\small[\href{http://arxiv.org/abs/0903.4152}{{arXiv:0903.4152
  {\small[astro-ph.SR]}}}]}.

\bibitem{deFreitasPacheco:2005ub}
de~Freitas~Pacheco, J.A., Regimbau, T., Vincent, S.  and Spallicci, A.,
  ``Expected coalescence rates of ns-ns binaries for laser beam
  interferometers'', {\em Int. J. Mod. Phys. D}, {\bf 15}, 235--250, (2006).
  {\small[\href{http://dx.doi.org/10.1142/S0218271806007699}{DOI}]},
  {\small[\href{http://adsabs.harvard.edu/abs/2006IJMPD..15..235D}{ADS}]},
  {\small[\href{http://arxiv.org/abs/astro-ph/0510727}{{arXiv:astro-ph/0510727
  {\small[astro-ph]}}}]}.

\bibitem{DeVilliers:2008ut}
De~Villiers, J.-P., ``Some First Steps Towards a Radiation GRMHD Code:
  Radiative Effects on Accretion Rate onto a Kerr Black Hole'', (2008).
  {\small[\href{http://adsabs.harvard.edu/abs/2008arXiv0802.0848D}{ADS}]},
  {\small[\href{http://arxiv.org/abs/0802.0848}{{arXiv:0802.0848
  {\small[astro-ph]}}}]}.

\bibitem{Demorest:2010bx}
Demorest, P., Pennucci, T., Ransom, S., Roberts, M.  and Hessels, J., ``Shapiro
  delay measurement of a two solar mass neutron star'', {\em Nature}, {\bf
  467}, 1081--1083, (2010).
  {\small[\href{http://dx.doi.org/10.1038/nature09466}{DOI}]},
  {\small[\href{http://adsabs.harvard.edu/abs/2010Natur.467.1081D}{ADS}]},
  {\small[\href{http://arxiv.org/abs/1010.5788}{{arXiv:1010.5788
  {\small[astro-ph.HE]}}}]}.

\bibitem{Dessart:2008zd}
Dessart, L., Ott, C., Burrows, A., Rosswog, S.  and Livne, E., ``Neutrino
  signatures and the neutrino-driven wind in Binary Neutron Star Mergers'',
  {\em Astrophys. J.}, {\bf 690}, 1681--1705, (2009).
  {\small[\href{http://dx.doi.org/10.1088/0004-637X/690/2/1681}{DOI}]},
  {\small[\href{http://adsabs.harvard.edu/abs/2009ApJ...690.1681D}{ADS}]},
  {\small[\href{http://arxiv.org/abs/0806.4380}{{arXiv:0806.4380
  {\small[astro-ph]}}}]}.

\bibitem{Douchin:2001sv}
Douchin, F.  and Haensel, P., ``A unified equation of state of dense matter and
  neutron star structure'', {\em Astron. Astrophys.}, {\bf 380}, 151--167,
  (2001). {\small[\href{http://dx.doi.org/10.1051/0004-6361:20011402}{DOI}]},
  {\small[\href{http://adsabs.harvard.edu/abs/2001A%26A...380..151D}{ADS}]},
  {\small[\href{http://arxiv.org/abs/astro-ph/0111092}{{arXiv:astro-ph/0111092
  {\small[astro-ph]}}}]}.

\bibitem{Duez:2009yy}
Duez, M.D., Foucart, F., Kidder, L.E., Ott, C.D.  and Teukolsky, S.A.,
  ``Equation of state effects in black hole-neutron star mergers'', {\em Class.
  Quantum Grav.}, {\bf 27}, 114106, (2010).
  {\small[\href{http://dx.doi.org/10.1088/0264-9381/27/11/114106}{DOI}]},
  {\small[\href{http://adsabs.harvard.edu/abs/2010CQGra..27k4106D}{ADS}]},
  {\small[\href{http://arxiv.org/abs/0912.3528}{{arXiv:0912.3528
  {\small[astro-ph.HE]}}}]}.

\bibitem{Duez:2005cj}
Duez, M.D., Liu, Y.T., Shapiro, S.L., Shibata, M.  and Stephens, B.C.,
  ``Collapse of magnetized hypermassive neutron stars in general relativity'',
  {\em Phys. Rev. Lett.}, {\bf 96}, 031101, (2006).
  {\small[\href{http://dx.doi.org/10.1103/PhysRevLett.96.031101}{DOI}]},
  {\small[\href{http://adsabs.harvard.edu/abs/2006PhRvL..96c1101D}{ADS}]},
  {\small[\href{http://arxiv.org/abs/astro-ph/0510653}{{arXiv:astro-ph/0510653
  {\small[astro-ph]}}}]}.

\bibitem{Duez:2005sf}
Duez, M.D., Liu, Y.T., Shapiro, S.L.  and Stephens, B.C., ``Relativistic
  magnetohydrodynamics in dynamical spacetimes: Numerical methods and tests'',
  {\em Phys. Rev. D}, {\bf 72}, 024028, (2005).
  {\small[\href{http://dx.doi.org/10.1103/PhysRevD.72.024028}{DOI}]},
  {\small[\href{http://adsabs.harvard.edu/abs/2005PhRvD..72b4028D}{ADS}]},
  {\small[\href{http://arxiv.org/abs/astro-ph/0503420}{{arXiv:astro-ph/0503420
  {\small[astro-ph]}}}]}.

\bibitem{Duez:2008rb}
{Duez}, M.~D., {Foucart}, F., {Kidder}, L.~E., {Pfeiffer}, H.~P., {Scheel},
  M.~A.  and {Teukolsky}, S.~A., ``Evolving black hole-neutron star binaries in
  general relativity using pseudospectral and finite difference methods'', {\em
  Phys. Rev. D}, {\bf 78}, 104015, (2008).
  {\small[\href{http://dx.doi.org/10.1103/PhysRevD.78.104015}{DOI}]},
  {\small[\href{http://adsabs.harvard.edu/abs/2008PhRvD..78j4015D}{ADS}]},
  {\small[\href{http://arxiv.org/abs/0809.0002}{{arXiv:0809.0002
  {\small[gr-qc]}}}]}.

\bibitem{East:2011xa}
{East}, W.~E., {Pretorius}, F.  and {Stephens}, B.~C., ``{Eccentric black
  hole-neutron star mergers: effects of black hole spin and equation of
  state}'', (2011).
  {\small[\href{http://adsabs.harvard.edu/abs/2011arXiv1111.3055E}{ADS}]},
  {\small[\href{http://arxiv.org/abs/1111.3055}{{arXiv:1111.3055
  {\small[astro-ph.HE]}}}]}.

\bibitem{East:2011aa}
{East}, W.~E., {Pretorius}, F.  and {Stephens}, B.~C., ``{Hydrodynamics in full
  general relativity with conservative AMR}'', (2011).
  {\small[\href{http://adsabs.harvard.edu/abs/2011arXiv1112.3094E}{ADS}]},
  {\small[\href{http://arxiv.org/abs/1112.3094}{{arXiv:1112.3094
  {\small[gr-qc]}}}]}.

\bibitem{einsteintoolkit_web}
``Einstein Toolkit''. URL (accessed 30 March 2012):
  \newline\url{http://www.einsteintoolkit.org}.

\bibitem{Etienne:2007jg}
Etienne, Z.B., Faber, J.A., Liu, Y.T., Shapiro, S.L., Taniguchi, K.  and
  Baumgarte, T.W., ``Fully General Relativistic Simulations of Black
  Hole-Neutron Star Mergers'', {\em Phys. Rev. D}, {\bf 77}, 084002, (2008).
  {\small[\href{http://dx.doi.org/10.1103/PhysRevD.77.084002}{DOI}]},
  {\small[\href{http://adsabs.harvard.edu/abs/2008PhRvD..77h4002E}{ADS}]},
  {\small[\href{http://arxiv.org/abs/0712.2460}{{arXiv:0712.2460
  {\small[astro-ph]}}}]}.

\bibitem{Etienne:2010ui}
Etienne, Z.B., Liu, Y.T.  and Shapiro, S.L., ``Relativistic
  magnetohydrodynamics in dynamical spacetimes: A new AMR implementation'',
  {\em Phys. Rev. D}, {\bf 82}, 084031, (2010).
  {\small[\href{http://dx.doi.org/10.1103/PhysRevD.82.084031}{DOI}]},
  {\small[\href{http://adsabs.harvard.edu/abs/2010PhRvD..82h4031E}{ADS}]},
  {\small[\href{http://arxiv.org/abs/1007.2848}{{arXiv:1007.2848
  {\small[astro-ph.HE]}}}]}.

\bibitem{Etienne:2008re}
Etienne, Z.B., Liu, Y.T., Shapiro, S.L.  and Baumgarte, T.W., ``General
  relativistic simulations of black-hole-neutron-star mergers: Effects of
  black-hole spin'', {\em Phys. Rev. D}, {\bf 79}, 044024, (2009).
  {\small[\href{http://dx.doi.org/10.1103/PhysRevD.79.044024}{DOI}]},
  {\small[\href{http://adsabs.harvard.edu/abs/2009PhRvD..79d4024E}{ADS}]},
  {\small[\href{http://arxiv.org/abs/0812.2245}{{arXiv:0812.2245
  {\small[astro-ph]}}}]}.

\bibitem{Etienne:2011ea}
{Etienne}, Z.~B., {Liu}, Y.~T., {Paschalidis}, V.  and {Shapiro}, S.~L.,
  ``{General relativistic simulations of black-hole-neutron-star mergers:
  Effects of magnetic fields}'', {\em Phys. Rev. D}, {\bf 85}(6), 064029,
  064029, (March 2012).
  {\small[\href{http://dx.doi.org/10.1103/PhysRevD.85.064029}{DOI}]},
  {\small[\href{http://adsabs.harvard.edu/abs/2012PhRvD..85f4029E}{ADS}]},
  {\small[\href{http://arxiv.org/abs/1112.0568}{{arXiv:1112.0568
  {\small[astro-ph.HE]}}}]}.

\bibitem{Etienne:2011re}
{Etienne}, Z.~B., {Paschalidis}, V., {Liu}, Y.~T.  and {Shapiro}, S.~L.,
  ``{Relativistic MHD in dynamical spacetimes: Improved EM gauge condition for
  AMR grids}'', {\em Phys. Rev. D}, {\bf 85}, 024013, (2012).
  {\small[\href{http://dx.doi.org/10.1103/PhysRevD.85.024013}{DOI}]},
  {\small[\href{http://adsabs.harvard.edu/abs/2012PhRvD..85b4013E}{ADS}]},
  {\small[\href{http://arxiv.org/abs/1110.4633}{{arXiv:1110.4633
  {\small[astro-ph.HE]}}}]}.

\bibitem{Faber:2005yg}
Faber, J.A., Baumgarte, T.W., Shapiro, S.L., Taniguchi, K.  and Rasio, F.A.,
  ``The dynamical evolution of black hole-neutron star binaries in general
  relativity: simulations of tidal disruption'', {\em Phys. Rev. D}, {\bf 73},
  024012, (2006).
  {\small[\href{http://dx.doi.org/10.1103/PhysRevD.73.024012}{DOI}]},
  {\small[\href{http://adsabs.harvard.edu/abs/2006PhRvD..73b4012F}{ADS}]},
  {\small[\href{http://arxiv.org/abs/astro-ph/0511366}{{arXiv:astro-ph/0511366
  {\small[astro-ph]}}}]}.

\bibitem{Faber:2003sb}
Faber, J.A., Grandcl{\'{e}}ment, P.  and Rasio, F.A., ``Mergers of irrotational
  neutron star binaries in conformally flat gravity'', {\em Phys. Rev. D}, {\bf
  D69}, 124036, (2004).
  {\small[\href{http://dx.doi.org/10.1103/PhysRevD.69.124036}{DOI}]},
  {\small[\href{http://adsabs.harvard.edu/abs/2004PhRvD..69l4036F}{ADS}]},
  {\small[\href{http://arxiv.org/abs/gr-qc/0312097}{{arXiv:gr-qc/0312097
  {\small[gr-qc]}}}]}.

\bibitem{Faber:2002zn}
Faber, J.A., Grandcl{\'{e}}ment, P., Rasio, F.A.  and Taniguchi, K.,
  ``Measuring neutron star radii with gravitational wave detectors'', {\em
  Phys. Rev. Lett.}, {\bf 89}, 231102, (2002).
  {\small[\href{http://dx.doi.org/10.1103/PhysRevLett.89.231102}{DOI}]},
  {\small[\href{http://adsabs.harvard.edu/abs/2002PhRvL..89w1102F}{ADS}]},
  {\small[\href{http://arxiv.org/abs/astro-ph/0204397}{{arXiv:astro-ph/0204397
  {\small[astro-ph]}}}]}.

\bibitem{Faber:1999gj}
Faber, J.A.  and Rasio, F.A., ``PostNewtonian SPH calculations of binary
  neutron star coalescence. 1. Method and first results'', {\em Phys. Rev. D},
  {\bf 62}, 064012, (2000).
  {\small[\href{http://dx.doi.org/10.1103/PhysRevD.62.064012}{DOI}]},
  {\small[\href{http://adsabs.harvard.edu/abs/2000PhRvD..62f4012F}{ADS}]},
  {\small[\href{http://arxiv.org/abs/gr-qc/9912097}{{arXiv:gr-qc/9912097
  {\small[gr-qc]}}}]}.

\bibitem{Faber:2002cg}
Faber, J.A.  and Rasio, F.A., ``PostNewtonian SPH calculations of binary
  neutron star coalescence. 3. Irrotational systems and gravitational wave
  spectra'', {\em Phys. Rev. D}, {\bf 65}, 084042, (2002).
  {\small[\href{http://dx.doi.org/10.1103/PhysRevD.65.084042}{DOI}]},
  {\small[\href{http://adsabs.harvard.edu/abs/2002PhRvD..65h4042F}{ADS}]},
  {\small[\href{http://arxiv.org/abs/gr-qc/0201040}{{arXiv:gr-qc/0201040
  {\small[gr-qc]}}}]}.

\bibitem{Faber:2000uf}
Faber, J.A., Rasio, F.A.  and Manor, J.B., ``PostNewtonian SPH calculations of
  binary neutron star coalescence. 2. Binary mass ratio, equation of state, and
  spin dependence'', {\em Phys. Rev. D}, {\bf 63}, 044012, (2001).
  {\small[\href{http://dx.doi.org/10.1103/PhysRevD.63.044012}{DOI}]},
  {\small[\href{http://adsabs.harvard.edu/abs/2001PhRvD..63d4012F}{ADS}]},
  {\small[\href{http://arxiv.org/abs/gr-qc/0006078}{{arXiv:gr-qc/0006078
  {\small[gr-qc]}}}]}.

\bibitem{Farhi:1984qu}
Farhi, E.  and Jaffe, R.L., ``Strange Matter'', {\em Phys. Rev. D}, {\bf 30},
  2379--2390, (1984).
  {\small[\href{http://dx.doi.org/10.1103/PhysRevD.30.2379}{DOI}]},
  {\small[\href{http://adsabs.harvard.edu/abs/1984PhRvD..30.2379F}{ADS}]}.

\bibitem{Farris:2008fe}
Farris, B.D., Li, T.K., Liu, Y.T.  and Shapiro, S.L., ``Relativistic Radiation
  Magnetohydrodynamics in Dynamical Spacetimes: Numerical Methods and Tests'',
  {\em Phys. Rev. D}, {\bf 78}, 024023, (2008).
  {\small[\href{http://dx.doi.org/10.1103/PhysRevD.78.024023}{DOI}]},
  {\small[\href{http://adsabs.harvard.edu/abs/2008PhRvD..78b4023F}{ADS}]},
  {\small[\href{http://arxiv.org/abs/0802.3210}{{arXiv:0802.3210
  {\small[astro-ph]}}}]}.

\bibitem{Flanagan:1998zt}
Flanagan, {\'{E}}.{\'{E}}., ``Possible explanation for star-crushing effect in
  binary neutron star simulations'', {\em Phys. Rev. Lett.}, {\bf 82},
  1354--1357, (1999).
  {\small[\href{http://dx.doi.org/10.1103/PhysRevLett.82.1354}{DOI}]},
  {\small[\href{http://adsabs.harvard.edu/abs/1999PhRvL..82.1354F}{ADS}]},
  {\small[\href{http://arxiv.org/abs/astro-ph/9811132}{{arXiv:astro-ph/9811132
  {\small[astro-ph]}}}]}.

\bibitem{Flanagan:2007ix}
Flanagan, {\'{E}}.{\'{E}}.  and Hinderer, T., ``Constraining neutron star tidal
  Love numbers with gravitational wave detectors'', {\em Phys. Rev. D}, {\bf
  77}, 021502, (2008).
  {\small[\href{http://dx.doi.org/10.1103/PhysRevD.77.021502}{DOI}]},
  {\small[\href{http://adsabs.harvard.edu/abs/2008PhRvD..77b1502F}{ADS}]},
  {\small[\href{http://arxiv.org/abs/0709.1915}{{arXiv:0709.1915
  {\small[astro-ph]}}}]}.

\bibitem{Font:2007zz}
Font, J.A., ``Numerical hydrodynamics and magnetohydrodynamics in general
  relativity'', {\em Living Rev. Relativity}, {\bf 11}, lrr-2008-7, (2008).
  {\small[\href{http://adsabs.harvard.edu/abs/2008LRR....11....7F}{ADS}]}. URL
  (accessed 30 March 2012):
  \newline\url{http://www.livingreviews.org/lrr-2008-7}.

\bibitem{Foucart:2010eq}
Foucart, F., Duez, M.D., Kidder, L.E.  and Teukolsky, S.A., ``Black
  hole-neutron star mergers: effects of the orientation of the black hole
  spin'', {\em Phys. Rev. D}, {\bf 83}, 024005, (2011).
  {\small[\href{http://dx.doi.org/10.1103/PhysRevD.83.024005}{DOI}]},
  {\small[\href{http://adsabs.harvard.edu/abs/2011PhRvD..83b4005F}{ADS}]},
  {\small[\href{http://arxiv.org/abs/1007.4203}{{arXiv:1007.4203
  {\small[astro-ph.HE]}}}]}.

\bibitem{Foucart:2011mz}
{Foucart}, F., {Duez}, M.~D., {Kidder}, L.~E., {Scheel}, M.~A., {Szilagyi}, B.
  and {Teukolsky}, S.~A., ``{Black hole-neutron star mergers for 10 solar mass
  black holes}'', {\em Phys. Rev. D}, {\bf 85}, 044015, (2012).
  {\small[\href{http://dx.doi.org/10.1103/PhysRevD.85.044015}{DOI}]},
  {\small[\href{http://adsabs.harvard.edu/abs/2012PhRvD..85d4015F}{ADS}]},
  {\small[\href{http://arxiv.org/abs/1111.1677}{{arXiv:1111.1677
  {\small[gr-qc]}}}]}.

\bibitem{Fox:2005kv}
Fox, D.B. {et~al.}, ``The afterglow of grb050709 and the nature of the
  short-hard gamma-ray bursts'', {\em Nature}, {\bf 437}, 845--850, (2005).
  {\small[\href{http://dx.doi.org/10.1038/nature04189}{DOI}]},
  {\small[\href{http://adsabs.harvard.edu/abs/2005Natur.437..845F}{ADS}]},
  {\small[\href{http://arxiv.org/abs/astro-ph/0510110}{{arXiv:astro-ph/0510110
  {\small[astro-ph]}}}]}.

\bibitem{Freiburghaus:1999ApJ...525L.121F}
Freiburghaus, C., Rosswog, S.  and Thielemann, F.-K., ``R-Process in Neutron
  Star Mergers'', {\em Astrophys. J. Lett.}, {\bf 525}, L121--L124, (1999).
  {\small[\href{http://dx.doi.org/10.1086/312343}{DOI}]},
  {\small[\href{http://adsabs.harvard.edu/abs/1999ApJ...525L.121F}{ADS}]}.

\bibitem{Friedrich:1985aa}
{Friedrich}, H., ``{On the hyperbolicity of Einstein's and other gauge field
  equations}'', {\em Commun. Math. Phys.}, {\bf 100}, 525--543, (December
  1985). {\small[\href{http://dx.doi.org/10.1007/BF01217728}{DOI}]},
  {\small[\href{http://adsabs.harvard.edu/abs/1985CMaPh.100..525F}{ADS}]}.

\bibitem{Friedrich:1998xt}
{Friedrich}, H.  and {Nagy}, G., ``{The Initial boundary value problem for
  Einstein's vacuum field equations}'', {\em Commun.Math.Phys.}, {\bf 201},
  619--655, (1999).
  {\small[\href{http://dx.doi.org/10.1007/s002200050571}{DOI}]},
  {\small[\href{http://adsabs.harvard.edu/abs/1999CMaPh.201..619F}{ADS}]}.

\bibitem{Galeazzi:2011nn}
Galeazzi, F., Yoshida, S.  and Eriguchi, Y., ``Differentially-rotating neutron
  star models with a parametrized rotation profile'', (2011).
  {\small[\href{http://adsabs.harvard.edu/abs/2011arXiv1101.2664G}{ADS}]},
  {\small[\href{http://arxiv.org/abs/1101.2664}{{arXiv:1101.2664
  {\small[astro-ph.SR]}}}]}.

\bibitem{Garfinkle:2001ni}
Garfinkle, D., ``{Harmonic coordinate method for simulating generic
  singularities}'', {\em Phys. Rev. D}, {\bf 65}, 044029, (2002).
  {\small[\href{http://dx.doi.org/10.1103/PhysRevD.65.044029}{DOI}]},
  {\small[\href{http://adsabs.harvard.edu/abs/2002PhRvD..65d4029G}{ADS}]},
  {\small[\href{http://arxiv.org/abs/gr-qc/0110013}{{arXiv:gr-qc/0110013
  {\small[gr-qc]}}}]}.

\bibitem{Giacomazzo:2009mp}
Giacomazzo, B., Rezzolla, L.  and Baiotti, L., ``Can magnetic fields be
  detected during the inspiral of binary neutron stars?'', {\em Mon. Not. R.
  Astron. Soc. Lett.}, {\bf 399}, L164--L168, (2009).
  {\small[\href{http://dx.doi.org/10.1111/j.1745-3933.2009.00745.x}{DOI}]},
  {\small[\href{http://adsabs.harvard.edu/abs/2009MNRAS.399L.164G}{ADS}]},
  {\small[\href{http://arxiv.org/abs/0901.2722}{{arXiv:0901.2722
  {\small[gr-qc]}}}]}.

\bibitem{Giacomazzo:2010bx}
Giacomazzo, B., Rezzolla, L.  and Baiotti, L., ``Accurate evolutions of
  inspiralling and magnetized neutron-stars: Equal-mass binaries'', {\em Phys.
  Rev. D}, {\bf 83}, 044014, (2011).
  {\small[\href{http://dx.doi.org/10.1103/PhysRevD.83.044014}{DOI}]},
  {\small[\href{http://adsabs.harvard.edu/abs/2011PhRvD..83d4014G}{ADS}]},
  {\small[\href{http://arxiv.org/abs/1009.2468}{{arXiv:1009.2468
  {\small[gr-qc]}}}]}.

\bibitem{Gingold:1977sh}
Gingold, R.A.  and Monaghan, J.J., ``Smoothed particle hydrodynamics: Theory
  and application to non-spherical stars'', {\em Mon. Not. R. Astron. Soc.},
  {\bf 181}, 375--389, (1977).
  {\small[\href{http://adsabs.harvard.edu/abs/1977MNRAS.181..375G}{ADS}]}.

\bibitem{Glendenning:1984jr}
Glendenning, N.K., ``Neutron Stars Are Giant Hypernuclei?'', {\em Astrophys.
  J.}, {\bf 293}, 470--493, (1985).
  {\small[\href{http://dx.doi.org/10.1086/163253}{DOI}]},
  {\small[\href{http://adsabs.harvard.edu/abs/1985ApJ...293..470G}{ADS}]}.

\bibitem{Glendenning:1997ak}
Glendenning, N.K.  and Schaffner-Bielich, J., ``First order kaon condensate'',
  {\em Phys. Rev. C}, {\bf 60}, 025803, (1999).
  {\small[\href{http://dx.doi.org/10.1103/PhysRevC.60.025803}{DOI}]},
  {\small[\href{http://adsabs.harvard.edu/abs/1999PhRvC..60b5803G}{ADS}]},
  {\small[\href{http://arxiv.org/abs/astro-ph/9810290}{{arXiv:astro-ph/9810290
  {\small[astro-ph]}}}]}.

\bibitem{Godunov:1959}
Godunov, S., ``{A Difference Scheme for Numerical Solution of Discontinuous
  Solution of Hydrodynamic Equations}'', {\em Math. Sbornik}, {\bf 47},
  271--306, (1959). {translated US Joint Publ. Res. Service, JPRS 7226, 1969}.

\bibitem{Gold:2011df}
{Gold}, R., {Bernuzzi}, S., {Thierfelder}, M., {Bruegmann}, B.  and
  {Pretorius}, F., ``{Eccentric binary neutron star mergers}'', (2011).
  {\small[\href{http://adsabs.harvard.edu/abs/2011arXiv1109.5128G}{ADS}]},
  {\small[\href{http://arxiv.org/abs/1109.5128}{{arXiv:1109.5128
  {\small[gr-qc]}}}]}.

\bibitem{Goriely:2011vg}
{Goriely}, S., {Bauswein}, A.  and {Janka}, H.-T., ``{R-Process Nucleosynthesis
  in Dynamically Ejected Matter of Neutron Star Mergers}'', {\em Ap. J. Lett.},
  {\bf 738}, L32, (September 2011).
  {\small[\href{http://dx.doi.org/10.1088/2041-8205/738/2/L32}{DOI}]},
  {\small[\href{http://adsabs.harvard.edu/abs/2011ApJ...738L..32G}{ADS}]},
  {\small[\href{http://arxiv.org/abs/1107.0899}{{arXiv:1107.0899
  {\small[astro-ph.SR]}}}]}.

\bibitem{Gourgoulhon:2000nn}
Gourgoulhon, E., Grandcl{\'{e}}ment, P., Taniguchi, K., Marck, J.-A.  and
  Bonazzola, S., ``Quasiequilibrium sequences of synchronized and irrotational
  binary neutron stars in general relativity: 1. Method and tests'', {\em Phys.
  Rev. D}, {\bf 63}, 064029, (2001).
  {\small[\href{http://dx.doi.org/10.1103/PhysRevD.63.064029}{DOI}]},
  {\small[\href{http://adsabs.harvard.edu/abs/2001PhRvD..63f4029G}{ADS}]},
  {\small[\href{http://arxiv.org/abs/gr-qc/0007028}{{arXiv:gr-qc/0007028
  {\small[gr-qc]}}}]}.

\bibitem{Grandclement:2003ck}
Grandcl{\'{e}}ment, P., Ihm, M., Kalogera, V.  and Belczynski, K., ``Searching
  for gravitational waves from the inspiral of precessing binary systems:
  Astrophysical expectations and detection efficiency of 'spiky' templates'',
  {\em Phys. Rev. D}, {\bf 69}, 102002, (2004).
  {\small[\href{http://dx.doi.org/10.1103/PhysRevD.69.102002}{DOI}]},
  {\small[\href{http://adsabs.harvard.edu/abs/2004PhRvD..69j2002G}{ADS}]},
  {\small[\href{http://arxiv.org/abs/gr-qc/0312084}{{arXiv:gr-qc/0312084
  {\small[gr-qc]}}}]}.

\bibitem{Grindlay:2005ym}
Grindlay, J., {Portegies Zwart}, S.  and McMillan, S., ``{Short gamma-ray
  bursts from binary neutron star mergers in globular clusters}'', {\em Nature
  Phys.}, {\bf 2}, 116--119, (February 2006).
  {\small[\href{http://dx.doi.org/10.1038/nphys214}{DOI}]},
  {\small[\href{http://adsabs.harvard.edu/abs/2006NatPh...2..116G}{ADS}]},
  {\small[\href{http://arxiv.org/abs/astro-ph/0512654}{{arXiv:astro-ph/0512654
  {\small[astro-ph]}}}]}.

\bibitem{Guetta:2008qw}
Guetta, D.  and Stella, L., ``Short $\gamma$-Ray Bursts and Gravitational Waves
  from Dynamically Formed Merging Binaries'', {\em Astron. Astrophys.}, {\bf
  498}, 329--333, (2008).
  {\small[\href{http://dx.doi.org/10.1051/0004-6361:200810493}{DOI}]},
  {\small[\href{http://adsabs.harvard.edu/abs/2009A%26A...498..329G}{ADS}]},
  {\small[\href{http://arxiv.org/abs/0811.0684}{{arXiv:0811.0684
  {\small[astro-ph]}}}]}.

\bibitem{Gundlach:2005eh}
Gundlach, C., Mart{\'{\i}}n-Garc{\'{\i}}a, J.M., Calabrese, G.  and Hinder, I.,
  ``Constraint damping in the Z4 formulation and harmonic gauge'', {\em Class.
  Quantum Grav.}, {\bf 22}, 3767--3774, (2005).
  {\small[\href{http://dx.doi.org/10.1088/0264-9381/22/17/025}{DOI}]},
  {\small[\href{http://adsabs.harvard.edu/abs/2005gr.qc.....4114G}{ADS}]},
  {\small[\href{http://arxiv.org/abs/gr-qc/0504114}{{arXiv:gr-qc/0504114
  {\small[gr-qc]}}}]}.

\bibitem{Gundlach:2006tw}
{Gundlach}, C.  and {Mart{\'{\i}}n-Garc{\'{\i}}a}, J.~M., ``{Well-posedness of
  formulations of the Einstein equations with dynamical lapse and shift
  conditions}'', {\em Phys. Rev. D}, {\bf 74}, 024016, (2006).
  {\small[\href{http://dx.doi.org/10.1103/PhysRevD.74.024016}{DOI}]},
  {\small[\href{http://adsabs.harvard.edu/abs/2006PhRvD..74b4016G}{ADS}]},
  {\small[\href{http://arxiv.org/abs/gr-qc/0604035}{{arXiv:gr-qc/0604035
  {\small[gr-qc]}}}]}.

\bibitem{Hachisu:1986ApJS...61..479H}
Hachisu, I., ``A versatile method for obtaining structures of rapidly rotating
  stars'', {\em Astrophys. J. Suppl. Ser.}, {\bf 61}, 479--507, (1986).
  {\small[\href{http://dx.doi.org/10.1086/191121}{DOI}]},
  {\small[\href{http://adsabs.harvard.edu/abs/1986ApJS...61..479H}{ADS}]}.

\bibitem{Hachisu:1986ApJS...62..461H}
Hachisu, I., ``A versatile method for obtaining structures of rapidly rotating
  stars. II - Three-dimensional self-consistent field method'', {\em Astrophys.
  J. Suppl. Ser.}, {\bf 62}, 461--499, (1986).
  {\small[\href{http://dx.doi.org/10.1086/191148}{DOI}]},
  {\small[\href{http://adsabs.harvard.edu/abs/1986ApJS...62..461H}{ADS}]}.

\bibitem{had_web}
``HAD: the hyper AMR driver''. URL (accessed 20 March 2012):
  \newline\url{http://relativity.phys.lsu.edu/\~matt/had.html}.

\bibitem{Hobbs:2005yx}
Hobbs, G., Lorimer, D.R., Lyne, A.G.  and Kramer, M., ``A Statistical study of
  233 pulsar proper motions'', {\em Mon. Not. R. Astron. Soc.}, {\bf 360},
  974--992, (2005).
  {\small[\href{http://dx.doi.org/10.1111/j.1365-2966.2005.09087.x}{DOI}]},
  {\small[\href{http://adsabs.harvard.edu/abs/2005MNRAS.360..974H}{ADS}]},
  {\small[\href{http://arxiv.org/abs/astro-ph/0504584}{{arXiv:astro-ph/0504584
  {\small[astro-ph]}}}]}.

\bibitem{Hotokezaka:2011dh}
Hotokezaka, K., Kyutoku, K., Okawa, H., Shibata, M.  and Kiuchi, K., ``Binary
  Neutron Star Mergers: Dependence on the Nuclear Equation of State'', {\em
  Phys. Rev. D}, {\bf 83}, 124008, (2011).
  {\small[\href{http://dx.doi.org/10.1103/PhysRevD.83.124008}{DOI}]},
  {\small[\href{http://adsabs.harvard.edu/abs/2011PhRvD..83l4008H}{ADS}]},
  {\small[\href{http://arxiv.org/abs/1105.4370}{{arXiv:1105.4370
  {\small[astro-ph.HE]}}}]}.

\bibitem{Hulse:1974eb}
Hulse, R.A.  and Taylor, J.H., ``Discovery of a pulsar in a binary system'',
  {\em Astrophys. J.}, {\bf 195}, L51--L53, (1975).
  {\small[\href{http://dx.doi.org/10.1086/181708}{DOI}]},
  {\small[\href{http://adsabs.harvard.edu/abs/1975ApJ...195L..51H}{ADS}]}.

\bibitem{Husa:2007hp}
Husa, S., Gonzalez, J.A., Hannam, M., Bruegmann, B.  and Sperhake, U.,
  ``Reducing phase error in long numerical binary black hole evolutions with
  sixth order finite differencing'', {\em Class. Quantum Grav.}, {\bf 25},
  105006, (2008).
  {\small[\href{http://dx.doi.org/10.1088/0264-9381/25/10/105006}{DOI}]},
  {\small[\href{http://adsabs.harvard.edu/abs/2008CQGra..25j5006H}{ADS}]},
  {\small[\href{http://arxiv.org/abs/0706.0740}{{arXiv:0706.0740
  {\small[gr-qc]}}}]}.

\bibitem{Isenberg:2007zg}
Isenberg, J.A., ``Waveless approximation theories of gravity'', {\em Int. J.
  Mod. Phys. D}, {\bf 17}, 265--273, (2008).
  {\small[\href{http://dx.doi.org/10.1142/S0218271808011997}{DOI}]},
  {\small[\href{http://adsabs.harvard.edu/abs/2008IJMPD..17..265I}{ADS}]},
  {\small[\href{http://arxiv.org/abs/gr-qc/0702113}{{arXiv:gr-qc/0702113
  {\small[gr-qc]}}}]}.

\bibitem{Janka:1999qu}
Janka, H.-T., Eberl, T., Ruffert, M.  and Fryer, C.L., ``Black hole: Neutron
  star mergers as central engines of gamma-ray bursts'', {\em Astrophys. J.
  Lett.}, {\bf 527}, L39--L42, (1999).
  {\small[\href{http://dx.doi.org/10.1086/312397}{DOI}]},
  {\small[\href{http://adsabs.harvard.edu/abs/1999ApJ...527L..39J}{ADS}]},
  {\small[\href{http://arxiv.org/abs/astro-ph/9908290}{{arXiv:astro-ph/9908290
  {\small[astro-ph]}}}]}.

\bibitem{Janka:1995cq}
Janka, H.-T.  and Ruffert, M., ``Can neutrinos from neutron star mergers power
  gamma-ray bursts?'', {\em Astron. Astrophys.}, {\bf 307}, L33--L36, (1996).
  {\small[\href{http://adsabs.harvard.edu/abs/1996A%26A...307L..33J}{ADS}]},
  {\small[\href{http://arxiv.org/abs/astro-ph/9512144}{{arXiv:astro-ph/9512144
  {\small[astro-ph]}}}]}.

\bibitem{Kalogera:2006uj}
Kalogera, V., Belczynski, K., Kim, C., O'Shaughnessy, R.W.  and Willems, B.,
  ``Formation of Double Compact Objects'', {\em Phys. Rept.}, {\bf 442},
  75--108, (2007).
  {\small[\href{http://dx.doi.org/10.1016/j.physrep.2007.02.008}{DOI}]},
  {\small[\href{http://adsabs.harvard.edu/abs/2007PhR...442...75K}{ADS}]},
  {\small[\href{http://arxiv.org/abs/astro-ph/0612144}{{arXiv:astro-ph/0612144
  {\small[astro-ph]}}}]}.

\bibitem{Kaplan10}
Kaplan, J., Ott, C., Muhlberger, C., Duez, M., Foucart, F.  and Scheel, M.,
  ``Simulations of Neutron-Star Binaries using the Spectral Einstein Code
  (SpEC)'', APS April Meeting 2010, p. 14005. American Physical Society,
  (2010).
  {\small[\href{http://adsabs.harvard.edu/abs/2010APS..APRP14005K}{ADS}]}.

\bibitem{Kettner:1994zs}
Kettner, C., Weber, F., Weigel, M.K.  and Glendenning, N.K., ``Structure and
  stability of strange and charm stars at finite temperatures'', {\em Phys.
  Rev. D}, {\bf 51}, 1440--1457, (1995).
  {\small[\href{http://dx.doi.org/10.1103/PhysRevD.51.1440}{DOI}]},
  {\small[\href{http://adsabs.harvard.edu/abs/1995PhRvD..51.1440K}{ADS}]}.

\bibitem{Kim:2002uw}
Kim, C., Kalogera, V.  and Lorimer, D.R., ``The probability distribution of
  binary pulsar coalescence rates. I. double neutron star systems in the
  galactic field'', {\em Astrophys. J.}, {\bf 584}, 985--995, (2003).
  {\small[\href{http://dx.doi.org/10.1086/345740}{DOI}]},
  {\small[\href{http://adsabs.harvard.edu/abs/2003ApJ...584..985K}{ADS}]},
  {\small[\href{http://arxiv.org/abs/astro-ph/0207408}{{arXiv:astro-ph/0207408
  {\small[astro-ph]}}}]}.

\bibitem{Kiuchi:2009jt}
Kiuchi, K., Sekiguchi, Y., Shibata, M.  and Taniguchi, K., ``Longterm general
  relativistic simulation of binary neutron stars collapsing to a black hole'',
  {\em Phys. Rev. D}, {\bf 80}, 064037, (2009).
  {\small[\href{http://dx.doi.org/10.1103/PhysRevD.80.064037}{DOI}]},
  {\small[\href{http://adsabs.harvard.edu/abs/2009PhRvD..80f4037K}{ADS}]},
  {\small[\href{http://arxiv.org/abs/0904.4551}{{arXiv:0904.4551
  {\small[gr-qc]}}}]}.

\bibitem{Kiuchi:2010ze}
Kiuchi, K., Sekiguchi, Y., Shibata, M.  and Taniguchi, K., ``Exploring
  binary-neutron-star-merger scenario of short-gamma-ray bursts by
  gravitational-wave observation'', {\em Phys. Rev. Lett.}, {\bf 104}, 141101,
  (2010).
  {\small[\href{http://dx.doi.org/10.1103/PhysRevLett.104.141101}{DOI}]},
  {\small[\href{http://adsabs.harvard.edu/abs/2010PhRvL.104n1101K}{ADS}]},
  {\small[\href{http://arxiv.org/abs/1002.2689}{{arXiv:1002.2689
  {\small[astro-ph.HE]}}}]}.

\bibitem{Kochanek:1992wk}
Kochanek, C.S., ``Coalescing binary neutron stars'', {\em Astrophys. J.}, {\bf
  398}, 234--247, (1992).
  {\small[\href{http://dx.doi.org/10.1086/171851}{DOI}]},
  {\small[\href{http://adsabs.harvard.edu/abs/1992ApJ...398..234K}{ADS}]}.

\bibitem{Komatsu:1989zz}
Komatsu, H., Eriguchi, Y.  and Hachisu, I., ``Rapidly rotating general
  relativistic stars -- I. Numerical method and its application to uniformly
  rotating polytropes'', {\em Mon. Not. R. Astron. Soc.}, {\bf 237}, 355--379,
  (1989).
  {\small[\href{http://adsabs.harvard.edu/abs/1989MNRAS.237..355K}{ADS}]}.

\bibitem{Kounine:2010js}
Kounine, A., ``Status of the AMS Experiment'', {\em arXiv}, e-print, (2010).
  {\small[\href{http://adsabs.harvard.edu/abs/2010arXiv1009.5349K}{ADS}]},
  {\small[\href{http://arxiv.org/abs/1009.5349}{{arXiv:1009.5349
  {\small[astro-ph.HE]}}}]}.

\bibitem{Kramer:2008ARA&A..46..541K}
Kramer, M.  and Stairs, I.H., ``The Double Pulsar'', {\em Annu. Rev. Astron.
  Astrophys.}, {\bf 46}, 541--572, (2008).
  {\small[\href{http://dx.doi.org/10.1146/annurev.astro.46.060407.145247}{DOI}%
]}, {\small[\href{http://adsabs.harvard.edu/abs/2008ARA%26A..46..541K}{ADS}]}.

\bibitem{Kreiss:2007zz}
Kreiss, H.O., Reula, O., Sarbach, O.  and Winicour, J., ``Well-posed
  initial-boundary value problem for the harmonic Einstein equations using
  energy estimates'', {\em Class. Quantum Grav.}, {\bf 24}, 5973--5984, (2007).
  {\small[\href{http://dx.doi.org/10.1088/0264-9381/24/23/017}{DOI}]},
  {\small[\href{http://adsabs.harvard.edu/abs/2007CQGra..24.5973K}{ADS}]},
  {\small[\href{http://arxiv.org/abs/0707.4188}{{arXiv:0707.4188
  {\small[gr-qc]}}}]}.

\bibitem{Kreiss:2006mi}
Kreiss, H.-O.  and Winicour, J., ``{Problems which are well-posed in a
  generalized sense with applications to the Einstein equations}'', {\em Class.
  Quant. Grav.}, {\bf 23}, S405--S420, (2006).
  {\small[\href{http://dx.doi.org/10.1088/0264-9381/23/16/S07}{DOI}]},
  {\small[\href{http://adsabs.harvard.edu/abs/2006CQGra..23S.405K}{ADS}]},
  {\small[\href{http://arxiv.org/abs/gr-qc/0602051}{{arXiv:gr-qc/0602051
  {\small[gr-qc]}}}]}.

\bibitem{Kuranov:2009pp}
Kuranov, A.G., Popov, S.B.  and Postnov, K.A., ``Pulsar spin-velocity alignment
  from single and binary neutron star progenitors'', {\em Mon. Not. R. Astron.
  Soc.}, {\bf 395}, 2087--2094, (2009).
  {\small[\href{http://dx.doi.org/10.1111/j.1365-2966.2009.14595.x}{DOI}]},
  {\small[\href{http://adsabs.harvard.edu/abs/2009MNRAS.395.2087K}{ADS}]},
  {\small[\href{http://arxiv.org/abs/0901.1055}{{arXiv:0901.1055
  {\small[astro-ph.SR]}}}]}.

\bibitem{Kyutoku:2011vz}
Kyutoku, K., Okawa, H., Shibata, M.  and Taniguchi, K., ``Gravitational waves
  from spinning black hole-neutron star binaries: dependence on black hole
  spins and on neutron star equations of state'', {\em Phys. Rev. D}, {\bf
  84}(6), 064018, 064018, (September 2011).
  {\small[\href{http://adsabs.harvard.edu/abs/2011PhRvD..84f4018K}{ADS}]},
  {\small[\href{http://arxiv.org/abs/1108.1189}{{arXiv:1108.1189
  {\small[astro-ph.HE]}}}]}.

\bibitem{Kyutoku:2010zd}
Kyutoku, K., Shibata, M.  and Taniguchi, K., ``Gravitational waves from
  nonspinning black hole-neutron star binaries: dependence on equations of
  state'', {\em Phys. Rev. D}, {\bf 82}, 044049, (2010).
  {\small[\href{http://dx.doi.org/10.1103/PhysRevD.82.044049}{DOI}]},
  {\small[\href{http://adsabs.harvard.edu/abs/2010PhRvD..82d4049K}{ADS}]},
  {\small[\href{http://arxiv.org/abs/1008.1460}{{arXiv:1008.1460
  {\small[astro-ph.HE]}}}]}.

\bibitem{Lai:1993ve}
Lai, D., Rasio, F.A.  and Shapiro, S.L., ``Ellipsoidal figures of equilibrium -
  Compressible models'', {\em Astrophys. J. Suppl. Ser.}, {\bf 88}, 205--252,
  (1993). {\small[\href{http://dx.doi.org/10.1086/191822}{DOI}]},
  {\small[\href{http://adsabs.harvard.edu/abs/1993ApJS...88..205L}{ADS}]}.

\bibitem{Lai:1993ApJ...406L..63L}
Lai, D., Rasio, F.A.  and Shapiro, S.L., ``Hydrodynamic instability and
  coalescence of close binary systems'', {\em Astrophys. J. Lett.}, {\bf 406},
  L63--L66, (1993). {\small[\href{http://dx.doi.org/10.1086/186787}{DOI}]},
  {\small[\href{http://adsabs.harvard.edu/abs/1993ApJ...406L..63L}{ADS}]}.

\bibitem{Lai:1993rs}
Lai, D., Rasio, F.A.  and Shapiro, S.L., ``Equilibrium, stability and orbital
  evolution of close binary systems'', {\em Astrophys. J.}, {\bf 423},
  344--370, (1994). {\small[\href{http://dx.doi.org/10.1086/173812}{DOI}]},
  {\small[\href{http://adsabs.harvard.edu/abs/1994ApJ...423..344L}{ADS}]},
  {\small[\href{http://arxiv.org/abs/astro-ph/9307032}{{arXiv:astro-ph/9307032
  {\small[astro-ph]}}}]}.

\bibitem{Lai:1993pa}
Lai, D., Rasio, F.A.  and Shapiro, S.L., ``Hydrodynamic instability and
  coalescence of binary neutron stars'', {\em Astrophys. J.}, {\bf 420},
  811--829, (1994). {\small[\href{http://dx.doi.org/10.1086/173606}{DOI}]},
  {\small[\href{http://adsabs.harvard.edu/abs/1994ApJ...420..811L}{ADS}]},
  {\small[\href{http://arxiv.org/abs/astro-ph/9304027}{{arXiv:astro-ph/9304027
  {\small[astro-ph]}}}]}.

\bibitem{Lai:1994hf}
Lai, D., Rasio, F.A.  and Shapiro, S.L., ``Hydrodynamics of rotating stars and
  close binary interactions: Compressible ellipsoid models'', {\em Astrophys.
  J.}, {\bf 437}, 742--769, (1994).
  {\small[\href{http://dx.doi.org/10.1086/175036}{DOI}]},
  {\small[\href{http://adsabs.harvard.edu/abs/1994ApJ...437..742L}{ADS}]},
  {\small[\href{http://arxiv.org/abs/astro-ph/9404031}{{arXiv:astro-ph/9404031
  {\small[astro-ph]}}}]}.

\bibitem{Lai:1994ke}
Lai, D.  and Shapiro, S.L., ``Gravitational radiation from rapidly rotating
  nascent neutron stars'', {\em Astrophys. J.}, {\bf 442}, 259--272, (1995).
  {\small[\href{http://dx.doi.org/10.1086/175438}{DOI}]},
  {\small[\href{http://adsabs.harvard.edu/abs/1995ApJ...442..259L}{ADS}]},
  {\small[\href{http://arxiv.org/abs/astro-ph/9408053}{{arXiv:astro-ph/9408053
  {\small[astro-ph]}}}]}.

\bibitem{Lai:2006pr}
Lai, D.  and Wu, Y., ``Resonant Tidal Excitations of Inertial Modes in
  Coalescing Neutron Star Binaries'', {\em Phys. Rev. D}, {\bf 74}, 024007,
  (2006). {\small[\href{http://dx.doi.org/10.1103/PhysRevD.74.024007}{DOI}]},
  {\small[\href{http://adsabs.harvard.edu/abs/2006PhRvD..74b4007L}{ADS}]},
  {\small[\href{http://arxiv.org/abs/astro-ph/0604163}{{arXiv:astro-ph/0604163
  {\small[astro-ph]}}}]}.

\bibitem{Lattimer:1991nc}
Lattimer, J.M.  and Swesty, F.D., ``A Generalized equation of state for hot,
  dense matter'', {\em Nucl. Phys. A}, {\bf 535}, 331--376, (1991).
  {\small[\href{http://dx.doi.org/10.1016/0375-9474(91)90452-C}{DOI}]}.

\bibitem{LaxWendroff:1960}
Lax, P.D.  and Wendroff, B., ``{Systems of conservation laws}'', {\em Commun.
  Pure Appl. Math.}, {\bf 13}, 217--237, (1960).
  {\small[\href{http://dx.doi.org/10.1002/cpa.3160130205}{DOI}]}.

\bibitem{Lazzati:2005vn}
Lazzati, D., Ghirlanda, G.  and Ghisellini, G., ``SGR giant flares in the BATSE
  short GRB catalogue: Constraints from spectroscopy'', {\em Mon. Not. R.
  Astron. Soc. Lett.}, {\bf 362}, L8--L12, (2005).
  {\small[\href{http://dx.doi.org/10.1111/j.1745-3933.2005.00062.x}{DOI}]},
  {\small[\href{http://adsabs.harvard.edu/abs/2005MNRAS.362L...8L}{ADS}]},
  {\small[\href{http://arxiv.org/abs/astro-ph/0504308}{{arXiv:astro-ph/0504308
  {\small[astro-ph]}}}]}.

\bibitem{Lee:1998qk}
Lee, W.H.  and Klu{\'{z}}niak, W., ``Newtonian hydrodynamics of the coalescence
  of black holes with neutron stars. 1. Tidally locked binaries with a stiff
  equation of state'', {\em Astrophys. J.}, {\bf 526}, 178--199, (1999).
  {\small[\href{http://dx.doi.org/10.1086/307958}{DOI}]},
  {\small[\href{http://adsabs.harvard.edu/abs/1999ApJ...526..178L}{ADS}]},
  {\small[\href{http://arxiv.org/abs/astro-ph/9808185}{{arXiv:astro-ph/9808185
  {\small[astro-ph]}}}]}.

\bibitem{Lee:1999kcb}
Lee, W.H.  and Klu{\'{z}}niak, W., ``Newtonian hydrodynamics of the coalescence
  of black holes with neutron stars. II. Tidally locked binaries with a soft
  equation of state'', {\em Mon. Not. R. Astron. Soc.}, {\bf 308}, 780--794,
  (1999).
  {\small[\href{http://dx.doi.org/10.1046/j.1365-8711.1999.02734.x}{DOI}]},
  {\small[\href{http://adsabs.harvard.edu/abs/1999MNRAS.308..780L}{ADS}]},
  {\small[\href{http://arxiv.org/abs/astro-ph/9904328}{{arXiv:astro-ph/9904328
  {\small[astro-ph]}}}]}.

\bibitem{Lee:2009ca}
{Lee}, W.~H., {Ramirez-Ruiz}, E.  and {van de Ven}, G., ``{Short gamma-ray
  bursts from dynamically-assembled compact binaries in globular clusters:
  pathways, rates, hydrodynamics and cosmological setting}'', {\em Astrophys.
  J.}, {\bf 720}, 953--975, (2010).
  {\small[\href{http://dx.doi.org/10.1088/0004-637X/720/1/953}{DOI}]},
  {\small[\href{http://adsabs.harvard.edu/abs/2010ApJ...720..953L}{ADS}]},
  {\small[\href{http://arxiv.org/abs/0909.2884}{{arXiv:0909.2884
  {\small[astro-ph.HE]}}}]}.

\bibitem{Lehner:2005vc}
{Lehner}, L., {Liebling}, S.~L.  and {Reula}, O., ``{AMR, stability and higher
  accuracy}'', {\em Class. Quant. Grav.}, {\bf 23}, S421--S446, (2006).
  {\small[\href{http://dx.doi.org/10.1088/0264-9381/23/16/S08}{DOI}]},
  {\small[\href{http://adsabs.harvard.edu/abs/2006CQGra..23S.421L}{ADS}]},
  {\small[\href{http://arxiv.org/abs/gr-qc/0510111}{{arXiv:gr-qc/0510111
  {\small[gr-qc]}}}]}.

\bibitem{Limousin:2004vc}
Limousin, F., Gondek-Rosinska, D.  and Gourgoulhon, E., ``Last orbits of binary
  strange quark stars'', {\em Phys. Rev. D}, {\bf 71}, 064012, (2005).
  {\small[\href{http://dx.doi.org/10.1103/PhysRevD.71.064012}{DOI}]},
  {\small[\href{http://adsabs.harvard.edu/abs/2005PhRvD..71f4012L}{ADS}]},
  {\small[\href{http://arxiv.org/abs/gr-qc/0411127}{{arXiv:gr-qc/0411127
  {\small[gr-qc]}}}]}.

\bibitem{Lindblom:2005qh}
Lindblom, L., Scheel, M.A., Kidder, L.E., Owen, R.  and Rinne, O., ``{A New
  generalized harmonic evolution system}'', {\em Class. Quant. Grav.}, {\bf
  23}, S447--S462, (2006).
  {\small[\href{http://dx.doi.org/10.1088/0264-9381/23/16/S09}{DOI}]},
  {\small[\href{http://adsabs.harvard.edu/abs/2006CQGra..23S.447L}{ADS}]},
  {\small[\href{http://arxiv.org/abs/gr-qc/0512093}{{arXiv:gr-qc/0512093
  {\small[gr-qc]}}}]}.

\bibitem{Liu:2008xy}
Liu, Y.T., Shapiro, S.L., Etienne, Z.B.  and Taniguchi, K., ``General
  relativistic simulations of magnetized binary neutron star mergers'', {\em
  Phys. Rev. D}, {\bf 78}, 024012, (2008).
  {\small[\href{http://dx.doi.org/10.1103/PhysRevD.78.024012}{DOI}]},
  {\small[\href{http://adsabs.harvard.edu/abs/2008PhRvD..78b4012L}{ADS}]},
  {\small[\href{http://arxiv.org/abs/0803.4193}{{arXiv:0803.4193
  {\small[astro-ph]}}}]}.

\bibitem{Livne:2003ai}
Livne, E., Burrows, A., Walder, R., Lichtenstadt, I.  and Thompson, T.A., ``Two
  - dimensional, time - dependent, multi-group, multi-angle radiation
  hydrodynamics test simulation in the core - collapse supernova context'',
  {\em Astrophys. J.}, {\bf 609}, 277--287, (2004).
  {\small[\href{http://dx.doi.org/10.1086/421012}{DOI}]},
  {\small[\href{http://adsabs.harvard.edu/abs/2004ApJ...609..277L}{ADS}]},
  {\small[\href{http://arxiv.org/abs/astro-ph/0312633}{{arXiv:astro-ph/0312633
  {\small[astro-ph]}}}]}.

\bibitem{Loffler:2006nu}
L\"offler, F., Rezzolla, L.  and Ansorg, M., ``Numerical evolutions of a black
  hole-neutron star system in full general relativity: Head-on collision'',
  {\em Phys. Rev. D}, {\bf 74}, 104018, (2006).
  {\small[\href{http://dx.doi.org/10.1103/PhysRevD.74.104018}{DOI}]},
  {\small[\href{http://adsabs.harvard.edu/abs/2006PhRvD..74j4018L}{ADS}]},
  {\small[\href{http://arxiv.org/abs/gr-qc/0606104}{{arXiv:gr-qc/0606104
  {\small[gr-qc]}}}]}.

\bibitem{Loffler:2011ay}
{L{\"o}ffler}, F. {et~al.}, ``{The Einstein Toolkit: A Community Computational
  Infrastructure for Relativistic Astrophysics}'', (2011).
  {\small[\href{http://adsabs.harvard.edu/abs/2011arXiv1111.3344L}{ADS}]},
  {\small[\href{http://arxiv.org/abs/1111.3344}{{arXiv:1111.3344
  {\small[gr-qc]}}}]}.

\bibitem{Lombardi:2010xh}
{Lombardi Jr.}, J.C., Holtzman, W., Dooley, K.L., Gearity, K., Kalogera, V.
  and Rasio, F., ``Twin Binaries: Studies of Stability, Mass Transfer, and
  Coalescence'', {\em Astrophys. J.}, {\bf 737}, 49, (2011).
  {\small[\href{http://dx.doi.org/10.1088/0004-637X/737/2/49}{DOI}]},
  {\small[\href{http://adsabs.harvard.edu/abs/2011ApJ...737...49L}{ADS}]},
  {\small[\href{http://arxiv.org/abs/1009.1300}{{arXiv:1009.1300
  {\small[astro-ph.SR]}}}]}.

\bibitem{Lombardi:1997aw}
Lombardi~Jr, J.C., Rasio, F.A.  and Shapiro, S.L., ``PostNewtonian models of
  binary neutron stars'', {\em Phys. Rev. D}, {\bf 56}, 3416--3438, (1997).
  {\small[\href{http://dx.doi.org/10.1103/PhysRevD.56.3416}{DOI}]},
  {\small[\href{http://adsabs.harvard.edu/abs/1997PhRvD..56.3416L}{ADS}]},
  {\small[\href{http://arxiv.org/abs/astro-ph/9705218}{{arXiv:astro-ph/9705218
  {\small[astro-ph]}}}]}.

\bibitem{lorene_web}
``LORENE: Langage Objet pour la RElativit\'{e} Num\'{e}riquE''. URL (accessed
  30 March 2012): \newline\url{http://www.lorene.obspm.fr}.

\bibitem{Lorimer:2008se}
Lorimer, D.R., ``Binary and Millisecond Pulsars'', {\em Living Rev.
  Relativity}, {\bf 11}, lrr-2008-8, (2008).
  {\small[\href{http://arxiv.org/abs/0811.0762}{{arXiv:0811.0762
  {\small[astro-ph]}}}]}. URL (accessed 30 March 2012):
  \newline\url{http://www.livingreviews.org/lrr-2008-8}.

\bibitem{Lousto:2007rj}
Lousto, C.O.  and Zlochower, Y., ``Foundations of multiple black hole
  evolutions'', {\em Phys. Rev. D}, {\bf 77}, 024034, (2008).
  {\small[\href{http://dx.doi.org/10.1103/PhysRevD.77.024034}{DOI}]},
  {\small[\href{http://adsabs.harvard.edu/abs/2008PhRvD..77b4034L}{ADS}]},
  {\small[\href{http://arxiv.org/abs/0711.1165}{{arXiv:0711.1165
  {\small[gr-qc]}}}]}.

\bibitem{Lovelace:2005gr}
Lovelace, R.V.E., Romanova, M.M.  and Bisnovatyi-Kogan, G.S., ``Screening of
  the magnetic field of disk accreting stars'', {\em Astrophys. J.}, {\bf 625},
  957--965, (2005). {\small[\href{http://dx.doi.org/10.1086/429532}{DOI}]},
  {\small[\href{http://adsabs.harvard.edu/abs/2005ApJ...625..957L}{ADS}]},
  {\small[\href{http://arxiv.org/abs/astro-ph/0508168}{{arXiv:astro-ph/0508168
  {\small[astro-ph]}}}]}.

\bibitem{Lucy:1977zz}
Lucy, L.B., ``A numerical approach to the testing of the fission hypothesis'',
  {\em Astron. J.}, {\bf 82}, 1013--1024, (1977).
  {\small[\href{http://dx.doi.org/10.1086/112164}{DOI}]},
  {\small[\href{http://adsabs.harvard.edu/abs/1977AJ.....82.1013L}{ADS}]}.

\bibitem{Madsen:2004vw}
Madsen, J., ``Strangelet propagation and cosmic ray flux'', {\em Phys. Rev. D},
  {\bf 71}, 014026, (2005).
  {\small[\href{http://dx.doi.org/10.1103/PhysRevD.71.014026}{DOI}]},
  {\small[\href{http://adsabs.harvard.edu/abs/2005PhRvD..71a4026M}{ADS}]},
  {\small[\href{http://arxiv.org/abs/astro-ph/0411538}{{arXiv:astro-ph/0411538
  {\small[astro-ph]}}}]}.

\bibitem{Markakis:2011vd}
Markakis, C., Read, J.S., Shibata, M., Ury{\=u}, K., Creighton, J.D.E.,
  Friedman, J.L.  and Lackey, B.D., ``{Neutron star equation of state via
  gravitational wave observations}'', {\em J. Phys. Conf. Ser.}, {\bf 189},
  012024, (2009).
  {\small[\href{http://dx.doi.org/10.1088/1742-6596/189/1/012024}{DOI}]},
  {\small[\href{http://adsabs.harvard.edu/abs/2009JPhCS.189a2024M}{ADS}]},
  {\small[\href{http://arxiv.org/abs/1110.3759}{{arXiv:1110.3759
  {\small[gr-qc]}}}]}.

\bibitem{Marronetti:1999ya}
Marronetti, P., Mathews, G.J.  and Wilson, J.R., ``Irrotational binary neutron
  stars in quasiequilibrium'', {\em Phys. Rev. D}, {\bf 60}, 087301, (1999).
  {\small[\href{http://dx.doi.org/10.1103/PhysRevD.60.087301}{DOI}]},
  {\small[\href{http://adsabs.harvard.edu/abs/1999PhRvD..60h7301M}{ADS}]},
  {\small[\href{http://arxiv.org/abs/gr-qc/9906088}{{arXiv:gr-qc/9906088
  {\small[gr-qc]}}}]}.

\bibitem{Marti:1991wi}
{Mart}, J.~M., {Ib{\'a}ez}, J.~M.  and {Miralles}, J.~A., ``{Numerical
  relativistic hydrodynamics: Local characteristic approach}'', {\em Phys. Rev.
  D}, {\bf 43}, 3794--3801, (1991).
  {\small[\href{http://dx.doi.org/10.1103/PhysRevD.43.3794}{DOI}]},
  {\small[\href{http://adsabs.harvard.edu/abs/1991PhRvD..43.3794M}{ADS}]}.

\bibitem{Mathews:1997pf}
Mathews, G.J., Marronetti, P.  and Wilson, J.R., ``Relativistic hydrodynamics
  in close binary systems: Analysis of neutron star collapse'', {\em Phys. Rev.
  D}, {\bf 58}, 043003, (1998).
  {\small[\href{http://dx.doi.org/10.1103/PhysRevD.58.043003}{DOI}]},
  {\small[\href{http://adsabs.harvard.edu/abs/1998PhRvD..58d3003M}{ADS}]},
  {\small[\href{http://arxiv.org/abs/gr-qc/9710140}{{arXiv:gr-qc/9710140
  {\small[gr-qc]}}}]}.

\bibitem{Mathews:1997vw}
Mathews, G.J.  and Wilson, J.R., ``Binary induced neutron star compression,
  heating, and collapse'', {\em Astrophys. J.}, {\bf 482}, 929--941, (1997).
  {\small[\href{http://dx.doi.org/10.1086/304166}{DOI}]},
  {\small[\href{http://adsabs.harvard.edu/abs/1997ApJ...482..929M}{ADS}]},
  {\small[\href{http://arxiv.org/abs/astro-ph/9701142}{{arXiv:astro-ph/9701142
  {\small[astro-ph]}}}]}.

\bibitem{Mathews:1999km}
Mathews, G.J.  and Wilson, J.R., ``Revised relativistic hydrodynamical model
  for neutron star binaries'', {\em Phys. Rev. D}, {\bf 61}, 127304, (2000).
  {\small[\href{http://dx.doi.org/10.1103/PhysRevD.61.127304}{DOI}]},
  {\small[\href{http://adsabs.harvard.edu/abs/2000PhRvD..61l7304M}{ADS}]},
  {\small[\href{http://arxiv.org/abs/gr-qc/9911047}{{arXiv:gr-qc/9911047
  {\small[gr-qc]}}}]}.

\bibitem{Miller:2003vc}
Miller, M.A., Gressman, P.  and Suen, W.-M., ``{Towards a realistic neutron
  star binary inspiral: Initial data and multiple orbit evolution in full
  general relativity}'', {\em Phys. Rev. D}, {\bf 69}, 064026, (2004).
  {\small[\href{http://dx.doi.org/10.1103/PhysRevD.69.064026}{DOI}]},
  {\small[\href{http://adsabs.harvard.edu/abs/2004PhRvD..69f4026M}{ADS}]},
  {\small[\href{http://arxiv.org/abs/gr-qc/0312030}{{arXiv:gr-qc/0312030
  {\small[gr-qc]}}}]}.

\bibitem{Misner:1973grav.book}
Misner, C.W., Thorne, K.S.  and Wheeler, J.A., {\em Gravitation}, (W.H.
  Freeman, San Francisco, 1973).
  {\small[\href{http://adsabs.harvard.edu/abs/1973grav.book.....M}{ADS}]}.

\bibitem{Monaghan:1992rr}
Monaghan, J.J., ``Smoothed particle hydrodynamics'', {\em Annu. Rev. Astron.
  Astrophys.}, {\bf 30}, 543--574, (1992).
  {\small[\href{http://dx.doi.org/10.1146/annurev.aa.30.090192.002551}{DOI}]},
  {\small[\href{http://adsabs.harvard.edu/abs/1992ARA%26A..30..543M}{ADS}]}.

\bibitem{Moncrief:1974am}
Moncrief, V., ``Gravitational perturbations of spherically symmetric systems.
  I. The exterior problem'', {\em Ann. Phys. (N.Y.)}, {\bf 88}, 323--342,
  (1974). {\small[\href{http://dx.doi.org/10.1016/0003-4916(74)90173-0}{DOI}]},
  {\small[\href{http://adsabs.harvard.edu/abs/1974AnPhy..88..323M}{ADS}]}.

\bibitem{Nakamura:1989jk}
Nakamura, T.  and Oohara, K., ``Gravitational Radiation From Coalescing Binary
  Neutron Stars. 2. Simulations Including Back Reaction Potential'', {\em Prog.
  Theor. Phys.}, {\bf 82}, 1066--1083, (1989).
  {\small[\href{http://dx.doi.org/10.1143/PTP.82.1066}{DOI}]},
  {\small[\href{http://adsabs.harvard.edu/abs/1989PThPh..82.1066N}{ADS}]}.

\bibitem{Nakamura:1991rt}
Nakamura, T.  and Oohara, K., ``Gravitational radiation from coalescing binary
  neutron stars. 4: Tidal disruption'', {\em Prog. Theor. Phys.}, {\bf 86},
  73--89, (1991). {\small[\href{http://dx.doi.org/10.1143/PTP.86.73}{DOI}]},
  {\small[\href{http://adsabs.harvard.edu/abs/1991PThPh..86...73N}{ADS}]}.

\bibitem{Nakar:2005bs}
Nakar, E., Gal-Yam, A.  and Fox, D.B., ``The Local Rate and the Progenitor
  Lifetimes of Short-Hard Gamma-Ray Bursts: Synthesis and Predictions for
  LIGO'', {\em Astrophys. J.}, {\bf 650}, 281--290, (2006).
  {\small[\href{http://dx.doi.org/10.1086/505855}{DOI}]},
  {\small[\href{http://adsabs.harvard.edu/abs/2006ApJ...650..281N}{ADS}]},
  {\small[\href{http://arxiv.org/abs/astro-ph/0511254}{{arXiv:astro-ph/0511254
  {\small[astro-ph]}}}]}.

\bibitem{Nakar:2005hs}
Nakar, E., Gal-Yam, A., Piran, T.  and Fox, D.B., ``The Distances of short-hard
  GRBs and the SGR connection'', {\em Astrophys. J.}, {\bf 640}, 849--853,
  (2006). {\small[\href{http://dx.doi.org/10.1086/498229}{DOI}]},
  {\small[\href{http://adsabs.harvard.edu/abs/2006ApJ...640..849N}{ADS}]},
  {\small[\href{http://arxiv.org/abs/astro-ph/0502148}{{arXiv:astro-ph/0502148
  {\small[astro-ph]}}}]}.

\bibitem{Nakar:2011cw}
Nakar, E.  and Piran, T., ``Radio Remnants of Compact Binary Mergers - the
  Electromagnetic Signal that will follow the Gravitational Waves'', {\em
  arXiv}, e-print, (2011).
  {\small[\href{http://adsabs.harvard.edu/abs/2011arXiv1102.1020N}{ADS}]},
  {\small[\href{http://arxiv.org/abs/1102.1020}{{arXiv:1102.1020
  {\small[astro-ph.HE]}}}]}.

\bibitem{New:1997xi}
New, K.C.B.  and Tohline, J.E., ``The Relative stability against merger of
  close, compact binaries'', {\em Astrophys. J.}, {\bf 490}, 311--327, (1997).
  {\small[\href{http://dx.doi.org/10.1086/304861}{DOI}]},
  {\small[\href{http://adsabs.harvard.edu/abs/1997ApJ...490..311N}{ADS}]},
  {\small[\href{http://arxiv.org/abs/gr-qc/9703013}{{arXiv:gr-qc/9703013
  {\small[gr-qc]}}}]}.

\bibitem{Newman:1961qr}
Newman, E.  and Penrose, R., ``An Approach to gravitational radiation by a
  method of spin coefficients'', {\em J. Math. Phys.}, {\bf 3}, 566--578,
  (1962). {\small[\href{http://dx.doi.org/10.1063/1.1724257}{DOI}]},
  {\small[\href{http://adsabs.harvard.edu/abs/1962JMP.....3..566N}{ADS}]}.

\bibitem{Noble:2005gf}
Noble, S.C., Gammie, C.F., McKinney, J.C.  and Del~Zanna, L., ``Primitive
  variable solvers for conservative general relativistic
  magnetohydrodynamics'', {\em Astrophys. J.}, {\bf 641}, 626--637, (2006).
  {\small[\href{http://dx.doi.org/10.1086/500349}{DOI}]},
  {\small[\href{http://adsabs.harvard.edu/abs/2006ApJ...641..626N}{ADS}]},
  {\small[\href{http://arxiv.org/abs/astro-ph/0512420}{{arXiv:astro-ph/0512420
  {\small[astro-ph]}}}]}.

\bibitem{Nunez:2009wn}
Nunez, D.  and Sarbach, O., ``Boundary conditions for the
  Baumgarte-Shapiro-Shibata-Nakamura formulation of Einstein's field
  equations'', {\em Phys. Rev. D}, {\bf 81}, 044011, (2010).
  {\small[\href{http://dx.doi.org/10.1103/PhysRevD.81.044011}{DOI}]},
  {\small[\href{http://adsabs.harvard.edu/abs/2010PhRvD..81d4011N}{ADS}]},
  {\small[\href{http://arxiv.org/abs/0910.5763}{{arXiv:0910.5763
  {\small[gr-qc]}}}]}.

\bibitem{O'Connor:2010tk}
O'Connor, E.  and Ott, C.D., ``Black Hole Formation in Failing Core-Collapse
  Supernovae'', {\em Astrophys. J.}, {\bf 730}, 70, (2011).
  {\small[\href{http://dx.doi.org/10.1088/0004-637X/730/2/70}{DOI}]},
  {\small[\href{http://adsabs.harvard.edu/abs/2011ApJ...730...70O}{ADS}]},
  {\small[\href{http://arxiv.org/abs/1010.5550}{{arXiv:1010.5550
  {\small[astro-ph.HE]}}}]}.

\bibitem{Oechslin:2006vt}
Oechslin, R.  and Janka, H.-T., ``Short Gamma-Ray Bursts from Binary Neutron
  Star Mergers'', {\em AIP Conf.Proc.}, {\bf 861}, 708--713, (2006).
  {\small[\href{http://dx.doi.org/10.1063/1.2399647}{DOI}]},
  {\small[\href{http://adsabs.harvard.edu/abs/2006AIPC..861..708O}{ADS}]},
  {\small[\href{http://arxiv.org/abs/astro-ph/0604562}{{arXiv:astro-ph/0604562
  {\small[astro-ph]}}}]}.

\bibitem{Oechslin:2005mw}
Oechslin, R.  and Janka, H.-T., ``Torus Formation in Neutron Star Mergers and
  Well-Localized Short Gamma-Ray Bursts'', {\em Mon. Not. R. Astron. Soc.},
  {\bf 368}, 1489--1499, (2006).
  {\small[\href{http://dx.doi.org/10.1111/j.1365-2966.2006.10238.x}{DOI}]},
  {\small[\href{http://adsabs.harvard.edu/abs/2006MNRAS.368.1489O}{ADS}]},
  {\small[\href{http://arxiv.org/abs/astro-ph/0507099}{{arXiv:astro-ph/0507099
  {\small[astro-ph]}}}]}.

\bibitem{Oechslin:2007gn}
Oechslin, R.  and Janka, H.-T., ``Gravitational waves from relativistic neutron
  star mergers with nonzero-temperature equations of state'', {\em Phys. Rev.
  Lett.}, {\bf 99}, 121102, (2007).
  {\small[\href{http://dx.doi.org/10.1103/PhysRevLett.99.121102}{DOI}]},
  {\small[\href{http://adsabs.harvard.edu/abs/2007PhRvL..99l1102O}{ADS}]},
  {\small[\href{http://arxiv.org/abs/astro-ph/0702228}{{arXiv:astro-ph/0702228
  {\small[astro-ph]}}}]}.

\bibitem{Oechslin:2006uk}
Oechslin, R., Janka, H.-T.  and Marek, A., ``Relativistic neutron star merger
  simulations with non-zero temperature equations of state. 1. Variation of
  binary parameters and equation of state'', {\em Astron. Astrophys.}, {\bf
  467}, 395--409, (2007).
  {\small[\href{http://dx.doi.org/10.1051/0004-6361:20066682}{DOI}]},
  {\small[\href{http://adsabs.harvard.edu/abs/2007A%26A...467..395O}{ADS}]},
  {\small[\href{http://arxiv.org/abs/astro-ph/0611047}{{arXiv:astro-ph/0611047
  {\small[astro-ph]}}}]}.

\bibitem{Oechslin:2002vy}
Oechslin, R., Poghosyan, G.S.  and Ury{\={u}}, K., ``Quark matter in neutron
  star mergers'', {\em Nucl. Phys. A}, {\bf 718}, 706--708, (2003).
  {\small[\href{http://dx.doi.org/10.1016/S0375-9474(03)00895-9}{DOI}]},
  {\small[\href{http://adsabs.harvard.edu/abs/2003NuPhA.718..706O}{ADS}]},
  {\small[\href{http://arxiv.org/abs/astro-ph/0210655}{{arXiv:astro-ph/0210655
  {\small[astro-ph]}}}]}.

\bibitem{Oechslin:2001km}
Oechslin, R., Rosswog, S.  and Thielemann, F.K., ``Conformally flat smoothed
  particle hydrodynamics: application to neutron star mergers'', {\em Phys.
  Rev. D}, {\bf 65}, 103005, (2002).
  {\small[\href{http://dx.doi.org/10.1103/PhysRevD.65.103005}{DOI}]},
  {\small[\href{http://adsabs.harvard.edu/abs/2002PhRvD..65j3005O}{ADS}]},
  {\small[\href{http://arxiv.org/abs/gr-qc/0111005}{{arXiv:gr-qc/0111005
  {\small[gr-qc]}}}]}.

\bibitem{Oechslin:2004yj}
Oechslin, R., Ury{\={u}}, K., Poghosyan, G.S.  and Thielemann, F.K., ``The
  Influence of quark matter at high densities on binary neutron star mergers'',
  {\em Mon. Not. R. Astron. Soc.}, {\bf 349}, 1469--1480, (2004).
  {\small[\href{http://dx.doi.org/10.1111/j.1365-2966.2004.07621.x}{DOI}]},
  {\small[\href{http://adsabs.harvard.edu/abs/2004MNRAS.349.1469O}{ADS}]},
  {\small[\href{http://arxiv.org/abs/astro-ph/0401083}{{arXiv:astro-ph/0401083
  {\small[astro-ph]}}}]}.

\bibitem{O'Leary:2008xt}
{O'Leary}, R.~M., {Kocsis}, B.  and {Loeb}, A., ``{Gravitational waves from
  scattering of stellar-mass black holes in galactic nuclei}'', {\em Mon. Not.
  R. Astron. Soc.}, {\bf 395}, 2127--2146, (June 2009).
  {\small[\href{http://dx.doi.org/10.1111/j.1365-2966.2009.14653.x}{DOI}]},
  {\small[\href{http://adsabs.harvard.edu/abs/2009MNRAS.395.2127O}{ADS}]},
  {\small[\href{http://arxiv.org/abs/0807.2638}{{arXiv:0807.2638
  {\small[astro-ph]}}}]}.

\bibitem{Oohara:1989cb}
Oohara, K.  and Nakamura, T., ``Gravitational Radiation From A Coalescing
  Binary Neutron Star'', {\em Prog. Theor. Phys.}, {\bf 82}, 535--554, (1989).
  {\small[\href{http://dx.doi.org/10.1143/PTP.82.535}{DOI}]},
  {\small[\href{http://adsabs.harvard.edu/abs/1989PThPh..82..535O}{ADS}]}.

\bibitem{Oohara:1990dp}
Oohara, K.  and Nakamura, T., ``Gravitational Radiation From Coalescing Binary
  Neutron Stars. 3. Simulations From Equilibrium Model'', {\em Prog. Theor.
  Phys.}, {\bf 83}, 906--940, (1990).
  {\small[\href{http://dx.doi.org/10.1143/PTP.83.906}{DOI}]},
  {\small[\href{http://adsabs.harvard.edu/abs/1990PThPh..83..906O}{ADS}]}.

\bibitem{Oohara:1992dm}
Oohara, K.  and Nakamura, T., ``Gravitational radiation from coalescing binary
  neutron stars. 5. PostNewtonian calculation'', {\em Prog. Theor. Phys.}, {\bf
  88}, 307--316, (1993).
  {\small[\href{http://dx.doi.org/10.1143/PTP.88.307}{DOI}]},
  {\small[\href{http://adsabs.harvard.edu/abs/1992PThPh..88..307O}{ADS}]}.

\bibitem{O'Shaughnessy:2007fb}
O'Shaughnessy, R.W., Kalogera, V.  and Belczynski, K., ``Short Gamma-Ray Bursts
  and Binary Mergers in Spiral and Elliptical Galaxies: Redshift Distribution
  and Hosts'', {\em Astrophys. J.}, {\bf 675}, 566--585, (2008).
  {\small[\href{http://dx.doi.org/10.1086/526334}{DOI}]},
  {\small[\href{http://adsabs.harvard.edu/abs/2008ApJ...675..566O}{ADS}]},
  {\small[\href{http://arxiv.org/abs/0706.4139}{{arXiv:0706.4139
  {\small[astro-ph]}}}]}.

\bibitem{O'Shaughnessy:2009ft}
O'Shaughnessy, R., Kalogera, V.  and Belczynski, K., ``Binary compact object
  coalescence rates: The role of elliptical galaxies'', {\em Astrophys. J.},
  {\bf 716}, 615--633, (2010).
  {\small[\href{http://dx.doi.org/10.1088/0004-637X/716/1/615}{DOI}]},
  {\small[\href{http://adsabs.harvard.edu/abs/2010ApJ...716..615O}{ADS}]},
  {\small[\href{http://arxiv.org/abs/0908.3635}{{arXiv:0908.3635
  {\small[astro-ph.CO]}}}]}.

\bibitem{O'Shaughnessy:2006wh}
O'Shaughnessy, R.W., Kim, C., Kalogera, V.  and Belczynski, K., ``Constraining
  population synthesis models via empirical binary compact object merger and
  supernovae rates'', {\em Astrophys. J.}, {\bf 672}, 479--488, (2008).
  {\small[\href{http://dx.doi.org/10.1086/523620}{DOI}]},
  {\small[\href{http://adsabs.harvard.edu/abs/2008ApJ...672..479O}{ADS}]},
  {\small[\href{http://arxiv.org/abs/astro-ph/0610076}{{arXiv:astro-ph/0610076
  {\small[astro-ph]}}}]}.

\bibitem{Ozel:2010fw}
Ozel, F., Baym, G.  and Guver, T., ``Astrophysical Measurement of the Equation
  of State of Neutron Star Matter'', {\em Phys. Rev. D}, {\bf 82}, 101301,
  (2010). {\small[\href{http://dx.doi.org/10.1103/PhysRevD.82.101301}{DOI}]},
  {\small[\href{http://adsabs.harvard.edu/abs/2010PhRvD..82j1301O}{ADS}]},
  {\small[\href{http://arxiv.org/abs/1002.3153}{{arXiv:1002.3153
  {\small[astro-ph.HE]}}}]}.

\bibitem{Ozel:2009da}
Ozel, F.  and Psaltis, D., ``Reconstructing the Neutron-Star Equation of State
  from Astrophysical Measurements'', {\em Phys. Rev. D}, {\bf 80}, 103003,
  (2009). {\small[\href{http://dx.doi.org/10.1103/PhysRevD.80.103003}{DOI}]},
  {\small[\href{http://adsabs.harvard.edu/abs/2009PhRvD..80j3003O}{ADS}]},
  {\small[\href{http://arxiv.org/abs/0905.1959}{{arXiv:0905.1959
  {\small[astro-ph.HE]}}}]}.

\bibitem{Pandharipande:1989nmhi.conf..103P}
Pandharipande, V.R.  and Ravenhall, D.G., ``Hot Nuclear Matter'', in Soyeur,
  M., Flocard, H., Tamain, B.  and Porneuf, M., eds., {\em Nuclear Matter and
  Heavy Ion Collisions}, Proceedings of a NATO Advanced Research Workshop, held
  February 7\,--\,16, 1989, in Les Houches, France, NATO ASI Series B,  205, p.
  103, (Plenum Press, New York, 1989).
  {\small[\href{http://adsabs.harvard.edu/abs/1989nmhi.conf..103P}{ADS}]}.

\bibitem{Pandharipande:1975NuPhA.237..507P}
Pandharipande, V.R.  and Smith, R.A., ``A model neutron solid with $\pi^0$
  condensate'', {\em Nucl. Phys. A}, {\bf 237}, 507--532, (1975).
  {\small[\href{http://dx.doi.org/10.1016/0375-9474(75)90415-7}{DOI}]},
  {\small[\href{http://adsabs.harvard.edu/abs/1975NuPhA.237..507P}{ADS}]}.

\bibitem{Pazos:2009vb}
Pazos, E., Tiglio, M., Duez, M.D., Kidder, L.E.  and Teukolsky, S.A.,
  ``Orbiting binary black hole evolutions with a multipatch high order
  finite-difference approach'', {\em Phys. Rev. D}, {\bf 80}, 024027, (2009).
  {\small[\href{http://dx.doi.org/10.1103/PhysRevD.80.024027}{DOI}]},
  {\small[\href{http://adsabs.harvard.edu/abs/2009PhRvD..80b4027P}{ADS}]},
  {\small[\href{http://arxiv.org/abs/0904.0493}{{arXiv:0904.0493
  {\small[gr-qc]}}}]}.

\bibitem{Perna:2005tv}
{Perna}, R., {Armitage}, P.~J.  and {Zhang}, B., ``{Flares in long and short
  gamma-ray bursts: a common origin in a hyperaccreting accretion disk}'', {\em
  Astrophys. J. Lett.}, {\bf 636}, L29--L32, (2005).
  {\small[\href{http://dx.doi.org/10.1086/499775}{DOI}]},
  {\small[\href{http://adsabs.harvard.edu/abs/2006ApJ...636L..29P}{ADS}]},
  {\small[\href{http://arxiv.org/abs/astro-ph/0511506}{{arXiv:astro-ph/0511506
  {\small[astro-ph]}}}]}.

\bibitem{Peters:1963ux}
Peters, P.C.  and Mathews, J., ``Gravitational radiation from point masses in a
  Keplerian orbit'', {\em Phys. Rev. D}, {\bf 131}, 435--439, (1963).
  {\small[\href{http://dx.doi.org/10.1103/PhysRev.131.435}{DOI}]},
  {\small[\href{http://adsabs.harvard.edu/abs/1963PhRv..131..435P}{ADS}]}.

\bibitem{Piro:2006ja}
{Piro}, A.~L.  and {Pfahl}, E., ``{Fragmentation of Collapsar Disks and the
  Production of Gravitational Waves}'', {\em Astrophys. J. Lett.}, {\bf 658},
  1173--1176, (April 2007).
  {\small[\href{http://dx.doi.org/10.1086/511672}{DOI}]},
  {\small[\href{http://adsabs.harvard.edu/abs/2007ApJ...658.1173P}{ADS}]},
  {\small[\href{http://arxiv.org/abs/astro-ph/0610696}{{arXiv:astro-ph/0610696
  {\small[astro-ph]}}}]}.

\bibitem{PortegiesZwart:1998xm}
Portegies~Zwart, S.F., ``Gamma-ray binaries: Stable mass transfer from neutron
  star to black hole'', {\em Astrophys. J. Lett.}, {\bf 503}, L53--L56, (1998).
  {\small[\href{http://dx.doi.org/10.1086/311522}{DOI}]},
  {\small[\href{http://adsabs.harvard.edu/abs/1998ApJ...503L..53P}{ADS}]},
  {\small[\href{http://arxiv.org/abs/astro-ph/9804296}{{arXiv:astro-ph/9804296
  {\small[astro-ph]}}}]}.

\bibitem{Postnov:2007jv}
Postnov, K.  and Yungelson, L., ``The Evolution of Compact Binary Star
  Systems'', {\em Living Rev. Relativity}, {\bf 9}, lrr-2006-6, (2006).
  {\small[\href{http://adsabs.harvard.edu/abs/2006LRR.....9....6P}{ADS}]},
  {\small[\href{http://arxiv.org/abs/astro-ph/0701059}{{arXiv:astro-ph/0701059
  {\small[astro-ph]}}}]}. URL (accessed 20 September 2011):
  \newline\url{http://www.livingreviews.org/lrr-2006-6}.

\bibitem{Prakash:1995uw}
Prakash, M., Cooke, J.R.  and Lattimer, J.M., ``Quark - hadron phase transition
  in protoneutron stars'', {\em Phys. Rev. D}, {\bf 52}, 661--665, (1995).
  {\small[\href{http://dx.doi.org/10.1103/PhysRevD.52.661}{DOI}]},
  {\small[\href{http://adsabs.harvard.edu/abs/1995PhRvD..52..661P}{ADS}]}.

\bibitem{Pretorius:2005gq}
Pretorius, F., ``Evolution of binary black hole spacetimes'', {\em Phys. Rev.
  Lett.}, {\bf 95}, 121101, (2005).
  {\small[\href{http://dx.doi.org/10.1103/PhysRevLett.95.121101}{DOI}]},
  {\small[\href{http://adsabs.harvard.edu/abs/2005PhRvL..95l1101P}{ADS}]},
  {\small[\href{http://arxiv.org/abs/gr-qc/0507014}{{arXiv:gr-qc/0507014
  {\small[gr-qc]}}}]}.

\bibitem{Pretorius:2004jg}
Pretorius, F., ``Numerical relativity using a generalized harmonic
  decomposition'', {\em Class. Quantum Grav.}, {\bf 22}, 425--452, (2005).
  {\small[\href{http://dx.doi.org/10.1088/0264-9381/22/2/014}{DOI}]},
  {\small[\href{http://adsabs.harvard.edu/abs/2005CQGra..22..425P}{ADS}]},
  {\small[\href{http://arxiv.org/abs/gr-qc/0407110}{{arXiv:gr-qc/0407110
  {\small[gr-qc]}}}]}.

\bibitem{Price:2006fi}
Price, D.  and Rosswog, S., ``Producing ultra-strong magnetic fields in neutron
  star mergers'', {\em Science}, {\bf 312}, 719--722, (2006).
  {\small[\href{http://dx.doi.org/10.1126/science.1125201}{DOI}]},
  {\small[\href{http://adsabs.harvard.edu/abs/2006Sci...312..719P}{ADS}]},
  {\small[\href{http://arxiv.org/abs/astro-ph/0603845}{{arXiv:astro-ph/0603845
  {\small[astro-ph]}}}]}.

\bibitem{Rasio:1992ApJ...401..226R}
Rasio, F.A.  and Shapiro, S.L., ``Hydrodynamical evolution of coalescing binary
  neutron stars'', {\em Astrophys. J.}, {\bf 401}, 226--245, (1992).
  {\small[\href{http://dx.doi.org/10.1086/172055}{DOI}]},
  {\small[\href{http://adsabs.harvard.edu/abs/1992ApJ...401..226R}{ADS}]}.

\bibitem{Rasio:1994bd}
Rasio, F.A.  and Shapiro, S.L., ``Hydrodynamics of binary coalescence. 1.
  Polytropes with stiff equations of state'', {\em Astrophys. J.}, {\bf 432},
  242--261, (1994). {\small[\href{http://dx.doi.org/10.1086/174566}{DOI}]},
  {\small[\href{http://adsabs.harvard.edu/abs/1994ApJ...432..242R}{ADS}]},
  {\small[\href{http://arxiv.org/abs/astro-ph/9401027}{{arXiv:astro-ph/9401027
  {\small[astro-ph]}}}]}.

\bibitem{Rasio:1994uw}
Rasio, F.A.  and Shapiro, S.L., ``Hydrodynamics of binary coalescence. II.
  Polytropes with $\Gamma = 5/3$'', {\em Astrophys. J.}, {\bf 438}, 887--903,
  (1995). {\small[\href{http://dx.doi.org/10.1086/175130}{DOI}]},
  {\small[\href{http://adsabs.harvard.edu/abs/1995ApJ...438..887R}{ADS}]},
  {\small[\href{http://arxiv.org/abs/astro-ph/9406032}{{arXiv:astro-ph/9406032
  {\small[astro-ph]}}}]}.

\bibitem{Read:2008iy}
Read, J.S., Lackey, B.D., Owen, B.J.  and Friedman, J.L., ``Constraints on a
  phenomenologically parameterized neutron-star equation of state'', {\em Phys.
  Rev. D}, {\bf 79}, 124032, (2009).
  {\small[\href{http://dx.doi.org/10.1103/PhysRevD.79.124032}{DOI}]},
  {\small[\href{http://adsabs.harvard.edu/abs/2009PhRvD..79l4032R}{ADS}]},
  {\small[\href{http://arxiv.org/abs/0812.2163}{{arXiv:0812.2163
  {\small[astro-ph]}}}]}.

\bibitem{Read:2009yp}
{Read}, J.~S., {Markakis}, C., {Shibata}, M., {Ury{\= u}}, K., {Creighton},
  J.~D.~E.  and {Friedman}, J.~L., ``Measuring the neutron star equation of
  state with gravitational wave observations'', {\em Phys. Rev. D}, {\bf 79},
  124033, (2009).
  {\small[\href{http://dx.doi.org/10.1103/PhysRevD.79.124033}{DOI}]},
  {\small[\href{http://adsabs.harvard.edu/abs/2009PhRvD..79l4033R}{ADS}]},
  {\small[\href{http://arxiv.org/abs/0901.3258}{{arXiv:0901.3258
  {\small[gr-qc]}}}]}.

\bibitem{Regge:1957td}
Regge, T.  and Wheeler, J.A., ``Stability of a Schwarzschild singularity'',
  {\em Phys. Rev.}, {\bf 108}, 1063--1069, (1957).
  {\small[\href{http://dx.doi.org/10.1103/PhysRev.108.1063}{DOI}]},
  {\small[\href{http://adsabs.harvard.edu/abs/1957PhRv..108.1063R}{ADS}]}.

\bibitem{Rezzolla:2010fd}
Rezzolla, L., Baiotti, L., Giacomazzo, B., Link, D.  and Font, J.A., ``Accurate
  evolutions of unequal-mass neutron-star binaries: properties of the torus and
  short GRB engines'', {\em Class. Quantum Grav.}, {\bf 27}, 114105, (2010).
  {\small[\href{http://dx.doi.org/10.1088/0264-9381/27/11/114105}{DOI}]},
  {\small[\href{http://adsabs.harvard.edu/abs/2010CQGra..27k4105R}{ADS}]},
  {\small[\href{http://arxiv.org/abs/1001.3074}{{arXiv:1001.3074
  {\small[gr-qc]}}}]}.

\bibitem{Rezzolla:2011da}
{Rezzolla}, L., {Giacomazzo}, B., {Baiotti}, L., {Granot}, J., {Kouveliotou},
  C.  and {Aloy}, M.~A., ``The missing link: Merging neutron stars naturally
  produce jet-like structures and can power short Gamma-Ray Bursts'', {\em
  Astrophys. J. Lett.}, {\bf 732}, L6, (2011).
  {\small[\href{http://dx.doi.org/10.1088/2041-8205/732/1/L6}{DOI}]},
  {\small[\href{http://adsabs.harvard.edu/abs/2011ApJ...732L...6R}{ADS}]},
  {\small[\href{http://arxiv.org/abs/1101.4298}{{arXiv:1101.4298
  {\small[astro-ph.HE]}}}]}.

\bibitem{Rinne:2008vn}
Rinne, O., Buchman, L.T., Scheel, M.A.  and Pfeiffer, H.P., ``Implementation of
  higher-order absorbing boundary conditions for the Einstein equations'', {\em
  Class. Quantum Grav.}, {\bf 26}, 075009, (2009).
  {\small[\href{http://dx.doi.org/10.1088/0264-9381/26/7/075009}{DOI}]},
  {\small[\href{http://adsabs.harvard.edu/abs/2009CQGra..26g5009R}{ADS}]},
  {\small[\href{http://arxiv.org/abs/0811.3593}{{arXiv:0811.3593
  {\small[gr-qc]}}}]}.

\bibitem{Rosswog:2006rh}
Rosswog, S., ``Fallback accretion in the aftermath of a compact binary
  merger'', {\em Mon. Not. R. Astron. Soc. Lett.}, {\bf 376}, L48--L51, (2007).
  {\small[\href{http://dx.doi.org/10.1111/j.1745-3933.2007.00284.x}{DOI}]},
  {\small[\href{http://adsabs.harvard.edu/abs/2007MNRAS.376L..48R}{ADS}]},
  {\small[\href{http://arxiv.org/abs/astro-ph/0611440}{{arXiv:astro-ph/0611440
  {\small[astro-ph]}}}]}.

\bibitem{Rosswog:2001fh}
Rosswog, S.  and Davies, M.B., ``High-resolution calculations of merging
  neutron stars. I: Model description and hydrodynamic evolution'', {\em Mon.
  Not. R. Astron. Soc.}, {\bf 334}, 481--497, (2002).
  {\small[\href{http://dx.doi.org/10.1046/j.1365-8711.2002.05409.x}{DOI}]},
  {\small[\href{http://adsabs.harvard.edu/abs/2002MNRAS.334..481R}{ADS}]},
  {\small[\href{http://arxiv.org/abs/astro-ph/0110180}{{arXiv:astro-ph/0110180
  {\small[astro-ph]}}}]}.

\bibitem{Rosswog:2000qm}
Rosswog, S., Freiburghaus, C.  and Thielemann, F.K., ``Nucleosynthesis
  calculations for the ejecta of neutron star coalescences'', {\em Nucl. Phys.
  A}, {\bf 688}, 344--348, (2001).
  {\small[\href{http://dx.doi.org/10.1016/S0375-9474(01)00724-2}{DOI}]},
  {\small[\href{http://adsabs.harvard.edu/abs/2001NuPhA.688..344R}{ADS}]},
  {\small[\href{http://arxiv.org/abs/astro-ph/0012046}{{arXiv:astro-ph/0012046
  {\small[astro-ph]}}}]}.

\bibitem{Rosswog:2003rv}
Rosswog, S.  and Liebend\"orfer, M., ``High resolution calculations of merging
  neutron stars. 2: Neutrino emission'', {\em Mon. Not. R. Astron. Soc.}, {\bf
  342}, 673--689, (2003).
  {\small[\href{http://dx.doi.org/10.1046/j.1365-8711.2003.06579.x}{DOI}]},
  {\small[\href{http://adsabs.harvard.edu/abs/2003MNRAS.342..673R}{ADS}]},
  {\small[\href{http://arxiv.org/abs/astro-ph/0302301}{{arXiv:astro-ph/0302301
  {\small[astro-ph]}}}]}.

\bibitem{Rosswog:2007ue}
Rosswog, S.  and Price, D., ``MAGMA: a 3D, Lagrangian magnetohydrodynamics code
  for merger applications'', {\em Mon. Not. R. Astron. Soc.}, {\bf 379},
  915--931, (2007).
  {\small[\href{http://dx.doi.org/10.1111/j.1365-2966.2007.11984.x}{DOI}]},
  {\small[\href{http://adsabs.harvard.edu/abs/2007MNRAS.379..915R}{ADS}]},
  {\small[\href{http://arxiv.org/abs/0705.1441}{{arXiv:0705.1441
  {\small[astro-ph]}}}]}.

\bibitem{Rosswog:2002rt}
Rosswog, S.  and Ramirez-Ruiz, E., ``Jets, winds and bursts from coalescing
  neutron stars'', {\em Mon. Not. R. Astron. Soc. Lett.}, {\bf 336}, L7--L11,
  (2002).
  {\small[\href{http://dx.doi.org/10.1046/j.1365-8711.2002.05898.x}{DOI}]},
  {\small[\href{http://adsabs.harvard.edu/abs/2002MNRAS.336L...7R}{ADS}]},
  {\small[\href{http://arxiv.org/abs/astro-ph/0207576}{{arXiv:astro-ph/0207576
  {\small[astro-ph]}}}]}.

\bibitem{Rosswog:2003ts}
Rosswog, S.  and Ramirez-Ruiz, E., ``On the diversity of short gamma-ray
  bursts'', {\em Mon. Not. R. Astron. Soc. Lett.}, {\bf 343}, L36--L40, (2003).
  {\small[\href{http://dx.doi.org/10.1046/j.1365-8711.2003.06889.x}{DOI}]},
  {\small[\href{http://adsabs.harvard.edu/abs/2003MNRAS.343L..36R}{ADS}]},
  {\small[\href{http://arxiv.org/abs/astro-ph/0306172}{{arXiv:astro-ph/0306172
  {\small[astro-ph]}}}]}.

\bibitem{Rosswog:2003tn}
Rosswog, S., Ramirez-Ruiz, E.  and Davies, M.B., ``High Resolution Calculations
  of Merging Neutron Stars. 3. Gamma-Ray Bursts'', {\em Mon. Not. R. Astron.
  Soc.}, 1077--1090, (2003).
  {\small[\href{http://dx.doi.org/10.1046/j.1365-2966.2003.07032.x}{DOI}]},
  {\small[\href{http://adsabs.harvard.edu/abs/2003MNRAS.345.1077R}{ADS}]},
  {\small[\href{http://arxiv.org/abs/astro-ph/0306418}{{arXiv:astro-ph/0306418
  {\small[astro-ph]}}}]}.

\bibitem{Ruffert:1998qg}
Ruffert, M.  and Janka, H.-T., ``Gamma-ray bursts from accreting black holes in
  neutron star mergers'', {\em Astron. Astrophys.}, {\bf 344}, 573--606,
  (1999).
  {\small[\href{http://adsabs.harvard.edu/abs/1999A%26A...344..573R}{ADS}]},
  {\small[\href{http://arxiv.org/abs/astro-ph/9809280}{{arXiv:astro-ph/9809280
  {\small[astro-ph]}}}]}.

\bibitem{Ruffert:2001gf}
Ruffert, M.  and Janka, H.-T., ``Coalescing neutron stars - a step towards
  physical models III. Improved numerics and different neutron star masses and
  spins s'', {\em Astron. Astrophys.}, {\bf 380}, 544--577, (2001).
  {\small[\href{http://dx.doi.org/10.1051/0004-6361:20011453}{DOI}]},
  {\small[\href{http://adsabs.harvard.edu/abs/2001A%26A...380..544R}{ADS}]},
  {\small[\href{http://arxiv.org/abs/astro-ph/0106229}{{arXiv:astro-ph/0106229%
}}]}.

\bibitem{Ruffert:1995fs}
Ruffert, M.H., Janka, H.-T.  and Sch\"afer, G., ``Coalescing neutron stars: A
  Step towards physical models. 1: Hydrodynamic evolution and gravitational
  wave emission'', {\em Astron. Astrophys.}, {\bf 311}, 532--566, (1996).
  {\small[\href{http://adsabs.harvard.edu/abs/1996A%26A...311..532R}{ADS}]},
  {\small[\href{http://arxiv.org/abs/astro-ph/9509006}{{arXiv:astro-ph/9509006
  {\small[astro-ph]}}}]}.

\bibitem{Ruffert:1996by}
Ruffert, M., Janka, H.-T., Takahashi, K.  and Sch\"afer, G., ``Coalescing
  neutron stars: A Step towards physical models. 2. Neutrino emission, neutron
  tori, and gamma-ray bursts'', {\em Astron. Astrophys.}, {\bf 319}, 122--153,
  (1997).
  {\small[\href{http://adsabs.harvard.edu/abs/1997A%26A...319..122R}{ADS}]},
  {\small[\href{http://arxiv.org/abs/astro-ph/9606181}{{arXiv:astro-ph/9606181
  {\small[astro-ph]}}}]}.

\bibitem{Ruffert:1996qu}
Ruffert, M., Rampp, M.  and Janka, H.-T., ``Coalescing neutron stars:
  Gravitational waves from polytropic models'', {\em Astron. Astrophys.}, {\bf
  321}, 991--1006, (1997).
  {\small[\href{http://adsabs.harvard.edu/abs/1997A%26A...321..991R}{ADS}]},
  {\small[\href{http://arxiv.org/abs/astro-ph/9611056}{{arXiv:astro-ph/9611056
  {\small[astro-ph]}}}]}.

\bibitem{Ruiz:2007hg}
{Ruiz}, M., {Rinne}, O.  and {Sarbach}, O., ``{Outer boundary conditions for
  Einstein's field equations in harmonic coordinates}'', {\em Class. Quant.
  Grav.}, {\bf 24}, 6349--6378, (2007).
  {\small[\href{http://dx.doi.org/10.1088/0264-9381/24/24/012}{DOI}]},
  {\small[\href{http://adsabs.harvard.edu/abs/2007CQGra..24.6349R}{ADS}]},
  {\small[\href{http://arxiv.org/abs/0707.2797}{{arXiv:0707.2797
  {\small[gr-qc]}}}]}.

\bibitem{Ruiz:2007yx}
Ruiz, M., Takahashi, R., Alcubierre, M.  and Nunez, D., ``Multipole expansions
  for energy and momenta carried by gravitational waves'', {\em Gen. Relativ.
  Gravit.}, {\bf 40}, 1705--1729, (2008).
  {\small[\href{http://dx.doi.org/10.1007/s10714-007-0570-8}{DOI}]},
  {\small[\href{http://adsabs.harvard.edu/abs/2008GReGr..40.1705R}{ADS}]},
  {\small[\href{http://arxiv.org/abs/0707.4654}{{arXiv:0707.4654
  {\small[gr-qc]}}}]}. erratum: 10.1007/s10714-008-0684-7.

\bibitem{Sadowski:2007dz}
Sadowski, A., Belczynski, K., Bulik, T., Ivanova, N., Rasio, F.A.  and
  O'Shaughnessy, R., ``The Total Merger Rate of Compact Object Binaries In The
  Local Universe'', {\em Astrophys. J.}, {\bf 676}, 1162--1169, (2008).
  {\small[\href{http://dx.doi.org/10.1086/528932}{DOI}]},
  {\small[\href{http://adsabs.harvard.edu/abs/2008ApJ...676.1162S}{ADS}]},
  {\small[\href{http://arxiv.org/abs/0710.0878}{{arXiv:0710.0878
  {\small[astro-ph]}}}]}.

\bibitem{Saijo:2000qt}
Saijo, M., Shibata, M., Baumgarte, T.W.  and Shapiro, S.L., ``Dynamical bar
  instability in rotating stars: Effect of general relativity'', {\em
  Astrophys. J.}, {\bf 548}, 919--931, (2001).
  {\small[\href{http://dx.doi.org/10.1086/319016}{DOI}]},
  {\small[\href{http://adsabs.harvard.edu/abs/2001ApJ...548..919S}{ADS}]},
  {\small[\href{http://arxiv.org/abs/astro-ph/0010201}{{arXiv:astro-ph/0010201
  {\small[astro-ph]}}}]}.

\bibitem{Schaefer:2003ra}
Sch\"afer, G.  and Gopakumar, A., ``A Minimal no radiation approximation to
  Einstein's field equations'', {\em Phys. Rev. D}, {\bf 69}, 021501, (2004).
  {\small[\href{http://dx.doi.org/10.1103/PhysRevD.69.021501}{DOI}]},
  {\small[\href{http://adsabs.harvard.edu/abs/2004PhRvD..69b1501S}{ADS}]},
  {\small[\href{http://arxiv.org/abs/gr-qc/0310041}{{arXiv:gr-qc/0310041
  {\small[gr-qc]}}}]}.

\bibitem{Scheel:2008rj}
{Scheel}, M.~A., {Boyle}, M., {Chu}, T., {Kidder}, L.~E., {Matthews}, K.~D.
  and {Pfeiffer}, H.~P., ``High-accuracy waveforms for binary black hole
  inspiral, merger, and ringdown'', {\em Phys. Rev. D}, {\bf 79}, 024003,
  (2009). {\small[\href{http://dx.doi.org/10.1103/PhysRevD.79.024003}{DOI}]},
  {\small[\href{http://adsabs.harvard.edu/abs/2009PhRvD..79b4003S}{ADS}]},
  {\small[\href{http://arxiv.org/abs/0810.1767}{{arXiv:0810.1767
  {\small[gr-qc]}}}]}.

\bibitem{Schnetter:2003rb}
Schnetter, E., Hawley, S.H.  and Hawke, I., ``Evolutions in 3-D numerical
  relativity using fixed mesh refinement'', {\em Class. Quantum Grav.}, {\bf
  21}, 1465--1488, (2004).
  {\small[\href{http://dx.doi.org/10.1088/0264-9381/21/6/014}{DOI}]},
  {\small[\href{http://adsabs.harvard.edu/abs/2004CQGra..21.1465S}{ADS}]},
  {\small[\href{http://arxiv.org/abs/gr-qc/0310042}{{arXiv:gr-qc/0310042
  {\small[gr-qc]}}}]}.

\bibitem{Sekiguchi:2011mc}
{Sekiguchi}, Y., {Kiuchi}, K., {Kyutoku}, K.  and {Shibata}, M., ``{Effects of
  hyperons in binary neutron star mergers}'', {\em Phys. Rev. Lett.}, {\bf
  107}, 211101, (2011).
  {\small[\href{http://dx.doi.org/10.1103/PhysRevLett.107.211101}{DOI}]},
  {\small[\href{http://adsabs.harvard.edu/abs/2011PhRvL.107u1101S}{ADS}]},
  {\small[\href{http://arxiv.org/abs/1110.4442}{{arXiv:1110.4442
  {\small[astro-ph.HE]}}}]}.

\bibitem{Sekiguchi:2011zd}
Sekiguchi, Y., Kiuchi, K., Kyutoku, K.  and Shibata, M., ``Gravitational waves
  and neutrino emission from the merger of binary neutron stars'', {\em Phys.
  Rev. Lett.}, {\bf 107}, 051102, (2011).
  {\small[\href{http://dx.doi.org/10.1103/PhysRevLett.107.051102}{DOI}]},
  {\small[\href{http://adsabs.harvard.edu/abs/2011PhRvL.107e1102S}{ADS}]},
  {\small[\href{http://arxiv.org/abs/1105.2125}{{arXiv:1105.2125
  {\small[gr-qc]}}}]}.

\bibitem{Setiawan:2005ah}
Setiawan, S., Ruffert, M.  and Janka, H.-T., ``Three-dimensional simulations of
  non-stationary accretion by remnant black holes of compact object mergers'',
  {\em Astron. Astrophys.}, {\bf 458}, 553--567, (2006).
  {\small[\href{http://dx.doi.org/10.1051/0004-6361:20054193}{DOI}]},
  {\small[\href{http://adsabs.harvard.edu/abs/2006A%26A...458..553S}{ADS}]},
  {\small[\href{http://arxiv.org/abs/astro-ph/0509300}{{arXiv:astro-ph/0509300
  {\small[astro-ph]}}}]}.

\bibitem{Shen:1998gq}
Shen, H., Toki, H., Oyamatsu, K.  and Sumiyoshi, K., ``Relativistic equation of
  state of nuclear matter for supernova and neutron star'', {\em Nucl. Phys.
  A}, {\bf 637}, 435--450, (1998).
  {\small[\href{http://dx.doi.org/10.1016/S0375-9474(98)00236-X}{DOI}]},
  {\small[\href{http://adsabs.harvard.edu/abs/1998NuPhA.637..435S}{ADS}]},
  {\small[\href{http://arxiv.org/abs/nucl-th/9805035}{{arXiv:nucl-th/9805035
  {\small[nucl-th]}}}]}.

\bibitem{Shen:1998by}
Shen, H., Toki, H., Oyamatsu, K.  and Sumiyoshi, K., ``Relativistic equation of
  state of nuclear matter for supernova explosion'', {\em Prog. Theor. Phys.},
  {\bf 100}, 1013--1031, (1998).
  {\small[\href{http://dx.doi.org/10.1143/PTP.100.1013}{DOI}]},
  {\small[\href{http://adsabs.harvard.edu/abs/1998PThPh.100.1013S}{ADS}]},
  {\small[\href{http://arxiv.org/abs/nucl-th/9806095}{{arXiv:nucl-th/9806095}}%
]}.

\bibitem{Shibata:1996ci}
Shibata, M., ``Instability of synchronized binary neutron stars in the first
  post-Newtonian approximation of general relativity'', {\em Prog. Theor.
  Phys.}, {\bf 96}, 317--325, (1996).
  {\small[\href{http://dx.doi.org/10.1143/PTP.96.317}{DOI}]},
  {\small[\href{http://adsabs.harvard.edu/abs/1996PThPh..96..317S}{ADS}]}.

\bibitem{Shibata:1995sb}
Shibata, M., ``Numerical study of synchronized binary neutron stars in the
  postNewtonian approximation of general relativity'', {\em Phys. Rev. D}, {\bf
  55}, 6019--6029, (1997).
  {\small[\href{http://dx.doi.org/10.1103/PhysRevD.55.6019}{DOI}]},
  {\small[\href{http://adsabs.harvard.edu/abs/1997PhRvD..55.6019S}{ADS}]}.

\bibitem{Shibata:1999va}
Shibata, M., ``3-D numerical simulation of black hole formation using
  collisionless particles: Triplane symmetric case'', {\em Prog. Theor. Phys.},
  {\bf 101}, 251--282, (1999).
  {\small[\href{http://dx.doi.org/10.1143/PTP.101.251}{DOI}]},
  {\small[\href{http://adsabs.harvard.edu/abs/1999PThPh.101..251S}{ADS}]}.

\bibitem{Shibata:1999wi}
Shibata, M., ``Fully general relativistic simulation of merging binary
  clusters: Spatial gauge condition'', {\em Prog. Theor. Phys.}, {\bf 101},
  1199--1233, (1999).
  {\small[\href{http://dx.doi.org/10.1143/PTP.101.1199}{DOI}]},
  {\small[\href{http://adsabs.harvard.edu/abs/1999PThPh.101.1199S}{ADS}]},
  {\small[\href{http://arxiv.org/abs/gr-qc/9905058}{{arXiv:gr-qc/9905058
  {\small[gr-qc]}}}]}.

\bibitem{Shibata:1998sg}
Shibata, M., Baumgarte, T.W.  and Shapiro, S.L., ``Stability of coalescing
  binary stars against gravitational collapse: Hydrodynamical simulations'',
  {\em Phys. Rev. D}, {\bf 58}, 023002, (1998).
  {\small[\href{http://dx.doi.org/10.1103/PhysRevD.58.023002}{DOI}]},
  {\small[\href{http://adsabs.harvard.edu/abs/1998PhRvD..58b3002S}{ADS}]},
  {\small[\href{http://arxiv.org/abs/gr-qc/9805026}{{arXiv:gr-qc/9805026
  {\small[gr-qc]}}}]}.

\bibitem{Shibata:2000jt}
Shibata, M., Baumgarte, T.W.  and Shapiro, S.L., ``The bar-mode instability in
  differentially rotating neutron stars: simulations in full general
  relativity'', {\em Astrophys. J.}, {\bf 542}, 453--463, (2000).
  {\small[\href{http://dx.doi.org/10.1086/309525}{DOI}]},
  {\small[\href{http://adsabs.harvard.edu/abs/2000ApJ...542..453S}{ADS}]},
  {\small[\href{http://arxiv.org/abs/astro-ph/0005378}{{arXiv:astro-ph/0005378
  {\small[astro-ph]}}}]}.

\bibitem{Shibata:2005mz}
Shibata, M., Duez, M.D., Liu, Y.T., Shapiro, S.L.  and Stephens, B.C.,
  ``Magnetized hypermassive neutron star collapse: A Central engine for short
  gamma-ray bursts'', {\em Phys. Rev. Lett.}, {\bf 96}, 031102, (2006).
  {\small[\href{http://dx.doi.org/10.1103/PhysRevLett.96.031102}{DOI}]},
  {\small[\href{http://adsabs.harvard.edu/abs/2006PhRvL..96c1102S}{ADS}]},
  {\small[\href{http://arxiv.org/abs/astro-ph/0511142}{{arXiv:astro-ph/0511142
  {\small[astro-ph]}}}]}.

\bibitem{Shibata:2009cn}
Shibata, M., Kyutoku, K., Yamamoto, T.  and Taniguchi, K., ``Gravitational
  waves from black hole-neutron star binaries I: Classification of waveforms'',
  {\em Phys. Rev. D}, {\bf 79}, 044030, (2009).
  {\small[\href{http://dx.doi.org/10.1103/PhysRevD.79.044030}{DOI}]},
  {\small[\href{http://adsabs.harvard.edu/abs/2009PhRvD..79d4030S}{ADS}]},
  {\small[\href{http://arxiv.org/abs/0902.0416}{{arXiv:0902.0416
  {\small[gr-qc]}}}]}.

\bibitem{Shibata:1995we}
Shibata, M.  and Nakamura, T., ``Evolution of three-dimensional gravitational
  waves: Harmonic slicing case'', {\em Phys. Rev. D}, {\bf 52}, 5428--5444,
  (1995). {\small[\href{http://dx.doi.org/10.1103/PhysRevD.52.5428}{DOI}]},
  {\small[\href{http://adsabs.harvard.edu/abs/1995PhRvD..52.5428S}{ADS}]}.

\bibitem{Shibata:1992py}
Shibata, M., Nakamura, T.  and Oohara, K., ``Coalescence of spinning binary
  neutron stars of equal mass 3-D numerical simulations'', {\em Prog. Theor.
  Phys.}, {\bf 88}, 1079--1096, (1992).
  {\small[\href{http://dx.doi.org/10.1143/PTP.88.1079}{DOI}]},
  {\small[\href{http://adsabs.harvard.edu/abs/1992PThPh..88.1079S}{ADS}]}.

\bibitem{Shibata:1997dr}
Shibata, M., Oohara, K.  and Nakamura, T., ``Numerical study on the
  hydrodynamic instability of binary stars in the first post-Newtonian
  approximation of general relativity'', {\em Prog. Theor. Phys.}, {\bf 98},
  1081--1098, (1997).
  {\small[\href{http://dx.doi.org/10.1143/PTP.98.1081}{DOI}]},
  {\small[\href{http://adsabs.harvard.edu/abs/1997PThPh..98.1081S}{ADS}]},
  {\small[\href{http://arxiv.org/abs/gr-qc/9710023}{{arXiv:gr-qc/9710023
  {\small[gr-qc]}}}]}.

\bibitem{Shibata:2011fj}
Shibata, M., Suwa, Y., Kiuchi, K.  and Ioka, K., ``Afterglow of a Binary
  Neutron Star Merger'', {\em Astrophys. J. Lett.}, {\bf 734}, L36, (2011).
  {\small[\href{http://dx.doi.org/10.1088/2041-8205/734/2/L36}{DOI}]},
  {\small[\href{http://adsabs.harvard.edu/abs/2011ApJ...734L..36S}{ADS}]},
  {\small[\href{http://arxiv.org/abs/1105.3302}{{arXiv:1105.3302
  {\small[astro-ph.HE]}}}]}.

\bibitem{Shibata:1997xn}
Shibata, M.  and Taniguchi, K., ``Solving the Darwin problem in the first
  postNewtonian approximation of general relativity: Compressible model'', {\em
  Phys. Rev. D}, {\bf 56}, 811--825, (1997).
  {\small[\href{http://dx.doi.org/10.1103/PhysRevD.56.811}{DOI}]},
  {\small[\href{http://adsabs.harvard.edu/abs/1997PhRvD..56..811S}{ADS}]},
  {\small[\href{http://arxiv.org/abs/gr-qc/9705028}{{arXiv:gr-qc/9705028
  {\small[gr-qc]}}}]}.

\bibitem{Shibata:2006nm}
Shibata, M.  and Taniguchi, K., ``Merger of binary neutron stars to a black
  hole: disk mass, short gamma-ray bursts, and quasinormal mode ringing'', {\em
  Phys. Rev. D}, {\bf 73}, 064027, (2006).
  {\small[\href{http://dx.doi.org/10.1103/PhysRevD.73.064027}{DOI}]},
  {\small[\href{http://adsabs.harvard.edu/abs/2006PhRvD..73f4027S}{ADS}]},
  {\small[\href{http://arxiv.org/abs/astro-ph/0603145}{{arXiv:astro-ph/0603145
  {\small[astro-ph]}}}]}.

\bibitem{Shibata:2007zm}
Shibata, M.  and Taniguchi, K., ``Merger of black hole and neutron star in
  general relativity: Tidal disruption, torus mass, and gravitational waves'',
  {\em Phys. Rev. D}, {\bf 77}, 084015, (2008).
  {\small[\href{http://dx.doi.org/10.1103/PhysRevD.77.084015}{DOI}]},
  {\small[\href{http://adsabs.harvard.edu/abs/2008PhRvD..77h4015S}{ADS}]},
  {\small[\href{http://arxiv.org/abs/0711.1410}{{arXiv:0711.1410
  {\small[gr-qc]}}}]}.

\bibitem{ST_LRR}
Shibata, M.  and Taniguchi, K., ``Coalescence of Black Hole-Neutron Star
  Binaries'', {\em Living Rev. Relativity}, {\bf 14}, lrr-2011-6, (2011).
  {\small[\href{http://adsabs.harvard.edu/abs/2011LRR....14....6S}{ADS}]}. URL
  (accessed 30 March 2012):
  \newline\url{http://www.livingreviews.org/lrr-2011-6}.

\bibitem{Shibata:2003ga}
Shibata, M., Taniguchi, K.  and Ury{\={u}}, K., ``Merger of binary neutron
  stars of unequal mass in full general relativity'', {\em Phys. Rev. D}, {\bf
  68}, 084020, (2003).
  {\small[\href{http://dx.doi.org/10.1103/PhysRevD.68.084020}{DOI}]},
  {\small[\href{http://adsabs.harvard.edu/abs/2003PhRvD..68h4020S}{ADS}]},
  {\small[\href{http://arxiv.org/abs/gr-qc/0310030}{{arXiv:gr-qc/0310030
  {\small[gr-qc]}}}]}.

\bibitem{Shibata:2005ss}
Shibata, M., Taniguchi, K.  and Ury{\={u}}, K., ``Merger of binary neutron
  stars with realistic equations of state in full general relativity'', {\em
  Phys. Rev. D}, {\bf 71}, 084021, (2005).
  {\small[\href{http://dx.doi.org/10.1103/PhysRevD.71.084021}{DOI}]},
  {\small[\href{http://adsabs.harvard.edu/abs/2005PhRvD..71h4021S}{ADS}]},
  {\small[\href{http://arxiv.org/abs/gr-qc/0503119}{{arXiv:gr-qc/0503119
  {\small[gr-qc]}}}]}.

\bibitem{Shibata:1999wm}
Shibata, M.  and Ury{\={u}}, K., ``Simulation of merging binary neutron stars
  in full general relativity: Gamma = two case'', {\em Phys. Rev. D}, {\bf 61},
  064001, (2000).
  {\small[\href{http://dx.doi.org/10.1103/PhysRevD.61.064001}{DOI}]},
  {\small[\href{http://adsabs.harvard.edu/abs/2000PhRvD..61f4001S}{ADS}]},
  {\small[\href{http://arxiv.org/abs/gr-qc/9911058}{{arXiv:gr-qc/9911058
  {\small[gr-qc]}}}]}.

\bibitem{Shibata:2002jb}
Shibata, M.  and Ury{\={u}}, K., ``Gravitational waves from the merger of
  binary neutron stars in a fully general relativistic simulation'', {\em Prog.
  Theor. Phys.}, {\bf 107}, 265--303, (2002).
  {\small[\href{http://dx.doi.org/10.1143/PTP.107.265}{DOI}]},
  {\small[\href{http://adsabs.harvard.edu/abs/2002PThPh.107..265S}{ADS}]},
  {\small[\href{http://arxiv.org/abs/gr-qc/0203037}{{arXiv:gr-qc/0203037
  {\small[gr-qc]}}}]}.

\bibitem{Shibata:2006ks}
Shibata, M.  and Ury{\={u}}, K., ``Merger of black hole-neutron star binaries:
  Nonspinning black hole case'', {\em Phys. Rev. D}, {\bf 74}, 121503, (2006).
  {\small[\href{http://dx.doi.org/10.1103/PhysRevD.74.121503}{DOI}]},
  {\small[\href{http://adsabs.harvard.edu/abs/2006PhRvD..74l1503S}{ADS}]},
  {\small[\href{http://arxiv.org/abs/gr-qc/0612142}{{arXiv:gr-qc/0612142
  {\small[gr-qc]}}}]}.

\bibitem{Shibata:2006bs}
Shibata, M.  and Ury{\={u}}, K., ``Merger of black hole-neutron star binaries
  in full general relativity'', {\em Class. Quantum Grav.}, {\bf 24},
  S125--S138, (2007).
  {\small[\href{http://dx.doi.org/10.1088/0264-9381/24/12/S09}{DOI}]},
  {\small[\href{http://adsabs.harvard.edu/abs/2007CQGra..24..125S}{ADS}]},
  {\small[\href{http://arxiv.org/abs/astro-ph/0611522}{{arXiv:astro-ph/0611522
  {\small[astro-ph]}}}]}.

\bibitem{Shibata:2004qz}
Shibata, M., Ury{\={u}}, K.  and Friedman, J.L., ``Deriving formulations for
  numerical computation of binary neutron stars in quasicircular orbits'', {\em
  Phys. Rev. D}, {\bf 70}, 044044, (2004).
  {\small[\href{http://dx.doi.org/10.1103/PhysRevD.70.044044,
  10.1103/PhysRevD.70.129901}{DOI}]},
  {\small[\href{http://adsabs.harvard.edu/abs/2004PhRvD..70d4044S}{ADS}]},
  {\small[\href{http://arxiv.org/abs/gr-qc/0407036}{{arXiv:gr-qc/0407036
  {\small[gr-qc]}}}]}.

\bibitem{Smith:2009bx}
Smith, J.R. (LIGO Scientific Collaboration), ``The Path to the enhanced and
  advanced LIGO gravitational-wave detectors'', {\em Class. Quantum Grav.},
  {\bf 26}, 114013, (2009).
  {\small[\href{http://dx.doi.org/10.1088/0264-9381/26/11/114013}{DOI}]},
  {\small[\href{http://adsabs.harvard.edu/abs/2009CQGra..26k4013S}{ADS}]},
  {\small[\href{http://arxiv.org/abs/0902.0381}{{arXiv:0902.0381
  {\small[gr-qc]}}}]}.

\bibitem{Stephens:2006cn}
Stephens, B.C., Duez, M.D., Liu, Y.T., Shapiro, S.L.  and Shibata, M.,
  ``Collapse and black hole formation in magnetized, differentially rotating
  neutron stars'', {\em Class. Quantum Grav.}, {\bf 24}, S207--S220, (2007).
  {\small[\href{http://dx.doi.org/10.1088/0264-9381/24/12/S14}{DOI}]},
  {\small[\href{http://adsabs.harvard.edu/abs/2007CQGra..24..207S}{ADS}]},
  {\small[\href{http://arxiv.org/abs/gr-qc/0610103}{{arXiv:gr-qc/0610103
  {\small[gr-qc]}}}]}.

\bibitem{Stephens:2011as}
Stephens, B.C., East, W.E.  and Pretorius, F., ``Eccentric Black Hole-Neutron
  Star Mergers'', {\em Astrophys. J. Lett.}, {\bf 737}, L5, (2011).
  {\small[\href{http://dx.doi.org/10.1088/2041-8205/737/1/L5}{DOI}]},
  {\small[\href{http://adsabs.harvard.edu/abs/2011ApJ...737L...5S}{ADS}]},
  {\small[\href{http://arxiv.org/abs/1105.3175}{{arXiv:1105.3175
  {\small[astro-ph.HE]}}}]}.

\bibitem{Stephens:2008hu}
{Stephens}, B.~C., {Shapiro}, S.~L.  and {Liu}, Y.~T., ``{Collapse of
  magnetized hypermassive neutron stars in general relativity: Disk evolution
  and outflows}'', {\em Phys. Rev. D}, {\bf 77}, 044001, (2008).
  {\small[\href{http://dx.doi.org/10.1103/PhysRevD.77.044001}{DOI}]},
  {\small[\href{http://adsabs.harvard.edu/abs/2008PhRvD..77d4001S}{ADS}]},
  {\small[\href{http://arxiv.org/abs/0802.0200}{{arXiv:0802.0200
  {\small[astro-ph]}}}]}.

\bibitem{Stergioulas:2011gd}
Stergioulas, N., Bauswein, A., Zagkouris, K.  and Janka, H.-T., ``Gravitational
  waves and nonaxisymmetric oscillation modes in mergers of compact object
  binaries'', {\em Mon. Not. R. Astron. Soc.}, {\bf 418}, 427--436, (November
  2011).
  {\small[\href{http://dx.doi.org/10.1111/j.1365-2966.2011.19493.x}{DOI}]},
  {\small[\href{http://adsabs.harvard.edu/abs/2011MNRAS.418..427S}{ADS}]},
  {\small[\href{http://arxiv.org/abs/1105.0368}{{arXiv:1105.0368
  {\small[gr-qc]}}}]}.

\bibitem{Surman:2008qf}
Surman, R., McLaughlin, G.C., Ruffert, M., Janka, H.-T.  and Hix, W.R.,
  ``r-Process Nucleosynthesis in Hot Accretion Disk Flows from Black Hole -
  Neutron Star Mergers'', {\em Astrophys. J.}, {\bf 679}, L117--L120, (2008).
  {\small[\href{http://dx.doi.org/10.1086/589507}{DOI}]},
  {\small[\href{http://adsabs.harvard.edu/abs/2008ApJ...679L.117S}{ADS}]},
  {\small[\href{http://arxiv.org/abs/0803.1785}{{arXiv:0803.1785
  {\small[astro-ph]}}}]}.

\bibitem{Swesty:1999ke}
Swesty, F.D., Wang, E.Y.M.  and Calder, A.C., ``Numerical models of binary
  neutron star system mergers. 1. Numerical methods and equilibrium data for
  Newtonian models'', {\em Astrophys. J.}, {\bf 541}, 937--958, (2000).
  {\small[\href{http://dx.doi.org/10.1086/309460}{DOI}]},
  {\small[\href{http://adsabs.harvard.edu/abs/2000ApJ...541..937S}{ADS}]},
  {\small[\href{http://arxiv.org/abs/astro-ph/9911192}{{arXiv:astro-ph/9911192
  {\small[astro-ph]}}}]}.

\bibitem{Taniguchi:1998wi}
Taniguchi, K., Asada, H.  and Shibata, M., ``Irrotational and incompressible
  ellipsoids in the first postNewtonian approximation of general relativity'',
  {\em Prog. Theor. Phys.}, {\bf 100}, 703--735, (1998).
  {\small[\href{http://dx.doi.org/10.1143/PTP.100.703}{DOI}]},
  {\small[\href{http://adsabs.harvard.edu/abs/1998PThPh.100..703T}{ADS}]},
  {\small[\href{http://arxiv.org/abs/gr-qc/9809039}{{arXiv:gr-qc/9809039
  {\small[gr-qc]}}}]}.

\bibitem{Taniguchi:2007xm}
Taniguchi, K., Baumgarte, T.W., Faber, J.A.  and Shapiro, S.L.,
  ``Quasiequilibrium black hole-neutron star binaries in general relativity'',
  {\em Phys. Rev. D}, {\bf 75}, 084005, (2007).
  {\small[\href{http://dx.doi.org/10.1103/PhysRevD.75.084005}{DOI}]},
  {\small[\href{http://adsabs.harvard.edu/abs/2007PhRvD..75h4005T}{ADS}]},
  {\small[\href{http://arxiv.org/abs/gr-qc/0701110}{{arXiv:gr-qc/0701110
  {\small[gr-qc]}}}]}.

\bibitem{Taniguchi:2007aq}
Taniguchi, K., Baumgarte, T.W., Faber, J.A.  and Shapiro, S.L., ``Relativistic
  black hole-neutron star binaries in quasiequilibrium: Effects of the black
  hole excision boundary condition'', {\em Phys. Rev. D}, {\bf 77}, 044003,
  (2008). {\small[\href{http://dx.doi.org/10.1103/PhysRevD.77.044003}{DOI}]},
  {\small[\href{http://adsabs.harvard.edu/abs/2008PhRvD..77d4003T}{ADS}]},
  {\small[\href{http://arxiv.org/abs/0710.5169}{{arXiv:0710.5169
  {\small[gr-qc]}}}]}.

\bibitem{Taniguchi:2005fr}
{Taniguchi}, K., {Baumgarte}, T.~W., {Faber}, J.~A.  and {Shapiro}, S.~L.,
  ``{Black hole-neutron star binaries in general relativity: Effects of neutron
  star spin}'', {\em Phys. Rev. D}, {\bf 72}, 044008, (2005).
  {\small[\href{http://dx.doi.org/10.1103/PhysRevD.72.044008}{DOI}]},
  {\small[\href{http://adsabs.harvard.edu/abs/2005PhRvD..72d4008T}{ADS}]},
  {\small[\href{http://arxiv.org/abs/astro-ph/0505450}{{arXiv:astro-ph/0505450
  {\small[astro-ph]}}}]}.

\bibitem{Taniguchi:2002ns}
Taniguchi, K.  and Gourgoulhon, E., ``Quasiequilibrium sequences of
  synchronized and irrotational binary neutron stars in general relativity. 3.
  Identical and different mass stars with gamma = 2'', {\em Phys. Rev. D}, {\bf
  66}, 104019, (2002).
  {\small[\href{http://dx.doi.org/10.1103/PhysRevD.66.104019}{DOI}]},
  {\small[\href{http://adsabs.harvard.edu/abs/2002PhRvD..66j4019T}{ADS}]},
  {\small[\href{http://arxiv.org/abs/gr-qc/0207098}{{arXiv:gr-qc/0207098
  {\small[gr-qc]}}}]}.

\bibitem{Taniguchi:2003hx}
Taniguchi, K.  and Gourgoulhon, E., ``Various features of quasiequilibrium
  sequences of binary neutron stars in general relativity'', {\em Phys. Rev.
  D}, {\bf 68}, 124025, (2003).
  {\small[\href{http://dx.doi.org/10.1103/PhysRevD.68.124025}{DOI}]},
  {\small[\href{http://adsabs.harvard.edu/abs/2003PhRvD..68l4025T}{ADS}]},
  {\small[\href{http://arxiv.org/abs/gr-qc/0309045}{{arXiv:gr-qc/0309045
  {\small[gr-qc]}}}]}.

\bibitem{Taniguchi:2010kj}
Taniguchi, K.  and Shibata, M., ``Binary Neutron Stars in Quasi-equilibrium'',
  {\em Astrophys. J. Suppl. Ser.}, {\bf 188}, 187--208, (2010).
  {\small[\href{http://dx.doi.org/10.1088/0067-0049/188/1/187}{DOI}]},
  {\small[\href{http://adsabs.harvard.edu/abs/2010ApJS..188..187T}{ADS}]},
  {\small[\href{http://arxiv.org/abs/1005.0958}{{arXiv:1005.0958
  {\small[astro-ph.SR]}}}]}.

\bibitem{Taylor:1989sw}
Taylor, J.H.  and Weisberg, J.M., ``Further experimental tests of relativistic
  gravity using the binary pulsar PSR 1913+16'', {\em Astrophys. J.}, {\bf
  345}, 434--450, (1989).
  {\small[\href{http://dx.doi.org/10.1086/167917}{DOI}]},
  {\small[\href{http://adsabs.harvard.edu/abs/1989ApJ...345..434T}{ADS}]}.

\bibitem{cactus_web}
``The Cactus Code''. URL (accessed 30 March 2012):
  \newline\url{http://www.cactuscode.org/}.

\bibitem{Thierfelder:2011yi}
Thierfelder, M., Bernuzzi, S.  and Bruegmann, B., ``Numerical relativity
  simulations of binary neutron stars'', {\em Phys. Rev. D}, {\bf 84}, 044012,
  (2011). {\small[\href{http://dx.doi.org/10.1103/PhysRevD.84.044012}{DOI}]},
  {\small[\href{http://adsabs.harvard.edu/abs/2011PhRvD..84d4012T}{ADS}]},
  {\small[\href{http://arxiv.org/abs/1104.4751}{{arXiv:1104.4751
  {\small[gr-qc]}}}]}.

\bibitem{Tichy:2011gw}
Tichy, W., ``{Initial data for binary neutron stars with arbitrary spins}'',
  {\em Phys. Rev. D}, {\bf 84}, 024041, (2011).
  {\small[\href{http://dx.doi.org/10.1103/PhysRevD.84.024041}{DOI}]},
  {\small[\href{http://adsabs.harvard.edu/abs/2011PhRvD..84b4041T}{ADS}]},
  {\small[\href{http://arxiv.org/abs/1107.1440}{{arXiv:1107.1440
  {\small[gr-qc]}}}]}.

\bibitem{Toth:2000JCoPh.161..605T}
T{\'o}th, G., ``The $\nabla\cdot B=0$ Constraint in Shock-Capturing
  Magnetohydrodynamics Codes'', {\em J. Comp. Phys.}, {\bf 161}, 605--652,
  (2000). {\small[\href{http://dx.doi.org/10.1006/jcph.2000.6519}{DOI}]},
  {\small[\href{http://adsabs.harvard.edu/abs/2000JCoPh.161..605T}{ADS}]}.

\bibitem{Uryu:1997zc}
Ury{\={u}}, K.  and Eriguchi, Y., ``Stationary states of irrotational binary
  neutron star systems and their evolution due to gravitational wave
  emission'', {\em Mon. Not. R. Astron. Soc. Lett.}, {\bf 296}, L1--L5, (1998).
  {\small[\href{http://dx.doi.org/10.1046/j.1365-8711.1998.01385.x}{DOI}]},
  {\small[\href{http://adsabs.harvard.edu/abs/1998MNRAS.296L...1U}{ADS}]},
  {\small[\href{http://arxiv.org/abs/astro-ph/9712203}{{arXiv:astro-ph/9712203
  {\small[astro-ph]}}}]}.

\bibitem{Uryu:1998kq}
Ury{\={u}}, K.  and Eriguchi, Y., ``Stationary structures of irrotational
  binary systems: Models for close binary systems of compact stars'', {\em
  Astrophys. J. Suppl. Ser.}, {\bf 118}, 563--587, (1998).
  {\small[\href{http://dx.doi.org/10.1086/313146}{DOI}]},
  {\small[\href{http://adsabs.harvard.edu/abs/1998ApJS..118..563U}{ADS}]},
  {\small[\href{http://arxiv.org/abs/astro-ph/9808118}{{arXiv:astro-ph/9808118
  {\small[astro-ph]}}}]}.

\bibitem{Uryu:1999uu}
Ury{\={u}}, K.  and Eriguchi, Y., ``A New numerical method for constructing
  quasiequilibrium sequences of irrotational binary neutron stars in general
  relativity'', {\em Phys. Rev. D}, {\bf 61}, 124023, (2000).
  {\small[\href{http://dx.doi.org/10.1103/PhysRevD.61.124023}{DOI}]},
  {\small[\href{http://adsabs.harvard.edu/abs/2000PhRvD..61l4023U}{ADS}]},
  {\small[\href{http://arxiv.org/abs/gr-qc/9908059}{{arXiv:gr-qc/9908059
  {\small[gr-qc]}}}]}.

\bibitem{Uryu:2010su}
Ury{\={u}}, K., Gourgoulhon, E.  and Markakis, C., ``Thermodynamics of
  magnetized binary compact objects'', {\em Phys. Rev. D}, {\bf 82}, 104054,
  (2010). {\small[\href{http://dx.doi.org/10.1103/PhysRevD.82.104054}{DOI}]},
  {\small[\href{http://adsabs.harvard.edu/abs/2010PhRvD..82j4054U}{ADS}]},
  {\small[\href{http://arxiv.org/abs/1010.4409}{{arXiv:1010.4409
  {\small[gr-qc]}}}]}.

\bibitem{Uryu:2005vv}
Ury{\={u}}, K., Limousin, F., Friedman, J.L., Gourgoulhon, E.  and Shibata, M.,
  ``Binary neutron stars in a waveless approximation'', {\em Phys. Rev. Lett.},
  {\bf 97}, 171101, (2006).
  {\small[\href{http://dx.doi.org/10.1103/PhysRevLett.97.171101}{DOI}]},
  {\small[\href{http://adsabs.harvard.edu/abs/2006PhRvL..97q1101U}{ADS}]},
  {\small[\href{http://arxiv.org/abs/gr-qc/0511136}{{arXiv:gr-qc/0511136
  {\small[gr-qc]}}}]}.

\bibitem{Uryu:2009ye}
Ury{\={u}}, K., Limousin, F., Friedman, J.L., Gourgoulhon, E.  and Shibata, M.,
  ``Non-conformally flat initial data for binary compact objects'', {\em Phys.
  Rev. D}, {\bf 80}, 124004, (2009).
  {\small[\href{http://dx.doi.org/10.1103/PhysRevD.80.124004}{DOI}]},
  {\small[\href{http://adsabs.harvard.edu/abs/2009PhRvD..80l4004U}{ADS}]},
  {\small[\href{http://arxiv.org/abs/0908.0579}{{arXiv:0908.0579
  {\small[gr-qc]}}}]}.

\bibitem{Uryu:2000dw}
Ury{\={u}}, K., Shibata, M.  and Eriguchi, Y., ``Properties of general
  relativistic, irrotational binary neutron stars in close quasiequilibrium
  orbits: Polytropic equations of state'', {\em Phys. Rev. D}, {\bf 62},
  104015, (2000).
  {\small[\href{http://dx.doi.org/10.1103/PhysRevD.62.104015}{DOI}]},
  {\small[\href{http://adsabs.harvard.edu/abs/2000PhRvD..62j4015U}{ADS}]},
  {\small[\href{http://arxiv.org/abs/gr-qc/0007042}{{arXiv:gr-qc/0007042
  {\small[gr-qc]}}}]}.

\bibitem{Usui:2002fx}
Usui, F.  and Eriguchi, Y., ``Quasiequilibrium sequences of synchronously
  rotating binary neutron stars with constant rest masses in general
  relativity: Another approach without using the conformally flat condition'',
  {\em Phys. Rev. D}, {\bf 65}, 064030, (2002).
  {\small[\href{http://dx.doi.org/10.1103/PhysRevD.65.064030}{DOI}]},
  {\small[\href{http://adsabs.harvard.edu/abs/2002PhRvD..65f4030U}{ADS}]},
  {\small[\href{http://arxiv.org/abs/astro-ph/0112571}{{arXiv:astro-ph/0112571
  {\small[astro-ph]}}}]}.

\bibitem{Usui:1999eu}
Usui, F., Ury{\={u}}, K.  and Eriguchi, Y., ``A New numerical scheme to compute
  3-D configurations of quasiequilibrium compact stars in general relativity:
  Application to synchronously rotating binary star systems'', {\em Phys. Rev.
  D}, {\bf 61}, 024039, (2000).
  {\small[\href{http://dx.doi.org/10.1103/PhysRevD.61.024039}{DOI}]},
  {\small[\href{http://adsabs.harvard.edu/abs/2000PhRvD..61b4039U}{ADS}]},
  {\small[\href{http://arxiv.org/abs/gr-qc/9906102}{{arXiv:gr-qc/9906102
  {\small[gr-qc]}}}]}.

\bibitem{VanDenBroeck:2009gd}
Van Den~Broeck, C., Brown, D.A., Cokelaer, T., Harry, I., Jones, G.,
  Sathyaprakash, B.S., Tagoshi, H.  and Takahashi, H., ``Template banks to
  search for compact binaries with spinning components in gravitational wave
  data'', {\em Phys. Rev. D}, {\bf 80}, 024009, (2009).
  {\small[\href{http://dx.doi.org/10.1103/PhysRevD.80.024009}{DOI}]},
  {\small[\href{http://adsabs.harvard.edu/abs/2009PhRvD..80b4009V}{ADS}]},
  {\small[\href{http://arxiv.org/abs/0904.1715}{{arXiv:0904.1715
  {\small[gr-qc]}}}]}.

\bibitem{vanderSluys:2008qx}
{van der Sluys}, M., {Raymond}, V., {Mandel}, I., {R{\"o}ver}, C.,
  {Christensen}, N., {Kalogera}, V., {Meyer}, R.  and {Vecchio}, A.,
  ``Parameter estimation of spinning binary inspirals using Markov-chain Monte
  Carlo'', {\em Class. Quantum Grav.}, {\bf 25}, 184011, (2008).
  {\small[\href{http://dx.doi.org/10.1088/0264-9381/25/18/184011}{DOI}]},
  {\small[\href{http://adsabs.harvard.edu/abs/2008CQGra..25r4011V}{ADS}]},
  {\small[\href{http://arxiv.org/abs/0805.1689}{{arXiv:0805.1689
  {\small[gr-qc]}}}]}.

\bibitem{vanMeter:2006vi}
van Meter, J.R., Baker, J.G., Koppitz, M.  and Choi, D.-I., ``{How to move a
  black hole without excision: Gauge conditions for the numerical evolution of
  a moving puncture}'', {\em Phys. Rev. D}, {\bf 73}, 124011, (2006).
  {\small[\href{http://dx.doi.org/10.1103/PhysRevD.73.124011}{DOI}]},
  {\small[\href{http://adsabs.harvard.edu/abs/2006PhRvD..73l4011V}{ADS}]},
  {\small[\href{http://arxiv.org/abs/gr-qc/0605030}{{arXiv:gr-qc/0605030
  {\small[gr-qc]}}}]}.

\bibitem{Voss:2003ep}
Voss, R.  and Tauris, T.M., ``Galactic distribution of merging neutron stars
  and black holes - Prospects for short gamma-ray burst progenitors and
  LIGO/VIRGO'', {\em Mon. Not. R. Astron. Soc.}, {\bf 342}, 1169--1184, (2003).
  {\small[\href{http://dx.doi.org/10.1046/j.1365-8711.2003.06616.x}{DOI}]},
  {\small[\href{http://adsabs.harvard.edu/abs/2003MNRAS.342.1169V}{ADS}]},
  {\small[\href{http://arxiv.org/abs/astro-ph/0303227}{{arXiv:astro-ph/0303227
  {\small[astro-ph]}}}]}.

\bibitem{Wang:2005jg}
Wang, C., Lai, D.  and Han, J.L., ``Neutron star kicks in isolated and binary
  pulsars: observational constraints and implications for kick mechanisms'',
  {\em Astrophys. J.}, {\bf 639}, 1007--1017, (2006).
  {\small[\href{http://dx.doi.org/10.1086/499397}{DOI}]},
  {\small[\href{http://adsabs.harvard.edu/abs/2006ApJ...639.1007W}{ADS}]},
  {\small[\href{http://arxiv.org/abs/astro-ph/0509484}{{arXiv:astro-ph/0509484
  {\small[astro-ph]}}}]}.

\bibitem{Weisberg:2010zz}
Weisberg, J.M., Nice, D.J.  and Taylor, J.H., ``Timing Measurements of the
  Relativistic Binary Pulsar PSR B1913+16'', {\em Astrophys. J.}, {\bf 722},
  1030--1034, (2010).
  {\small[\href{http://dx.doi.org/10.1088/0004-637X/722/2/1030}{DOI}]},
  {\small[\href{http://adsabs.harvard.edu/abs/2010ApJ...722.1030W}{ADS}]},
  {\small[\href{http://arxiv.org/abs/1011.0718}{{arXiv:1011.0718
  {\small[astro-ph.GA]}}}]}.

\bibitem{Wilson:1989fnr..book..306W}
Wilson, J.R.  and Mathews, G.J., ``Relativistic hydrodynamics'', in Evans,
  C.R., Finn, L.S.  and Hobill, D.W., eds., {\em Frontiers in Numerical
  Relativity}, Proceedings of the International Workshop on Numerical
  Relativity, University of Illinois at Urbana-Champaign, USA, 9\,--\,13 May
  1988, pp. 306--314, (Cambridge University Press, Cambridge; New York, 1989).
  {\small[\href{http://adsabs.harvard.edu/abs/1989fnr..book..306W}{ADS}]}.

\bibitem{Wilson:1995uh}
Wilson, J.R.  and Mathews, G.J., ``Instabilities in Close Neutron Star
  Binaries'', {\em Phys. Rev. Lett.}, {\bf 75}, 4161--4164, (1995).
  {\small[\href{http://dx.doi.org/10.1103/PhysRevLett.75.4161}{DOI}]},
  {\small[\href{http://adsabs.harvard.edu/abs/1995PhRvL..75.4161W}{ADS}]}.

\bibitem{Wilson:1996ty}
Wilson, J.R., Mathews, G.J.  and Marronetti, P., ``Relativistic numerical model
  for close neutron star binaries'', {\em Phys. Rev. D}, {\bf 54}, 1317--1331,
  (1996). {\small[\href{http://dx.doi.org/10.1103/PhysRevD.54.1317}{DOI}]},
  {\small[\href{http://adsabs.harvard.edu/abs/1996PhRvD..54.1317W}{ADS}]},
  {\small[\href{http://arxiv.org/abs/gr-qc/9601017}{{arXiv:gr-qc/9601017
  {\small[gr-qc]}}}]}.

\bibitem{Winicour:2009dr}
Winicour, J., ``Disembodied boundary data for Einstein's equations'', {\em
  Phys. Rev. D}, {\bf 80}, 124043, (2009).
  {\small[\href{http://dx.doi.org/10.1103/PhysRevD.80.124043}{DOI}]},
  {\small[\href{http://adsabs.harvard.edu/abs/2009PhRvD..80l4043W}{ADS}]},
  {\small[\href{http://arxiv.org/abs/0909.1989}{{arXiv:0909.1989
  {\small[gr-qc]}}}]}.

\bibitem{Xing:1994ak}
Xing, Z.-G., Centrella, J.M.  and McMillan, S.L.W., ``Gravitational radiation
  from coalescing binary neutron stars'', {\em Phys. Rev. D}, {\bf 50},
  6247--6261, (1994).
  {\small[\href{http://dx.doi.org/10.1103/PhysRevD.50.6247}{DOI}]},
  {\small[\href{http://adsabs.harvard.edu/abs/1994PhRvD..50.6247Z}{ADS}]},
  {\small[\href{http://arxiv.org/abs/gr-qc/9411029}{{arXiv:gr-qc/9411029
  {\small[gr-qc]}}}]}.

\bibitem{Xing:1996sr}
Xing, Z.-G., Centrella, J.M.  and McMillan, S.L.W., ``Gravitational radiation
  from the coalescence of binary neutron stars: Effects due to the equation of
  state, spin, and mass ratio'', {\em Phys. Rev. D}, {\bf 54}, 7261--7277,
  (1996). {\small[\href{http://dx.doi.org/10.1103/PhysRevD.54.7261}{DOI}]},
  {\small[\href{http://adsabs.harvard.edu/abs/1996PhRvD..54.7261Z}{ADS}]},
  {\small[\href{http://arxiv.org/abs/gr-qc/9610039}{{arXiv:gr-qc/9610039
  {\small[gr-qc]}}}]}.

\bibitem{Yamamoto:2008js}
Yamamoto, T., Shibata, M.  and Taniguchi, K., ``Simulating coalescing compact
  binaries by a new code SACRA'', {\em Phys. Rev. D}, {\bf 78}, 064054, (2008).
  {\small[\href{http://dx.doi.org/10.1103/PhysRevD.78.064054}{DOI}]},
  {\small[\href{http://adsabs.harvard.edu/abs/2008PhRvD..78f4054Y}{ADS}]},
  {\small[\href{http://arxiv.org/abs/0806.4007}{{arXiv:0806.4007
  {\small[gr-qc]}}}]}.

\bibitem{York:1998hy}
York~Jr, J.W., ``Conformal `thin sandwich' data for the initial-value
  problem'', {\em Phys. Rev. Lett.}, {\bf 82}, 1350--1353, (1999).
  {\small[\href{http://dx.doi.org/10.1103/PhysRevLett.82.1350}{DOI}]},
  {\small[\href{http://adsabs.harvard.edu/abs/1999PhRvL..82.1350Y}{ADS}]},
  {\small[\href{http://arxiv.org/abs/gr-qc/9810051}{{arXiv:gr-qc/9810051
  {\small[gr-qc]}}}]}.

\bibitem{Yoshida:2006cn}
Yoshida, S., Bromley, B.C., Read, J.S., Ury{\={u}}, K.  and Friedman, J.L.,
  ``Models of helically symmetric binary systems'', {\em Class. Quantum Grav.},
  {\bf 23}, S599--S614, (2006).
  {\small[\href{http://dx.doi.org/10.1088/0264-9381/23/16/S16}{DOI}]},
  {\small[\href{http://adsabs.harvard.edu/abs/2006CQGra..23S.599Y}{ADS}]},
  {\small[\href{http://arxiv.org/abs/gr-qc/0605035}{{arXiv:gr-qc/0605035
  {\small[gr-qc]}}}]}.

\bibitem{Zerilli:1971wd}
Zerilli, F.J., ``Gravitational field of a particle falling in a schwarzschild
  geometry analyzed in tensor harmonics'', {\em Phys. Rev. D}, {\bf 2},
  2141--2160, (1970).
  {\small[\href{http://dx.doi.org/10.1103/PhysRevD.2.2141}{DOI}]},
  {\small[\href{http://adsabs.harvard.edu/abs/1970PhRvD...2.2141Z}{ADS}]}.

\bibitem{Zhang:2007nw}
Zhang, W.-Q., Woosley, S.E.  and Heger, A., ``Fallback and Black Hole
  Production in Massive Stars'', {\em Astrophys. J.}, {\bf 679}, 639--654,
  (2007). {\small[\href{http://dx.doi.org/10.1086/526404}{DOI}]},
  {\small[\href{http://adsabs.harvard.edu/abs/2008ApJ...679..639Z}{ADS}]},
  {\small[\href{http://arxiv.org/abs/astro-ph/0701083}{{arXiv:astro-ph/0701083
  {\small[astro-ph]}}}]}.

\bibitem{Zink:2007xn}
Zink, B., Schnetter, E.  and Tiglio, M., ``Multi-patch methods in general
  relativistic astrophysics. 1. Hydrodynamical flows on fixed backgrounds'',
  {\em Phys. Rev. D}, {\bf 77}, 103015, (2008).
  {\small[\href{http://dx.doi.org/10.1103/PhysRevD.77.103015}{DOI}]},
  {\small[\href{http://adsabs.harvard.edu/abs/2008PhRvD..77j3015Z}{ADS}]},
  {\small[\href{http://arxiv.org/abs/0712.0353}{{arXiv:0712.0353
  {\small[astro-ph]}}}]}.

\bibitem{Zorotovic:2010da}
Zorotovic, M., Schreiber, M.R., G{\"a}nsicke, B.T.  and Nebot
  G{\'o}mez-Mor{\'a}n, A., ``Post-common-envelope binaries from SDSS. IX:
  Constraining the common-envelope efficiency'', {\em Astron. Astrophys.}, {\bf
  520}, A86, (2010).
  {\small[\href{http://dx.doi.org/10.1051/0004-6361/200913658}{DOI}]},
  {\small[\href{http://adsabs.harvard.edu/abs/2010A%26A...520A..86Z}{ADS}]},
  {\small[\href{http://arxiv.org/abs/1006.1621}{{arXiv:1006.1621
  {\small[astro-ph.SR]}}}]}.

\end{thebibliography}

\end{document}